\documentclass[a4paper,11pt]{book}
\usepackage[T1]{fontenc}
\usepackage[utf8]{inputenc}
\usepackage{lmodern}
\usepackage{hyperref}
\usepackage{xcolor}
\usepackage{graphicx}
\usepackage[sort, compress]{cite} % 使用基础 cite 包
\usepackage[english]{babel}
\usepackage{geometry} % add by zlFu
\usepackage{booktabs} % add by zlFu
\usepackage{amsmath}% add by zlFu
\makeatletter
\renewcommand{\@chapapp}{}% Not necessary...

\makeatother

\title{\Huge \textbf{Foundation of}  \\ \textbf{\huge Affective Computing \& Interaction}
}
\author{\textsc{Changzeng Fu}\thanks{\url{https://cz26.github.io/CZ-HP/}} \\ \\ \\ \\ HACI Lab \\ SSTC \\ Northeastern University}

\begin{document}
\frontmatter
\maketitle

%%%%%%%%%%%%%%%%%%%%%%%%%%%%%%%%%%%%%%%%%%%%%%%%%%%%%%%%%%%%%%%
% Add a dedication paragraph to dedicate your book to someone %
%%%%%%%%%%%%%%%%%%%%%%%%%%%%%%%%%%%%%%%%%%%%%%%%%%%%%%%%%%%%%%%
% \begin{dedication}
% xxx
% \end{dedication}
\section*{Copyright}
Copyright \copyright \ 2025 Changzeng Fu (All Rights Reserved)\\
HACI Lab, Sydney Smart Technolegy College, Northeastern University
\\
\\
\\
Licensed under the Creative Commons Attribution-NonCommercial 4.0 Unported License (the “License”). 
You may not use this file except in compliance with the License. 
You may obtain a copy of the License at \url{http://creativecommons.org/licenses/by-nc/4.0.} 
Unless required by applicable law or agreed to in writing, software distributed under the License is distributed on an “AS IS” BASIS, 
WITHOUT WARRANTIES OR CONDITIONS OF ANY KIND, eitherexpress or implied. 
See the License for the specific language governing permissions and limitations under the License.

%%%%%%%%%%%%%%%%%%%%%%%%%%%%%%%%%%%%%%%%%%%%%%%%%%%%%%%%%%%%%%%%%%%%%%%%
% Auto-generated table of contents, list of figures and list of tables %
%%%%%%%%%%%%%%%%%%%%%%%%%%%%%%%%%%%%%%%%%%%%%%%%%%%%%%%%%%%%%%%%%%%%%%%%
\tableofcontents
\listoffigures
\listoftables

\mainmatter

% ---------- Part level ----------
\part{Basic Theory and Technical Framework}

% ---------- Chapter level ----------
\chapter{History and Development of Human–Computer Interaction}

% ---------- Section level ----------
\section{The Evolution of Human–Computer Interaction}

% ---------- Subsection level ----------
\subsection{Definition of Human–Computer Interaction}
Human-Computer Interaction (HCI) refers to the process of information exchange between people and computers using a certain dialogue language and a certain interactive method to complete a certain task. It includes the design and optimization of hardware equipment, the development and improvement of software interfaces, and the psychological and behavioral characteristics of users to achieve an efficient, natural and friendly interactive experience. Research in this field is an interdisciplinary field that explores the information exchange process between people and computers. It not only focuses on the design and implementation of computer systems, but also pays more attention to how users communicate and operate computers effectively. With the rapid development of information technology, human-computer interaction has become an indispensable and important part of computer science. Its research content covers multiple disciplines such as computer science, psychology, cognitive science, and design, and plays a vital role in improving the usability, efficiency and user satisfaction of computer systems.
\subsection{Importance of Human-Computer Interaction}
\begin{itemize}
    \item Improve work efficiency: Good human-computer interaction design can help users complete tasks more quickly and accurately, reduce operational errors, and thus improve overall work efficiency. For example, human-computer interaction technology can automate a variety of repetitive tasks, effectively reduce the number of operating steps, and improve processing efficiency. Gmail's smart mailbox function can not only automatically classify emails, but also provide intelligent reply suggestions. Users can complete the reply with just one click. E-commerce platforms such as Taobao and JD.com widely use intelligent customer service robots to automatically handle common problems and only transfer complex matters to manual customer service, significantly improving response speed. Design software has also introduced intelligent interaction mechanisms that can identify users' operating modes and actively recommend more efficient tools or shortcuts to speed up the pace of task completion.
\end{itemize}
% fig
\begin{figure}[h]
    \centering
    \includegraphics[width=0.5\textwidth]{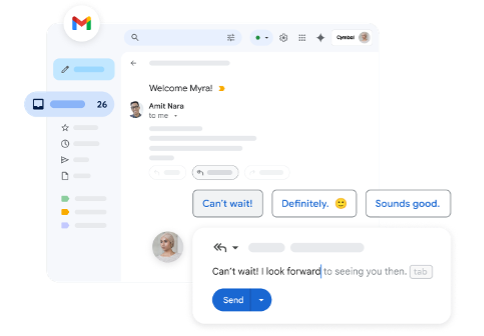}
    \caption{ Gmail smart reply suggestion diagram}
\end{figure}
\begin{itemize}
    \item Improve user experience: By optimizing the interactive interface and method, the computer system can be made more in line with the user's usage habits and psychological expectations, thereby enhancing user satisfaction and loyalty to the product. For example, human-computer interaction technology supports personalized customization of interfaces and functions, making system operations more in line with individual needs. Taking the Windows operating system as an example, users can set up multiple virtual desktops according to task types, and customize the order and interface layout of programs in the taskbar to create a more comfortable and efficient working environment, thereby improving user experience. The Windows operating system allows users to set up multiple virtual desktops according to different tasks, and customize the order and interface layout of fixed programs in the taskbar, thereby creating an efficient personal working environment.    
\end{itemize}
\begin{figure}[h]
    \centering
    \includegraphics[width=0.5\textwidth]{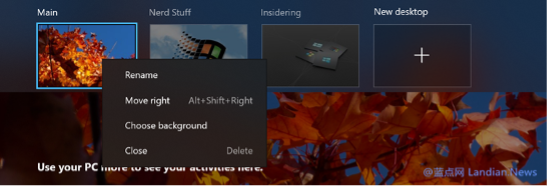}
    \caption{Windows Virtual Desktop Diagram}
\end{figure}
\begin{itemize}
    \item Promote technology popularization: Simple and easy-to-use interactive methods can lower the threshold for using computer technology, allowing more people to easily access and use computers and related equipment, and promote the widespread application and popularization of information technology. For example, the "elderly mode" widely used in smart device systems significantly reduces the difficulty of operation for middle-aged and elderly users by enlarging fonts, simplifying interface structures, and enhancing voice interaction and navigation prompts, thereby effectively promoting the popularization of information technology among the elderly.
\end{itemize}
\begin{figure}[h]
    \centering
    \includegraphics[width=0.5\textwidth]{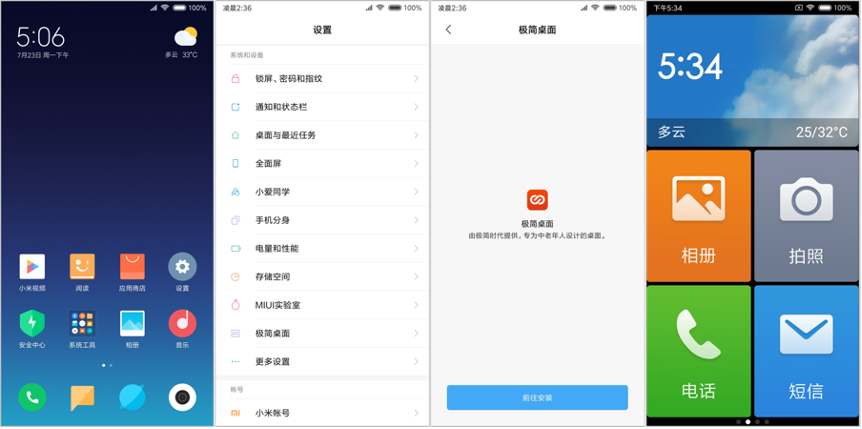}
    \caption{Xiaomi Elder Mode Icon}
\end{figure}
\begin{itemize}
    \item Promoting social development: The progress of human-computer interaction technology has not only changed people's lives and working methods, but also brought innovation and changes to various fields of society, such as education, medical care, entertainment, transportation, etc., and promoted the information and intelligent development of society. For example, the voice control system in smart cars helps drivers concentrate on driving and complete multiple operations at the same time, improving traffic safety and travel experience.
\end{itemize}
\begin{figure}
    \centering
    \includegraphics[width=0.5\textwidth]{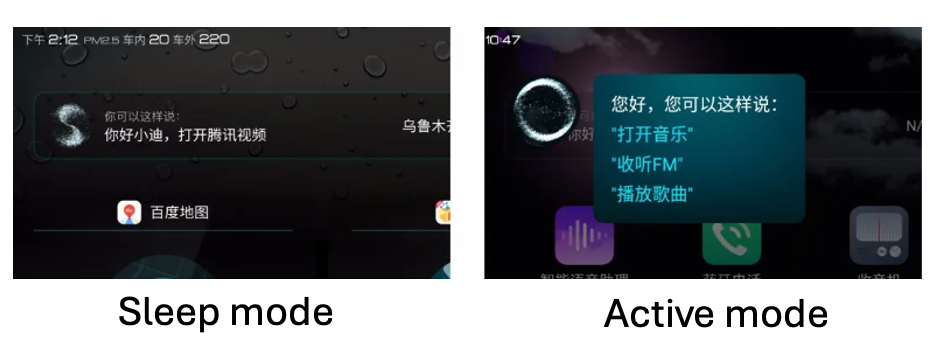}
    \caption{BYD's AI intelligent voice system}
\end{figure}
\section{The Evolution of Human-Computer Interaction}
The development of human-computer interaction reflects the leap from mechanical instructions to natural interaction. Its evolution process can be roughly divided into the following stages:
\begin{figure}[h]
    \centering
    \includegraphics[width=0.5\textwidth]{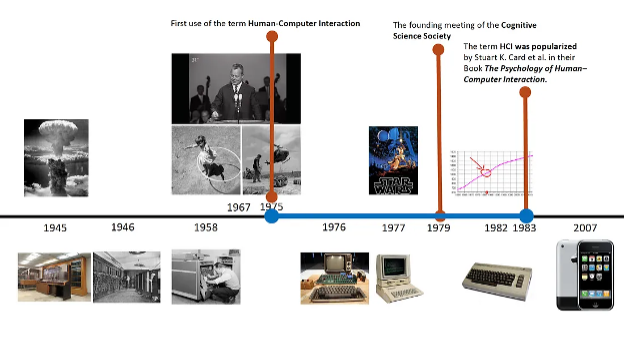}
    \caption{Schematic diagram of the development of human-computer interaction related events}
\end{figure}

\subsection{Early manual work stage}
In the early stages of computer development, the way of human-computer interaction was extremely primitive and clumsy. Computers at that time were large in size and limited in performance. Operators needed to manually write binary codes and input them into the computer before the computer could perform the corresponding operations. This interactive method has extremely high technical requirements for operators. They need to have deep computer expertise and programming skills to communicate effectively with the computer. Due to the cumbersome and complicated operation process and the slow processing speed of the computer, the efficiency of human-computer interaction is low, which greatly limits the popularity and application scope of computers. Human-computer interaction at this stage is mainly about people adapting to computers. Users must operate according to the requirements and rules of the computer. The computer can hardly provide users with any convenient and friendly interactive experience.
\begin{figure}
    \centering
    \includegraphics[width=0.5\textwidth]{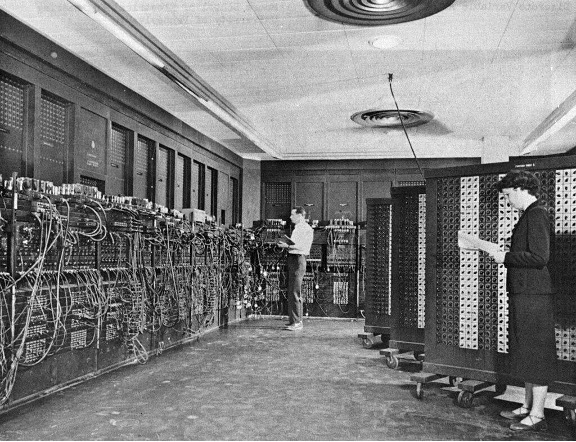}
    \caption{The world’s first general-purpose computer “ENIAC”}
\end{figure}

\subsection{Job Control Language and Interactive Command Language Stage}
In the mid-1960s, with the gradual development of computer technology, job control languages and interactive command languages emerged. Computer users at this stage were mainly professional programmers who could interact with computers by writing batch job languages or interactive command languages. Although it was still necessary to memorize a large number of commands and master keyboard operations, the interactive method had been significantly improved compared to the early manual operation stage. Programmers could debug programs and understand the execution of computers more conveniently, thereby improving work efficiency. This interactive method gradually expanded the scope of computer use, extending from professional scientific research institutions and large enterprises to more fields and user groups. However, this command-line-based interactive method still has certain limitations. It requires users to have certain computer expertise and programming basics, and there is still a high threshold for ordinary users.
\begin{figure}
    \centering
    \includegraphics[width=0.5\textwidth]{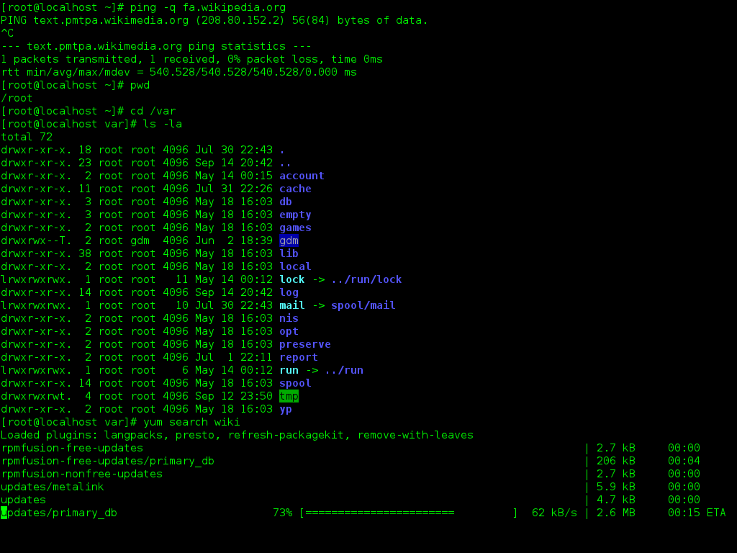}
    \caption{GNOME Terminal3 (Fedora 15)Example inBashScreen capture of the session}
\end{figure}

\subsection{Graphical User Interface (GUI) Stage}
In the 1980s, the emergence of the Graphical User Interface (GUI) marked a major breakthrough in human-computer interaction technology. The main features of GUI are desktop metaphor, WIMP technology (windows, icons, menus, pointers), direct manipulation and "What You See Is What You Get (WYSIWYG)". This interactive method, with its simple and easy-to-learn, intuitive and convenient characteristics, greatly reduces the difficulty of using computers, allowing ordinary users who do not understand computer expertise to easily operate computers. The emergence of GUI not only changed the way users interact with computers, but also promoted the rapid development of computer software and applications, and promoted the popularization and widespread application of computers. For example, the Macintosh computer launched by Apple in 1984 was the first personal computer to widely adopt a graphical user interface. It was widely welcomed by users for its friendly user interface and convenient operation, and became an important milestone in the history of human-computer interaction. Subsequently, Microsoft's Windows operating system also gradually became popular, further consolidating the dominant position of GUI in the field of human-computer interaction.
\begin{figure}
    \centering
    \includegraphics[width=0.5\textwidth]{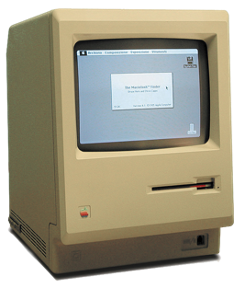}
    \caption{The original Macintosh 128k}
\end{figure}

\subsection{Web User Interface Stage}
In the 1990s, with the rise and development of the Internet, the web user interface came into being. Web browsers based on Hypertext Markup Language (HTML) and Hypertext Transfer Protocol (HTTP) became the representative of web user interfaces. The emergence of web user interfaces enabled users to access information resources worldwide through browsers, greatly enriching the content and form of human-computer interaction. Human-computer interaction technology developed rapidly during this period, and new technologies and applications continued to emerge, such as search engines, network acceleration, multimedia animation, and chat tools. Web user interfaces not only provided users with a more convenient way to obtain and communicate information, but also promoted the rapid development of emerging fields such as e-commerce, online education, and telecommuting. For example, the emergence of Google search engines greatly improved the efficiency of users in obtaining information and became one of the indispensable tools in people's daily lives; and instant messaging tools such as QQ and WeChat changed people's communication methods, allowing people to communicate and exchange conveniently anytime and anywhere. The development of web user interfaces has made human-computer interaction no longer limited to local computers, but has expanded to the global cyberspace, bringing a broader development space for human-computer interaction.

\subsection{Multi-channel and multimedia intelligent human-computer interaction stage}
Entering the 21st century, with the continuous development and integration of computer technology, communication technology, sensor technology, etc., human-computer interaction technology has entered a new stage of development - the multi-channel, multimedia intelligent human-computer interaction stage. The main feature of this stage is to use multiple sensory channels and action channels (such as voice, handwriting, posture, sight, expression, etc.) of people for input, and interact with the computer environment in a parallel and imprecise way. This interaction method is more natural, efficient and intelligent, and can better meet the diverse needs of users in different scenarios. For example, the emergence of virtual reality (VR) and augmented reality (AR) technology has brought a new immersive experience to human-computer interaction. Users can enter the virtual three-dimensional space by wearing devices such as VR helmets or AR glasses, and interact naturally with virtual objects, as if they were in a real virtual world. This immersive interaction method not only enhances the user's sense of participation and interactivity, but also brings new application models and development opportunities to games, education, training, design and other fields. At the same time, the continuous advancement of speech recognition technology has also made voice interaction an important interaction method. Users can communicate with computers through voice commands without manually entering text, which greatly improves the efficiency and convenience of interaction. For example, the emergence of intelligent voice assistants such as Apple's Siri, Microsoft's Cortana, and Amazon's Alexa allows users to query information, set reminders, control devices, and other operations through voice, which has brought great convenience to users' lives and work. In addition, interactive technologies such as gesture recognition, expression recognition, and eye control are also constantly developing and improving, providing richer and more diverse means of interaction for human-computer interaction. The development of these technologies has promoted the transformation of human-computer interaction from traditional two-dimensional plane interaction to three-dimensional space interaction, and from single-modal interaction to multi-modal fusion interaction, making human-computer interaction gradually develop in a more natural, intelligent, and humanized direction.
\begin{figure}
    \centering
    \includegraphics[width=0.5\textwidth]{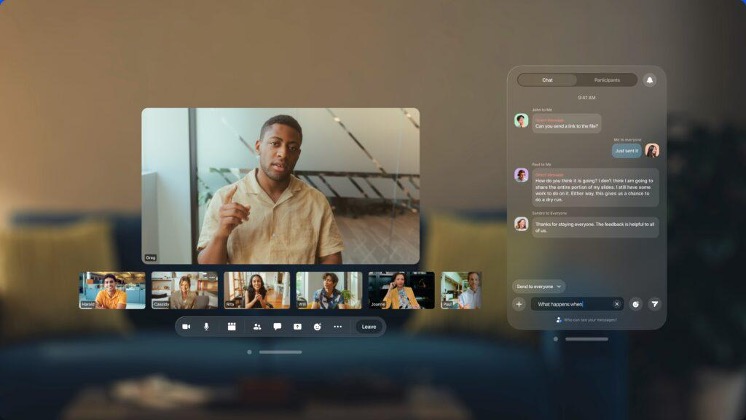}
    \caption{Using Zoom with Apple Vision Pro}
\end{figure}

\section{Key technologies and equipment for human-computer interaction}
The development of human-computer interaction is inseparable from the support of various key technologies and equipment. They have played an important role in different stages of development and continuously promoted the advancement of human-computer interaction technology.

\subsection{Development of input devices}
Input devices are an important part of human-computer interaction, and their development history is closely related to the progress of human-computer interaction technology. Early input devices were mainly keyboards and mice, which provided users with basic input functions, but the operation was relatively cumbersome and not intuitive enough. With the development of technology, the emergence of touch screen technology has changed the traditional input method. Users can directly interact with the content on the screen by clicking, sliding, zooming, etc. on the touch screen with their fingers, which greatly improves the naturalness and convenience of interaction. For example, the popularity of smartphones and tablets has made touch screens one of the most commonly used input devices in people's daily lives. In addition to touch screens, the progress of voice recognition technology has also brought new changes to input devices. Voice input devices can convert users' voice signals into text or instructions that can be recognized by computers. Users do not need to enter text manually, but can complete various operations by voice, such as voice search, voice input text, voice control devices, etc. This input method not only improves input efficiency, but also provides a more convenient interaction method for users who are not convenient to use keyboards or touch screens. In addition, new input devices such as gesture recognition devices, eye tracking devices, and brain-computer interface devices are also emerging and developing. Gesture recognition devices can capture user gestures through cameras or sensors and convert them into corresponding operation instructions. Users can interact with computers through simple gestures, such as waving, pointing, rotating, etc. Eye tracking devices can monitor the user's line of sight and gaze point in real time. Users can control the cursor or select content on the screen through their eyes. This input method is of great significance for some special user groups, such as people with disabilities or users who need to operate with both hands. Brain-computer interface devices are more cutting-edge input devices that can directly read the brain's neural signals and convert them into instructions that the computer can understand, thereby realizing direct interaction between users and computers. Although brain-computer interface technology is still in the development stage, its future development prospects are broad and it is expected to bring more revolutionary changes to human-computer interaction.
communication methods.
\begin{figure}
    \centering
    \includegraphics[width=0.5\textwidth]{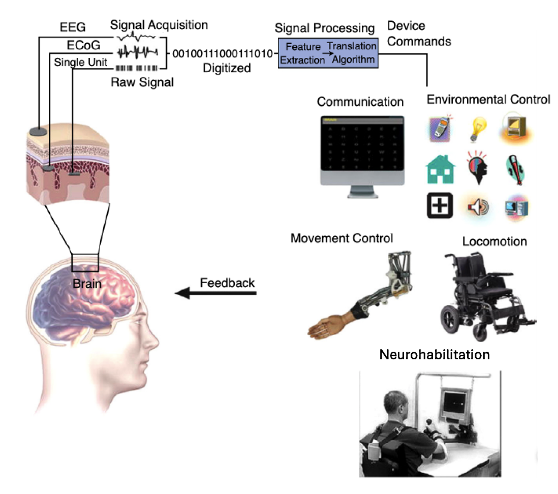}
    \caption{Typical BCI system components and their communication methods}
\end{figure}

\subsection{Evolution of output devices}
Output devices play a vital role in human-computer interaction, and their development history also witnesses the continuous progress of human-computer interaction technology. From the early simple display devices to today's diversified output methods, the evolution of output devices has not only improved the effect of information presentation, but also greatly enriched the user's interactive experience.

\subsubsection{Development of display devices}
Display devices are the core components of output devices, and their development history is closely related to the progress of human-computer interaction technology.

Early display devices: In the early stages of computer development, display devices were mainly simple cathode ray tube (CRT) displays. These displays had low resolution and rough display effects, and could only present information to users in the form of text or simple graphics. For example, the early DOS system interface only displayed content in the form of characters, and users needed to enter instructions through the command line to operate the computer, and the display effect was extremely limited.

\begin{figure}
    \centering
    \includegraphics[width=0.5\textwidth]{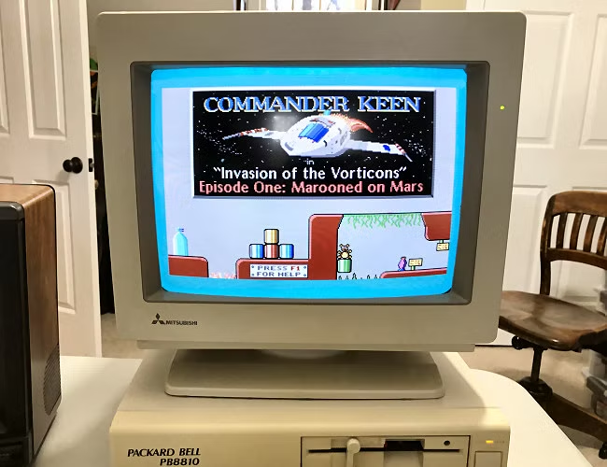}
    \caption{CRT display}
\end{figure}

Promotion of graphical user interface: With the emergence of graphical user interface (GUI), display devices have ushered in major changes. The Macintosh computer launched by Apple in 1984 was the first to widely adopt the graphical user interface. Its supporting display can present information in the form of graphics, icons, windows, etc., which greatly improves the readability and ease of use of information. Subsequently, liquid crystal displays (LCDs) gradually replaced CRT monitors and became the mainstream display device. LCD monitors have the advantages of small size, low power consumption, and clear display effects. They can support higher resolution and richer color expression, providing users with a more intuitive and beautiful visual experience.

\begin{figure}
    \centering
    \includegraphics[width=0.5\textwidth]{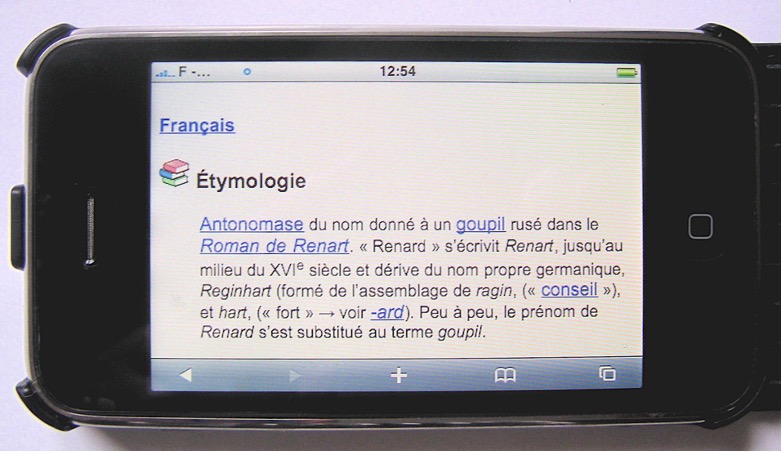}
    \caption{LCD screen in the first generation iPhone}
\end{figure}

Popularization of touch screens: Entering the 21st century, with the rise of mobile devices, touch screen technology has been widely used. The touch screen is not only an input device, but also an output device. It integrates display functions with input functions. Users can operate by directly touching the content on the screen, realizing a more natural and convenient way of interaction. For example, the touch screens of smartphones and tablets can display images, videos and text content with high resolution and high contrast, and support multi-touch gestures. Users can interact with the content on the screen through operations such as clicking, sliding, and zooming, which greatly enhances the user's interactive experience.

Virtual reality and augmented reality display devices: In recent years, the rise of virtual reality (VR) and augmented reality (AR) technologies has brought a new visual experience to human-computer interaction. Devices such as VR helmets and AR glasses create immersive three-dimensional virtual environments or augmented reality scenes for users through high-resolution displays and advanced optical technology. Users can move and interact freely in the virtual environment, as if they were in a real virtual world. For example, VR helmets can provide a 360-degree panoramic view, allowing users to experience virtual scenes immersively; AR glasses can overlay virtual information such as navigation instructions and virtual objects in the real world, providing users with a richer and more intuitive way of presenting information. The development of these display devices has not only promoted the advancement of human-computer interaction technology, but also brought new application models and development opportunities to games, education, training, design and other fields.

\subsubsection{Development of audio output devices}
Audio output devices also play an important role in human-computer interaction, and their development history has also witnessed the continuous advancement of human-computer interaction technology.

Early audio output devices: In the early stages of computer development, audio output devices were mainly simple speakers and headphones. These devices were mainly used to play system prompts or simple audio files. They had relatively simple functions and relatively rough sound quality. For example, in the early DOS system, the computer could only emit a simple beep to prompt the user that an operation was completed or an error occurred.

Development of multimedia audio: With the rise of multimedia technology, audio output devices have been significantly improved. Computers began to support high-fidelity audio playback, and the sound quality of speakers and headphones has also been greatly improved. Users can listen to music, watch movies, make voice calls, etc. through audio output devices. Audio output devices have become an indispensable part of human-computer interaction. For example, high-quality headphones can provide an immersive audio experience, allowing users to better enjoy multimedia content; and high-fidelity speakers can be used in home theater systems to bring users shocking audio effects.

The promotion of voice interaction: In recent years, the rapid development of speech recognition and speech synthesis technology has made voice interaction an important way of interaction. The emergence of intelligent voice assistants such as Apple's Siri, Microsoft's Cortana, Amazon's Alexa, etc., allows users to have natural conversations and interactions with devices through voice. Audio output devices play a key role in voice interaction. They can not only clearly play the responses of voice assistants, but also help users better understand and operate devices through voice prompts. For example, users can query information, set reminders, control devices, etc. through voice commands, and voice output devices can feedback the results to users in the form of natural language, greatly improving the efficiency and convenience of interaction.

The rise of spatial audio technology: With the development of virtual reality and augmented reality technologies, spatial audio technology has emerged. Spatial audio technology can simulate the sound propagation effect in the real world, allowing users to feel the direction, distance and spatial sense of sound in a virtual environment, further enhancing the user's sense of immersion and interactive experience. For example, in VR games, users can hear the sounds of virtual objects coming from different directions, so that they can feel the game environment more realistically; in AR applications, spatial audio technology can provide users with more intuitive navigation prompts and voice interaction experience. The development of spatial audio technology not only brings a richer auditory experience to human-computer interaction, but also provides a more realistic immersive environment for virtual reality and augmented reality applications.

\begin{figure}
    \centering
    \includegraphics[width=0.5\textwidth]{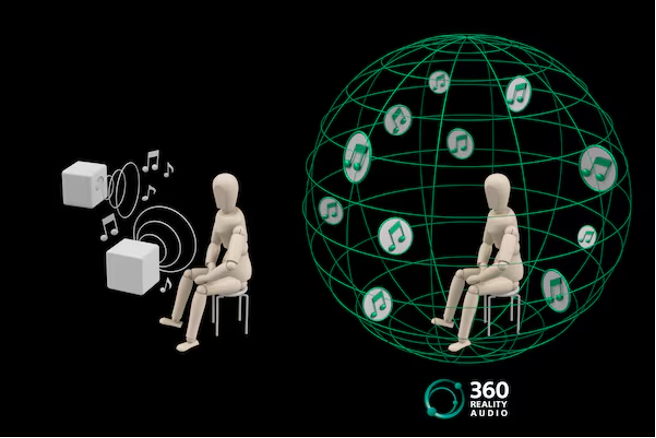}
    \caption{Schematic diagram of speaker sound effects under Sony's 360 Reality Audio technology}
\end{figure}

\subsubsection{Innovation of other output devices}
In addition to display devices and audio output devices, other types of output devices also play an important role in human-computer interaction and are constantly being innovated.

Haptic feedback devices: Haptic feedback devices provide users with a richer and more intuitive interactive experience by simulating real tactile sensations. For example, the vibration feedback function in a game controller allows users to feel the collision and movement of virtual objects during the game, enhancing the immersion of the game; the tactile feedback function in smart watches and mobile phones can remind users to receive notifications or complete operations through vibrations or slight tactile prompts. In recent years, tactile feedback technology has continued to develop, and more advanced tactile feedback devices such as tactile feedback gloves and tactile feedback clothing have emerged. These devices can simulate the touch, pressure, and texture of the real world through precise tactile feedback, allowing users to experience real tactile experiences in virtual environments. For example, tactile feedback gloves allow users to feel the shape and texture of virtual objects when touching and grabbing them in virtual reality, as if they were really touching the real thing.
\begin{figure}
    \centering
    \includegraphics[width=0.5\textwidth]{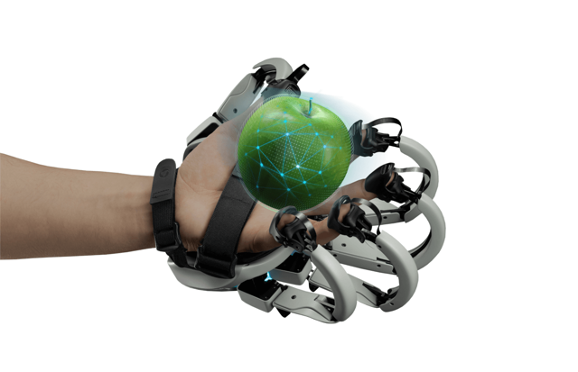}
    \caption{Tactile feedback gloves TESLAGLOVE}
\end{figure}

Smell output device: Smell output device is an emerging output device that enhances the user's interactive experience by releasing different smells. Although the smell output device is still in the development stage, it has broad application prospects. For example, in virtual reality and augmented reality applications, smell output devices can be combined with visual and audio output devices to provide users with a more realistic immersive experience. Users can not only see and hear the virtual scene in the virtual environment, but also smell the smell associated with the scene.For example, Sony launched the Scent Output Module, a programmable scent output device that releases preset scent molecules through a micro-cartridge. The device can control the type, intensity and release timing of the scent according to the scene, and achieve synchronization with vision and hearing. In application scenarios such as immersive multimedia displays, virtual reality experiences, and retail displays, the Scent Output Module significantly enhances the user's sensory immersion and situational awareness.Such as the fragrance of flowers in the forest, the smell of sea breeze at the seaside, etc., further enhancing the user's sense of immersion and participation. In addition, odor output devices can also be used in medical rehabilitation, psychotherapy and other fields, helping patients relieve stress and improve mood by releasing specific odors.

\begin{figure}
    \centering
    \includegraphics[width=0.5\textwidth]{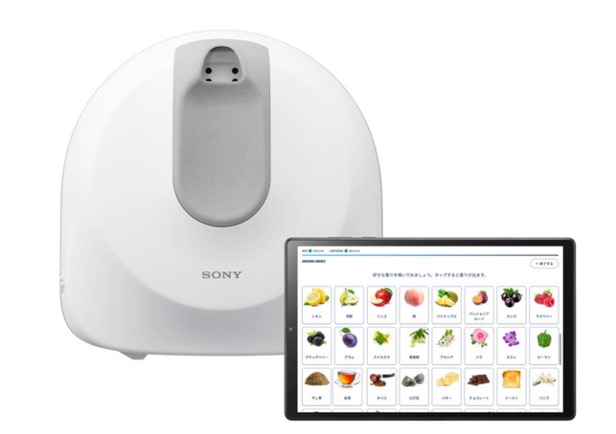}
    \caption{Customized ink cartridge for Sony's olfactory measurement system}
\end{figure}

Brain-computer interface output device: Brain-computer interface (BCI) technology is a technology that directly reads brain neural signals and interacts with computers. Although brain-computer interface technology is currently mainly focused on input devices, its application in the field of output devices is gradually gaining attention. For example, through brain-computer interface technology, computers can directly control external devices or generate corresponding outputs based on the user's brain signals, such as controlling the movement of virtual objects, generating voice or text information, etc. This output method can not only provide a more convenient means of interaction for people with disabilities, but may also bring a more natural and efficient interactive experience to future virtual reality and augmented reality applications. Although brain-computer interface output devices still face many technical challenges, their future development prospects are broad and are expected to bring more revolutionary changes to human-computer interaction.

\begin{figure}
    \centering
    \includegraphics[width=0.5\textwidth]{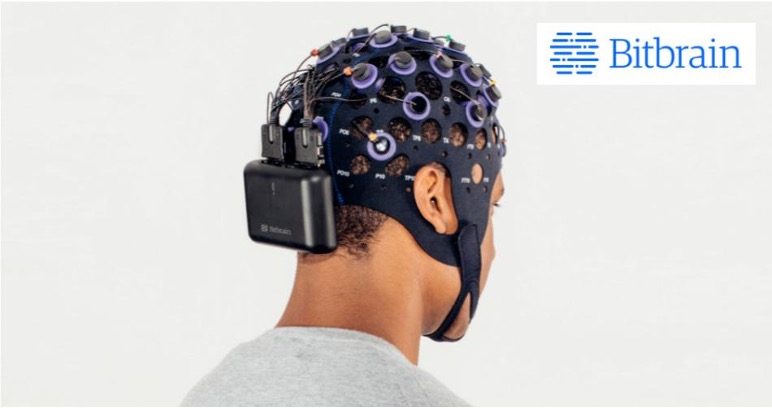}
    \caption{Brain-computer interface device from Bitbrain}
\end{figure}

\section{Overview of Modern Interaction Technologies}
In today's digital age, human-computer interaction technology is developing at an astonishing speed, showing diversified and intelligent characteristics. These technologies have not only greatly changed the way we interact with computer devices, but also profoundly affected our daily life, work and entertainment. The following will elaborate on modern interaction technology from four aspects: voice interaction, visual interaction, brain-computer interface and tactile feedback.

\subsection{Voice interaction: opening a new era of natural conversation}

\subsubsection{Technical principles and core architecture}
The core of voice interaction technology lies in speech recognition and natural language processing. Speech recognition is the process of converting the user's voice signal into text, while natural language processing is responsible for understanding the semantics of these texts and generating corresponding responses. In recent years, the rapid development of deep learning technology, especially the emergence of the Transformer model, has brought a qualitative leap in voice interaction technology. Based on the attention mechanism, the Transformer model can efficiently process sequence data, thereby achieving high-precision command understanding. It is trained through a large amount of voice data, learns the characteristics of speech and the laws of language, and greatly improves the accuracy of speech recognition. It can accurately understand user commands even in complex contexts and accents.

\subsubsection{Application scenarios and advantages}
The application scenarios of voice interaction technology are extremely wide, covering smart home, smart office, smart driving and other fields. In the smart home environment, users can control various devices in the home, such as lights, curtains, air conditioners, etc., through voice commands to achieve intelligent home control. For example, the user only needs to say "turn on the lights in the living room", and the smart speaker can accurately recognize and execute the command, greatly improving the convenience of life. In the field of smart office, voice interaction technology can help users quickly query information, arrange meetings, send emails, etc., to improve work efficiency. In smart driving, voice interaction technology allows drivers to perform navigation queries, make phone calls and other operations without leaving the steering wheel, improving driving safety.

\begin{figure}
    \centering
    \includegraphics[width=0.5\textwidth]{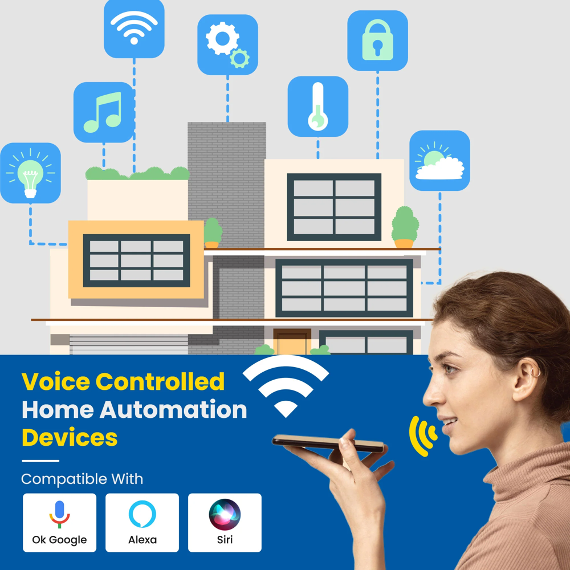}
    \caption{Schematic diagram of voice-controlled smart furniture}
\end{figure}

The advantage of voice interaction lies in its naturalness and convenience. Users do not need to learn complex operating procedures, they can complete various operations by simply expressing their needs in natural language. This interactive method not only saves time, but also lowers the user's usage threshold, allowing more people to easily use smart devices.

\subsubsection{Future development trends and challenges}
In the future, voice interaction technology will continue to develop in a more intelligent, personalized and emotional direction. With the continuous advancement of artificial intelligence technology, voice assistants will be able to better understand users' emotions and intentions and provide more intimate and personalized services. For example, voice assistants can adjust the tone and content of their responses based on the user's emotional state, providing users with a warmer and more humane interactive experience.

However, voice interaction technology also faces some challenges. First, the accuracy of voice recognition still needs to be improved in some complex environments, such as in noisy environments or when facing multiple accents. Secondly, the privacy and security issues of voice interaction have also attracted people's attention. The user's voice data contains a lot of personal information, and how to ensure the security and privacy of this data is an urgent problem to be solved. In addition, the popularization of voice interaction technology still needs to overcome some technical difficulties, such as multi-language support, dialect recognition, etc., to meet the needs of different regions and user groups.

\subsection{Visual Interaction: Building an Immersive Visual Experience}

\subsubsection{Technical principles and core architecture}
Visual interaction technology mainly relies on computer vision and image processing technology. The user's visual information is obtained through RGB-D sensors (color cameras and depth sensors), and then the Convolutional Neural Network (CNN) is used to process and analyze this information to achieve recognition of facial expressions, body movements, etc. CNN is a deep learning model that can automatically learn features in images and is trained with a large amount of image data to enable it to have powerful image recognition capabilities. For example, in facial expression recognition, CNN can identify the user's emotional state such as happiness, anger, sadness, and joy; in body movement recognition, it can capture the user's gestures, body postures and other action information.

\subsubsection{Application scenarios and advantages}
Visual interaction technology has been widely used in games, virtual reality, education and other fields. In the field of games, visual interaction technology allows players to interact with characters in the game through body movements, which enhances the immersion and fun of the game. For example, in some somatosensory games, players can control the actions of game characters by waving their arms, jumping and other actions, as if they were in the game world. For example, Kinect developed by Microsoft uses depth cameras and computer vision technology to capture and recognize players' full-body movements in real time. Players can interact naturally with the virtual game environment without holding a controller, which reflects the typical application of visual interaction in somatosensory games. In virtual reality applications, visual interaction technology can capture users' movements and expressions in real time, allowing users to interact naturally with virtual objects in a virtual environment, such as grabbing virtual objects with their hands and controlling virtual interfaces with their eyes. In the field of education, visual interaction technology can create more vivid and intuitive teaching scenes and improve students' learning interest and effectiveness. For example, through body movement recognition technology, students can interact with virtual teaching models and understand complex knowledge concepts more intuitively.

\begin{figure}
    \centering
    \includegraphics[width=0.5\textwidth]{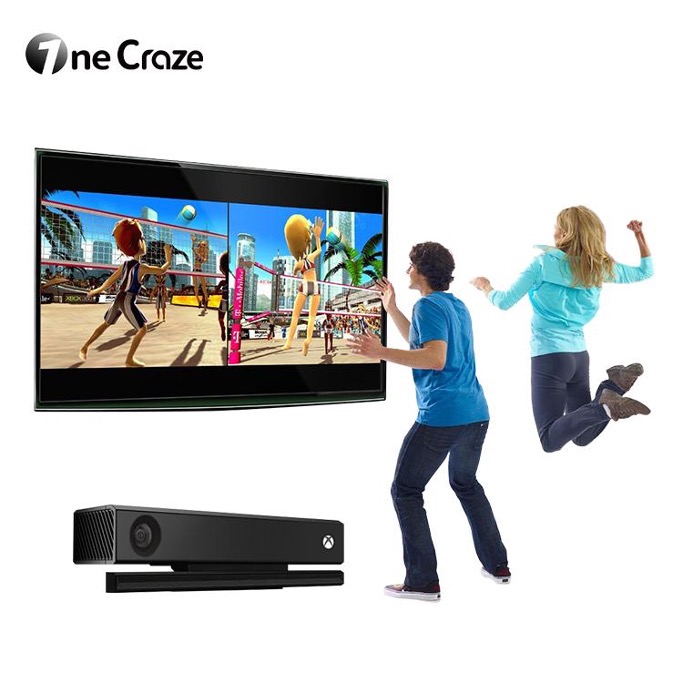}
    \caption{Schematic diagram of somatosensory game interaction based on visual recognition - screenshot of Kinect game scene}
\end{figure}

The advantage of visual interaction lies in its intuitiveness and immersion. Users can interact with devices through natural visual movements without complicated operation procedures. This interaction method is more in line with human natural behavior habits. At the same time, visual interaction technology can provide users with a richer visual experience, enhance users' sense of participation and immersion, and make users feel more pleasant and real during use.

\subsubsection{Future development trends and challenges}
In the future, visual interaction technology will develop in the direction of being more intelligent, natural, and multimodal. With the continuous advancement of computer vision technology, visual interaction systems will be able to more accurately identify and understand users' visual information, and achieve more complex and natural interaction scenarios. For example, future visual interaction systems will be able to identify subtle visual information such as users' eyes and micro-expressions, providing users with a more accurate and personalized interaction experience. At the same time, visual interaction technology will be deeply integrated with other interaction technologies (such as voice interaction, tactile feedback, etc.) to build a richer and more natural multimodal interaction environment.

However, visual interaction technology also faces some challenges. First, the accuracy and robustness of visual interaction systems still need to be improved in some complex environments, such as under poor lighting conditions or complex user movements. Second, the privacy and security issues of visual interaction technology have also attracted people's attention. The user's visual information contains a lot of personal privacy, and how to ensure the security and privacy of this information is an important research direction. In addition, the popularization of visual interaction technology also requires solving some technical problems, such as equipment costs, computing resources, etc., in order to lower its application threshold and enable it to be more widely used in various scenarios.

\subsection{Brain-computer interface: exploring direct dialogue between the human brain and the machine}

\subsubsection{Technical principles and core architecture}
Brain-Computer Interface (BCI) technology is a technology that directly reads brain neural signals and interacts with computers. Its core principle is to collect neural electrical activity signals from the cerebral cortex through electroencephalogram (EEG) sensors, and then use signal processing and decoding algorithms to convert these signals into instructions that the computer can understand. EEG signals are a non-invasive way to collect brain signals. By placing multiple electrodes on the scalp, the brain's neural electrical activity can be monitored in real time. After these signals are pre-processed by amplification and filtering, they are then feature extracted and classified by machine learning algorithms to decode the user's intentions. For example, users can imagine actions such as "moving forward" or "turning left" to make the computer or external device operate according to the corresponding instructions.

\subsubsection{Application scenarios and advantages}
Brain-computer interface technology has important application value in the field of medical rehabilitation. For disabled people and users with limited mobility, brain-computer interface technology provides them with a new way of interaction, allowing them to directly control prosthetic limbs, wheelchairs and other equipment through brain signals, and realize autonomous life and activities. For example, some brain-computer interface systems can help paralyzed patients control mechanical arms to perform operations such as grasping and moving through brain signals, greatly improving their quality of life and self-care ability. In addition, brain-computer interface technology also has broad application prospects in the fields of neuroscience research and psychotherapy. By monitoring the brain's neural activity in real time, researchers can better understand the brain's working mechanism and cognitive process, providing important support for the development of neuroscience.

\begin{figure}
    \centering
    \includegraphics[width=0.5\textwidth]{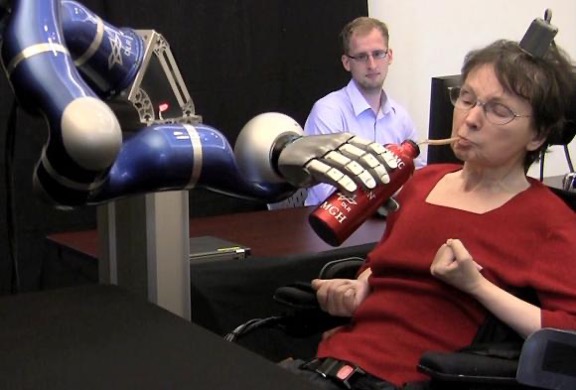}
    \caption{A paralyzed person uses a brain-computer interface to control a robotic arm}
\end{figure}

The advantages of brain-computer interface lie in its directness and efficiency. It can directly read the brain's neural signals without going through traditional intermediate links such as muscle movement or speech, thus realizing direct dialogue between the human brain and the machine. This interactive method not only improves the efficiency of interaction, but also provides new possibilities for users who cannot interact with the outside world through traditional methods, and has important social significance and application value.

\subsubsection{Future development trends and challenges}
In the future, brain-computer interface technology will develop in a more accurate, stable and convenient direction. With the continuous advancement of neuroscience and signal processing technology, brain-computer interface systems will be able to decode brain signals more accurately and achieve more complex and natural interaction scenarios. For example, future brain-computer interface technology will be able to achieve real-time decoding and feedback of brain signals, allowing users to interact with computers more naturally, just like using their own limbs. At the same time, brain-computer interface technology will be deeply integrated with other technologies (such as virtual reality, augmented reality, etc.) to provide users with a richer and more immersive interactive experience.

However, brain-computer interface technology also faces some challenges. First, the accuracy and stability of the brain-computer interface system still need to be improved, especially in complex environments and under long-term use. Secondly, the privacy and security issues of brain-computer interface technology have also attracted people's attention. Brain signals contain a lot of personal privacy and sensitive information. How to ensure the security and privacy of this information is an urgent problem to be solved. In addition, the popularization of brain-computer interface technology also needs to overcome some technical difficulties, such as equipment cost, wearing comfort, etc., in order to improve its application feasibility and user experience.

\subsection{Haptic feedback: giving machines delicate tactile expression capabilities}

\subsubsection{Technical principles and core architecture}
Touch Tactile feedback technology is a technology that enhances the user's interactive experience by simulating real tactile sensations. Its core principle is to use piezoelectric materials and force control algorithms to convert electrical signals into mechanical vibrations or force feedback, thereby simulating different tactile effects. Piezoelectric materials are materials that can convert electrical energy into mechanical energy. When an electrical signal is applied, the piezoelectric material will produce corresponding mechanical deformation, thereby generating vibration or force feedback. The force control algorithm can accurately control the deformation degree and vibration mode of the piezoelectric material according to different application scenarios and user needs, thereby achieving a variety of tactile effects. For example, in virtual reality (VR) and augmented reality (AR) applications, the force control algorithm can adjust the intensity, frequency and duration of tactile feedback in real time according to the characteristics of objects in the virtual scene and the user's interactive actions, allowing users to feel the texture and shape of virtual objects. In the field of robotics, force control algorithms can be used to control the robot's tactile perception and action response, allowing the robot to interact with humans more naturally and flexibly.

\subsubsection{Application scenarios and advantages}
Haptic feedback technology plays an important role in many fields. In the field of consumer electronics, such as smartphones, tablets and gaming devices, vibration and force feedback are used to enhance the user's sense of reality and immersion. In the automotive field, haptic feedback technology is used to improve driving safety and user experience, such as alerting drivers to potential dangers through tactile signals. In the medical field, haptic feedback technology is used in surgical simulation and rehabilitation training to help doctors gain a real operating experience and improve surgical accuracy. In the field of education, haptic feedback technology helps students better understand complex concepts and improve learning outcomes by simulating the touch of real objects.

\begin{figure}
    \centering
    \includegraphics[width=0.5\textwidth]{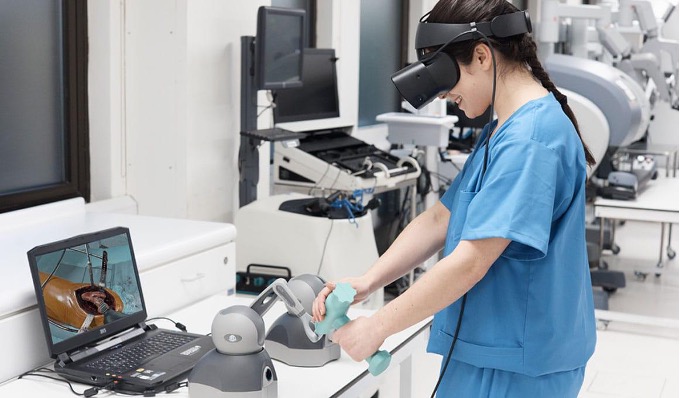}
    \caption{VR simulation provides tactile feedback to surgeons}
\end{figure}

Haptic feedback technology significantly enhances the user experience, making interactions more realistic and immersive. It improves operational precision, especially in scenarios that require delicate manipulation, such as medical surgery and industrial assembly. In addition, haptic feedback technology enriches the interaction methods, adds the dimension of touch, and makes the interaction more diverse and natural. In areas such as automobiles, haptic feedback technology improves safety through timely tactile signal reminders.

\subsubsection{Future development trends and challenges}
Haptic feedback technology is developing towards higher precision and realism, and will provide a more delicate tactile experience in the future, such as more realistic temperature and texture feedback. Multi-sensory integrated interactive design is another important trend. Tactile feedback will be integrated with other sensory technologies such as vision and hearing to form a more immersive multimodal interactive experience. In addition, tactile feedback technology will be combined with artificial intelligence to achieve intelligent and adaptive control, and adjust the parameters of tactile feedback in real time according to user behavior and feedback to provide a more personalized experience. Wearable and portable are also the future development direction. Tactile feedback devices will be lighter and more comfortable, making them easier for users to use in daily life.

Although tactile feedback technology has broad application prospects, it also faces some challenges. Individual differences and adaptability issues in tactile perception are one of them. Different users perceive and respond to tactile stimuli differently, which requires tactile feedback devices to adapt to the preferences of different users. The portability and cost issues of tactile feedback devices also need to be addressed. High-precision tactile feedback devices are usually expensive, which limits their popularity. In addition, tactile feedback technology still has limitations in simulating real tactile sensations, and the accuracy and realism need to be further improved. Interdisciplinary research and cooperation are also important aspects of promoting the development of tactile feedback technology, involving multiple fields such as materials science, control engineering, and computer science. Finally, with the widespread application of tactile feedback technology, user privacy and security protection have become important issues, and relevant technologies and management measures need to be strengthened.

\section{Concept and goal of human-machine integration}
In today's era of rapid technological development, human-machine integration has become an important area of concern. Human-machine integration emphasizes the deep integration of robots and humans at multiple levels. This integration is not just a simple interaction, but a full-scale, in-depth collaboration and cooperation, aiming to build a more harmonious and efficient human-machine relationship and promote the progress and development of human society.

\subsection{Definition of human-machine integration}
Human-machine integration covers the integration of three key levels: physical, cognitive, and emotional. From a physical perspective, robots share the same physical space with humans. They need to be able to accurately perceive the surrounding environment and the physical state and movements of humans, and be able to physically contact and interact with humans in a safe and coordinated manner. For example, on an industrial production line, collaborative robots need to work with human workers to complete assembly tasks, which requires robots to accurately perceive the position and movements of human workers to avoid collisions, and to cooperate according to the operating rhythm of human workers to achieve efficient collaborative work.

At the cognitive level, human-machine collaboration requires robots to understand human intentions, knowledge, and thinking processes. Robots need to have certain cognitive abilities to accurately interpret and understand human instructions, and to reason and make decisions based on actual conditions. At the same time, robots also need to be able to convey their working status, intentions, and knowledge to humans, so that humans can understand the robot's behavior and decision-making basis, thereby achieving cognitive synchronization and collaboration between the two parties. For example, in the medical field, surgical robots need to be able to understand the doctor's surgical intentions and operating steps, and be able to make corresponding decisions and adjustments based on the real-time situation during the operation. At the same time, they need to promptly feedback the progress of the operation and possible problems to the doctor, assisting the doctor in completing high-precision and difficult surgical operations.

Emotional integration is an important part of human-machine integration. Robots need to have the ability to perceive and express emotions, be able to perceive human emotional states, and be able to respond to human emotional needs in an appropriate manner. This emotional resonance can enhance the trust and affinity between humans and machines, making humans more willing to cooperate and communicate with robots. For example, in the field of education, educational robots can adjust teaching methods and rhythms by sensing students' emotional states, such as tension, anxiety, excitement, etc., and give students appropriate emotional support and encouragement to improve students' learning effects and enthusiasm.

\subsection{Core Goals of Human-Machine Integration}

\subsubsection{Dynamic adaptability}
Dynamic adaptability is one of the important goals of human-machine integration. Through multimodal perception technology, robots can obtain various information about humans in real time, including vision, voice, physiological signals, etc., so as to accurately understand human intentions and needs. Visual perception enables robots to observe human movements, expressions and surroundings, voice perception enables robots to understand human language instructions and communication content, and physiological signal perception can reveal human physical state and emotional changes, such as heart rate, brain waves, etc. The fusion of these multimodal information enables robots to fully and accurately grasp the state of humans, so as to better adapt to human behavior and needs.

For example, in a smart home environment, robots can perceive the activities of family members through vision, such as whether they are at home, in which room, and what activities they are doing; perceive the needs of family members through voice, such as the need to play music and adjust the indoor temperature; and perceive the physical condition of family members through physiological signals, such as whether they are tired or nervous. Based on this information, robots can dynamically adjust their behavior and services to provide family members with a personalized comfort experience. If a family member shows physiological signals of fatigue while working, the robot can automatically adjust the indoor light and temperature and play soothing music to create a relaxing working environment for family members; if a family member shows excitement while entertaining, the robot can recommend corresponding entertainment content and adjust the sound and lighting effects to enhance the entertainment atmosphere.

In the industrial field, dynamic adaptability is also crucial. Robots need to adjust quickly according to the real-time situation on the production line and the operating needs of workers. For example, when the product model on the production line changes or the operating speed of workers changes, the robot can obtain this information in time through multimodal perception, and adjust its working mode and operating speed accordingly to ensure the smooth progress of the production process. In addition, collaborative robots (cobots) use different safety interaction modes, such as safety monitoring stop, manual guidance, speed and distance monitoring, and force and power limitation. These modes can ensure the safety and efficiency of human-machine cooperation in a dynamically changing working environment. This dynamic adaptability not only improves production efficiency, but also ensures the stability of product quality. At the same time, it also reduces the workload of workers, allowing them to focus more on the operation of key links.

\begin{figure}
    \centering
    \includegraphics[width=0.5\textwidth]{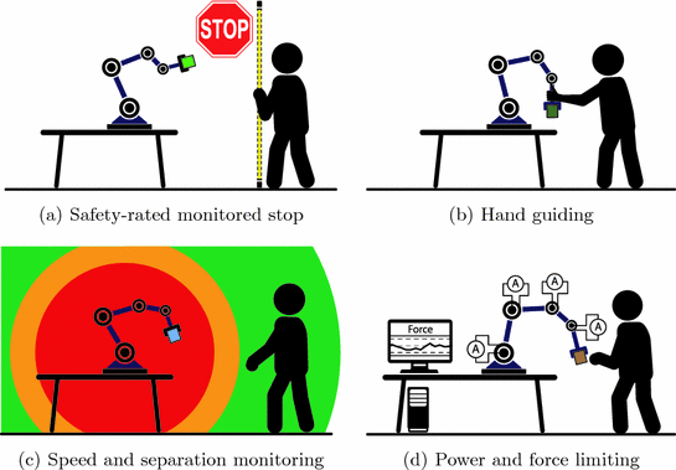}
    \caption{Four common safe interaction modes used by collaborative robots (cobots) when working with humans}
\end{figure}

\subsubsection{Two-way learning mechanism}
The two-way learning mechanism of human-machine integration promotes collaboration and learning between humans and machines, and achieves common growth and progress for both parties. Human guidance of robot task execution is an important aspect of the two-way learning mechanism. With rich experience and professional knowledge, humans can provide robots with clear task instructions and operating specifications to help robots quickly learn and master new tasks. In this process, humans pass on their skills and knowledge to robots through direct interaction with robots, such as demonstration operations and verbal guidance. Robots learn and imitate by sensing human actions and instructions, and continuously optimize their own behavior patterns.

For example, in the agricultural field, farmers can guide agricultural robots to perform operations such as sowing, fertilizing, and weeding. Farmers demonstrate the correct operation methods and techniques to the robots, such as the depth and spacing of sowing, the amount and location of fertilization, etc. The robots perceive the farmers' operation actions through vision and sensors, store and analyze this information, and gradually learn and master these agricultural skills. As the learning progresses, the robots can autonomously adjust the operation parameters according to different farmland environments and crop growth conditions, thereby improving agricultural production efficiency and quality.

At the same time, robots can also assist humans in optimizing their decision-making. With their powerful computing and data analysis capabilities, robots can quickly process and analyze large amounts of data and provide decision-making support for humans. For example, in the financial field, intelligent robots can analyze massive amounts of financial market data, including stock prices, trading volumes, macroeconomic indicators, etc. Through machine learning algorithms and data analysis models, they can predict market trends and risks and provide decision-making advice to financial analysts. Financial analysts can combine the robot's analysis results with their own professional knowledge to make more scientific and reasonable investment decisions, reduce investment risks, and increase investment returns.

In the medical field, robots can also provide doctors with auxiliary decision-making support. For example, in terms of disease diagnosis, robots can analyze patients' medical records, examination reports, imaging data and other data, and combine medical knowledge and clinical experience to provide doctors with possible disease diagnosis suggestions and treatment plans. Doctors can refer to the robot's suggestions and combine their own clinical judgment to make more accurate diagnosis and treatment decisions, thereby improving medical quality and efficiency.

\subsubsection{Emotional resonance}
Emotional resonance is one of the important goals of human-machine integration. Building an empathy model based on affective computing technology enables robots to perceive and respond to human emotional states, enhancing emotional connection and trust between humans and machines. Affective computing is an important branch of artificial intelligence, which involves the recognition, understanding and expression of human emotions. Through affective computing technology, robots can use various sensors and algorithms to monitor and analyze human facial expressions, voice intonation, physiological signals, etc. in real time, so as to accurately perceive human emotional states, such as happiness, sadness, anger, fear, etc.

For example, the "Emotional Gloves" project of MIT Media Lab is a typical application of emotional computing technology. This glove has multiple sensors built in, which can monitor the wearer's hand movements, pressure, temperature and other information in real time, and identify the wearer's emotional state through data analysis and machine learning algorithms. When the wearer is in a negative emotional state such as tension and anxiety, the gloves can give the wearer emotional feedback through vibration, light and other means to help the wearer relieve emotional pressure. At the same time, the gloves can also transmit emotional information to the robot that interacts with it, so that the robot can perceive the wearer's emotional state and make corresponding emotional responses.

\begin{figure}
    \centering
    \includegraphics[width=0.5\textwidth]{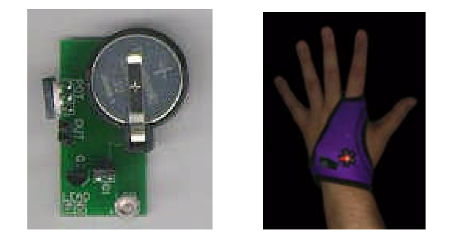}
    \caption{MIT Media Lab’s “emotional gloves” and their circuit boards}
\end{figure}

In the field of education, emotional resonance is also of great significance. Educational robots can sense students' emotional states through emotional computing technology, such as whether they feel confused, frustrated, excited, etc. during the learning process, and adjust teaching methods and content according to students' emotional states. When students encounter difficulties in learning and show frustrated emotions, educational robots can give students encouragement and comfort, help students overcome difficulties through more vivid and interesting teaching methods, and stimulate students' interest and enthusiasm in learning. This emotional resonance can enhance the emotional connection between students and educational robots, making students more willing to accept the help and guidance of educational robots and improve learning effects.

In the service industry, emotional resonance can also improve service quality. For example, in hotel services, service robots can perceive guests' needs and emotional states through emotional computing technology. When guests show tired emotions, robots can take the initiative to provide services such as luggage handling and room guidance, and communicate with guests in a warm and friendly language, making guests feel cared for and comfortable. This kind of emotional resonance can enhance guests' satisfaction and loyalty to hotel services and enhance the hotel's brand image and competitiveness.

The concept and goal of human-machine integration have drawn a beautiful blueprint for the harmonious coexistence and coordinated development of humans and machines in the future. By achieving goals such as dynamic adaptability, two-way learning mechanisms and emotional resonance, robots will be able to better integrate into human society, become effective assistants and partners in human life and work, and jointly promote the progress and development of human society. However, to achieve these goals, we still need to conduct in-depth research and exploration in many aspects such as technology, ethics, and society, and solve a series of challenges and problems, such as the improvement of human-machine interaction technology, the ethical norms of robots, and the social acceptance of human-machine integration. Only in this way can we truly realize the beautiful vision of human-machine integration and create a new era of harmonious coexistence between humans and machines.

%-----------------------------------------------------

\chapter{Scientific Basis of Affective Computing}

\section{Definition and Model of Affective Computing}
Affective computing is an important concept proposed by Rosalind Picard in 1997. Its core is to quantify and process emotional information through algorithms so that computer systems can recognize, understand and express human emotions. Research in this field aims to bridge the emotional gap between humans and machines, so that machines can interact with humans more naturally. With the continuous advancement of science and technology, affective computing has shown great application potential in many fields such as human-computer interaction, intelligent customer service, and mental health monitoring, and has become an important research direction in the field of artificial intelligence.

\subsection{Definition of Affective Computing}
The research content of affective computing covers many aspects of emotion, including the recognition, understanding and expression of emotion. Emotion recognition refers to the extraction of emotional features from human behavior and physiological data through various technical means, such as facial expression recognition, voice emotion recognition, physiological signal recognition, etc., to determine the individual's current emotional state. For example, by analyzing features such as the shape of eyebrows and the curvature of the mouth in facial expressions, it is possible to identify whether the individual is in an emotional state such as happiness, anger or sadness. Emotional understanding goes a step further and requires the computer system to understand the meaning of emotions and the reasons behind them, that is, to be able to infer the events or situations that trigger emotions based on the emotional state. For example, when a person is identified as being in an angry state, the system can analyze that it may be caused by frustration at work, interpersonal conflicts, etc. Emotional expression refers to the ability of computer systems to express emotions in a natural and human-understandable way, such as through voice intonation, facial expressions, body language, etc., to convey emotional information to humans. For example, when communicating with users, intelligent customer service systems can adjust their voice intonation according to the user's emotional state, express emotions such as sympathy, comfort or happiness, and enhance the user's emotional experience.

The realization of emotional computing requires the help of a variety of technical means, including machine learning, deep learning, computer vision, speech processing, physiological signal processing, etc. The development of these technologies provides strong support for emotional computing, which continuously improves the performance of emotional computing. For example, deep learning algorithms have achieved remarkable results in emotion recognition tasks, and can automatically learn emotional features from a large amount of data to improve the accuracy of emotion recognition. Computer vision technology can realize real-time capture and analysis of facial expressions and capture subtle changes in human emotional expression. Speech processing technology can extract emotional features in speech, such as intonation, speech speed, volume, etc., to realize speech emotion recognition. Physiological signal processing technology can monitor the physiological signals of the human body, such as heart rate, skin conductance, etc. These physiological signals are closely related to emotional state and provide important physiological basis for emotional computing.

\begin{figure}
    \centering
    \includegraphics[width=0.5\textwidth]{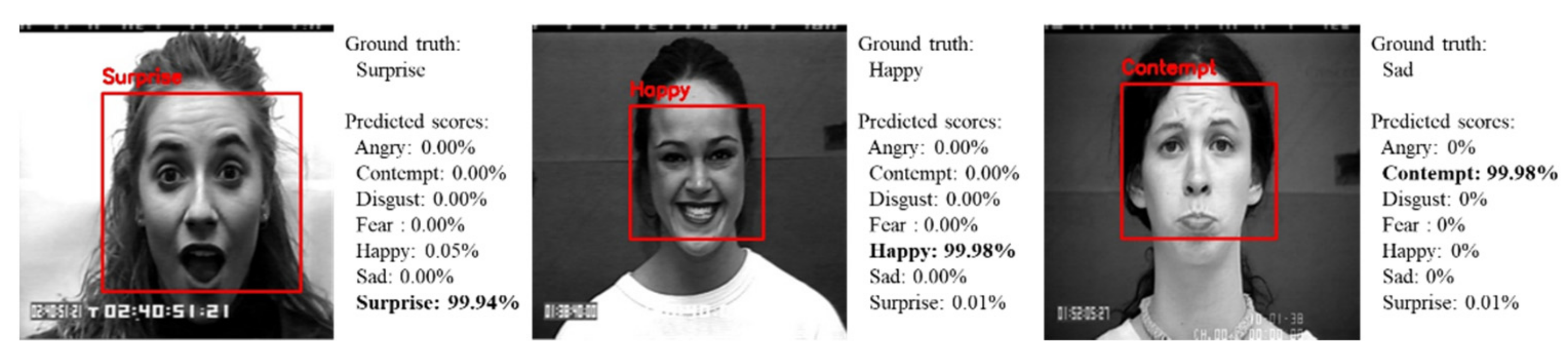}
    \caption{Classification results of emotion recognition based on facial key points through directed graph neural network}
\end{figure}

\subsection{Development of Affective Computing}
The development of affective computing can be traced back to the 1990s. With the continuous development of artificial intelligence and computer technology, affective computing has gradually become a research hotspot. In the early days, the research on affective computing focused on the recognition and expression of emotions, and realized basic emotional interactions through simple rules and models. For example, some early chatbots were able to judge the user's emotional state based on the keywords entered by the user and give corresponding emotional responses. However, these early affective computing systems were relatively simple in function, and the accuracy of emotion recognition and the naturalness of expression needed to be improved.

With the development of machine learning and deep learning technologies, the performance of emotional computing has been significantly improved. Machine learning algorithms can learn emotional features from large amounts of data, build emotional classification models, and improve the accuracy of emotion recognition. Deep learning algorithms further improve the performance of emotional computing, and can automatically extract high-level features from data to achieve more accurate emotion recognition and understanding. For example, convolutional neural networks (CNNs) have achieved remarkable results in facial expression recognition tasks, and can automatically extract features from facial images to achieve high-accuracy emotion recognition. Recurrent neural networks (RNNs) and their variants, such as long short-term memory networks (LSTMs), have performed well in speech emotion recognition and emotion sequence modeling, and can handle long-term dependencies in sequence data and improve the effect of emotion recognition.

In recent years, the research on affective computing has gradually developed towards multimodal affective computing, that is, integrating data from multiple modalities, such as facial expressions, voice, physiological signals, text, etc., to achieve more comprehensive and accurate emotion recognition and understanding. Multimodal affective computing can make full use of the advantages of different modal data to improve the performance of affective computing. For example, facial expressions can provide intuitive emotional information, voice can convey emotional characteristics such as tone and speech speed, physiological signals can reflect the physiological changes of emotions, and text can express the content and semantics of emotions. By integrating these multimodal data, affective computing systems can more accurately identify and understand human emotional states and achieve more natural and efficient human-computer interaction.

\begin{figure}
    \centering
    \includegraphics[width=0.5\textwidth]{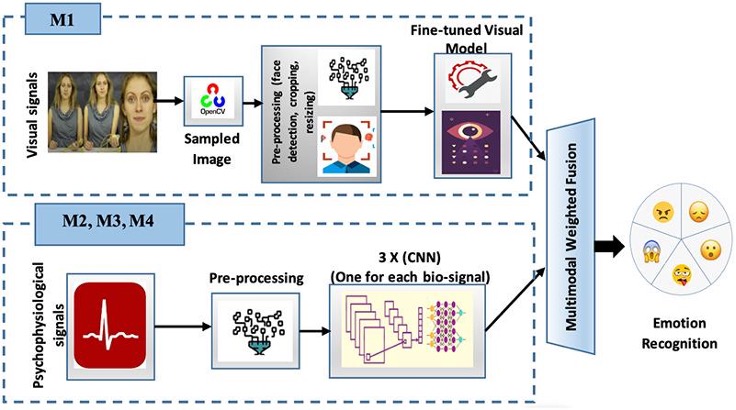}
    \caption{System framework of multimodal emotion recognition method (MEmoR) based on emotion biomarkers}
\end{figure}

The "Emotion Engine" framework proposed by Huawei in 2022 is an important progress in the field of emotional computing. The framework integrates multimodal data, including facial expressions, voice, physiological signals, etc., and achieves an emotional reasoning accuracy of 89.2\% through advanced algorithms and models. This achievement shows that through the integration of multimodal data and the application of deep learning technology, the performance of emotional computing has been significantly improved, providing strong support for the practical application of emotional computing.

The core of Huawei's "Emotion Engine" framework is multimodal data fusion and deep learning algorithms. Multimodal data fusion can fully utilize the advantages of different modal data to improve the accuracy and robustness of emotion recognition. For example, facial expressions can provide intuitive emotional information, voice can convey emotional characteristics such as tone and speed, and physiological signals can reflect the physiological changes of emotions. By fusing these multimodal data, the emotion engine can more comprehensively and accurately identify and understand the emotional state of human beings.

Deep learning algorithms play a key role in Huawei's "emotion engine" framework. By building deep neural network models, such as convolutional neural networks (CNNs) and recurrent neural networks (RNNs), the emotion engine can automatically learn emotional features from multimodal data to classify and predict emotional states. For example, in facial expression recognition, CNN can automatically extract features from facial images to achieve high-accuracy emotion recognition; in speech emotion recognition, RNN can process sequence information in speech signals to classify and predict emotional states.

Huawei's "emotion engine" framework has shown broad application prospects in multiple fields. In the field of intelligent customer service, the emotion engine can identify the user's emotional state in real time, adjust the customer service strategy according to the user's emotional state, provide more intimate and personalized services, and improve user satisfaction. In the field of mental health monitoring, the emotion engine can monitor the user's facial expressions, voice, physiological signals and other data to understand the user's mental health state in real time, discover psychological problems in a timely manner, and provide support for mental health intervention. In the field of human-computer interaction, the emotion engine can realize emotional interaction between humans and machines, enabling machines to better understand human emotional needs and provide a more natural and efficient human-computer interaction experience.

\subsection{Integration and Innovation of Affective Computing Models}

With the continuous development of affective computing technology, the integration and innovation of affective computing models have become a hot topic of research. The integration of multimodal affective computing models can make full use of the advantages of different modal data and improve the performance of affective computing. For example, the integration of multiple modal data such as facial expressions, voice, and physiological signals can achieve more accurate and comprehensive emotion recognition and understanding. The development of deep learning algorithms provides strong support for the integration of multimodal affective computing models. By constructing a deep neural network model, emotional features can be automatically learned from multimodal data to achieve the classification and prediction of emotional states.

In addition to the integration of multimodal data, the innovation of emotional computing models is also constantly advancing. For example, some studies have begun to explore the integration of emotional computing and cognitive computing, trying to build a more intelligent and more humane emotional computing model. Cognitive computing can simulate human cognitive processes, including perception, memory, thinking, decision-making, etc. Combining cognitive computing with emotional computing can enable emotional computing models to not only recognize and understand emotions, but also perform emotional reasoning and emotional decision-making, achieving more intelligent and more humane emotional interactions.

\begin{figure}
    \centering
    \includegraphics[width=0.5\textwidth]{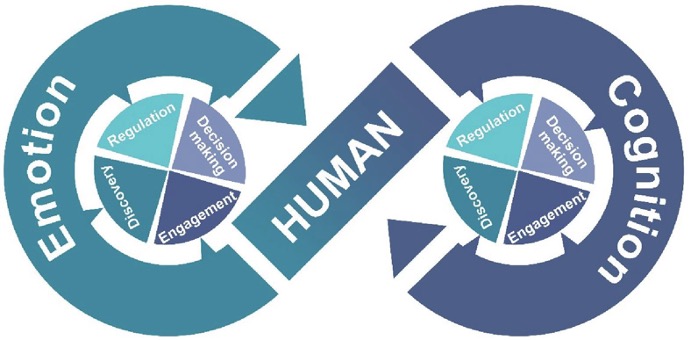}
    \caption{Interaction model of emotion and cognition}
\end{figure}

In addition, the interpretability and transparency of affective computing models have also become important research directions. With the deepening application of affective computing in various fields, the interpretability and transparency of affective computing models have become increasingly important. Interpretable affective computing models can explain the results and processes of affective computing to users, enhancing users' trust and acceptance of affective computing. For example, in the medical field, interpretable affective computing models can explain the results of affective computing to doctors, helping doctors better understand the emotional state of patients and make more accurate diagnosis and treatment decisions.

As an important branch of artificial intelligence, the development of affective computing is of great significance to the advancement and application of artificial intelligence technology. Through affective computing, computer systems can better understand human emotional needs, provide a more natural and efficient human-computer interaction experience, and bring more convenience and well-being to human life and work. In the future, with the continuous development of affective computing technology, affective computing will demonstrate its application value in more fields and become an important driving force for the development of artificial intelligence technology.

\section{Psychological Basis of Emotion}

\subsection{Discrete Emotion Theory}
The discrete emotion theory was proposed by the famous psychologist Paul Ekman and has a profound impact in the field of affective computing. Ekman believes that basic emotions are common to all humans and are universal across cultures. He classifies basic emotions into six categories: happiness, sadness, anger, surprise, disgust, and fear.

\begin{figure}
    \centering
    \includegraphics[width=0.5\textwidth]{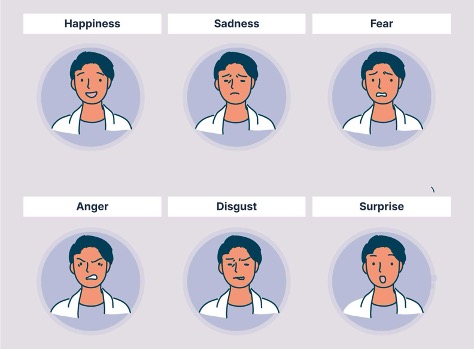}
    \caption{Illustration of Ekman's six basic emotion models}
\end{figure}

Happiness is a positive emotional experience, usually manifested by features such as relaxed facial muscles and upturned corners of the mouth. In social interactions, happy emotional expressions can promote connections and cooperation between people. For example, when people are at a party or celebration, happy expressions and laughter can create a relaxed and pleasant atmosphere and enhance the emotional bond between each other.

Sadness is a negative emotional experience, and its facial expression features are usually retracted eyebrows and downturned mouth corners. The emotional expression of sadness helps individuals gain sympathy and support from others. For example, when a person encounters setbacks or loses a loved one, sad expressions and crying can attract the attention and comfort of people around them, helping the individual to overcome difficulties.

Anger is a strong emotional response, and its facial expressions are characterized by frowning, glaring, pursed lips or clenched teeth, etc. The emotional expression of anger can, to a certain extent, protect the interests of individuals and maintain fairness and justice. For example, when faced with unfair treatment or infringement, angry expressions and words can convey dissatisfaction and warnings to the other party, prompting the other party to change their behavior.

Surprise is a short and intense emotional experience, and its facial expression features include wide eyes, raised eyebrows, and open mouth. The emotional expression of surprise can attract the attention and interest of others and promote the exchange and sharing of information. For example, when hearing surprising news, people will express their feelings through surprised expressions and words, thus attracting the attention and discussion of others.

Disgust is an emotional response to something unpleasant or offensive, with facial features such as wrinkled nose, pursed lips or raised upper lip. The emotional expression of disgust helps individuals avoid harmful or unpleasant things and protect their physical and mental health. For example, when faced with disgusting food or smell, disgust expressions and reactions can remind individuals to stay away from these things.

Fear is an emotional response to potential danger or threat, and its facial expression features include wide eyes, raised eyebrows, and slightly open mouth. The emotional expression of fear can arouse the vigilance and protection of others and help individuals deal with potential dangers. For example, when encountering dangerous situations, expressions of fear and screaming can attract the attention and help of others, increasing the individual's chance of survival.

These basic emotions are expressed in a variety of ways, such as facial expressions and voice intonation, providing a clear classification basis for emotion recognition in affective computing. By recognizing and understanding these basic emotions, computers can better perceive and respond to human emotional states, achieving more natural and efficient human-computer interaction.

\subsection{Dimensional Model}
The dimensional model was proposed by psychologist James Russell. Unlike discrete emotion theory, the dimensional model regards emotion as a continuous, multi-dimensional space. He constructed an emotion space coordinate system using valence and arousal.

\begin{figure}
    \centering
    \includegraphics[width=0.5\textwidth]{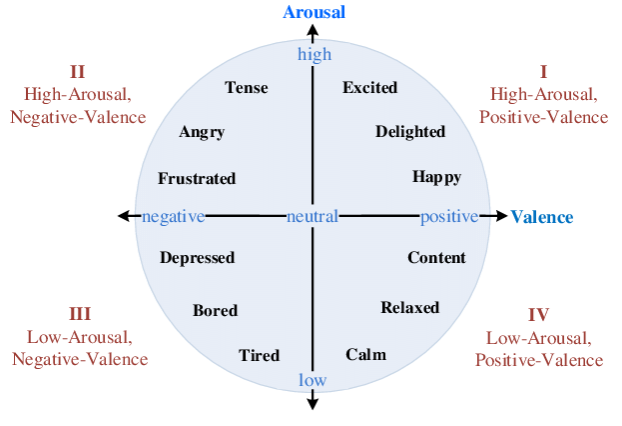}
    \caption{Two-dimensional Valence-Arousal model}
\end{figure}

Valence indicates the degree of positivity or negativity of an emotion, ranging from very pleasant to very unpleasant. Emotions with positive valence are usually associated with experiences such as pleasure, satisfaction, and happiness, while emotions with negative valence are associated with experiences such as pain, dissatisfaction, and anger. For example, when a person achieves success or reward, he or she will be in an emotional state with high positive valence; and when encountering failure or frustration, he or she will be in an emotional state with low negative valence.

Arousal indicates the level of emotional activation, ranging from calm to excitement. Low arousal emotional states are usually characterized by calmness, relaxation, sleepiness, etc., while high arousal emotional states are characterized by excitement, tension, excitement, etc. For example, a person would be in a low arousal emotional state when resting in a quiet environment, and in a high arousal emotional state when participating in intense sports or activities.

This model can describe and quantify emotional states in more detail, providing a more flexible and comprehensive framework for affective computing. By measuring and analyzing valence and arousal, computers can more accurately identify and understand human emotional states, thereby achieving more personalized and precise emotional interactions. For example, in an intelligent customer service system, by analyzing the user's emotional dimensions, the system can provide more intimate and appropriate services and improve user satisfaction.

\section{Neuroscience Basis of Emotion}

\subsection{Amygdala and emotional response}

In the field of neuroscience, studies have found that the amygdala plays a key role in emotional processing. Located in the medial temporal lobe of the brain, the amygdala is an almond-shaped neural nucleus that receives information from the senses and quickly activates emotion-related physiological responses, such as accelerated heartbeat and shortness of breath.

\begin{figure}
    \centering
    \includegraphics[width=0.5\textwidth]{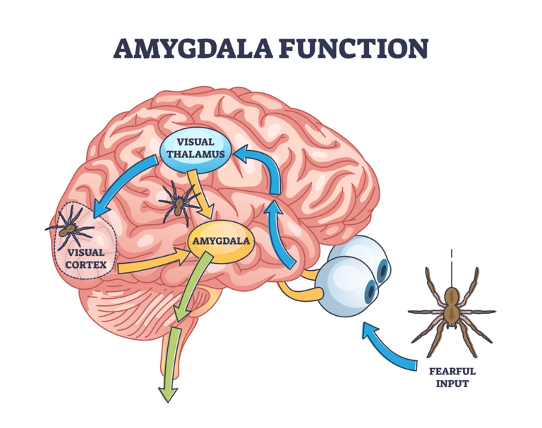}
    \caption{Amygdala function and response mechanism during threat perception}
\end{figure}

When an individual faces a potential threat or danger, the amygdala will react quickly and activate the "fight or flight" response. For example, in the event of a sudden dangerous situation, such as a wild animal attack or a traffic accident, the amygdala will quickly activate the body's stress response, causing the individual's heart rate to speed up, blood pressure to rise, and muscles to tense, so that the individual can quickly respond, such as running away or defending himself.

In addition, the amygdala is also involved in learning and remembering emotional stimuli. Through interactions with other areas of the brain, the amygdala is able to associate emotional experiences with specific stimuli or situations to form emotional memories. For example, when an individual experiences a traumatic event in childhood, such as being bitten by a dog, the amygdala will associate this fear emotion with the image of a dog, forming a fear memory. Later in life, when the individual sees a dog again, the amygdala will quickly activate the fear response, causing the individual to escape or avoid the dog.

\subsection{Prefrontal cortex and emotional cognition}
The prefrontal cortex is the area of the brain responsible for higher cognitive functions, including the evaluation, expression, and control of emotions. It is involved in the analysis and processing of emotional information, helping individuals understand the meaning and consequences of emotions and make appropriate emotional responses.

In terms of emotional assessment, the prefrontal cortex is able to perform complex analysis and judgment of emotional stimuli. For example, when an individual faces a complex social situation, such as a conflict with a colleague at work, the prefrontal cortex will assess the nature, severity, and possible consequences of the conflict, helping the individual to make appropriate emotional responses, such as anger, grievance, or understanding.

In terms of emotional expression, the prefrontal cortex controls the way and degree of emotional expression of an individual. It can regulate emotional expression signals such as facial expressions and voice intonation, allowing individuals to express their emotions appropriately according to different social situations and communication partners. For example, in formal business situations, the prefrontal cortex will inhibit individuals from expressing overly excited or emotional emotions, allowing individuals to maintain a calm and professional image; while in communication with close friends, the prefrontal cortex allows individuals to express their emotions more freely and authentically.

In terms of emotional control, the prefrontal cortex can help individuals regulate and control their emotional responses and avoid excessive or inappropriate emotional expression. For example, when an individual faces stress or frustration, the prefrontal cortex will activate the self-regulation mechanism to help the individual calm down, stay calm and rational, and thus better cope with the problem.

\subsection{Synergistic effect of amygdala and prefrontal cortex}
The interaction between the amygdala and the prefrontal cortex is crucial for human emotional experience and behavioral responses. Working together, they enable humans to have rich and diverse emotional experiences and to make appropriate emotional responses in different situations.

\begin{figure}
    \centering
    \includegraphics[width=0.3\textwidth]{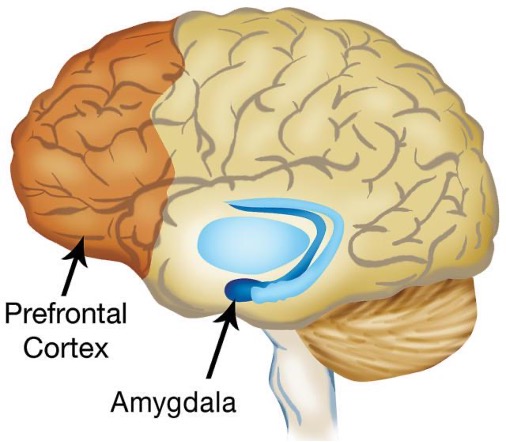}
    \caption{Location of the prefrontal cortex and amygdala in the brain}
\end{figure}

The amygdala plays a leading role in rapid emotional reactions. When an individual faces sudden danger or threat, the amygdala will quickly activate emotion-related physiological reactions, enabling the individual to respond in a timely manner. For example, when encountering a sudden dangerous situation, an individual may have a fear reaction in an instant, with an accelerated heartbeat and rapid breathing, which is the result of the amygdala's rapid reaction.

The prefrontal cortex plays an important role in the processing and regulation of complex emotions. It can deeply analyze and process emotional information, help individuals understand the meaning and consequences of emotions, and make appropriate emotional responses. For example, when facing complex interpersonal relationships or social situations, individuals need to use the cognitive ability of the prefrontal cortex to evaluate the nature and impact of emotions, regulate their own emotional expression and behavioral responses, and maintain good interpersonal relationships and social order.

This synergy provides important biological inspiration for the design of affective computing models. The affective computing model can draw on the collaborative working mechanism of the amygdala and prefrontal cortex in the human brain to design algorithms and models that can quickly respond to and complexly process emotional information. For example, in intelligent security systems, a rapid response mechanism similar to the amygdala can be used to monitor and warn potential dangerous situations in real time; at the same time, combined with the cognitive and regulatory capabilities of the prefrontal cortex, complex social situations and emotional information can be analyzed and processed to improve the intelligence level and accuracy of the system.

The psychological and neuroscience foundations of emotion provide rich theoretical support and practical guidance for affective computing. Through in-depth research and application of these basic theories, affective computing can better simulate the human emotion processing mechanism, achieve more natural and efficient human-computer emotional interaction, and bring more convenience and well-being to human life and work.

\section{Mechanisms of Emotion Recognition and Expression}
Emotion recognition and expression are two important links in affective computing, which together constitute the core mechanism of affective computing. Through emotion recognition, computer systems can perceive and understand human emotional states; through emotion expression, computer systems can convey emotional information in a way that humans can understand, thus achieving emotional interaction between humans and machines.

\subsection{Emotion Recognition Mechanism}

\subsubsection{Facial Expression Recognition}
Facial expression is one of the important ways for humans to express emotions. Through the movement and changes of facial muscles, rich emotional information can be conveyed. Facial expression recognition technology aims to identify the emotional state of an individual by analyzing facial images or video data.The changes in facial expressions not only involve the basic recognition of emotions (such as anger, joy, sadness, etc.), but also include the intensity and subtle changes of emotions (such as a slight smile or a tight mouth corner). These subtle changes in expression can usually reveal the individual's psychological state in a specific situation and can more accurately understand the user's emotional response.

Facial key point extraction

Facial key point extraction is a basic step in facial expression recognition, and its purpose is to locate the key feature points of the face, such as the position and shape of the eyes, eyebrows, nose, mouth, etc. These key points contain rich facial expression information, and by extracting and analyzing these key points, subtle changes in facial expressions can be captured.

For example, when a person smiles, the corners of the mouth will rise and wrinkles will appear at the corners of the eyes; when a person is angry, the eyebrows will be wrinkled and the corners of the mouth will be closed. By extracting the position and changes of these key points, the emotional state of the individual can be preliminarily judged.

Modern facial expression recognition systems are not limited to 2D images, but also use deep learning technology combined with 3D models to accurately recognize facial expressions under complex lighting and angle conditions.The 3D-CNN (Convolutional Neural Network) algorithm is a convolutional neural network algorithm suitable for processing three-dimensional data. It can automatically extract features from facial images or videos and perform emotion classification. Through structures such as convolutional layers, pooling layers, and fully connected layers, the 3D-CNN algorithm can effectively capture the temporal and spatial characteristics of facial expressions and improve the accuracy of emotion recognition.

For example, the OpenFace toolkit detects key points of the face, such as the position and movement of the eyes, eyebrows, and mouth, and analyzes and models these key points using the 3D-CNN algorithm to identify different emotional expressions. This method can accurately capture subtle changes in facial expressions and has a high degree of accuracy in identifying emotions.

\begin{figure}
    \centering
    \includegraphics[width=0.5\textwidth]{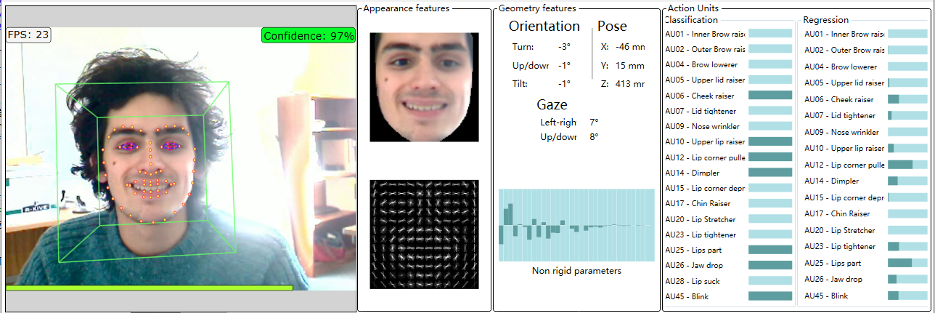}
    \caption{ OpenFace detects key points of the face}
\end{figure}

Facial expression recognition technology has broad application prospects in many fields. In the field of human-computer interaction, through facial expression recognition, the computer system can perceive the user's emotional state in real time, and adjust the interaction method according to the user's emotional state, providing a more natural and efficient human-computer interaction experience. For example, in an intelligent customer service system, through facial expression recognition, the system can judge the user's emotional state, such as anger, dissatisfaction, etc., and adjust the service strategy in time to improve user satisfaction.

In the field of mental health monitoring, facial expression recognition technology can be used to monitor the mental health status of individuals. For example, by analyzing changes in facial expressions, emotional abnormalities in patients with mental illnesses such as depression and anxiety can be discovered in a timely manner, providing support for mental health intervention.

\subsubsection{Speech Emotion Recognition}
Speech is another important way for humans to express emotions. Speech characteristics such as intonation, speaking speed, and volume are closely related to emotional states. Speech emotion recognition technology aims to identify an individual's emotional state by analyzing speech signals.

MFCC (Mel-Frequency Cepstral Coefficient) is a parameter that can better reflect the spectral characteristics of speech signals and can capture the emotional information in speech. MFCC features simulate the perceptual characteristics of the human auditory system and convert speech signals into a series of characteristic coefficients. These characteristic coefficients contain information such as the frequency and amplitude of speech and can effectively reflect the emotional characteristics of speech.In addition to MFCC, modern speech emotion recognition has also introduced more audio processing technologies, such as linear predictive coding (LPC) and speech enhancement technology, which can improve the recognition accuracy in noisy environments and thus improve the robustness of emotion recognition.

\begin{figure}
    \centering
    \includegraphics[width=0.5\textwidth]{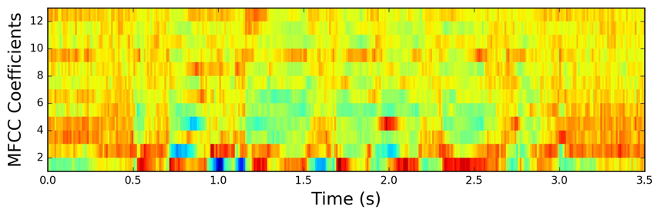}
    \caption{MFCC (Mel-Frequency Cepstral Coefficient) graph}
\end{figure}

For example, when a person is in a happy emotional state, the voice is usually higher in pitch, faster in speed, and louder in volume; and when a person is in a sad emotional state, the voice is usually lower in pitch, slower in speed, and lower in volume. By extracting MFCC features, these emotional features can be captured, providing a basis for speech emotion recognition.

LSTM (Long Short-Term Memory) network is a special neural network that can process long-term dependencies in sequence data and is suitable for speech emotion classification tasks. By introducing memory units and gating mechanisms, LSTM network can effectively solve the gradient vanishing and gradient exploding problems of traditional neural networks when processing sequence data, and improve the accuracy of emotion classification.

\begin{figure}
    \centering
    \includegraphics[width=0.5\textwidth]{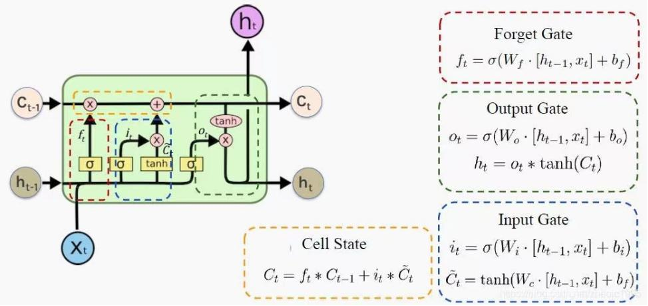}
    \caption{LSTM architecture diagram}
\end{figure}

For example, by inputting MFCC features into the LSTM network for training and classification, accurate recognition of speech emotions can be achieved. The LSTM network can automatically learn the emotional features in speech signals and classify emotions based on these features, such as happiness, sadness, anger, etc.Furthermore, LSTM networks can also be combined with other deep learning architectures such as convolutional neural networks (CNNs) to form hybrid models to improve the performance of speech emotion recognition, especially the recognition accuracy in complex environments.

Speech emotion recognition technology has important application value in many fields. In the field of intelligent customer service, through speech emotion recognition, the system can perceive the user's emotional state in real time, and adjust the service strategy according to the user's emotional state to provide more intimate and personalized services. For example, when the user shows anger or dissatisfaction, the system can adjust the service attitude in time, appease the user's emotions, and improve user satisfaction.In addition, in applications such as smart homes and in-car voice assistants, speech emotion recognition not only helps improve user experience, but also enhances the system's ability to respond to complex situations. For example, in an in-car voice assistant, when the driver feels tired or anxious, the system can provide help, such as playing relaxing music or suggesting a break.

In the field of human-computer interaction, speech emotion recognition technology can realize emotional interaction between humans and machines, enabling machines to better understand human emotional needs. For example, in voice assistant applications, through speech emotion recognition, voice assistants can provide corresponding emotional responses based on the user's emotional state, such as comfort, encouragement or congratulations, to enhance the user's emotional experience.

\subsubsection{Physiological Signal Recognition}

hysiological signals are an objective reflection of human emotional states. For example, there is a close mapping relationship between physiological signals such as electrocardiogram (ECG) and galvanic skin response (GSR) and emotional states. Physiological signal recognition technology aims to identify an individual's emotional state by collecting and analyzing physiological signals.

\begin{figure}
    \centering
    \includegraphics[width=0.5\textwidth]{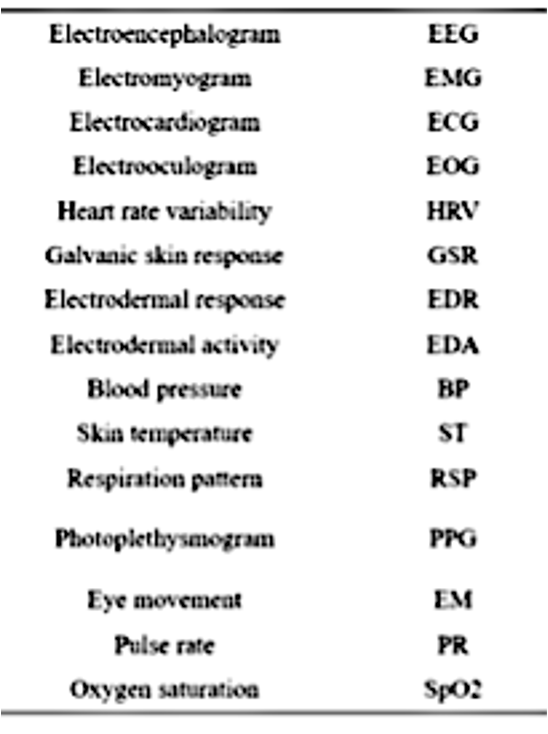}
    \caption{Physiological signal recognition process}
\end{figure}

ECG signals reflect the electrical activity of the heart and are closely related to emotional states. When an individual is in different emotional states, the ECG signal will show different characteristics. For example, when an individual is in an emotional state such as tension or anxiety, the ECG signal will show characteristics such as faster heart rate and reduced heart rate variability; while when an individual is in an emotional state such as relaxation or calmness, the ECG signal will show characteristics such as slower heart rate and increased heart rate variability.

The GSR signal reflects the change in skin conductivity and is closely related to the emotional state. When an individual is in different emotional states, the GSR signal will show different characteristics. For example, when an individual is in an emotional state such as tension or anxiety, the GSR signal will show characteristics such as increased skin conductivity; and when an individual is in an emotional state such as relaxation or calmness, the GSR signal will show characteristics such as decreased skin conductivity.

Physiological signal recognition technology has important application prospects in many fields. In the field of mental health monitoring, by collecting and analyzing ECG and GSR signals, it is possible to monitor the mental health status of individuals in real time, detect psychological problems in a timely manner, and provide support for mental health intervention. For example, through physiological signal recognition technology, the emotional state of patients with mental illnesses such as depression and anxiety can be monitored, providing important diagnostic basis for doctors.

In the field of human-computer interaction, physiological signal recognition technology can realize emotional interaction between humans and machines, enabling machines to better understand human emotional needs. For example, in intelligent security systems, through physiological signal recognition technology, the system can monitor the emotional state of security personnel in real time. When security personnel are in an emotional state such as tension or fatigue, the system can issue an alarm in time to remind security personnel to pay attention to safety.

\subsection{The Mechanism of Emotional Expression}
The mechanism of emotion expression is an important part of affective computing, which aims to express emotions through robots or computer systems and realize emotional interaction between humans and machines. Emotional expression can make robots communicate with humans more naturally and vividly, and enhance the emotional connection between humans and machines. The mechanism of emotion expression mainly includes two aspects: robot expression simulation and emotional speech synthesis.

\subsubsection{Robot Expression Simulation}

Robot expression simulation technology conveys emotional information through the robot's facial expressions, realizing emotional interaction between humans and machines. Pneumatic artificial muscle technology is an advanced robot expression simulation technology that realizes the changes in the robot's facial expressions through pneumatic artificial muscles. Pneumatic artificial muscles are soft and flexible, and can simulate the movement of human facial muscles, allowing robots to produce rich and diverse expression changes. For example, the "Geminoid" robot of Hiroshi Ishiguro's laboratory in Japan realizes micro-expression simulation through pneumatic artificial muscles, which can reproduce human facial expressions more realistically. This type of robot has broad application prospects in the fields of human-computer interaction, emotional companionship, etc. Through pneumatic artificial muscle technology, robots can realize a variety of expressions such as smiling, frowning, and surprise, allowing robots to express emotions more naturally and vividly.

\begin{figure}
    \centering
    \includegraphics[width=0.5\textwidth]{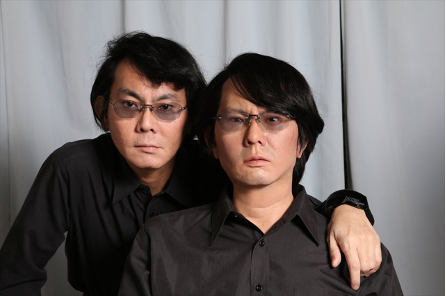}
    \caption{“Geminoid” robot}
\end{figure}
Robot expression simulation technology has important application value in many fields. In the field of human-computer interaction, through robot expression simulation, robots can express emotions more naturally and vividly, and enhance the emotional connection between humans and machines. For example, in the field of education, educational robots can interact with students emotionally through expression simulation to improve students' interest and enthusiasm in learning. Educational robots can encourage students through expressions such as smiles and nods, and enhance students' motivation to learn.

In the field of emotional companionship, robot expression simulation technology can provide emotional companionship for groups such as the elderly and children. For example, emotional companion robots can interact with users through expression simulation, alleviate users' loneliness and anxiety, and improve their quality of life. Emotional companion robots can interact with users through expressions such as smiling and blinking to enhance users' emotional experience.

In addition, robot expression simulation technology also has important applications in the field of medical rehabilitation. For example, rehabilitation robots can interact with patients through expression simulation, help patients relieve psychological pressure during rehabilitation and improve rehabilitation effects. Rehabilitation robots can enhance patients' confidence in rehabilitation through encouraging expressions and language.

\subsubsection{Avatar Expression Generation}
Avatar expression generation technology aims to give virtual characters (Avatars) rich, natural and infectious expressions through advanced computer graphics and artificial intelligence algorithms. The core of this technology is to map the captured human expression data or expression parameters generated by the algorithm to the facial model of the virtual character, so as to achieve real-time driving and generation of expressions. Its key technologies include facial feature extraction, expression parameterization, expression animation generation, etc. Facial feature extraction technology can accurately locate and track key points of the face from images or videos, such as the corners of the eyes, the corners of the mouth, eyebrows, etc. The movement and changes of these key points contain rich emotional information. Expression parameterization converts the movement information of these key points into quantifiable parameters, such as the degree of stretching of facial muscles, the upward angle of the corners of the mouth, etc. These parameters can be recognized and processed by computer systems. Expression animation generation technology uses these parameters to drive the facial model of the virtual character and generate corresponding expression animations. For example, when the parameter indicates that the corners of the mouth are raised, the corners of the mouth of the virtual character will rise accordingly, showing a smiling expression. Through the collaborative work of these technologies, Avatar expression generation technology can make the expressions of virtual characters more realistic and vivid, and enhance their interactivity and immersion in the virtual environment.

\begin{figure}
    \centering
    \includegraphics[width=0.5\textwidth]{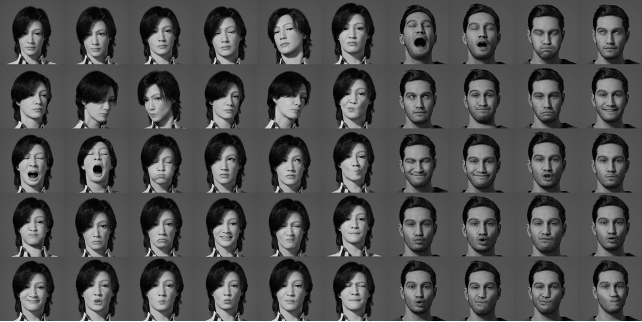}
    \caption{Avatar expression generation}
\end{figure}

Currently, the most widely used Avatar expression generation technologies include ZEGO Avatar SDK, PhotoCore SDK, and Avatarify:

ZEGO Avatar SDK is a powerful avatar expression generation tool that provides two ways to drive the expression and mouth shape of virtual people in real time: voice drive and expression follow-up. In terms of voice drive, ZEGO Avatar SDK captures and analyzes the user's voice information, and uses deep learning algorithms to convert voice signals into virtual people's mouth shape animation. For example, in online education scenarios, virtual teachers can use voice drive to adjust their mouth shape and expression in real time according to the content of the explanation, making the teaching process more vivid and interesting, and improving students' learning interest and attention. In terms of expression follow-up, ZEGO Avatar SDK uses a camera to capture the user's facial expression, and uses face recognition and expression recognition technology to map the user's expression to the virtual character in real time. For example, in social live broadcasts, the anchor can use expression follow-up to let the virtual image imitate his or her own expression in real time, enhance the interactivity and affinity with the audience, and improve the fun and viewing of the live broadcast.

\begin{figure}
    \centering
    \includegraphics[width=0.5\textwidth]{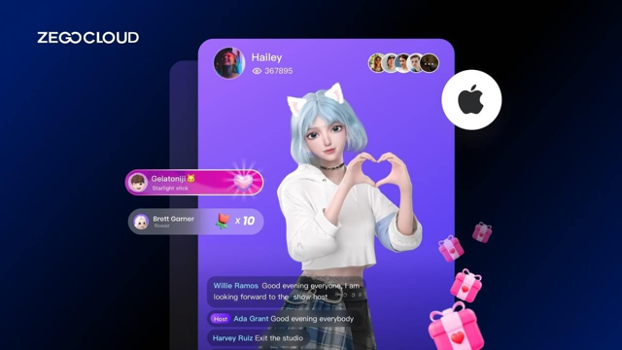}
    \caption{Creating a virtual human using ZEGO Avatar SDK}
\end{figure}
Xiangxin SDK is another advanced Avatar expression generation tool, which provides two ways to drive the expressions of virtual people: expression capture and voice drive. In terms of expression capture, Xiangxin SDK uses the camera to capture the user's facial expressions, and generates corresponding expression animations by detecting and analyzing the user's facial feature points. For example, in the virtual makeup trial application, users can use expression capture technology to see the effects of different makeup attempts on the virtual image in real time, helping users to better choose the makeup style that suits them. In terms of voice drive, Xiangxin SDK can generate expressions and lip animations that match the voice content based on the user's voice input. For example, in the intelligent customer service system, the virtual customer service can adjust the expression and lip shape in real time according to the user's voice problems to provide more intimate and personalized services.

\begin{figure}
    \centering
    \includegraphics[width=0.5\textwidth]{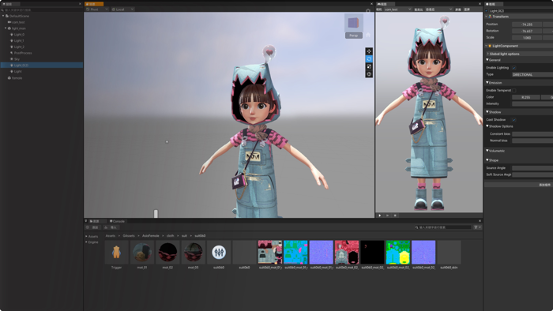}
    \caption{Avatarify real-time facial animation generation tool}
\end{figure}

Avatarify is a real-time facial animation generation tool based on deep learning, which allows users to map their own or other people's facial expressions to virtual characters, static images or videos in real time. Users can capture their facial expressions through the camera, and then select any static image or virtual character, and Avatarify will synchronize the user's facial expressions to the selected virtual avatar in real time. For example, in a virtual meeting, users can use Avatarify to map their expressions to virtual characters to increase the fun and interactivity of the meeting. In addition, Avatarify also supports compatibility with common virtual meeting software and live broadcast platforms. Users can use Avatarify on these platforms to convert their facial expressions in real time and generate high-quality virtual image videos.

\begin{figure}
    \centering
    \includegraphics[width=0.5\textwidth]{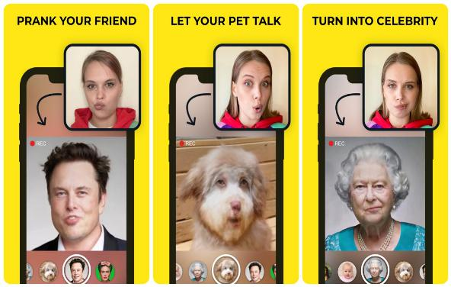}
    \caption{Avatarify real-time facial animation generation tool}
\end{figure}

\subsubsection{Emotional Speech Synthesis}

Emotional speech synthesis technology generates emotional speech through speech synthesis to achieve emotional interaction between humans and machines. The WaveNet algorithm is a speech synthesis model based on deep learning, which can learn the temporal characteristics and emotional characteristics of speech signals and generate emotional speech. The WaveNet algorithm can effectively improve the naturalness and emotional expression ability of speech synthesis by introducing technologies such as attention mechanism and generative adversarial network. For example, using the WaveNet algorithm to generate natural intonation is an advanced technology for emotional speech synthesis. Through the WaveNet algorithm, speech with emotional colors, such as happiness, sadness, anger, etc., can be generated, making the robot's speech more natural and vivid. The WaveNet algorithm can capture the emotional characteristics in speech and generate speech intonation that matches the emotional state.

Emotional speech synthesis technology has important application value in many fields. In the field of human-computer interaction, emotional speech synthesis can make the robot's voice more natural and vivid, and enhance the emotional connection between humans and machines. For example, in the intelligent customer service system, through emotional speech synthesis, the system can adjust the voice tone according to the user's emotional state and provide more intimate and personalized services. When the user shows anger or dissatisfaction, the system can respond to the user with a soothing tone to ease the user's emotions. In the field of education, emotional speech synthesis technology can be used for speech synthesis of educational robots. For example, educational robots can interact with students emotionally through emotional speech synthesis to improve students' learning interest and enthusiasm. Educational robots can praise students' progress with encouraging tones to enhance students' learning motivation. In the medical field, emotional speech synthesis technology can be used for speech synthesis of medical robots. For example, medical robots can interact with patients emotionally through emotional speech synthesis to relieve patients' anxiety and tension and improve patients' treatment effects. Medical robots can comfort patients with gentle tones to enhance patients' confidence in treatment. In addition, emotional speech synthesis technology also has important applications in the entertainment field. For example, virtual idols can interact with fans emotionally through emotional speech synthesis to enhance fans' emotional experience. Virtual idols can interact with fans in vivid tones, enhancing fans' sense of participation and loyalty.

% ---------------------------------------------------------

\chapter{Visual Interaction Technology}

\section{Principles of visual perception}
Vision is an extremely amazing ability that supports many behaviors and cognitive activities of organisms. About a century ago, Ramon y Cajal used the Golgi method to conduct the first in-depth study of the retina. Because the retina has a well-defined and orderly structure and the direction of flow of nerve signals is easy to identify, Cajal was able to come up with important conclusions about the basic organizational principles of the nervous system. These discoveries directly promoted his establishment of the neuron theory and the theory of "dynamic polarization" of nerve cells, laying the foundation for our understanding of the modern neuronal information processing function.

Over the past century, the retina, as an extension of the brain, has remained a focus of visual research. Scientists have conducted extensive research on its anatomical structure and functional properties, as well as its role in visual data processing. Most early studies focused on the compound eyes of invertebrates because of technical challenges in making similar measurements on vertebrate eyes. However, in recent years, important progress has been made in the study of animals such as cats and monkeys, revealing many information processing mechanisms that are common across species. A deep understanding of the basic operating principles of the nervous system is crucial for future research in neurobionics and visual simulation.¬¬

\begin{figure}
    \centering
    \includegraphics[width=0.5\textwidth]{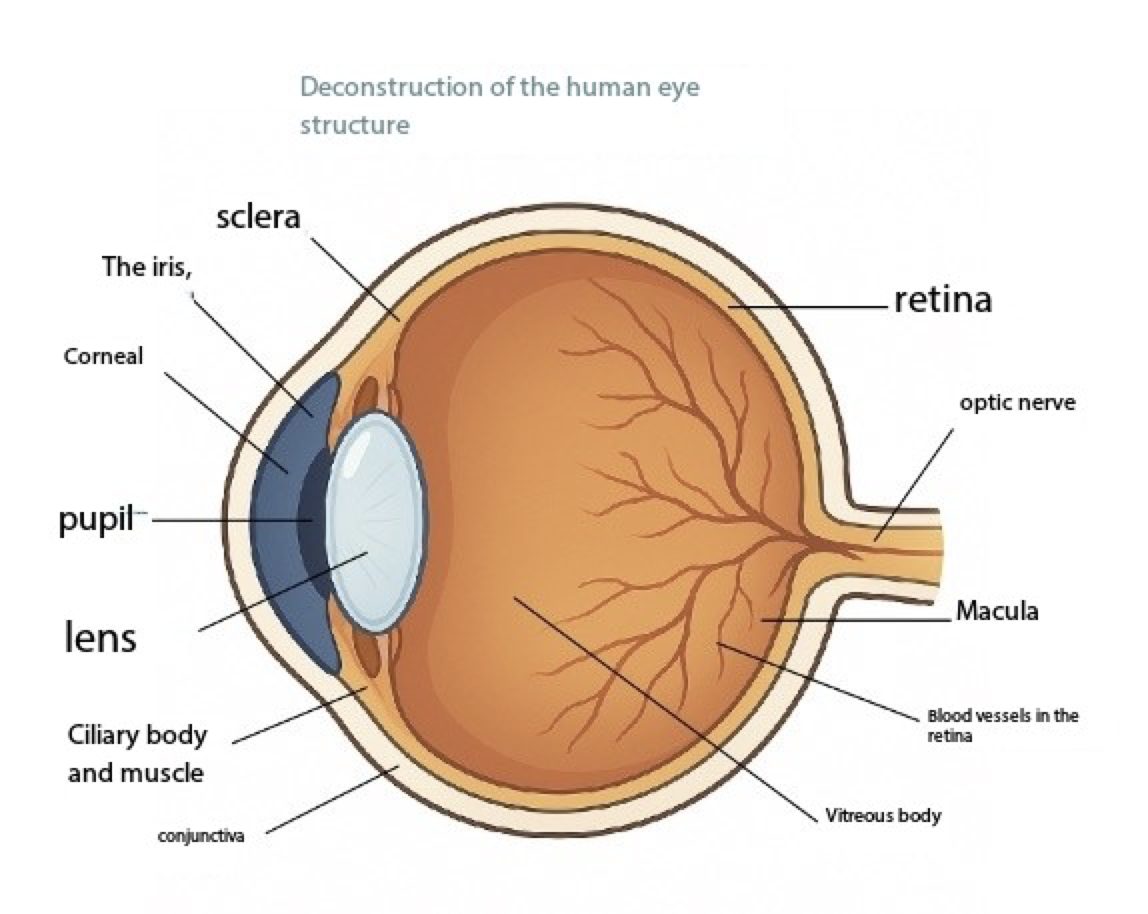}
    \caption{Structure of the human eye}
\end{figure}

In 1932, Hartline and Graham made a groundbreaking discovery in their study of the photoreceptors of the horseshoe crab's compound eyes. They observed that the output signal of the photoreceptors was linearly related to the logarithm of the incident light intensity. This logarithmic compression mechanism allows horseshoe crabs to perceive visually when the input light intensity spans 6 to 7 orders of magnitude, while the dynamic range of their neurons only covers 2 to 3 orders of magnitude. In addition, the output activity of the photoreceptors gradually adapts from the initial peak of the stimulus to a lower stable value, and the final output value of this process is also related to the logarithm of the light intensity. Due to this property, slowly changing light in the environment, such as the alternation of day and night, does not significantly affect visual perception. Similar studies in 1956 further revealed the processing mechanism of spatial information in the visual system. Scientists found that the retina can effectively ignore slow spatial changes in light patterns. This function is mainly achieved by two neural mechanisms: mutual lateral inhibition and self-inhibition. Lateral inhibition enhances edges and contrast through spatial high-pass filtering, while self-inhibition suppresses slow changes in light intensity in the environment through temporal high-pass filtering. These mechanisms ensure that the visual system can prioritize attention to dynamic and salient features in a scene.

Compared to the compound eyes of invertebrates, the arrangement of neurons in the vertebrate retina is more complex. The retina is usually composed of three layers of neural cell bodies, each separated by two layers of synaptic networks. The outermost layer contains photoreceptor cells, including rods and cones. Rods are mainly responsible for night vision in low light conditions, while cones provide color vision and detailed vision in bright environments. The middle layer includes bipolar cells, horizontal cells, and Aklin cells. Bipolar cells connect photoreceptors to the ganglion cells of the retina and transmit light-sensitive information to the latter. Horizontal cells and Aklin cells form lateral connections and are responsible for regulating and integrating local neural activity. Ganglion cells are the output units of the retina. Their axons pass through the inner surface of the retina and converge at the optic disc to form the optic nerve, which transmits signals to the lateral geniculate body (LGN) of the brain for further processing.

The number and distribution of visual receptors play a key role in visual perception. The human eye has about 120 million rods and 6.5 million cones. Rods are extremely sensitive to photons, allowing us to perceive the environment in low light conditions, while cones provide the ability to recognize colors and details under normal lighting. The high concentration of cones in the fovea gives the human eye extremely high visual acuity at the point of fixation. Recent studies have further revealed the functional differentiation between the fovea and the peripheral retina. The fovea focuses on high-resolution color vision and fine vision, while the peripheral retina plays an important role in motion detection and visual perception in low light conditions. By combining visual psychophysics experiments and neurophysiological studies, scientists continue to deepen their understanding of the retina's perception mechanism and the brain's way of processing visual information. Such research not only deepens the understanding of the visual system, but also provides valuable inspiration for the design of artificial visual systems.

Diffraction in the pupil and aberrations in the eye's optics cause blurring of the retinal image. In the foveal region, these blurred images are sampled by a set of cones that form a precise triangular grid. The inner segments of the cones (i.e., the active sensing elements) occupy most of the space between them, a property that has a dual effect: on the one hand, it maximizes quantum capture and thus reduces photon noise; on the other hand, the integration of light over the cone aperture reduces the spatial frequency response. Under normal vision conditions, moiré fringes are not evident in the foveal region, indicating that the eye's optical system effectively filters out high spatial frequencies in the retinal image to avoid image distortion. Diffraction and aberrations, rather than receptor density, thus become the main limiting factors for spatial resolution. The morphology of the eye strikes a balance between reducing diffraction but increasing aberrations at larger pupil diameters, or reducing aberrations but being limited by diffraction at smaller pupil diameters. Interestingly, it is actually the size and spacing of the cones that may ultimately determine the spatial resolution. The optical system of the eye appears to have evolved to not significantly interfere with the resolution achievable with the smallest cone spacing. Even in very small eyes, cone spacings less than 2 micrometers are rare. Cone diameters show remarkable consistency across eye sizes, likely to prevent optical crosstalk between receptor outer segments. As short optical waveguides, cone outer segments have limited photon retention capabilities, and if cone spacing is too small, captured photons may trigger photopolymerization in adjacent cones, affecting detection accuracy. Therefore, eye structures, including pupil diameter and optical quality, may be more about ensuring optical isolation of cone outer segments rather than limiting cone spacing due to poor optical quality. The matching relationship between the resolution limit set by cone spacing and the highest frequency passed by the eye's optical system has long been verified. Recent studies have further revealed which factor, between eye optics and cone separation, plays a more significant role in determining actual spatial resolution. This finding provides a new perspective for understanding the biological significance of spatial resolution.

Whatever the limiting factors of the eye’s resolution, it is certain that evolutionary pressures have driven primates to develop a visual system that operates near physical and biochemical limits. With the exception of some birds of prey such as eagles, few organisms have eyes with higher resolution. Remarkably, the compound eyes of many insects also show similar capabilities in terms of detection resolution. This suggests that even organisms with relatively simple nervous systems can gain significant survival advantages from the large amount of data and interactive information brought by high-resolution visual systems. In addition, the optical quality of a single-chamber eye is generally lower than that of a single-lens camera. In such an optical system, the distance between receptors is pushed to its biochemical limit, requiring not only powerful data processing capabilities of the retina and brain, but also the extraction of effective information from images with limited optical and geometric quality. This highlights the adaptability of the nervous system. Although the retina shows a high degree of specialization in visual processing, many information processing modules of the cerebral cortex still show a high degree of flexibility and follow a few general information processing principles. Existing neural architectures mainly determine the rough mapping between different brain regions, but these mappings can be reshaped, thereby further revealing the universal principles of neural information processing.

When exploring the impact of evolution on the visual system, we should be cautious in using evolution as a theoretical basis. As Crick pointed out, we should not over-speculate on the specific constraints imposed by evolution unless in a broad context. At the same time, we should avoid assuming that evolution will always choose mathematically optimal solutions. By directly experimentally verifying these hypotheses, researchers can more accurately understand which mechanisms are possible and which are not. Finally, further experiments have shown that the limitations of neural processing on visual resolution are far less significant than optical blur. For example, by directly constructing interference fringes on the retina to measure the contrast sensitivity of the fovea, the results showed that under optimal conditions, the degree of neural blur is comparable to optical blur. The experiment showed that observers only needed 8\% contrast to detect interference fringes with a spatial frequency of 60c/°. This result supports the ability of the retina to perform complex data compression and lateral interactions, and also shows that the retina has not yet reached the limit of its neural processing capabilities. This also further verifies the hypothesis that the center of the receptive field of some ganglion cells may be supplied by a single cone cell, providing a new perspective for understanding the fine processing capabilities of the retina.

In the 1960s and 1970s, the study of vertebrate retinas was mainly focused on cats. By the end of the 1970s, scientists had a relatively in-depth understanding of the overall structure of the cat retina and began to apply these research results to computational models of computer vision. The importance of this development stage lies in that it provides a biological basis for visual computing, allowing the design of artificial visual systems to benefit from the optimization mechanism of the natural visual system. The study of the cat retina has revealed its unique structural and functional characteristics. As the first information processing checkpoint of the visual system, the retina undertakes the initial conversion from light signals to neural signals. The regulation of light adaptation and dark adaptation capabilities enables the retina to work effectively in different lighting environments. Rods and cones are responsible for night and daytime visual functions, respectively, and transmit information through three independent neural pathways to adapt to changes from strong light to dim light. This information transmission pathway can be described by the receptive field characteristics of bipolar cells. In the dark-adapted retina, the receptive field of bipolar cells is mainly composed of an excitatory center, while in the light-adapted environment, the receptive field presents a concentric structure, including an excitatory center and an inhibitory surrounding area, which enhances the detection ability of local contrast.

Studies have shown that cats' B-type horizontal cells are mainly responsible for the formation of the center of the bipolar cell receptive field, while the larger A-type horizontal cells jointly determine the surrounding structure of the receptive field through electronic interactions. This mechanism ensures that the retina can effectively process light signals and optimize the transmission of visual information. Bipolar cells take on the role of linear integration of light signals in this process, and as the information enters the ganglion cell layer, the processing becomes more complicated. Studies on CB bj bipolar cells have shown that they show opposite polarity responses to light and dark stripes in the center of the receptive field, while showing the same polarity in response to light and dark stripes around the receptive field. This characteristic indicates that there are significant differences in the information processing mode of the retina under light adaptation and dark adaptation conditions, and the visual system will adjust its own operating mechanism under different lighting environments to optimize the effect of information acquisition.

Ganglion cells are key nodes for retinal information transmission. In the cat retina, ganglion cells can be divided into two major categories: Alpha cells and Beta cells. Alpha cells have large cell bodies and sparsely branched dendrites, forming a wide receptive field, while Beta cells have medium-sized cell bodies, small dendrites but dense branches, and a relatively small receptive field range. The difference between X-type cells (Beta cells) and Y-type cells (Alpha cells) lies in the way they integrate input signals. X/Beta cells mainly collect synaptic inputs from a small number of cone bipolar cells, which have large receptive field overlaps, while Y/Alpha cells receive inputs from a large number of cone bipolar cells, but have less receptive field overlaps. These connection patterns determine the differences in signal processing between the two types of ganglion cells. The former is mainly responsible for fine visual processing, while the latter is good at transmitting dynamic information.

\begin{figure}
    \centering
    \includegraphics[width=0.5\textwidth]{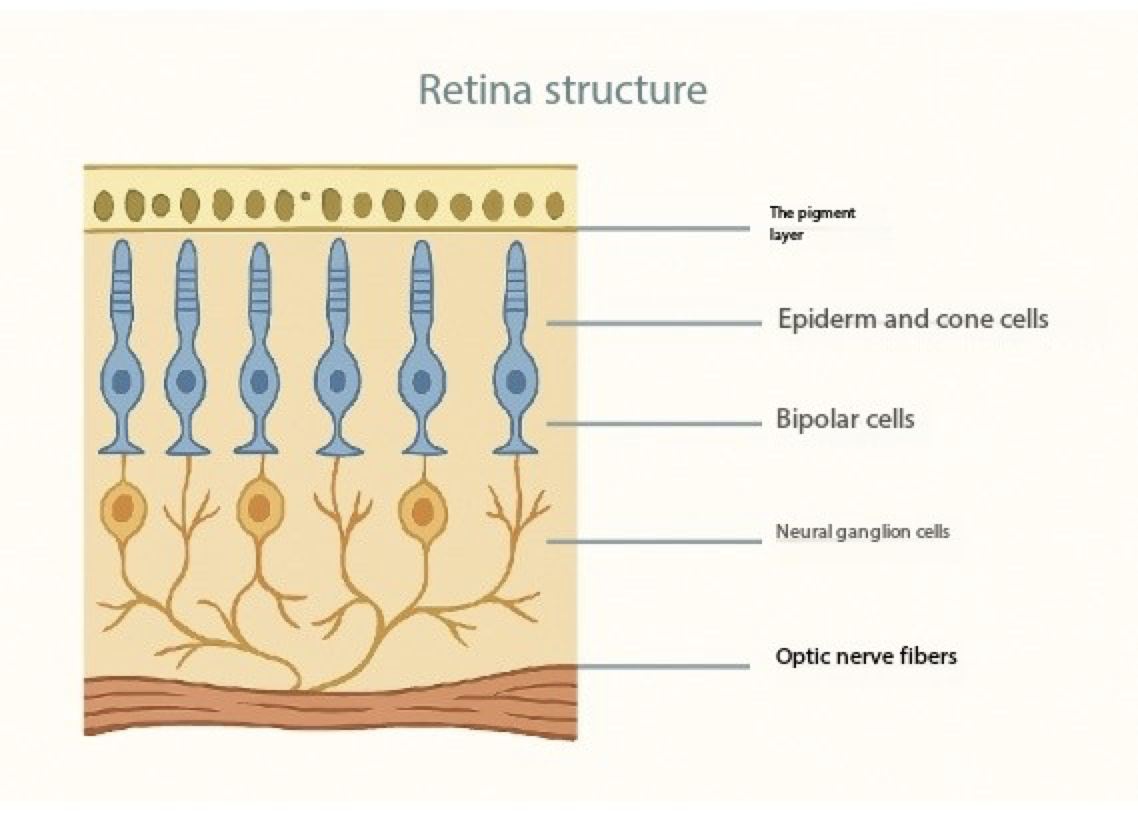}
    \caption{Retinal structure}
\end{figure}

The information processing of the retina is shaped by evolutionary pressures to maximize the dynamic range while minimizing the effects of noise. The gain regulation mechanism of the retina helps optimize the transmission of visual signals and improve the adaptability of the visual system in complex environments. For example, the connection pattern between core bipolar cells and X/Beta cells constitutes a signal enhancement mechanism similar to a "push-pull" amplifier, making the contrast of visual information more prominent. In addition, "ON" cells and "OFF" cells synapse in different sublayers of the inner plexiform layer of the retina and receive input from different types of bipolar cells, thereby maintaining the stability of information transmission under different lighting conditions.

These studies not only provide rich biological inspiration for modern visual sensor technology and visual information processing, but also promote the innovative development of computer vision. For example, bionic sensors based on retinal structure are being widely studied to improve the adaptability of computer vision systems to changes in lighting. In addition, researchers also draw on the information encoding mechanism of biological retina to optimize image processing algorithms, so that artificial vision systems can extract image features more efficiently. With the development of artificial intelligence and computer vision, imitating the structure and function of biological visual systems has become an important direction for improving machine vision capabilities. We will explore in depth the development of visual sensor technology and its application in visual information processing to further understand the nature of visual perception.

\subsection{Visual Sensor Technology}

From one or more light sources, light reflects off one or more surfaces in the world, passes through the camera's optics (lens), and eventually reaches an imaging sensor. How are the photons that reach this sensor converted into the digital (R, G, B) values that we observe when we view a digital image?

When light strikes an imaging sensor, it is typically received by an active sensing area, accumulated for an exposure time (usually expressed as a fraction of a second, e.g., 1/125, 1/60, 1/30), and then passed to a set of sensing amplifiers. There are two main types of sensors used in digital still and video cameras today: charge-coupled devices (CCDs) and complementary metal-oxide semiconductors (CMOS).

In a CCD, photons accumulate in each active light trap until the exposure ends. Then, during the transfer phase, charges are transferred from one light trap to another in a “bucket relay” fashion until they are passed to the sensor amplifier, which amplifies the signal and feeds it into an analog-to-digital converter (ADC) [10]. Early CCD sensors were prone to "blooming," where the charge from an overexposed pixel spills over into adjacent pixels, but most newer CCDs have adopted anti-blooming techniques (“troughs” are used to hold the excess charge).

In CMOS, photons striking the sensor directly affect the conductivity (or gain) of the photodetector, which can be selectively gated to control exposure duration and amplified locally using a multiplexing scheme when read out. Traditionally, CCD sensors have been preferred over CMOS in high-quality sensitive applications such as DSLRs, while CMOS is better suited for low-power applications, but most digital cameras today use CMOS.

The main factors affecting digital image sensor performance include shutter speed, sampling pitch, fill factor, chip size, analog gain, sensor noise, and the resolution (and quality) of the analog-to-digital converter. The actual values of many of these parameters can be read from the EXIF tags embedded in the digital image, while others can be obtained from camera manufacturers’ specification sheets or camera review and calibration websites [11].

Shutter speed (exposure time) directly controls the amount of light that reaches the sensor, thus determining whether the image is overexposed or underexposed. (For bright scenes, photographers sometimes use neutral density filters to allow the use of large apertures or slow shutter speeds to achieve shallow depth of field or motion blur.) For dynamic scenes, shutter speed also determines the amount of motion blur in the final image. In general, higher shutter speeds (less motion blur) make subsequent analysis easier (see Section 10.3 for techniques to remove such blur). However, when capturing and displaying video, some motion blur may be desirable to avoid flickering effects.

The sampling pitch is the physical spacing between adjacent sensing cells on an imaging chip. Sensors with a smaller sampling pitch have a higher sampling density and therefore provide higher resolution (measured in pixels) for a given active chip area. However, a smaller pitch also means that each sensor has a smaller area to accumulate as many photons; this makes it less photosensitive and more susceptible to noise.

The fill factor is the ratio of the active sensing area area to the theoretically available sensing area area (i.e., the product of the horizontal and vertical sampling pitches). Higher fill factors are generally preferred because they capture more light and reduce aliasing (see Section 2.3.1). However, this must be balanced against the need to place additional electronics between the active sensing areas. The fill factor of a camera can be experimentally determined through a photometric camera calibration process.

\begin{figure}
    \centering
    \includegraphics[width=0.9\textwidth]{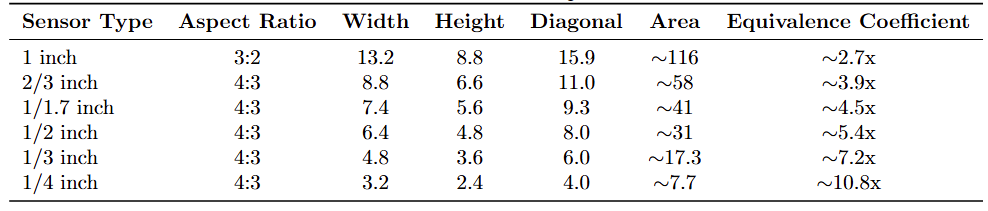}
    \caption{Sensor Size Comparison}
\end{figure}

Video and point-and-shoot cameras have traditionally used small chip areas (1/4-inch to 1/2-inch sensors), while digital SLR cameras strive to approach the size of the traditional 35mm film frame [13]. When overall device size is not important, larger chip sizes are more desirable because each sensor cell can have a higher light sensitivity. (For compact cameras, smaller chips mean that all optical components can be scaled down.) However, larger chips are more expensive to produce, not only because fewer chips can be cut from each wafer, but also because the probability of a chip defect increases linearly with chip area.

Prior to analog-to-digital conversion, the sensed signal is typically amplified by a sensor amplifier. In video cameras, the gain of these amplifiers has traditionally been controlled by automatic gain control (AGC) logic, which adjusts the gain value to achieve a good overall exposure. In newer digital still cameras, users now have some additional control over gain via the ISO setting, which is usually expressed in ISO standard units such as 100, 200, or 400. Since the automatic exposure control of most cameras also adjusts aperture and shutter speed, manually setting ISO is like manually specifying aperture and shutter speed, removing a degree of freedom from the camera's control. In theory, higher gain allows the camera to perform better in low light conditions (since the exposure time is already at a maximum, reducing motion blur). In practice, however, higher ISO settings often amplify sensor noise.

Throughout the sensing process, noise can come from multiple sources, which may include fixed pattern noise, dark current noise, photon noise, amplifier noise, and quantization noise (Healey and Kondepudy 1994; Tsin, Ramesh, and Kanade 2001). The final amount of noise in the sampled image depends on all of these factors, as well as the incident light (controlled by scene radiance and aperture), exposure time, and sensor gain. In addition, in low-light conditions, noise may be due to low photon counts, in which case a Poisson noise model may be more appropriate than a Gaussian noise model.

As discussed in more detail in Section 10.1.1, Liu, Szeliski, Kang, et al. (2008) use this model, combined with an empirical database of camera response functions (CRFs) obtained from Grossberg and Nayar (2004), to estimate the noise level function (NLF) for a given image, which predicts the variance of the noise at a given pixel as a function of its brightness (a separate NLF is estimated for each color channel). An alternative approach is to precalibrate the NLF (McCamy, Marcus, and Davidson 1976) by photographing a scene containing a variety of colors and brightnesses before taking the picture (e.g., the McCamy-White card shown in Figure 10.3b). When estimating the variance, be sure to discard or downweight pixels with large gradients, since small changes between exposures will affect the sensed values of these pixels. Unfortunately, due to the complex interactions occurring within the sensing system, the precalibration process may need to be repeated at different exposure times and gain settings.

In practice, most computer vision algorithms (such as image denoising, edge detection, and stereo matching) can benefit from a preliminary estimate of the noise level. If it is not possible to pre-calibrate the camera or take multiple photos of the same scene, the simplest approach is to find areas with nearly constant brightness and estimate the noise variance in these areas (Liu, Szeliski, Kang et al. 2008).

The final step in the analog processing chain in an imaging sensor is the analog-to-digital conversion (ADC). While a variety of techniques can be used to implement this process, two important quantities are the resolution of the process (how many bits it produces) and its noise level (how many of those bits are useful in practice). For most cameras, the number of bits typically stated (8 bits for compressed JPEG images, a nominal 16 bits for the RAW format offered by some DSLRs) exceeds the number of bits actually available. The way to tell is by calibrating the noise of a given sensor, for example, by taking multiple photos of the same scene and plotting the estimated noise versus brightness.

Once the irradiance values received by the sensor are converted into digital bits, most cameras perform various digital signal processing (DSP) operations to enhance the image and then compress and store the pixel values. These operations include color filter array (CFA) demosaicing, white point setting, and mapping brightness values through a gamma function to increase the perceived dynamic range of the signal.

\subsection{Visual information processing}
Now that we have delved into how images are formed through the complex interaction of 3D scene elements, lighting conditions, and camera optics and sensors, it is time to turn our attention to a core part of visual information processing: image preprocessing. This stage plays a vital role in the vast majority of computer vision applications, and its purpose is to transform raw images into a form suitable for subsequent analysis through a series of image processing techniques. Whether it is adjusting exposure and color balance to optimize visual effects, reducing image noise, enhancing sharpness, or even correcting image orientation through rotation, these operations are intended to lay a solid foundation for further processing. Although some may question whether image processing belongs to the core of computer vision, it is undeniable that many practical applications - from computational photography to object recognition - are inseparable from carefully designed image preprocessing steps. Only through these steps can we ensure that the subsequent analysis can obtain satisfactory results.

The core of image processing is to use a series of operators to map the pixel values of one image to another image, so as to achieve specific visual adjustments or enhancements. These techniques are usually taught as an extension of the basic signal processing course in the field of electrical engineering, and their importance is self-evident. There are many classic image processing textbooks on the market, covering a wide range of content from basic theory to advanced applications, providing learners with rich resources. This article will start with the most basic image transformation and gradually explore more complex processing methods in depth, striving to provide readers with a comprehensive and clear perspective.

\begin{equation}
    g(x, y) = T(f(x, y)) 
\end{equation}

We first focus on the simplest image transformations, namely point operators. These operators act on each pixel independently and do not rely on its neighborhood information, so they are also called point processes. As shown in Formula 3-1, f(x,y) represents the pixel value of the input image at the coordinate (x,y), g(x,y) represents the pixel value of the output image at the same coordinate (x,y), and T() represents a point operator. Typical examples include brightness and contrast adjustment, color correction, and simple image synthesis operations. Imagine that by slightly adjusting the brightness scaling or performing image addition, we can significantly improve the visibility of an image; or by color transformation, make the image present a more natural tone. These operations seem simple, but they play an indispensable role in computational photography and graphics applications. In addition, image matting and synthesis techniques also rely on such point operators to achieve background separation or foreground fusion by manipulating pixel values. To further optimize the appearance of the image, we will also introduce a more global method called histogram equalization. By analyzing the tonal distribution of the image and remapping the pixel values, histogram equalization can significantly improve the contrast and visual appeal of the image without introducing too many artificial traces. For example, in underexposed photos, this technology can effectively brighten dark details and make the picture look brand new.

However, relying solely on point operators is often not enough to cope with more complex image processing needs. Therefore, we turn to neighborhood-based operators, also known as local operators. The characteristic of these operators is that the value of each output pixel depends not only on itself, but also on the values of its neighboring pixels. Local adaptive histogram equalization is a typical example, which dynamically adjusts the hue by analyzing the area around the pixel to optimize image details under different lighting conditions. In addition to hue adjustment, neighborhood operators are also widely used in image filtering tasks, such as smoothing images through soft blur, sharpening details to highlight features, emphasizing edges to enhance contours, or removing noise to improve clarity. Among them, linear filters are the most common type of neighborhood operators, whose output pixel values are the weighted sum of the input pixel values, and the weights are determined by the filter function. This process is usually called a convolution operation, which is favored for its mathematical simplicity and practicality.

Let's take a look at some specific linear filters to better understand how they work. The simplest is the moving average filter, also known as the box filter. It smoothes an image by averaging the pixel values within a fixed-size window (e.g., a K×K region). This operation is essentially equivalent to convolving the image with a convolution kernel of all 1s and scaling the result appropriately. For larger convolution kernels, direct calculations can be resource-intensive, and separable filters provide an efficient alternative. This approach significantly reduces the computational complexity by sliding the window horizontally and vertically, respectively, and gradually updating the running sum. This technique is also similar to the concept of a cumulative area table, which will be discussed further in the following content.

To achieve a smoother filter, we can convolve the image with a piecewise linear "tent" function, also known as a Bartlett filter. The two-dimensional form of this filter, such as a 3×3 bilinear kernel, is formed by the outer product of two first-order spline functions, and its smoothing effect is more natural than that of a simple box filter. If the linear tent function is convolved with itself, a cubic approximation spline is obtained, which is often regarded as an approximation of the Gaussian kernel. This kernel is particularly important in the construction of image pyramids because it can effectively smooth images in multi-scale analysis while preserving key features. By repeatedly convolving with a box filter, we can also approximate the effect of a Gaussian kernel, although in applications that require strict rotational symmetry, a carefully designed sampled Gaussian filter is still a better choice. Linear filtering has uses far beyond smoothing. It is also a good preprocessing tool for tasks such as edge extraction and interest point detection. For example, the Sobel operator combines horizontal central difference and vertical tent filtering to emphasize horizontal edges while smoothing noise, showing the powerful flexibility of linear filtering.

In addition to linear filtering, nonlinear filters can also play a unique role in certain scenarios. For example, edge-preserving median filters and bilateral filters can effectively remove noise while preserving image edge details as much as possible. Morphological operators are very useful in binary image processing, and are used to perform operations such as dilation and erosion; while semi-global operators, such as distance transforms and connected component analysis, further expand the scope of application of neighborhood operators. Together, these tools form a rich ecosystem of image processing technology, providing diverse solutions for different needs.

In more complex scenarios, neighborhood operators can also be cascaded to form image pyramids and wavelet transforms. These structures not only improve processing efficiency by analyzing images at multiple resolutions, but also provide the possibility of accelerating certain operations. For example, in target detection or image compression, the pyramid structure can quickly capture multi-scale features. In addition, Fourier transform, as a powerful analysis tool, can help us understand the frequency characteristics of neighborhood operators and implement efficient large-kernel convolution through fast Fourier transform (FFT). This method decouples computational complexity from the size of the convolution kernel, making it possible to process large filters.

Geometric transformations are another important class of global operators, including operations such as rotation, shearing, and perspective deformation. These transformations are particularly critical when registering images, correcting distortions, or building panoramas. For example, with perspective transformation, we can correct an obliquely shot document image to a frontal view, greatly improving its readability. Finally, we will explore global optimization methods, such as minimizing energy functionals or using Bayesian Markov random field models for optimal estimation. These methods take image processing to a new level and can integrate global information to solve complex visual problems.

From simple point operators to complex global optimization, image processing techniques provide a solid foundation for visual information processing. By flexibly applying these tools, we can not only improve image quality, but also pave the way for subsequent computer vision tasks. Whether it is smoothing noise, extracting edges, or adjusting geometric structures, each step adds value to the final analysis and understanding. As technology continues to advance, the application prospects of these methods will be broader, opening a door for us to the deeper mysteries of the visual world.

\section{Facial Expression Recognition}

Facial expressions are the dynamic changes in facial muscles driven by inner emotional states, intentions or social communication. This phenomenon has become a long-standing research topic in the field of behavioral science since Darwin first systematically expounded its importance in 1872. As early as 1978, scholars tried to automatically analyze facial expressions by tracking the movement of specific calibration points in image sequences. This pioneering work marked the beginning of the technicalization of facial expression research. Since then, with the rapid development of computer technology, many systems have been designed and applied to understand and interpret this natural form of human communication. These efforts not only reveal the complex mechanisms behind facial expressions, but also provide us with valuable tools for applications in psychology, computer science, and human-computer interaction.

The core of facial expression analysis is to automatically detect and identify facial movements and their characteristic changes through visual information. This process is usually completed by a computer system, aiming to capture the subtle movements of facial muscles and the meaning they convey. However, facial expression analysis and emotion analysis are often confused in the field of computer vision, although there are significant differences between the two. Emotion analysis requires a deeper understanding. It not only involves facial expressions, but also requires a comprehensive judgment based on multiple factors such as context, body posture, voice intonation, individual differences, and cultural background. For example, the same facial movement may express completely different meanings in different situations: it may be an expression of emotion, a manifestation of intention, or even closely related to cognitive processes or physical states. In contrast, computerized facial expression analysis systems focus more on the measurement and recognition of facial movements themselves, and usually do not directly consider these external factors. Nevertheless, advances in psychological research, face detection technology, face tracking methods, and human motion analysis have laid a solid foundation for automatic facial expression analysis, which has shown broad application prospects in many fields such as emotional communication, clinical psychology, pain assessment, intelligent environment design, and multimodal human-computer interaction.

\begin{figure}
    \centering
    \includegraphics[width=0.5\textwidth]{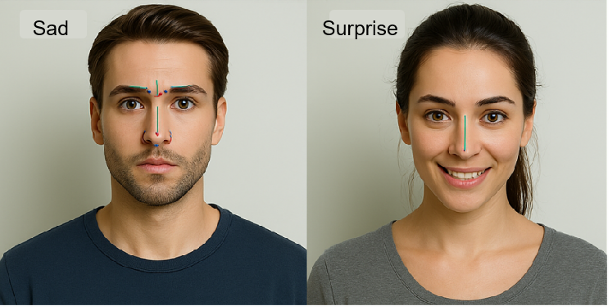}
    \caption{Expression analysis}
\end{figure}

The implementation of automatic facial expression analysis usually relies on a clear workflow. The first is the face acquisition stage, which aims to automatically locate the face area from the input image or video sequence. In order to cope with large head movements, the system may need to integrate head detection, tracking, and pose estimation techniques. For example, in video analysis, the accuracy of the analysis can be ensured by detecting the face in each frame or continuously tracking after locating the first frame. Next is the facial data extraction and representation stage, which focuses on capturing the changes in facial features caused by changes in expression. There are two main types of extraction methods: geometric feature-based methods and appearance feature-based methods. The former constructs feature vectors to characterize the geometric morphology of the face by analyzing the shape and position of facial components (such as mouth, eyes, eyebrows, and nose); the latter uses image filtering techniques (such as Gabor wavelets) to process the entire face or specific areas to extract appearance features. In order to improve the robustness of the analysis, facial normalization is usually performed before feature extraction or expression recognition to eliminate the influence of head rotation or facial scale differences.

Finally, facial expression recognition, as the core of the entire system, converts the extracted facial changes into understandable expressions. These changes may be identified as specific facial action units, or further classified as typical emotional expressions, such as happiness, surprise, or sadness. Depending on whether the time dimension is introduced in the analysis, the recognition methods can be divided into static analysis based on a single frame and dynamic analysis based on a sequence. The former is suitable for static images, while the latter is more suitable for capturing the evolution of expressions in videos. It is worth noting that the recognition of facial expressions not only relies on technical means, but is also closely related to the understanding of the classification and coding of expressions themselves. Research in this field provides a theoretical basis for the subsequent classification and coding of facial expressions, and also promotes the continuous optimization of related algorithms.

As an important part of human behavior, the importance of facial expressions has been fully verified in behavioral science. Studies have shown that in information transmission, language content accounts for only 7\%, voice and tone contribute 38\%, and facial expressions have an impact on communication effectiveness of up to 55\%. This finding highlights the key role of facial expression analysis in understanding human behavior and emotional communication. Since Darwin, this topic has attracted the attention of many psychologists and computer scientists and has developed into a vibrant research direction in the field of pattern recognition. The rise of automatic facial expression analysis is due to the pioneering contributions of many scholars, who gradually built a framework for analyzing basic emotional expressions through visual clues in images and videos. These efforts have not only enriched academic research, but also opened up new possibilities for application scenarios such as cross-language communication and human-computer interaction.

\begin{figure}
    \centering
    \includegraphics[width=0.4\textwidth]{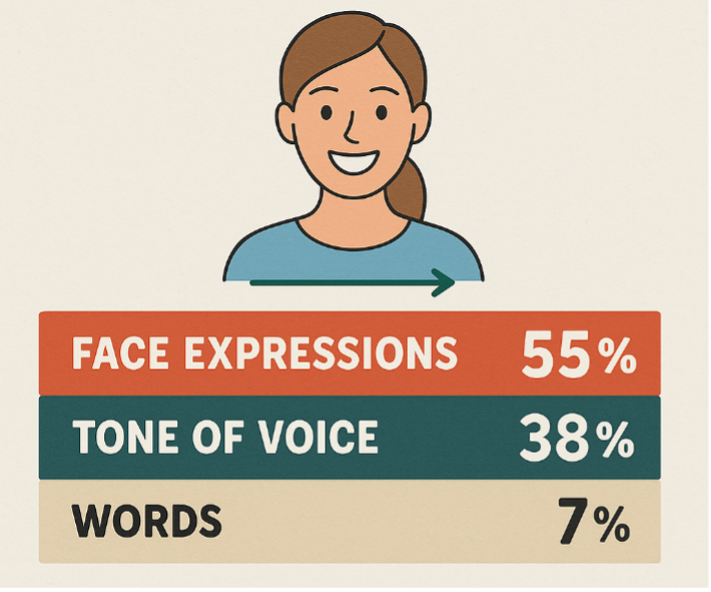}
    \caption{Human information transmission}
\end{figure}

Before delving into the technical details of facial expression analysis, it is necessary to briefly introduce the classification and encoding of facial expressions. Research in this field is largely inspired by the basic emotion theory, which holds that humans have six universally recognizable emotional expressions: happiness, surprise, disgust, fear, anger, and sadness. These expressions can be systematically described by facial action units, each of which corresponds to a specific combination of muscle movements. This encoding system provides a standardized reference framework for subsequent recognition algorithms. However, the impact of cultural factors on expression encoding cannot be ignored. In different cultural backgrounds, the same action unit may carry different emotional meanings, which brings new challenges to algorithm design.

Based on the above background, the design and optimization of facial expression recognition algorithms have become the focus of research. Common algorithms are usually based on basic emotion theory, and identify the corresponding expression categories by analyzing facial features in static images or video sequences. For example, algorithms based on geometric features may focus on the displacement of facial key points, while algorithms based on appearance pay more attention to the extraction of texture changes. In addition, in order to improve the adaptability of the algorithm, researchers have begun to explore dynamic sequence analysis methods to capture the evolution of expressions using time information. The continuous development of these algorithms has not only improved the recognition accuracy, but also provided technical support for cross-cultural expression analysis. For example, by extracting facial muscle movement areas through optical flow method and mapping them to action units, the differences in expression patterns in different cultures can be further analyzed. This method, combined with machine learning techniques such as decision trees, can generate classification rules for specific cultures, thereby significantly improving the performance of the algorithm.

Facial expression analysis has undergone a profound evolution from early theoretical exploration to today's technical application. Its core lies in capturing and interpreting subtle changes in human faces through computer systems, and the realization of this process is inseparable from the in-depth understanding of expression classification and encoding and the continuous innovation of recognition algorithms. In the following content, we will further explore the specific implementation methods of these key elements and their application potential in practical scenarios.

\subsection{Classification and encoding of facial expressions}

Facial expression is one of the important ways to express emotions. It transmits emotional information through the movement of different facial muscles. In the classification and coding research of facial expressions, the Facial Action Coding System (FACS) is one of the most widely used frameworks. The system was proposed by Ekman and aims to provide human observers with a set of standardized tools to describe and analyze facial activities.

FACS characterizes facial expressions by identifying visual action units (AUs) of facial muscles. It is a comprehensive coding system based on anatomy that can effectively describe six basic emotional expressions, including joy, sadness, anger, surprise, fear, and disgust. Ekman et al. established a set of AUs containing 44 basic action units and numbered them according to a standardized list. Among these AUs, the researchers selected the 18 most distinctive action units to accurately describe the characteristics of the six basic emotional expressions. The FACS coding system not only allows researchers to record facial expressions in a systematic way, but also describes more complex emotional states through AU combinations.

In the field of facial expression recognition, the FACS encoding program provides a method based on language description to make the encoding of facial expressions more standardized and systematic. Each facial expression is composed of one or more AUs, which may show different degrees of emotional intensity. For example, the classic Emotion Facial Action System (Emotion FACS, EMFACS) maps facial expressions to specific emotion categories through AU combinations based on FACS rules. This method has made facial expression analysis widely used in many fields such as psychology, computer vision, and human-computer interaction.

Most Automatic Facial Expression Analysis (AFEA) systems focus on recognizing a small set of typical emotional expressions, such as disgust, fear, joy, surprise, sadness, and anger. This classification method is influenced by Darwin's theory of emotional expression and is supported by the research of Ekman and Friesen and Izard et al., who believe that each emotion has a corresponding typical facial expression. However, in daily life, facial expressions are often more subtle, and the expression of emotions relies more on subtle changes in specific areas of the face. For example, the lips may be closed in anger, the corners of the mouth may droop slightly in sadness, and the eyebrows may be raised in surprise. Small movements of the eyebrows, eyelids, or mouth play an important role in emotional expression and paralinguistic communication. For example, a slight raise of the eyebrows may indicate doubt or concern.

In order to capture these subtle emotional changes more accurately, researchers have proposed more fine-grained facial expression recognition methods. FACS, as a system based on manual observation, can detect and encode these subtle changes in facial movements. Professionally trained observers can manually encode various facial expression features by analyzing video images in slow motion. The FACS system consists of 44 action units, 30 of which are related to the contraction anatomy of specific facial muscles, and the remaining 14 action units are classified as miscellaneous actions. These AUs can appear symmetrically or asymmetrically. During the FACS encoding process, researchers usually use a five-level scale to assess the intensity of muscle contraction to more finely portray changes in facial expressions.

Although Ekman and Friesen pointed out that certain AU combinations can represent typical expressions of emotions, FACS itself is not an emotion classification system, but a purely descriptive coding system. In order to convert facial expression codes into specific emotion categories, researchers have developed derivative systems such as EMFACS, which allow facial expressions to be mapped to specific emotion categories (such as joy or anger), or even further summarized into broader emotion categories (such as positive or negative emotions).

The shape, texture, color, and hair distribution of the face vary significantly depending on the individual's gender, ethnic background, and age. For example, infants' skin is usually smoother and less textured, and their eyebrows and hair are relatively sparse or even absent. At the same time, different ethnic groups also differ significantly in facial features. For example, Asians and Northern Europeans differ significantly in the shape of the eye opening and the contrast between the iris and sclera. These differences may affect the robustness of eye tracking and facial feature analysis. In addition, items such as beards, glasses, or jewelry may obscure facial features, further complicating facial analysis. However, despite the important impact these individual differences in appearance may have on facial analysis, research on their impact is still relatively limited.

Facial analysis studies on people of different ages have shown that many traditional algorithms perform poorly when applied to infants. For example, the optical flow and high gradient component detection algorithms optimized for young adults are significantly less applicable to infants. This difference mainly comes from the lack of skin texture on infants' faces, more fat tissue, underdeveloped facial structures, and the lack of transient wrinkles. All of these factors lead to significant differences in the performance of infants and adults in facial analysis tasks. Therefore, establishing robust algorithms that can adapt to different ages, genders, and ethnic backgrounds is a key challenge in facial expression research.

In addition to individual differences in appearance, facial expressions also have individual characteristics. These differences are manifested in facial plasticity, specific morphology, intensity of expression, and frequency and rate of overall expression. Studies have shown that these individual characteristics can not only be used for identity recognition, but also as biometric features to enhance the accuracy of facial recognition algorithms. These expressive differences are particularly evident in individuals with damage to the facial nerves or central nervous system. Therefore, when developing facial expression analysis algorithms, it is necessary to cover a wide range of sample groups, including people of different nationalities, ages, and genders, while also considering the presence or absence of facial hair, the wearing of jewelry or glasses, and the expression characteristics of normal people and clinically impaired individuals.

Traditional facial expression analysis usually assumes that expressions are unitary and start and end in a neutral state. However, in reality, the dynamic changes of facial expressions are much more complicated than this assumption, especially at the level of action units (AUs). Action units often appear in combinations and may show serial dependencies. The transition from one action unit or action combination to another action unit does not necessarily go through a neutral state. Therefore, in order to build a robust facial analysis system, it is necessary to parse the behavioral flow of expressions and include enough training data of dynamically combined action units. The combinations of these action units may be additive or non-additive.

\begin{figure}
    \centering
    \includegraphics[width=0.5\textwidth]{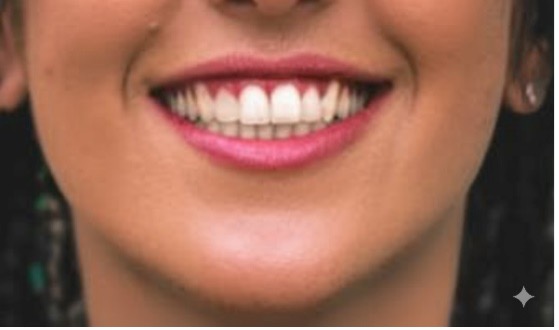}
    \caption{Action unit AU12}
\end{figure}

A typical example of an additive combination is a smile (AU 12) with an open mouth. With different degrees of lip separation and jaw lowering, this combination can be encoded as AU 12 + 25, AU 12 + 26, or AU 12 + 27. For example, in the case of AU 12 + 27, the facial analysis system needs to be able to detect different levels of mouth opening and continue to recognize AU 12 despite different intensity changes. Detection of such dynamic changes is critical to improving the accuracy of facial analysis.

In contrast, non-additive combinations are more complex and often involve co-articulation effects. A typical example is the combination of AU 12+15, which often appears in embarrassing situations. Although AU 12 acts to lift the cheeks and corners of the mouth, the downward pressure of AU 15 affects the movement of the corners of the mouth, resulting in the final facial appearance change being highly dependent on the timing relationship. If the actions of AU 12 and AU 15 occur simultaneously, the visual effect will be different from when they occur sequentially, which is usually more common. Therefore, in order to improve the accuracy of facial expression analysis, the database should include separate action units, as well as cover additive and non-additive action combinations, especially those involving co-articulation effects.

The intensity variation of facial movements is also an important dimension of expression analysis. Manual FACS coding systems usually use a 3-point or 5-point intensity scale to describe the intensity variation of action units. For example, in the eye region, AU 41 (slightly droopy eyelids), AU 42 (moderate droopy eyelids), AU 43 (completely closed eyelids), and AU 45 (blinking) together constitute a series of different degrees of eye closure. Computer vision researchers have also explored automated intensity representation methods, such as Essa and Pentland using optical flow to analyze the intensity variation of smiles, and Kimura and Yachida quantifying the intensity variation of specific emotional expressions. In addition, Lien et al. proposed an intensity quantification method for action units, while Tian et al. successfully distinguished different intensities of eye closure through Gabor features and artificial neural networks, with reliability comparable to that of human coders. Further studies, such as the work of Bartlett and colleagues, clarified the relationship between marginal changes in classifier output and expression intensity by analyzing expression dynamics, while Yang et al. transformed the intensity estimation problem into a ranking problem and used the RankBoost model for intensity grading. These research results show that it is feasible to automatically identify the intensity of facial movements using computer vision technology and can help improve the accuracy of facial expression analysis.

In the study of facial expression recognition, the distinction between deliberate and spontaneous expressions is also crucial. Most facial expression databases are collected by asking subjects to perform specific expression tasks, and the facial movements under these controlled conditions may differ significantly in appearance and timing from spontaneous expressions in daily life. Neurophysiological studies have shown that deliberate facial movements are mainly controlled by pyramidal motor pathways, while spontaneous expressions involve extrapyramidal motor pathways. This difference in control mechanism leads to the fact that deliberate expressions are usually inferior to spontaneous expressions in terms of motor refinement and symmetry. For example, most people are able to spontaneously raise their outer eyebrows while keeping their inner eyebrows still, but few people can consciously complete this action. Similarly, spontaneous drooping of the corners of the mouth (AU 15) and the lifting and narrowing of the inner eyebrows (AU 1 + 4) are typical characteristics of sadness, but it is difficult for untrained individuals to actively make these expressions. Therefore, the lack of spontaneous expression data in the training data of facial expression analysis systems may lead to reduced robustness of the model in practical applications. Furthermore, since many pattern recognition methods (e.g., hidden Markov models) rely on the temporal dynamics of expressions, expression databases need to cover both spontaneous and intentional expressions to ensure that the algorithms can adapt to different modes of expression.

In addition to the changes in action units and intensity, facial expression analysis is also affected by external environmental factors. Among them, facial orientation, head posture, scene complexity and interactive environment are all important factors affecting facial recognition and expression analysis. In the field of facial recognition, a lot of research has been done on facial postures at different angles. For example, the FERET database covers frontal and oblique angles to develop recognition methods that are robust to posture changes. However, in the field of facial expression analysis, multi-view research is relatively rare. Most existing methods assume that the face only rotates slightly within the plane, and do not fully consider the large three-dimensional rotation of the head. In fact, many facial expressions are usually accompanied by changes in head posture. For example, Kraut and Johnson's research found that smiles often appear when people turn their heads to face others, while Camras et al.'s research shows that babies' surprised expressions are usually accompanied by the movement of tilting their heads back. Therefore, to build a posture-invariant facial expression analysis system, it is necessary to introduce image data containing significant head posture changes and combine it with a three-dimensional model or deep learning method to improve the system's adaptability to different perspectives.

In addition, the complexity of the background in the real environment is also a major challenge for facial expression analysis. Many current databases have images with a single background and usually only one individual, while in real application scenarios, the interaction between people is a factor that cannot be ignored. For example, in social scenarios, facial expressions may be affected by other individuals, and may even present different visual characteristics due to environmental factors (such as lighting changes, occlusion, etc.). Therefore, in order to improve the practicality of the automatic expression recognition system, it is necessary to cover a variety of environmental variables in the database construction process, including different lighting conditions, background complexity, and multi-person interaction scenarios, to ensure that the model can adapt to the diverse needs of real-world applications.

\subsection{Facial expression recognition algorithm}

In the facial expression recognition algorithm, the preprocessing algorithm undertakes many basic tasks such as data cleaning, image enhancement and feature highlighting. Its role in the entire system is like a cornerstone, which directly determines the effect of subsequent feature extraction and pattern recognition. Facial image acquisition is often affected by various factors such as illumination, noise, posture change and occlusion. Therefore, a series of preprocessing operations must be performed on the original image at the data input stage to improve the image quality and the robustness of the system. The preprocessing algorithm must first denoise the input image. Since various noises such as salt and pepper noise and Gaussian noise are easily introduced during camera equipment, ambient lighting and image transmission, it is necessary to use filtering algorithms to smooth and reduce noise on the image. Commonly used median filtering, mean filtering and Gaussian filtering play an important role in balancing noise removal and detail retention. After denoising, there may still be local brightness imbalance problems in the image due to uneven ambient lighting or overexposure. At this time, histogram equalization or adaptive histogram equalization technology is needed to adjust the contrast of the image, so that facial features are more obvious and reduce recognition errors caused by different lighting conditions.

After image denoising and enhancement, the next step is to locate and correct the facial area. Facial detection technology plays a key role here. The facial area in the image is located through the detection algorithm, eliminating background interference and providing clear area information for subsequent processing. The detection method can use traditional cascade classifiers based on template matching or Haar features, or modern deep learning techniques such as convolutional neural networks to quickly and accurately locate the facial area. After the face is detected, correction operations are essential due to the diversity of different shooting angles and facial postures. By performing geometric transformations on the facial image, such as affine transformation or perspective transformation, the facial image is rotated, scaled and normalized, so that each facial organ (eyes, nose, mouth, etc.) is in a standardized position, so that the subsequent feature extraction process can more effectively capture the subtle differences in expression changes.

\begin{equation}
    I_{\text{resized}} = \text{Resize}(I_{\text{original}}, W, H)
\end{equation}

Normalization is an important part of the preprocessing stage. Since there are large differences in factors such as size, resolution, and shooting distance of different images, unifying the images to a standard size helps improve the robustness of the system. The commonly used normalization method not only includes the unification of geometric dimensions such as formula (3-2), but also covers the standardization of grayscale values and color channels. Grayscale normalization technology can reduce grayscale fluctuations caused by lighting conditions, while color space conversion can convert image information from RGB space to a color space that is more suitable for describing facial details, such as HSV or YCbCr, which has obvious advantages for the subsequent extraction of expression features.

In addition to the above basic preprocessing operations, in recent years, more and more studies have introduced deep learning-based preprocessing modules to address the special challenges in facial expression recognition. Such modules can not only achieve more sophisticated processing based on traditional image enhancement and correction, but also adaptively learn the optimal preprocessing parameters from the data. Through end-to-end training, the preprocessing network and the subsequent feature extraction network can be jointly optimized, so that the entire system can effectively suppress noise and environmental interference while retaining key information. Such a joint training strategy greatly improves the system's adaptability in complex scenes, especially under low illumination and background clutter, and can significantly improve the accuracy and robustness of facial expression recognition.

In practical applications, the preprocessing algorithm must also consider the issue of real-time performance. Facial expression recognition systems are often used in fields such as human-computer interaction, intelligent monitoring, and emotional computing, and have high requirements for response speed. Therefore, under the premise of ensuring the preprocessing effect, the algorithm designer needs to optimize the algorithm to reduce the computational complexity. For example, on an embedded platform with limited hardware resources, a lightweight filtering algorithm and a simple and efficient facial detection model can be used, combined with region cropping and multi-threaded processing strategies to achieve real-time processing requirements. This type of optimization requires not only theoretical proof of the effectiveness of the algorithm, but also a large number of experimental verifications to ensure that the system can work stably in different scenarios.

In addition, the preprocessing algorithm also plays an important role in data enhancement. Facial expression recognition datasets often have problems with insufficient sample numbers or uneven sample distribution. In the preprocessing stage, data can be augmented through operations such as image rotation, scaling, translation, and flipping, thereby enriching the training data and improving the generalization ability of the model. Data enhancement can not only alleviate the overfitting problem, but also make the model more robust when facing various angles, postures, and expression changes. It is worth mentioning that in the process of data enhancement, it is very important to maintain the authenticity and coherence of the facial structure, otherwise it may introduce noise that does not conform to the actual situation and affect the training effect of the model.

In response to different scenarios and application requirements, the preprocessing algorithm may also fuse multiple information for collaborative optimization. For example, in a video sequence, changes in facial expressions are often continuous and sequential. Therefore, based on the preprocessing of a single-frame image, fusing time domain information to jointly process consecutive frames can effectively reduce the information loss caused by single-frame processing. At the same time, combined with motion estimation and optical flow analysis technology, facial micro-expression changes can be captured more accurately, thereby improving the sensitivity and accuracy of recognition. This preprocessing method based on spatiotemporal information provides a new idea and technical path for solving the problem that static images cannot reflect the true emotional state.

Although the preprocessing module is at the forefront, its complexity and diversity make it an indispensable part of the entire facial expression recognition system. From image acquisition, noise suppression, illumination correction, area positioning, geometric transformation, normalization processing to data enhancement, each step directly affects the final recognition effect. Researchers are constantly exploring new algorithms and technical means to cope with increasingly complex application scenarios. With the continuous development of deep learning and computer vision technology, the preprocessing algorithm will develop in a more intelligent, automated and efficient direction in the future. It will not only play a role in traditional image processing, but also form a seamless connection with subsequent feature extraction and classification modules, and jointly promote the widespread application of facial expression recognition technology in intelligent interaction, emotional computing and other fields.

In the facial expression recognition system, feature extraction algorithm plays a core role. Its task is to extract the most discriminative facial information from the preprocessed image and provide a reliable basis for subsequent classification and decision-making. With the continuous development of computer vision and machine learning technology, traditional manual feature extraction methods and modern automatic feature extraction methods based on deep learning have achieved remarkable results in their respective fields. Traditional methods mainly rely on well-designed feature descriptors to characterize facial expression changes by capturing local texture, shape, edge and other information. For example, Local Binary Pattern (LBP) is widely used in facial expression recognition as an effective texture feature descriptor. This method converts subtle grayscale changes in the image into digital symbols by binary encoding the local area, thereby capturing subtle expression changes. At the same time, Gabor wavelet transform has also become an important means of feature extraction due to its sensitivity to local spatial frequency and directional information. Using Gabor filter groups to decompose facial images in multiple scales and directions can obtain rich texture and edge information, providing strong support for expression classification. Another widely studied method is the Histogram of Oriented Gradients (HOG), which captures the geometric structure and edge features of the face by counting the distribution information of the gradient direction in the image. Although traditional feature extraction methods perform well under certain conditions, they usually rely on expert knowledge and manual design, have poor adaptability to environmental changes, lighting fluctuations, and posture changes, and have certain limitations when processing large-scale data.

\begin{figure}
    \centering
    \includegraphics[width=0.5\textwidth]{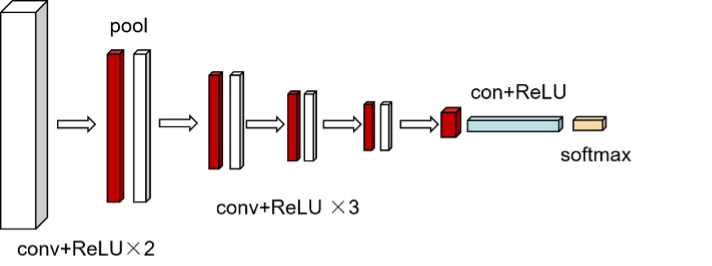}
    \caption{VGG neural network}
\end{figure}

With the rapid development of deep learning technology, feature extraction methods based on convolutional neural networks have gradually become mainstream. Deep learning methods can be trained end-to-end on large-scale data sets, and automatically learn feature representations suitable for facial expression recognition through a multi-layer network structure. Each layer in the network can abstract and transform the input image at different levels, thereby capturing the gradual features from low-level texture information to high-level semantic information. Using a large amount of annotated data for training, deep learning models can not only automatically adapt to changes in lighting and posture, but also optimize the synergy between feature extraction and classification modules through joint training to improve the overall recognition accuracy. In recent years, many researchers have begun to focus on the fusion method of deep features and traditional manual features, and achieve better recognition results by complementing the advantages of the two. This method utilizes the ability of deep networks to learn big data to a certain extent, while retaining the advantages of traditional methods in capturing local details, making the system more robust and generalizable when dealing with complex emotional expressions and micro-expression recognition.

In practical applications, the features of facial expressions are not limited to the features of static images, but also include dynamic information and temporal features. Since the changes in facial expressions are often continuous and dynamic, how to extract effective spatiotemporal features from continuous frames has become an important direction of current research. Researchers have tried to build a hybrid network model that can capture the characteristics of time evolution by combining the methods of convolutional neural networks and recurrent neural networks. This type of model can not only extract spatial features in a single frame image, but also reveal the dynamic process of expression changes through time series modeling, thereby more accurately reflecting the subtle changes in emotions. At the same time, optical flow analysis technology has also been introduced into dynamic feature extraction. By calculating the pixel motion information between continuous frames, it helps to reveal the subtle motion changes in facial micro-expressions and provide richer information support for subsequent emotion recognition. The extraction of dynamic features not only improves the system's ability to respond to rapid changes in expression, but also provides a new perspective for capturing complex emotional states.

On the other hand, the research on feature extraction algorithms also needs to consider the problem of feature dimensionality reduction and selection. In high-dimensional data, there is a lot of redundant information and noise. Directly using high-dimensional features for classification often leads to increased computational complexity and even causes the dimensionality disaster. In order to solve this problem, dimensionality reduction methods such as principal component analysis and linear discriminant analysis are widely used in facial expression recognition. By transforming and screening high-dimensional features, the most discriminative low-dimensional feature representation can be extracted, thereby reducing computational complexity and improving the generalization ability of the classifier. In recent years, with the development of kernel methods and sparse coding technology, some nonlinear dimensionality reduction methods have also been proposed. These methods can better capture the nonlinear relationship between features and provide more refined feature descriptions for complex expression recognition. Feature selection plays a vital role in the entire recognition system. Only by retaining the features that are most sensitive to emotional states can high-precision recognition effects be achieved in practical applications.

In addition, in order to deal with the sample imbalance problem in facial expression recognition, the feature extraction algorithm also needs to combine data enhancement technology to expand the training data. Data enhancement is not limited to geometric transformations such as image rotation, scaling, and translation, but also involves the adjustment of color, contrast, and lighting conditions, so that the model can maintain a high recognition accuracy in a variety of environments. Using data enhancement to generate diverse training samples can effectively alleviate the overfitting problem caused by insufficient samples or uneven category distribution, thereby improving the robustness of the entire system. In the feature extraction stage, combining data enhancement technology with feature extraction algorithms not only helps to improve the generalization ability of the model, but also provides more solid data support for the subsequent recognition of complex emotional expressions.

In facial expression analysis systems, the final step is to recognize facial expressions based on the extracted features. To accomplish this task, researchers have explored and applied a variety of classification algorithms, including neural networks, support vector machines, linear discriminant analysis, K-nearest neighbors, multinomial logistic regression, hidden Markov models, tree-enhanced naive Bayes, and RankBoost. In addition, some systems rely solely on rule-based classification methods, which make inferences based on the definition of facial action units. Depending on how the data is processed, expression recognition methods can be roughly divided into frame-based recognition methods and sequence-based recognition methods.

Frame-based recognition methods only use the information of the current frame for classification. They can analyze a single frame independently or with the help of a reference image (such as a neutral facial image) for comparison. This method does not rely on time information, so the computational complexity is low and it is suitable for real-time applications. In contrast, sequence-based recognition methods use time series information to analyze the evolution of expressions across multiple frames, which can more accurately capture the dynamic characteristics of facial expression changes. This type of method is usually combined with time series processing techniques such as hidden Markov models and recurrent neural networks to improve recognition accuracy.

Frame-based expression recognition methods do not consider temporal information, only use the information of the current input image, and independently process a single frame or a frame in the entire sequence. This method has been widely used in facial expression recognition tasks and has been extensively studied in the literature. Several research teams have used different classification algorithms for expression recognition, including neural networks, support vector machines, linear discriminant analysis, Bayesian networks, and rule-based classifiers.

\begin{figure}
    \centering
    \includegraphics[width=0.9\textwidth]{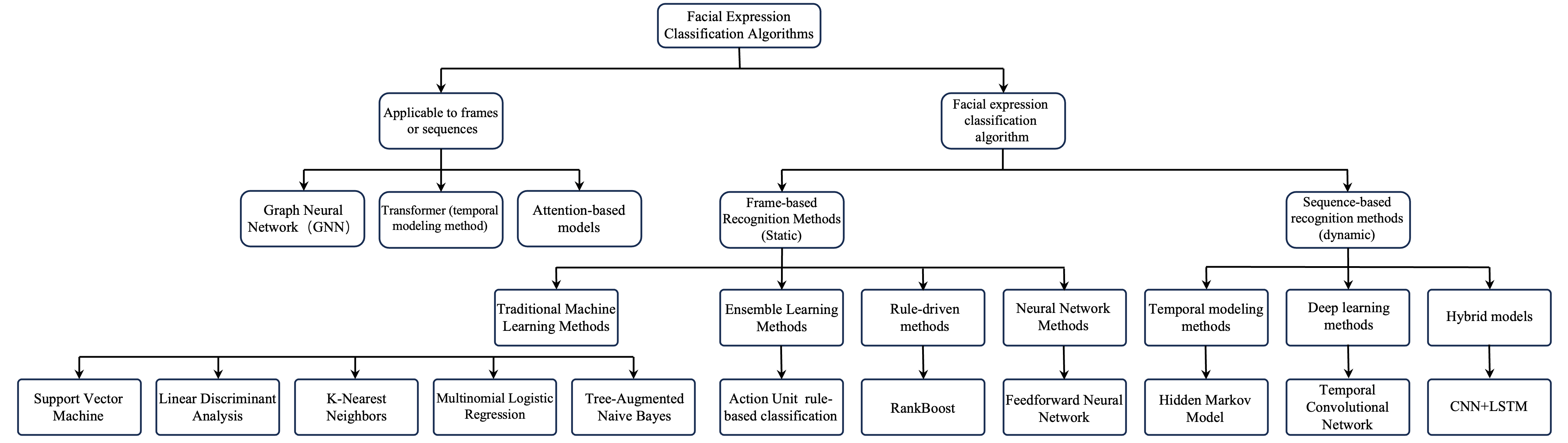}
    \caption{Facial expression classification algorithm}
\end{figure}

For example, Tian et al. used a neural network-based approach to recognize facial action units. They used a three-layer neural network with one hidden layer and trained it using a standard back-propagation method. The system is able to process expressions in the upper and lower parts of the face, and can simultaneously process normalized geometric features, appearance features, or a combination of both. The recognition results are output in the form of facial action units, and multiple facial action unit combinations can be recognized simultaneously, which gives it an advantage over traditional methods. In experiments, the system achieved an overall recognition rate of 95.5\% when recognizing neutral expressions and 16 facial action unit combinations.

Support vector machines have also been applied to facial expression recognition. For example, a study used a two-stage classifier to identify neutral expressions and six basic emotions. First, they trained a set of pairwise classifiers so that each support vector machine could distinguish between two different emotions. Subsequently, the researchers tested different methods, such as nearest neighbor classification, simple voting, and multinomial logistic regression, to convert the representations produced in the first stage into probability distributions to ultimately determine the expression category. Among them, the multinomial logistic regression method achieved the best recognition rate of 91.5\% in independent tests. In addition, Wen and Huang used another two-stage classification method. They first used a neural network to classify expressions into neutral and non-neutral categories, and then used a Gaussian mixture model to further subdivide non-neutral expressions. Experiments showed that the average recognition rate of this method in independent tests of people was 71\%. At the same time, Yang et al. used the RankBoost method for expression recognition and used L1 regularization technology to optimize model performance. On the Cohn-Kanade database, this method achieved a recognition rate of 88% and was able to estimate the intensity of the expression, further enhancing the classification effect.

Unlike frame-based methods, sequence-based expression recognition methods use the temporal dynamic information of facial expressions for classification. Since the natural change of expression is often a continuous process, sequence-based methods can more accurately characterize the evolution of expression and reduce errors caused by single-frame misjudgment. To this end, researchers have adopted technologies such as hidden Markov models and recurrent neural networks to improve the accuracy of recognition. For example, some studies use hidden Markov models for facial expression recognition, which can effectively model the temporal changes of expression and show high accuracy in facial action unit recognition tasks. Especially in datasets involving spontaneous expressions, hidden Markov models can better cope with the challenges caused by head movement, illumination changes, and different speeds of expression changes due to their powerful temporal modeling capabilities. In addition to hidden Markov models, recurrent neural networks are also widely used in sequence-based expression recognition tasks. For example, variants such as long short-term memory networks can effectively model long-term dependencies. Experimental results on multiple datasets show that this method can achieve high accuracy when dealing with expression recognition tasks.

In addition, rule-based classifiers are also used for sequence-based expression recognition tasks. This method combines the definition of facial action units, analyzes time series data, and infers expression categories based on established rules. Although this method performs well in some cases, it is gradually replaced by data-driven methods in recent years because the formulation of rules is complex and difficult to adapt to the changing actual application scenarios. The researchers' rule-based method achieved high-precision recognition of blinks, blink tremors, and non-blinks, with an overall accuracy of 98\%. Among them, blink tremors are defined as two or more rapidly repeated blinking behaviors, during which the eyes are only partially open. In the classification task of blinking and non-blinking, the system achieved 100\% accuracy.

However, the classification task of the eyebrow region is quite challenging. For the recognition of the three cases of eyebrow raising, eyebrow lowering and no eyebrow movement, the system's accuracy is only 57\%. This low accuracy is mainly attributed to the small number of samples of eyebrow lowering, which makes it impossible to make reliable point estimates. If the eyebrow lowering category is removed from the analysis, the recognition accuracy can be improved to 80\%. It is worth noting that human FACS coders also show similar difficulties in the same task, with only about 50\% consistency in the recognition of eyebrow lowering. Factors such as low frequency of eyebrow lowering, occlusion of glasses, and forward head movement together increase the complexity of coding and recognition.

Among machine learning methods, support vector machines combined with Gabor feature representation are widely used for facial expression recognition. Gabor features can effectively capture the changes in local facial texture and provide rich information for expression classification. A system first uses SVM for preliminary classification, and then uses hidden Markov model (HMM) to further model the dynamic changes of AU (action unit). There are two main ways to apply HMM: one is to directly use Gabor features as input, and the other is to use the output of SVM as input. After using principal component analysis (PCA) to reduce the Gabor features to 100 dimensions, the system trained two HMMs for the classification of blinking and non-blinking, and used leave-one-out cross-validation. Under the configuration of five states and three Gaussian models, the HMM directly based on Gabor features achieved a recognition rate of 95.7\%. When using SVM output as input, the recognition rate is further improved to 98.1\%.

For the recognition task of the eyebrow region, HMM combined with Gabor features after PCA dimension reduction and HMM combined with SVM output achieved accuracies of 70.1\% and 66.9\% respectively. When the category of eyebrow descent was excluded, the recognition accuracies increased to 90.9\% and 89.5\% respectively. This shows that in the classification of eyebrow movement, the performance of existing methods still has room for further improvement, especially the optimization of sample imbalance and feature extraction.

In addition to rule-based and SVM-based methods, Bayesian network classifier is also an important facial expression recognition method. This method provides probabilistic inference capabilities for different expressions by modeling the dependencies between features. A study evaluated multiple Bayesian network classifiers for the first time, including Gaussian Naive Bayes, Cauchy Naive Bayes (NB-Cauchy), and Tree-augmented Naive Bayes (TAN). These classifiers performed differently in frame-based expression recognition. In particular, the TAN classifier performed best in the classification of neutral expressions and six emotion-specific expressions due to its ability to model feature dependency structures. In an individual-independent test on the Cohn-Kanade database, the TAN classifier achieved a recognition rate of 73.2\%.

\section{Body language recognition}
Humans interact with others and the surrounding environment through multiple senses such as vision, hearing and touch. Among these perception modes, the visual modality is particularly important in modern human-computer interaction and has become the core input and output channel of the next generation of intelligent systems. With the development of technology, the recognition and understanding of human body language has gradually received attention. Body language, as a natural non-verbal communication method, is widely present in human daily life. From gestures when talking, postures when thinking, to movements when working, body language is not only an important medium for information transmission, but also carries rich emotions and intentions.

In recent years, the gaming and entertainment industry has become one of the main driving forces for the advancement of human-computer interaction technology. Many innovative interaction methods abandon traditional physical controllers and rely on natural human movements to achieve interaction. For example, the Kinect project allows users to interact with game consoles through gestures and voice commands. This method not only provides an immersive entertainment experience, but also provides an important application scenario for the development of body language recognition technology.

In addition to the field of games, body language also plays a key role in emotional expression and social communication. Psychological research shows that human emotional states are often conveyed through multiple modalities such as facial expressions, voice intonation, and body movements. Among these modalities, body movements are an important clue for identifying emotions because of their intuitiveness and authenticity. Researchers have found that body movements are often accompanied by specific emotional changes, such as exaggerated gestures when excited and drooping postures when sad. By analyzing these body clues, the emotional state of an individual can be effectively inferred.

However, despite the rich research results accumulated in the field of emotion recognition in the field of psychology, automated body language recognition technology still faces many challenges. Past research has focused on emotion recognition of facial expressions and voice, but relatively less attention has been paid to body movements. In addition, most existing automatic emotion analysis systems are mainly based on static features, ignoring the dynamic characteristics of emotional expression. The expression of emotions usually goes through stages such as onset, peak, and decay, and each stage contains rich information. How to accurately capture these temporal dynamics has become a major problem in automated body language recognition.

In order to solve these problems, computer scientists and human-computer interaction researchers have proposed a variety of methods in recent years to try to achieve automatic analysis of body language. The recognition of body language includes key links such as action detection, feature extraction and semantic parsing. In the action detection stage, the system captures the movement trajectory of the human body through sensors such as cameras to generate raw data. In the feature extraction stage, representative motion features are extracted from the raw data using technologies such as spatiotemporal interest points and skeleton tracking. Finally, these features are analyzed and modeled through methods such as deep learning to achieve semantic parsing of the action.

\begin{figure}
    \centering
    \includegraphics[width=0.3\textwidth]{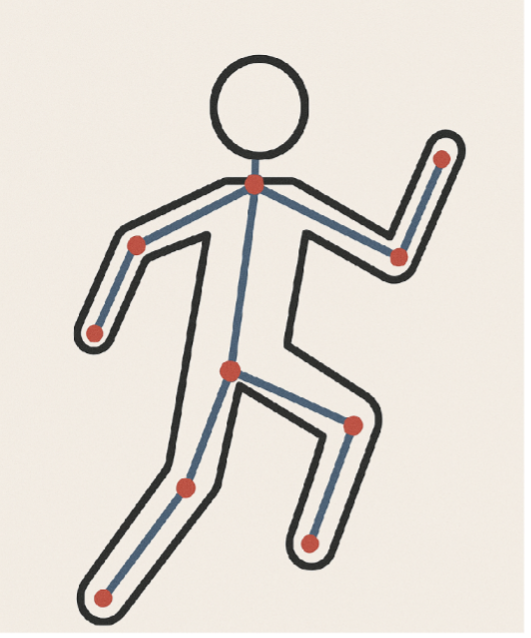}
    \caption{Action recognition}
\end{figure}
Semantic parsing of body movements is the core of body language recognition. It involves mapping low-level motion data to high-level semantic labels, such as identifying whether an action is waving, pointing, or hugging. In addition, the semantics of actions are often affected by the environment and context. The same gesture may have completely different meanings in different scenarios, so accurate semantic parsing requires comprehensive judgment in combination with contextual information. In the application of automated body language recognition, emotion recognition is a typical scenario. For example, by analyzing a person's body movements, their current emotional state can be inferred. In practical applications, such technologies have been widely used in mental health monitoring, human-computer interaction optimization and other fields. For example, by building a multimodal emotion recognition system and integrating the emotional features of facial expressions, sounds and body movements, the accuracy of emotion recognition can be significantly improved.

This chapter will discuss the core issues of body language recognition. First, it will explore the importance of body language in emotional expression and review the relevant research progress. Then, it will deeply analyze the semantic parsing methods of body movements, including key technologies such as action feature extraction, model training and context fusion. Finally, this chapter will introduce the technical progress and application cases in the field of body language recognition in recent years, and further demonstrate the wide application of body language recognition in emotion analysis and its future development direction.

\subsection{Semantic parsing of body movements}
Body language is an important part of human communication, covering a variety of non-verbal behaviors such as facial expressions, body posture, gestures, eye movements, touch, and the use of personal space. Through these subtle movements and gestures, a person's inner state is revealed. From the dilation of the iris and the direction of the gaze to the position of the hands and feet, the way you sit, walk, stand or lie, all of these can convey emotions and psychological states.

Hands play a particularly prominent role in body language. The posture and position of the hands often reflect a person's honesty. For example, when the palms are facing the other person, it usually conveys honesty and openness; on the contrary, hiding the hands behind the back may indicate concealment or insincerity. In the communication process, using open gestures often increases credibility, which is a communication strategy often used by speakers, debaters and politicians. Research shows that people who use open gestures are more likely to be seen as friendly and reliable by the audience.

Head posture can also reveal a wealth of emotional information. Listeners nod to encourage speakers to continue speaking, and this simple action can significantly enhance the fluency of communication. The frequency and amplitude of nodding can also convey different messages: gentle nodding usually indicates patience and understanding, while quick or mechanical nodding may indicate impatience. When the head is kept in a neutral position, eye contact between people is more natural; when the chin is raised, it may show confidence, superiority or even arrogance. On the other hand, exposing the neck may be a signal of submissiveness, especially in social situations. Women often tilt their heads unconsciously when expressing interest, and this posture often conveys gentleness and affinity. A lowered chin may indicate negative emotions or an aggressive attitude.

In contrast, the torso contains less information in body language, but its direction and angle are equally important. Facing others directly is often seen as a direct and confident gesture, but in some cases it can also appear aggressive. Slightly tilting the body and maintaining a certain angle is often interpreted as a sign of friendliness and relaxation. Especially when combined with positive signals such as nodding and smiling, a forward-leaning posture often represents curiosity and interest. To accurately interpret body language, it is necessary to consider all parts of the body comprehensively. The body language recognition system is based on this principle, and infers the emotional state of an individual by analyzing different body signals. This system is widely used in psychological research, behavioral analysis, intelligent human-computer interaction and other fields.

It is worth noting that the meaning of gestures often has strong cultural differences. With the development of globalization and the widespread dissemination of mass media, certain gestures have gradually shown cross-cultural commonality, especially among the younger generation. However, the meaning of many traditional gestures still has significant cultural differences. For example, the thumbs-up gesture is commonly used to represent the number "1" in Europe, "5" in Japan, and may be regarded as an offensive insult in Australia and Greece. Nevertheless, with the mutual influence of global cultures, this gesture has become a universal symbol of approval and recognition in many regions.

Facial expressions are a common way for humans to express emotions, and their cultural differences are relatively small. Studies have found that American and Japanese babies show striking similarities in expressing emotions, indicating that facial expressions of certain emotions may have a biological basis. This phenomenon also applies to many basic postures and body movements. Nevertheless, further research is still needed to deeply understand the specific impact of culture and context on body language. Scholars usually focus on body language in a specific activity or scene in order to more clearly identify behavioral differences in different cultural backgrounds. For example, shaking hands symbolizes respect and friendliness in many cultures, but in some cultures, physical contact is not common, and people may choose to nod or put their hands together to greet. Understanding these cultural differences not only helps cross-cultural communication, but also provides an important reference for the design of human-computer interaction systems.

\subsection{Body language recognition technology}
In this section, we will introduce the core components of a system called emotional body gesture recognition. This system is designed to recognize and infer emotional states by analyzing human gestures. To show the overall architecture of the system more clearly, we can refer to the relevant diagram.

\begin{figure}
    \centering
    \includegraphics[width=0.5\textwidth]{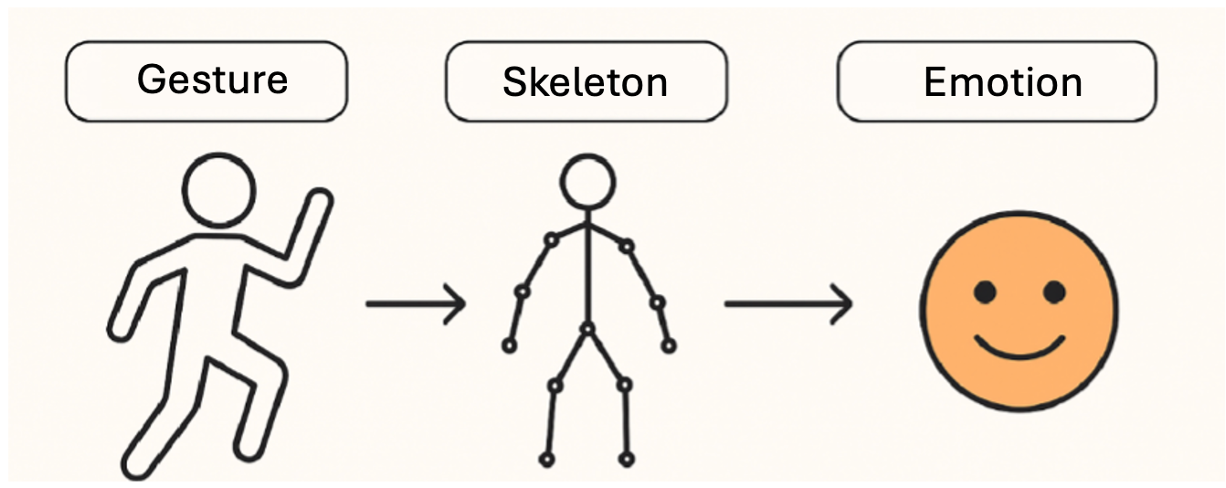}
    \caption{Emotional body posture recognition system}
\end{figure}
The first step of the system is to perform input modeling, that is, to reasonably characterize human posture and emotional targets. This link is crucial because it directly determines the design direction of the subsequent automated pipeline. Depending on different application scenarios and needs, you can choose to use a public database for training, or you can collect a dedicated data set to improve the pertinence of the model. The choice of model type affects the performance, accuracy, and real-time performance of the system, so multiple factors need to be weighed. In addition, the various components in the system should be compatible with each other, and they need to work together from data preprocessing to final emotion recognition to ensure the optimization of the overall performance.

Although different emotional body gesture recognition systems vary in their specific implementations, they generally follow the same basic steps. First, people need to be detected from images or videos. The goal of human detection is to segment the human region from the background, which is the basis for subsequent posture analysis and emotion recognition. Usually, this task is accomplished by determining the bounding box of the human body in the image. However, this task is very challenging due to the non-rigid nature of the human body and the significant differences in posture, clothing, and body shape. In natural environments, illumination changes and occlusion issues further increase the difficulty of detection. Human detection usually follows a standard object detection pipeline, including candidate region extraction, feature representation, classification, and post-processing steps. The feature extraction stage aims to capture the key features of the human body so that the model can accurately distinguish between the human body and the background. With depth information, the search space can be effectively narrowed down and the process of background segmentation can be significantly simplified.

In the early human detection methods, Viola and Jones proposed a detection method based on a cascade structure. This method was originally used for face detection and achieved efficient detection performance through cascade classifiers. It uses an adaptive enhancement algorithm to automatically select the best features to reduce the amount of calculation. Although this method performs well in simple scenes, it has certain limitations when facing complex backgrounds and occlusions. Subsequently, gradient-based features became an important development in the field of human detection. The introduction of the oriented gradient histogram feature provides a robust feature representation for object detection tasks. By capturing the distribution of gradient directions in the image, this feature effectively overcomes the impact of illumination changes. Since then, variants of the oriented gradient histogram feature have been widely used in various detection algorithms and have become an important cornerstone of human detection.

Modeling of human body structure is also an important direction of detection methods. The deformable component model is a widely used technology that solves the problem of human posture diversity by modeling local parts of the human body and their relative positions. In this model, the appearance of local parts is easier to model than the global appearance, and the training efficiency is improved by sharing component features. Although the deformable component model performs well in dealing with occlusion problems, its complexity may lead to performance bottlenecks in some scenarios.

In recent years, with the development of deep neural networks, human detection technology has made significant progress. Through training on large-scale datasets, deep learning models have shown excellent performance in image feature extraction and pattern recognition. In some tasks, the detection accuracy of deep neural networks has surpassed traditional methods. However, deep learning methods usually have high computational complexity, especially in sliding window detection, which will lead to a decrease in detection speed. To solve this problem, a common optimization strategy is to adopt a cascade detection architecture. First, a shallow network is used to preliminarily screen the candidate regions, thereby greatly reducing the number of candidate regions. Then, the deeper network only performs further fine detection on high-confidence regions. This method greatly improves the detection speed while ensuring detection accuracy. In addition, some studies have proposed a way to introduce complexity penalties in the optimization objectives to achieve a balance between detection accuracy and computational cost. By considering complexity loss in the learning process, the system can select lower-cost features in the early stages and postpone more expensive features to later stages. This strategy has achieved significant performance improvements in applications such as pedestrian detection.

Research has shown that head and hand trajectories play a significant role in emotion recognition. This trajectory representation can divide the emotional valence and arousal space into four quadrants, namely high valence and high arousal (such as entertainment, pride), high valence and low arousal (such as pleasure, relief, interest), low valence and high arousal (such as anger, fear, despair), and low valence and low arousal (such as cold anger, anxiety, sadness). The accuracy of emotion classification can be further improved by clustering the trajectory representation into compact emotion groups.

In the study of emotion classification based on body posture, some scholars used a simple representation of the upper body and divided body postures into six main emotion categories. These categories are often combinations of emotional states, such as anger-disgust, anger-fear, anger-happiness, fear-sadness-surprise, uncertainty-fear-surprise, and uncertainty-surprise. Although the amount of data and the number of participants in this type of research are limited, by training multiple classifiers, the results show that the Bayesian network classifier performs best in terms of accuracy.

In addition, for the recognition of dynamic body postures, researchers have compared the performance of multiple classification algorithms, including dynamic time warping, decision trees, and hidden naive Bayes. These algorithms effectively identify emotional states such as anger, happiness, pleasure, and sadness by modeling and classifying the time series of body postures. Among them, the combination of the dynamic time warping algorithm and the nearest neighbor classifier showed a higher accuracy.

Other studies have identified five basic emotional states, including anger, fear, happiness, sadness, and relaxation, through skeletal geometric features. In the comparison of different classifiers, the ensemble tree algorithm performed well in terms of accuracy. The use of skeletal features reduces the interference of environmental noise and enhances the robustness of the model. In some studies, independent modeling methods of body parts are also widely used in motion analysis. By decomposing and modeling the movements of parts such as arms, head, and torso, specific emotional expressions can be captured more accurately. In contrast, the structured body model describes human activities as a tree structure based on body parts, with each node representing a specific activity. This method has good interpretability and accuracy in identifying complex emotional states. At the same time, some researchers have further introduced background information and contextual features to enhance the effect of emotion recognition. By using convolutional neural networks to extract features from the body and background and fuse them, a variety of emotional states including calmness, affection, fatigue, and pain can be identified. In addition, this type of method can also output the intensity values of valence, arousal, and dominance, thereby providing more delicate emotion analysis.

Although body posture plays an important role in emotion recognition, it usually serves as a complementary signal to facial expression, voice, or context. Psychological research shows that humans are affected by body posture and situational factors when recognizing facial expressions. Therefore, combining language and non-verbal communication channels can form a more complete emotion recognition system. Multimodal fusion is one of the important development directions in the field of body language recognition in recent years. In some studies, researchers have explored the interaction between voice and body posture, revealing the impact of emotional state on these two communication channels. By extracting speech features such as prosody and Mel-frequency cepstral coefficients and combining them with the movement features of the head, lower body, and upper body, the interactive dynamics of speech and posture can be effectively modeled. Experimental results show that the multimodal recognition system significantly outperforms the single-modal system in the accuracy of emotion classification.

Another type of research uses the fusion of gesture and speech for emotion recognition. In this method, the gesture recognition module uses video and three-dimensional accelerometers to obtain motion data, while the speech recognition module uses open source speech recognition software for analysis. The advantages of the multimodal method are experimentally verified by using weighted criteria and majority voting to fuse the recognition results of the two modalities. In the study of fusing gesture and facial expression, the dual-modal emotion recognition method also achieved good results. The researchers extracted features from the dual video streams of the body and face and fused them at the feature level and decision level to achieve better recognition performance than a single modality. Further research also combined facial, body and speech data, adopted early and late fusion strategies, and used statistical classification methods for recognition, which significantly improved the recognition accuracy. In addition, some multimodal late fusion structures have also been proposed to cope with the challenge of emotion recognition in noisy environments. This method combines facial action units and high-dimensional body posture features through stacked generalization technology to recognize emotional states such as surprise, happiness, anger, sadness and fear. The results show that the multimodal method can still maintain high accuracy under noisy conditions.

% --------------------------------

\chapter{Voice Interaction Technology}

\section{Speech Recognition and Synthesis}
Speech is not just about words, it also carries a wealth of emotional information. Research shows that when expressing emotions and attitudes, speech itself conveys only a small part of the information, while non-verbal sound attributes and facial expressions together undertake a larger task of information transmission. Non-verbal sound attributes, including intonation, pitch, speaking speed, and timbre, can effectively reflect the speaker's emotional state. This feature makes speech play a vital role in daily communication. In the context of the continuous development of speech technology, emotional speech conversion has become a research field that has attracted much attention. The goal of emotional speech conversion is to convert the emotional state of speech from one emotion to another while maintaining the language content and the speaker's identity. This technology has a wide range of application scenarios, including film and television dubbing, virtual assistants, robot interaction, and emotional healing. By injecting emotions into synthesized speech, the system can present a more natural and human voice experience.

\begin{figure}
    \centering
    \includegraphics[width=0.5\textwidth]{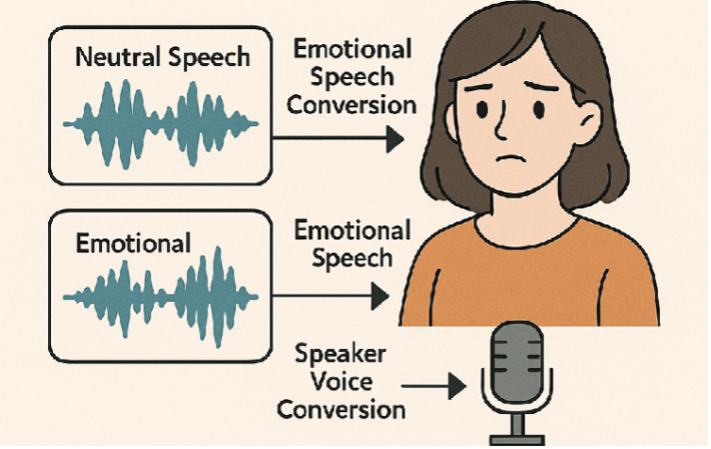}
    \caption{ Emotional speech conversion technology}
\end{figure}

In recent years, speech synthesis technology has made significant progress in generating natural and fluent speech. However, many existing systems still have shortcomings in terms of emotional expression. Although some speech synthesis systems can generate speech with simple emotion annotations, the emotional dimension and subtlety expressed in real scenes are far less than that of humans. To solve this problem, emotional speech conversion technology has emerged, which achieves accurate conversion of emotions by fine-tuning the spectral and prosodic features of speech. At the same time, speaker voice conversion technology has also made important progress. The core goal of speaker voice conversion is to convert the speaker identity of the source speech into the sound characteristics of the target speaker while keeping the semantic content unchanged. Early studies used statistical methods such as Gaussian mixture models and partial least squares regression to establish a mapping relationship between source features and target features. With the rise of deep learning, neural network-based methods have gradually become mainstream, including deep neural networks, recurrent neural networks, generative adversarial networks, and sequence-to-sequence models with attention mechanisms. These methods not only improve the conversion quality, but also achieve excellent performance in cross-speaker and cross-emotion tasks.

It is worth noting that both emotional speech conversion and speaker speech conversion involve the core technology of speech signal processing. Speech signal processing includes steps such as speech feature extraction, noise reduction, and signal enhancement, which provide a basis for subsequent speech modeling and conversion. In emotional speech conversion, in addition to the conversion of spectral features, the adjustment of prosodic features is particularly important. Prosodic features include pitch curve, speech rate change, and energy distribution, which directly affect the emotional expression of speech. In contrast, speaker speech conversion pays more attention to the personalized characteristics of sound, such as formant frequency and vocal tract length. In recent years, research on emotional speech conversion based on non-parallel training data has also made significant progress. Non-parallel data refers to data sets where the source emotion and the target emotion do not have a direct correspondence. Researchers have solved the problem of traditional methods' dependence on parallel data through methods such as domain translation, multi-task learning, and speaker disentanglement. In addition, the introduction of deep generative models such as generative adversarial networks and autoencoders also provides powerful modeling capabilities for emotional speech conversion.

In practical applications, emotional speech conversion not only needs to make breakthroughs in the accuracy of emotion conversion, but also needs to meet application requirements in terms of naturalness and real-time performance. For example, in human-computer interaction scenarios, the system needs to quickly respond to the user's emotional changes and achieve an emotion-driven conversation experience. In the future, with the further development of speech recognition and synthesis technology, emotional speech conversion systems that combine multimodal emotion recognition and natural language understanding will gradually become a reality. The following content will further explore the basics of speech signal processing and the latest progress in speech recognition technology in recent years. By gaining a deep understanding of these key technologies, we can more comprehensively grasp the core principles and application prospects of speech emotion conversion and speaker voice conversion.

\subsection{Basics of Speech Signal Processing}
Speech recognition is a technology that integrates knowledge from multiple fields such as linguistics, computer science, and electrical engineering. Its core goal is to convert human speech into text or instructions, and it is widely used in scenarios such as voice assistants, smart homes, healthcare, and customer service systems. The implementation of speech recognition involves a series of complex signal processing and machine learning technologies, including speech acquisition, preprocessing, feature extraction, acoustic modeling, language modeling, and ultimately speech-to-text conversion.

The processing of speech signals usually starts with the conversion of analog signals to digital signals. Speech is a continuous analog signal that needs to be converted into digital form through sampling and quantization techniques. The sampling rate determines the time resolution of the signal, while the number of quantization bits affects the amplitude resolution of the signal. Generally, the sampling rate of speech signals is 16kHz or 44.1kHz, and the quantization accuracy is usually 16 bits. After signal conversion, speech preprocessing is an indispensable link. The main goal of preprocessing is to improve the quality of the signal and reduce the interference of noise. Common preprocessing steps include noise suppression, silence detection, endpoint detection, and signal enhancement. Noise suppression technology can remove background noise through frequency domain or time domain methods, while silence detection and endpoint detection are used to locate the start and end positions of speech to ensure the effectiveness of subsequent processing.

Feature extraction of speech signals is a key step in speech recognition systems. The purpose of feature extraction is to convert the original speech waveform into a compact and representative feature vector. The most commonly used feature extraction methods include Mel-frequency cepstral coefficients (MFCC) and Mel-frequency cepstral coefficients. These methods extract spectral features by simulating the auditory characteristics of the human ear and provide a speech representation that is more in line with human auditory perception.

\begin{figure}
    \centering
    \includegraphics[width=0.5\textwidth]{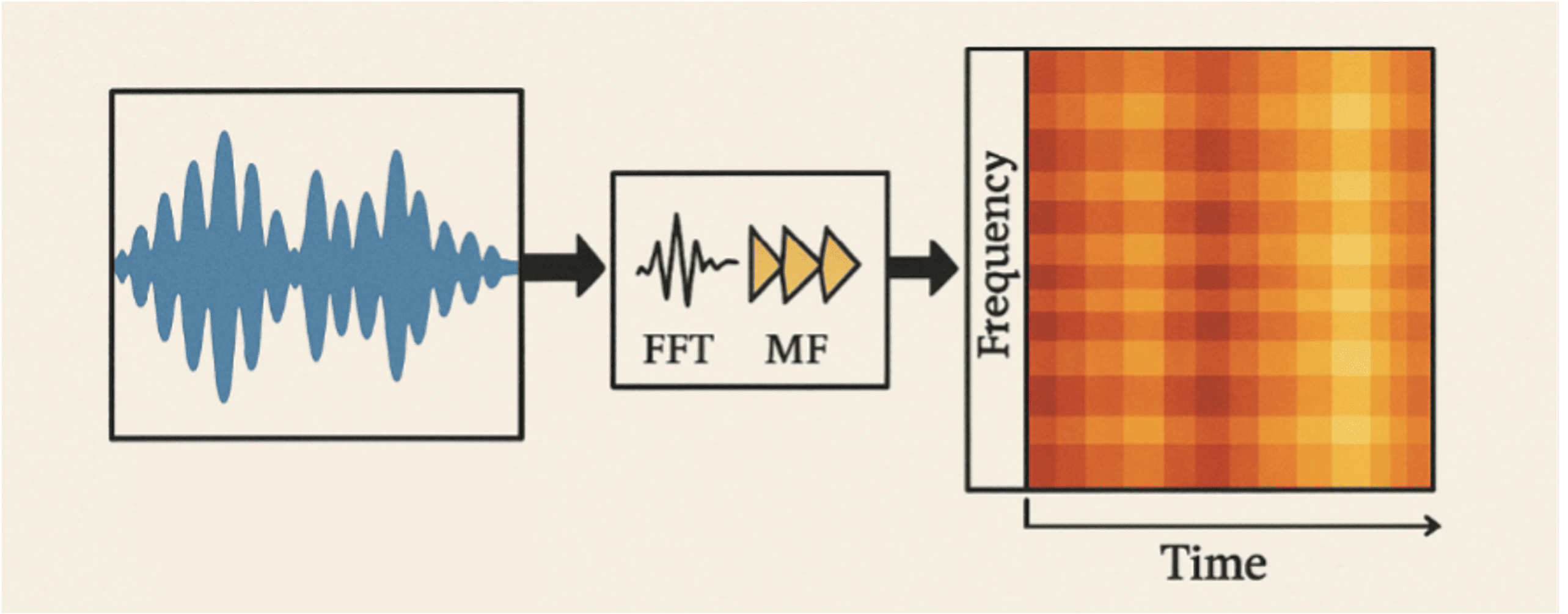}
    \caption{Feature extraction in speech processing}
\end{figure}

After the feature vectors are generated, the acoustic model maps these features to the corresponding speech units. The task of the acoustic model is to capture the relationship between phonemes and features. Traditional acoustic models mainly rely on hidden Markov models to model time series data, describing the dynamic changes of speech signals through state transition probabilities and observation probabilities. With the development of deep learning, deep neural networks have gradually become the mainstream method for acoustic modeling. Deep neural networks learn complex speech feature representations through multi-layer nonlinear mapping, significantly improving the recognition accuracy of the model.

Language model also plays a vital role in speech recognition. It uses grammatical rules and contextual information to predict the possibility of specific words or sentences. Language models based on statistical methods, such as n-gram models, were once widely used, while neural network language models in recent years have performed well in accuracy and robustness by learning language features from large-scale corpora. With the rapid development of deep learning and big data, speech recognition technology has made great progress. Traditional hybrid models, such as the combination of Gaussian mixture models and hidden Markov models, have gradually been replaced by deep neural network models. End-to-end deep neural network models no longer rely on manually designed features, but directly learn the mapping relationship between input speech and target text, which simplifies the system architecture and improves recognition accuracy.

As speech recognition technology matures and becomes more popular, voice assistants and voice interaction devices have become an important part of people's daily lives. From Apple's Siri to Amazon's Alexa, these intelligent voice assistants understand and respond to voice commands by integrating speech recognition and natural language processing technologies. In addition, open source tools such as Kaldi and LibriSpeech provide researchers and developers with a powerful speech recognition research platform, promoting the continued development of this field.

\subsection{Advances in speech recognition technology}

In the past decade, significant progress has been made in the neural network architecture in the field of speech recognition. From early basic models to current advanced deep learning networks, researchers are constantly exploring new methods to improve the accuracy and efficiency of speech recognition. Different types of neural network models have been applied, including recurrent neural networks, convolutional neural networks, and the Transformer network, which has received much attention in recent years. The introduction and improvement of these architectures have brought revolutionary improvements to speech recognition systems.

Early speech recognition research was limited by computing power and storage resources, and the lack of large-scale multilingual datasets was also a major obstacle. Researchers mainly relied on traditional statistical models, such as hidden Markov models based on Gaussian mixture models. Although these models modeled acoustic signals through mathematical reasoning and achieved certain results at the time, their performance in complex environments was still greatly limited. In 2010, the launch of the Google Voice Search application marked a major breakthrough in the field of speech recognition. Millions of users interacted with speech recognition technology through this application, generating a huge amount of speech data. This data was not only used to optimize the speech recognition model, but also laid the foundation for the launch of subsequent voice assistants such as Siri. At that time, the Google Voice Search system had stored more than 230 billion words from user searches. This data-driven approach greatly promoted the performance of the model.

In 2011, a study on large-vocabulary continuous speech recognition further promoted the progress of speech recognition technology. The researchers proposed a hybrid system that combines hidden Markov models and deep belief networks. This model replaced the traditional Gaussian mixture model and significantly reduced the recognition error rate. In tests, the accuracy of the system was significantly higher than that of the system based on the Gaussian mixture model. The success of this study has led researchers to pay extensive attention to the combination of neural networks and hidden Markov models, promoting the widespread application of neural networks in speech recognition. Soon after, another study published in 2012 further demonstrated the great potential of deep neural networks in speech recognition. By replacing the Gaussian mixture model with a deep neural network, the researchers successfully reduced the word error rate of the speech recognition system significantly. In multiple speech recognition tasks, the deep neural network model showed performance far superior to traditional methods. Even with limited training time and computing resources, deep neural networks still showed excellent performance. This discovery quickly sparked widespread interest in the application of deep learning in speech recognition in the research community.

With the continuous development of deep learning technology, researchers continue to explore ways to optimize the training of deep neural networks. The performance of deep neural networks has been significantly improved by properly initializing weights, using faster hardware devices, and adopting more efficient training algorithms. In addition, the increase in network depth and the use of context-dependent output units have further improved the accuracy of the model. At the same time, researchers have also developed a variety of methods to optimize the hyperparameters of neural networks, improve speech preprocessing processes, and achieve multi-dialect and multi-language speech recognition. The advantages of deep neural networks are not only reflected in the field of speech recognition, but have also achieved great success in tasks such as image classification and natural language processing. One of its key features is its powerful feature learning ability, which can autonomously extract relevant features from raw data, reducing reliance on manual feature engineering. With the development of hardware and deep learning frameworks, the training and deployment of deep neural networks have become more efficient, making this technology widely used.

However, deep neural networks also have certain limitations. Training large-scale deep neural networks usually requires a large amount of labeled data and computing resources, which poses a considerable challenge to small and medium-sized research teams. In addition, the overfitting problem is particularly prominent when data is insufficient, resulting in a decrease in the generalization performance of the model on unseen data. To address these problems, researchers have introduced new architectures such as recurrent neural networks and convolutional neural networks. Recurrent neural networks perform well in processing sequence data and are particularly suitable for capturing time-dependent information. They use a cyclic connection structure to associate the current output with the previous state, thereby effectively modeling the temporal characteristics of speech signals in speech recognition tasks. Subsequently, convolutional neural networks have also been widely used in speech recognition tasks due to their advantages in extracting local features. Convolutional neural networks can extract the time-frequency characteristics of audio signals through convolution operations, further improving the accuracy of speech recognition.

\begin{figure}
    \centering
    \includegraphics[width=0.5\textwidth]{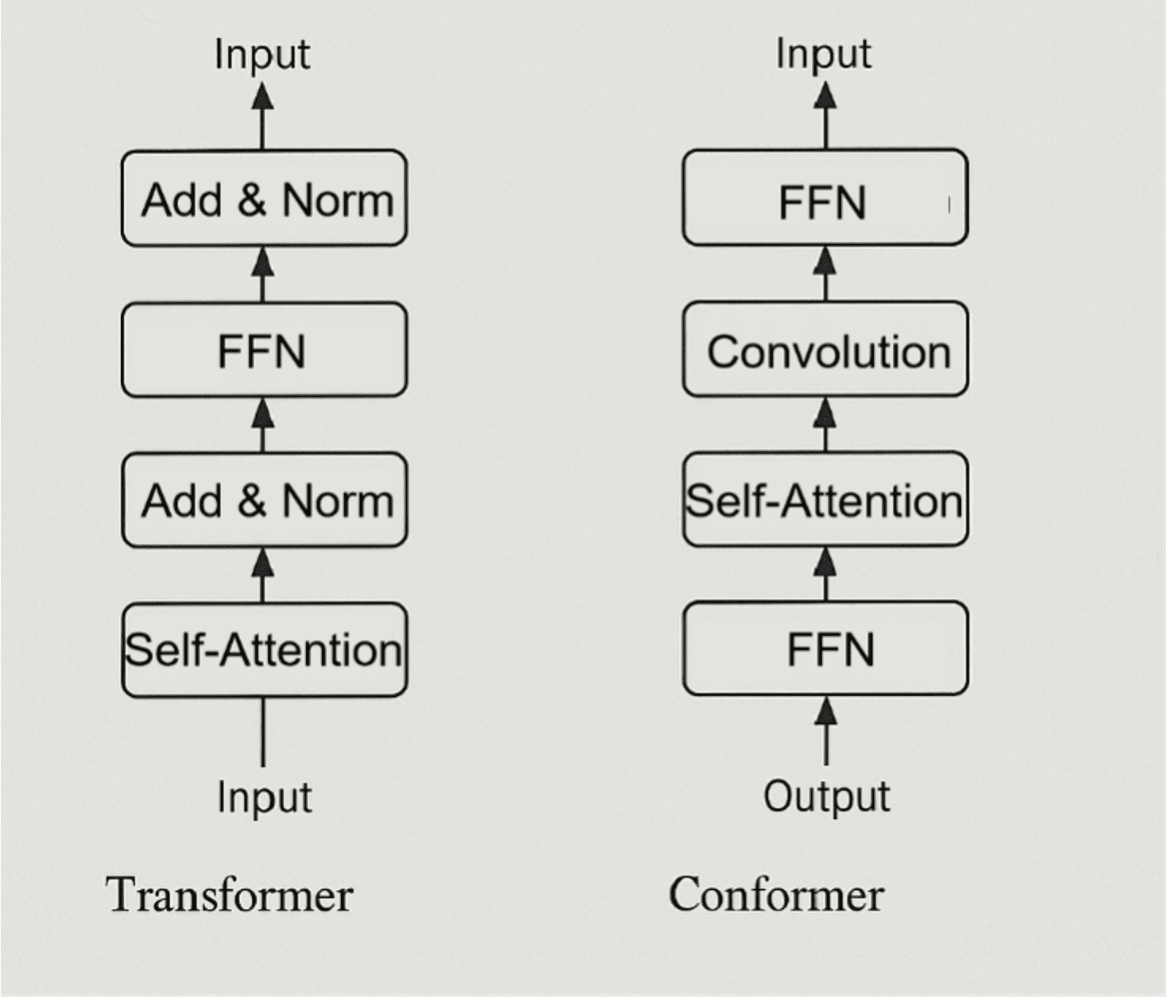}
    \caption{Transformer and Conformer network architecture}
\end{figure}

In recent years, the emergence of Transformer and Conformer networks has further promoted the development of speech recognition technology. Transformer networks have attracted attention for their excellent performance in capturing long-distance dependencies with their self-attention mechanism. They can process sequence data in parallel, greatly improving training speed and model performance. Conformer networks combine the local feature extraction capabilities of convolutional neural networks with Transformer, taking into account the capture of long-distance dependencies and local features, bringing significant performance improvements to speech recognition.

\section{Emotional Speech Analysis}
As an important research direction in the field of speech signal processing and pattern recognition, emotional speech analysis has received extensive attention in recent years. With the continuous development of artificial intelligence technology, the acquisition and understanding of emotional information in human-computer interaction scenarios is particularly important. Emotional speech not only carries the speaker's language information, but more importantly, it conveys rich emotions and psychological states. By deeply exploring the emotional components contained in speech signals, we can better understand the expression mechanism of human emotions, thereby improving the performance of intelligent systems in application fields such as speech recognition, emotional computing, and human-computer interaction. In the research process of emotional speech analysis, there is both the construction of theoretical models and experimental verification based on actual data. This requires researchers to have interdisciplinary knowledge reserves, master the basic principles of acoustic signals, and understand the laws of emotional expression in psychology and linguistics. When faced with complex and changeable emotional information, researchers need to consider how to extract effective emotional features and associate these features with emotional categories through reasonable algorithms. This involves both traditional signal processing technology and the support of modern machine learning methods.

In the overall framework of emotional speech analysis, data preprocessing, feature extraction and emotion classification constitute the basic process. Among them, data preprocessing mainly involves noise reduction, framing, windowing and other processing of the original speech signal to ensure that the subsequent feature extraction process can be carried out on the basis of relatively clean data. In fact, as a non-stationary signal, the inherent information content of the speech signal is often interfered by various factors such as the recording environment and individual differences of the speaker. Therefore, in the preprocessing stage, how to balance the retention of information and the removal of noise has become a problem that must be faced in emotional speech analysis. At the same time, the emotional information in the speech signal is usually manifested as changes in multiple levels such as spectrum, time domain and energy distribution. A single feature is often difficult to fully characterize the complexity of the emotional state, which requires researchers to conduct a comprehensive analysis of the speech signal from multiple dimensions.

In order to solve this problem, academia and industry have proposed a variety of feature extraction methods for emotional speech. By extracting a variety of acoustic features including fundamental frequency, resonance peaks, spectral envelopes, energy curves, and speech speed, a multidimensional feature space describing emotional changes is constructed. These features can not only reflect the physical characteristics of speech signals, but also reveal the emotional tendencies of speakers to a certain extent. For example, in emotional states such as excitement or anger, the energy and fundamental frequency of speech signals are often higher, while in calm or depressed emotions, they may show relatively low frequency and energy levels. Based on this, different emotional states will show their own unique distribution characteristics in the feature space, thus providing a basis for subsequent emotional classification. In this context, how to efficiently and accurately extract and utilize these features has become a core issue in the study of emotional speech analysis.

On the other hand, the analysis methods of emotional speech are also constantly evolving, from the initial rule-based methods to the deep learning models that have emerged in recent years, and the technical means are constantly enriched. Traditional methods mainly rely on expert knowledge to build emotional models and achieve emotion recognition through statistical methods or pattern matching; the development of deep neural networks in recent years has brought revolutionary changes to emotional speech analysis, especially the advantages of convolutional neural networks and recurrent neural networks in processing time series data, making automatic feature extraction and emotion classification possible. In practical applications, these methods can often overcome the limitations of traditional algorithms that rely too much on feature design, directly learn high-level emotional representations from raw speech signals, and significantly improve the accuracy and robustness of emotion recognition. However, deep learning methods also pose new challenges to the demand for large amounts of labeled data and computing resources. In scenarios where data is insufficient or the labels are inconsistent, their generalization ability still needs to be further improved.

Emotional speech analysis not only involves the technical implementation of signal processing and pattern recognition, but also involves the cross-integration of multiple disciplines such as psychology and cognitive science. From theory to practice, every step needs to fully consider the complexity of the speech signal itself and the diversity of emotional expression. Current research results have verified the effectiveness of the solution based on the combination of acoustic features and data-driven methods in emotion recognition to a certain extent, and also laid a solid foundation for subsequent system optimization and application expansion. In actual engineering applications, whether it is intelligent customer service, voice assistant or emotional computing system, higher requirements are put forward for emotional speech analysis, which promotes the continuous development of this field.

In the following content of this chapter, the feature extraction method and analysis method of emotional speech will be further discussed in detail. First, the theoretical basis, extraction algorithm and practical application effect of various acoustic features will be deeply analyzed to explore how to separate feature information that is sensitive to emotions from complex speech signals; secondly, various emotional speech analysis methods that have emerged in recent years will be introduced in detail, including the latest progress based on traditional machine learning models and deep learning models, and their advantages and disadvantages in practical applications will be discussed. Through the analysis of these key technologies, we hope to provide readers with a comprehensive and systematic emotional speech analysis framework that can not only reflect the latest progress of current research, but also provide theoretical support and practical guidance for future development directions. The entire chapter strives to build a bridge between theory and practice, so that the field of emotional speech analysis is not only forward-looking in academia, but also shows strong vitality and broad prospects in practical applications.

\subsection{Feature extraction of emotional speech}
Speech features are the most critical part of emotional speech recognition. Researchers have explored and used many features for emotional speech recognition. However, a universally applicable standard solution for speech feature extraction and specific classifiers has not yet been formed. In order to achieve better results in emotional speech recognition, some studies have introduced discriminant information to retain local information. The extraction method of speech features depends on actual needs and is mainly divided into local features, global features, or both. Local features reflect the dynamic changes of speech signals in a short period of time, also known as short-term features or segmental features; while global features characterize the overall characteristics of the entire speech segment through statistical methods, including parameters such as minimum, maximum, mean, and standard deviation. Such features are also called long-term features or suprasegmental features.

In emotional speech recognition, local and global features are mainly divided into four categories: spectral features, prosodic features, Teager energy operator features, and speech quality features. Spectral features are determined by the characteristics of the vocal tract and can well characterize speech signals in the frequency domain. The time domain signal can be converted into a frequency domain signal through Fourier transform, thereby extracting spectral features. One of the most commonly used spectral features is the Mel frequency cepstral coefficient. The conversion relationship between the Mel frequency and the actual frequency can be expressed by the following formula:

\begin{equation}
    m = 2595 \log_{10} \left( \frac{f}{700} + 1 \right)
\end{equation}
Before converting to the frequency domain, speech signals are usually divided into several frames. Mel-frequency cepstral coefficient features can then be extracted through inverse Fourier transform. In addition, linear prediction cepstral coefficients are also an important spectral feature that can effectively capture the resonance characteristics of the vocal tract and thus obtain emotional information in speech. Similarly, gammatone frequency cepstral coefficients are also widely used in emotion recognition. Studies have shown that different combinations of spectral features such as linear prediction cepstral coefficients, Mel-frequency cepstral coefficients, cepstral features, etc. can further improve recognition performance.

Prosodic features are another important type of features in emotion recognition, which usually include parameters such as pitch, duration and energy. The maximum, minimum, variance, range, mean and standard deviation of pitch can all be used as powerful features to extract emotional information. The changes in prosodic features are directly affected by the vibration of the vocal cords and the air flow, so these features carry a lot of emotion-related information. Researchers generally divide prosodic features into pitch features, intensity features and intonation features. Statistical parameters such as mean, standard deviation, skewness, kurtosis, etc. further enrich the representation of emotional features. Since humans are highly dependent on prosodic features when perceiving emotions, these features have shown significant advantages in emotion recognition tasks.

The Teager energy operator feature is also widely used in emotional speech recognition. It is based on the principle of energy detection and simulates the perception of speech energy by the human auditory system. The nonlinear formula of the Teager energy operator is as follows:

\begin{equation}
    U[f(n)] = f^2(n) - f(n-1)f(n+1)
\end{equation}

Where u is the Teager energy operator and f(n) represents the speech signal. The Teager energy operator feature is particularly effective under stress conditions because it can capture key frequency bands and frequency changes in the speech production process. Under stress conditions, changes in harmonic distribution lead to significant differences in speech energy distribution, and the Teager energy operator feature can identify emotional states through these changes.

Finally, speech quality features are an important class of parameters used to describe the characteristics of individual voices. Speech quality features play a vital role in many speech processing fields, such as speaker identification and emotion recognition. Glottal source characteristics determine speech quality features, including parameters such as formant frequency, bandwidth, glottal parameters, harmonic noise ratio, jitter, and shimmer. These features have shown excellent ability in capturing the emotional content in speech. Speech quality features not only reflect the physiological characteristics of speech, but also reflect the emotional state of individuals.

\subsection{Emotional speech analysis methods}

The application of traditional machine learning methods in the field of speech signal processing mainly focuses on the combination of feature extraction and classification models. Researchers have extensively evaluated a variety of classifiers to improve the accuracy of speech emotion recognition. Common traditional classifiers include support vector machines, Gaussian mixture models, hidden Markov models, artificial neural networks, and k-nearest neighbor algorithms. In order to optimize feature selection, some studies have introduced methods based on Fisher's criterion and correlation analysis, and conducted in-depth tests on different data sets. For example, in one study, researchers used extreme learning machines, support vector machines, back propagation neural networks, and k-nearest neighbor classifiers, and achieved average accuracies of 89.9\%, 87.2\%, 82.3\%, and 80.7\% on the CASIA data set, respectively. In addition, the application of support vector machines and hidden Markov models on the SUSAS database has also achieved considerable recognition results.

Many studies have further explored the impact of different feature extraction methods on emotion recognition. For example, linear prediction analysis and Mel-frequency cepstral coefficients were used for feature extraction, combined with generalized fuzzy neural networks and support vector machines for classification, achieving 98\% and 82\% accuracy on the EMODB database, respectively. In addition, some studies have used hybrid methods to conduct experiments on multiple databases, using support vector machine classifiers to identify four emotions: anger, neutrality, sadness, and happiness, achieving an average accuracy of 73\%. Some researchers have also proposed a method based on Fourier parameters and tested it on multiple databases, achieving the best accuracy of 71\%. Overall, traditional methods mainly rely on manually designed feature extraction and classification models, and their performance is limited to a certain extent by feature selection.

With the rapid development of deep learning technology, the field of speech emotion recognition has ushered in new breakthroughs. Deep learning is considered a subset of machine learning. It realizes the automatic learning of complex features through multi-layer neural networks. Compared with traditional methods, deep learning has stronger feature expression ability and can automatically learn more comprehensive features, thereby improving the accuracy of emotion recognition. In recent years, emotion recognition models based on deep learning have achieved remarkable results, showing excellent flexibility, scalability and high recognition accuracy.

Current deep learning methods are mainly divided into three categories: methods based on automatically learned features, methods based on manual features, and methods based on spectrograms. Among them, convolutional neural networks are widely used in emotion recognition research. For example, a study proposed a spatiotemporal and frequency cascade network, which achieved an average accuracy of 71.98\%, 82.1\%, 75.6\%, and 54.75\% on the IEMOCAP, EMODB, eNTERFACE, and SAVEE databases, respectively. In addition, a deep emotion classification network for emotion recognition was proposed to correct the errors of traditional methods and improve the recognition effect. Long short-term memory networks, as a deep learning model suitable for sequence data, have also been widely used in emotion recognition. For example, some researchers evaluated models based on long short-term memory networks on the RAVDESS database and achieved an improvement in recognition accuracy. In addition, one-dimensional convolutional neural networks have also been used for emotion classification, further improving the generalization ability of the model. Many studies have also converted speech signals into spectrograms and applied convolutional neural networks for emotion recognition. This method has achieved good results on multiple databases. For example, on the SAVEE database, the spectrogram feature extraction method combined with a deep neural network achieved the best accuracy of 81.7%.

\begin{figure}
    \centering
    \includegraphics[width=0.5\textwidth]{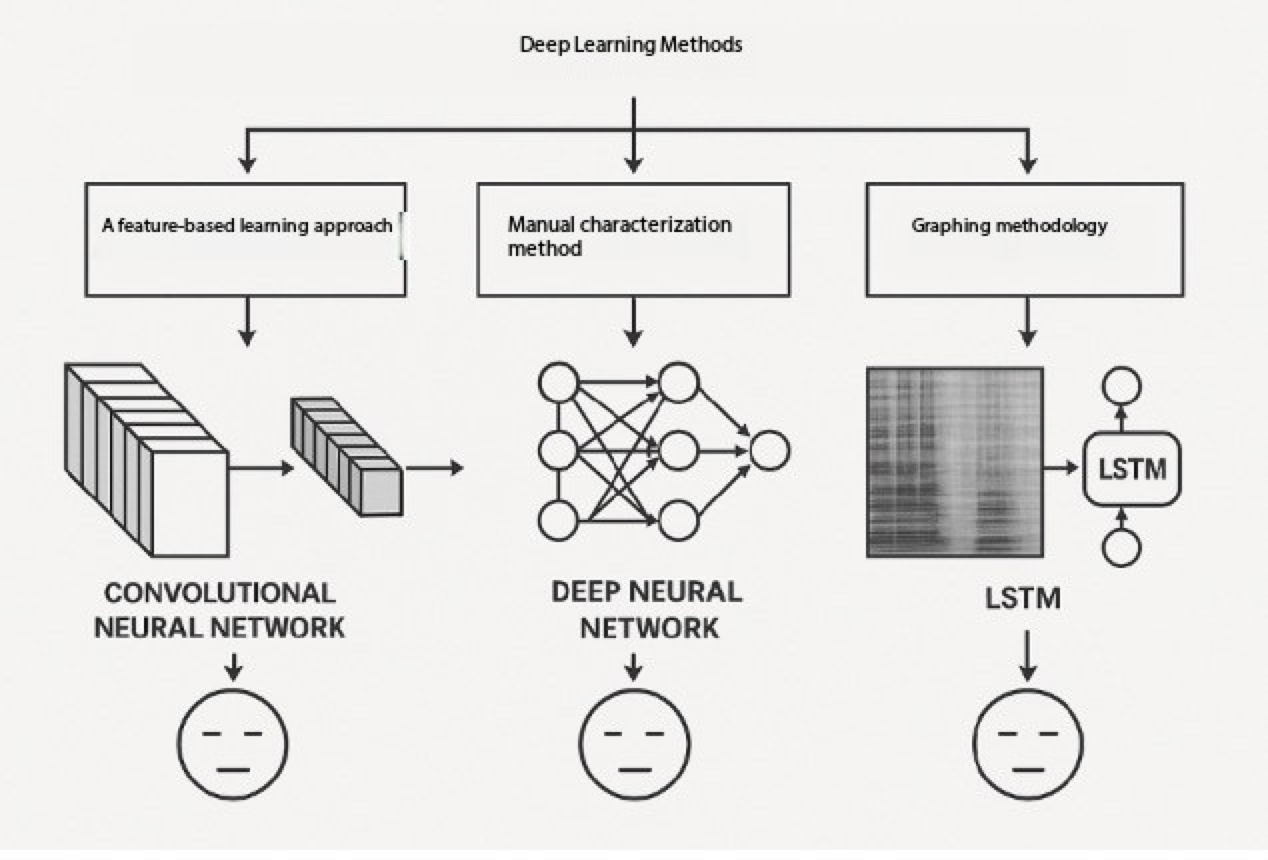}
    \caption{Main types of deep learning}
\end{figure}

Hybrid deep learning models show high robustness in emotion recognition. For example, researchers have proposed a method of combining convolutional neural networks with other classifiers to improve the accuracy of emotion recognition. In addition, models such as GoogleNet, three-dimensional convolutional neural networks, restricted Boltzmann machines, and deep belief networks have also been used for emotion recognition and tested on multiple databases. The introduction of models such as autoencoders, recurrent neural networks, and bidirectional long short-term memory networks has further enhanced the performance of deep learning methods in the field of speech emotion recognition. For example, a study used a bidirectional long short-term memory network and achieved a high accuracy of 93.97\% on multiple databases.

In addition, some researchers have combined manual features with deep neural networks for emotion recognition and tested them on multiple databases. For example, a classifier based on a deep belief network combined with the eGEMAPS feature extraction method improves the accuracy of emotion recognition. Another study used the openSMILE tool to extract features and combined recurrent neural networks with long short-term memory networks to achieve a high recognition rate. The method of automatically learning features through convolutional neural networks is also widely used in emotion recognition, such as extracting Mel frequency cepstral coefficient features and combining them with recurrent neural networks for classification, which achieved good recognition results on the EMODB database. .

From the overall research trend, few researchers adopted deep learning methods in the field of emotional speech recognition until 2013. Most researchers tended to use deep belief networks as classifiers. In 2014, convolutional neural networks gradually became a research hotspot, especially the convolutional neural network method based on spectrograms received widespread attention. By 2015, with the application of recursive neural networks, researchers achieved a best recognition accuracy of 81\% on the RECOLA database. In 2016, researchers further explored the performance of multiple classifiers such as recursive neural networks, deep belief networks, convolutional neural networks, and deep neural networks. In 2017, the performance of deep belief networks on the CASEC database was particularly outstanding, achieving a best accuracy of 94.60\%. In 2018, the accuracy of the model based on deep convolutional neural networks on the EMODB dataset increased to 92.71\%.

In 2019, the research direction showed a diversified trend, and researchers began to try to integrate different models to improve the accuracy of emotion recognition. For example, the combination of a two-dimensional convolutional neural network and a long short-term memory network achieved the best accuracy of 95.89\% and 89.16\% on the EMODB dataset and IEMOCAP dataset, respectively. In 2020, convolutional neural networks remained the mainstream method of research, especially on the EMODB dataset, where the best recognition accuracy of 95\% was achieved. In 2021, researchers continued to explore convolutional neural networks and hybrid methods in depth, further improving the performance of the model.

In terms of the use of research databases, most researchers chose IEMOCAP, EMODB, RAVDESS, and SAVEE databases for model evaluation. Among them, the IEMOCAP database is the most commonly used, accounting for 48\%; followed by the EMODB database, accounting for 38\%; RAVDESS and SAVEE databases account for 17\% and 10\% respectively. These databases cover a variety of emotion categories and scenarios, providing rich speech samples.

In terms of method selection, convolutional neural networks are the most commonly used deep learning method by researchers, followed by long short-term memory networks and recurrent neural networks. According to the application method, there are three typical ways to use deep learning methods in emotional speech recognition: the first way is to extract speech features and then input them into a deep learning model for emotion recognition. This method relies on a manually designed feature set, such as Mel-frequency cepstral coefficients and prosodic features. The second way is to convert speech signals into spectrograms and then use deep learning models for recognition. Convolutional neural networks are particularly outstanding in this field. The third way is to automatically learn speech features through deep learning models, eliminating the process of feature engineering and greatly simplifying model training and optimization. For example, a study used a hybrid model of convolutional neural networks and long short-term memory networks to extract features from spectrograms, achieving the best accuracy of 89.16\% and 95.89\% on the IEMOCAP and EMODB databases, respectively. In another study, long short-term memory networks were used to automatically learn speech features, achieving the best accuracy of 84.30\% on the RAVDESS database. Additionally, in a study on the SAVEE database, researchers used convolutional neural networks to extract features from spectrograms and reported a best accuracy of 81.05\%.

Although convolutional neural networks and long short-term memory networks dominate, some researchers have also explored methods such as autoencoders, deep belief networks, and deep Boltzmann machines. Although these methods are relatively less used, they also show certain advantages in specific scenarios. From the research trends, it can be seen that the application of deep learning methods in the field of emotional speech recognition has increased year by year since 2013. Currently, most of the proposed models use deep learning methods and have achieved significant improvements in recognition accuracy and computational cost.

% ---------------------------------------------

\chapter{Brain-Computer Interface Technology}

\section{Working principle of brain-computer interface}
Brain-computer interface is a revolutionary communication technology that enables people to interact with external devices by thinking alone by directly capturing and analyzing brain signals, without relying on traditional neuromuscular output pathways such as peripheral nerves or muscle activity. Its core working principle is to use high-precision sensors to monitor various physiological changes in the brain when performing specific mental activities. These activities may include imagining hand movements, trying to express language, or focusing on a task. During these processes, the brain generates electrical signals, blood flow changes, or other measurable indicators. Then, through advanced signal processing algorithms and machine learning models, these complex and changing signals are converted into clear and actionable instructions. These instructions can drive a variety of external devices, such as spelling tools, mobile robots, exoskeleton systems, prostheses, and even virtual reality systems and smart home devices, thus realizing an amazing vision of "mind control". This technology is unique because it breaks through the traditional human-computer interaction dependence on physical actions and establishes an unprecedented way for humans to connect with the digital world. It not only expands the boundaries of technology applications, but also redefines the relationship between humans and machines.

\begin{figure}
    \centering
    \includegraphics[width=0.5\textwidth]{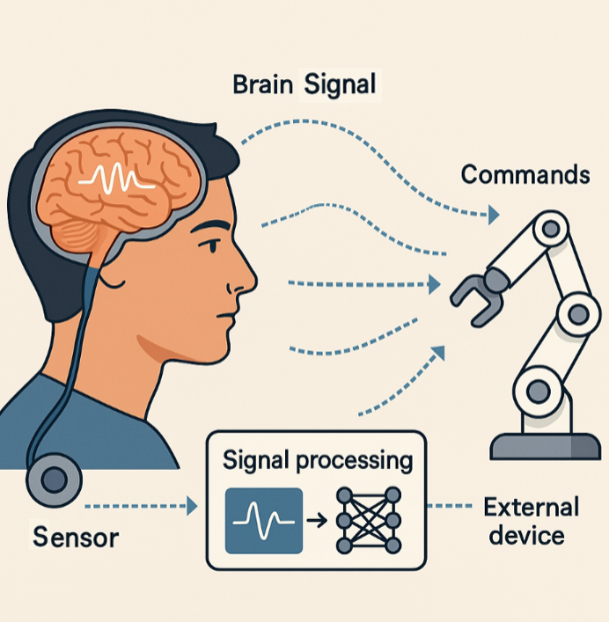}
    \caption{Schematic diagram of the working principle of brain-computer interface}
\end{figure}

The origin and development of brain-computer interface technology are deeply driven by medical needs, especially for patients with severe movement disorders caused by diseases or accidents. For example, individuals with amyotrophic lateral sclerosis (ALS), brainstem stroke, spinal cord injury or quadriplegia often lose the ability to communicate with the outside world through body movements. Brain-computer interface provides them with an alternative means of interaction. Imagine that a patient can gradually select letters, spell out complete words, and communicate with family members or medical staff by focusing on the picture of hand movements in his mind or staring at the flashing signals on the screen. This "alternative" function makes brain-computer interface an important tool to compensate for the loss of physical function. Its typical applications include controlling the precise movement of prostheses, manipulating the direction of movement of wheelchairs, or interacting with computer interfaces through brain signals. These functions not only help patients regain a certain degree of independence, but also significantly improve their quality of life. At the same time, another great value of brain-computer interface lies in the "restoration" function. With the help of the natural plasticity of the brain and combined with systematic rehabilitation training, this technology can help patients gradually activate damaged neural pathways. For example, stroke patients may gradually recover the basic ability to grasp objects, stand, and even walk after receiving brain-computer interface-assisted training several times a week for several weeks under the guidance of professional therapists. These successful practices in the medical field not only verify the great potential of brain-computer interfaces in the intersection of neuroscience and engineering, but also lay a solid foundation for its expansion into a wider range of application scenarios.

With the rapid progress of cognitive neuroscience, brain imaging technology, microelectronics technology and artificial intelligence, the performance of brain-computer interface systems has been significantly improved in the past decade. Early systems mainly relied on simpler EEG signal acquisition methods, with low signal resolution and slow processing speed. For example, it may take a patient several seconds to select a letter through brain signals, and the function seems to be single and the efficiency is limited. Today, thanks to the promotion of technological innovation, modern brain-computer interface systems are able to capture brain activity through more precise equipment. For example, some invasive technologies can directly record the activity of neurons by implanting microelectrodes in the brain; while non-invasive devices can achieve more accurate monitoring of brain signals through improved electroencephalography technology or near-infrared spectroscopy technology. These advances enable signal decoding to be carried out in near real time. For example, the latest research shows that patients can directly generate text output by imagining themselves writing specific words. This "mental spelling" technology greatly improves the efficiency of communication. At the same time, the reduction of hardware costs, the miniaturization of equipment size and the optimization of software algorithms have also significantly enhanced the flexibility and practicality of brain-computer interfaces in practical applications. These technological breakthroughs not only push brain-computer interfaces from the laboratory to reality, but also provide more possibilities for us to deeply understand their working mechanisms.

From the perspective of technological evolution, the development of brain-computer interfaces has roughly gone through three stages. Initially, it was the proof-of-concept stage, where researchers demonstrated the basic ability to extract signals from the brain and drive devices. Although the functions were limited, they were enough to inspire people's imagination of the future. Next came the simulation stage, during which the technology began to simulate the functions of traditional input devices, such as using brain signals to replace the mouse or keyboard to complete the selection, but the overall experience has not yet brought about a disruptive change. Today, brain-computer interfaces are moving towards maturity, and researchers are beginning to tap their unique potential and design new experiences that go beyond traditional interaction modes. For example, future systems may dynamically adjust the difficulty of games or work environments by monitoring emotional states in real time; or adaptively optimize interface design based on the user's cognitive load. These innovations have expanded the application of brain-computer interfaces from simple functional replacement or restoration to a new dimension of enhancing human capabilities and optimizing daily life.

To fully understand the working principle of brain-computer interface, we need to analyze from the source of the signal. First of all, the acquisition of EEG signals is the cornerstone of the entire system, which determines whether brain activity can be accurately captured and converted into usable data. Then, the transmission link of EEG signals efficiently transmits these weak signals to the processing unit to ensure the real-time and integrity of the information. On this basis, the classification and application of brain-computer interfaces show the diversified development direction of technology, from medical rehabilitation to consumer entertainment, presenting a broad prospect. Below, we will explore these key parts one by one to reveal how brain-computer interfaces transform from subtle fluctuations in the brain into a powerful force to change lives.

\subsection{EEG signal acquisition}
EEG signals are an important part of biological signals. By recording the electrical activity of neurons in the cerebral cortex, they can reflect the functional state of the brain and have a wide range of applications in the biomedical field. EEG signals are not only of great value in brain-computer interfaces, diagnosis of neurological diseases (such as epilepsy, sleep disorders), cognitive research, etc., but are also used for daily health monitoring. As a voltage-type biological signal, its typical amplitude ranges from a few microvolts to hundreds of microvolts, and its frequency is mainly distributed between 0.5 Hz and 100 Hz. The signal amplitude is low and is susceptible to noise interference, especially 1/f noise (flicker noise) and environmental noise (such as 50/60 Hz power frequency noise). Therefore, the EEG signal acquisition system needs to have high precision, low noise and strong anti-interference capabilities to ensure that the signal quality can meet the needs of subsequent analysis and application.

EEG signal acquisition mainly relies on analog front-end circuits and system technology, and its design goals include improving signal fidelity, reducing power consumption, reducing chip area, and enhancing system integration. A typical EEG signal acquisition system consists of electrodes, amplifiers, multiplexers, and analog-to-digital converters. The electrodes are used to capture weak electrical signals on the surface of the scalp or cortex, and then the signals are amplified, filtered, and digitized by the analog front-end circuit to form data that can be analyzed by the digital signal processing unit. The performance of the acquisition system directly determines the signal quality, and the key indicators include input reference noise (usually less than 5 microvolts root mean square value), common mode rejection ratio (usually required to exceed 100 decibels), input impedance (needed to be greater than 1 gigahertz to reduce the impact of the electrode-skin interface), and electrode offset tolerance (ranging from ±50 millivolts to ±350 millivolts, depending on the specific application). These indicators determine the sensitivity and anti-interference ability of the system, which is crucial for the accurate acquisition of EEG signals.

\begin{figure}
    \centering
    \includegraphics[width=0.5\textwidth]{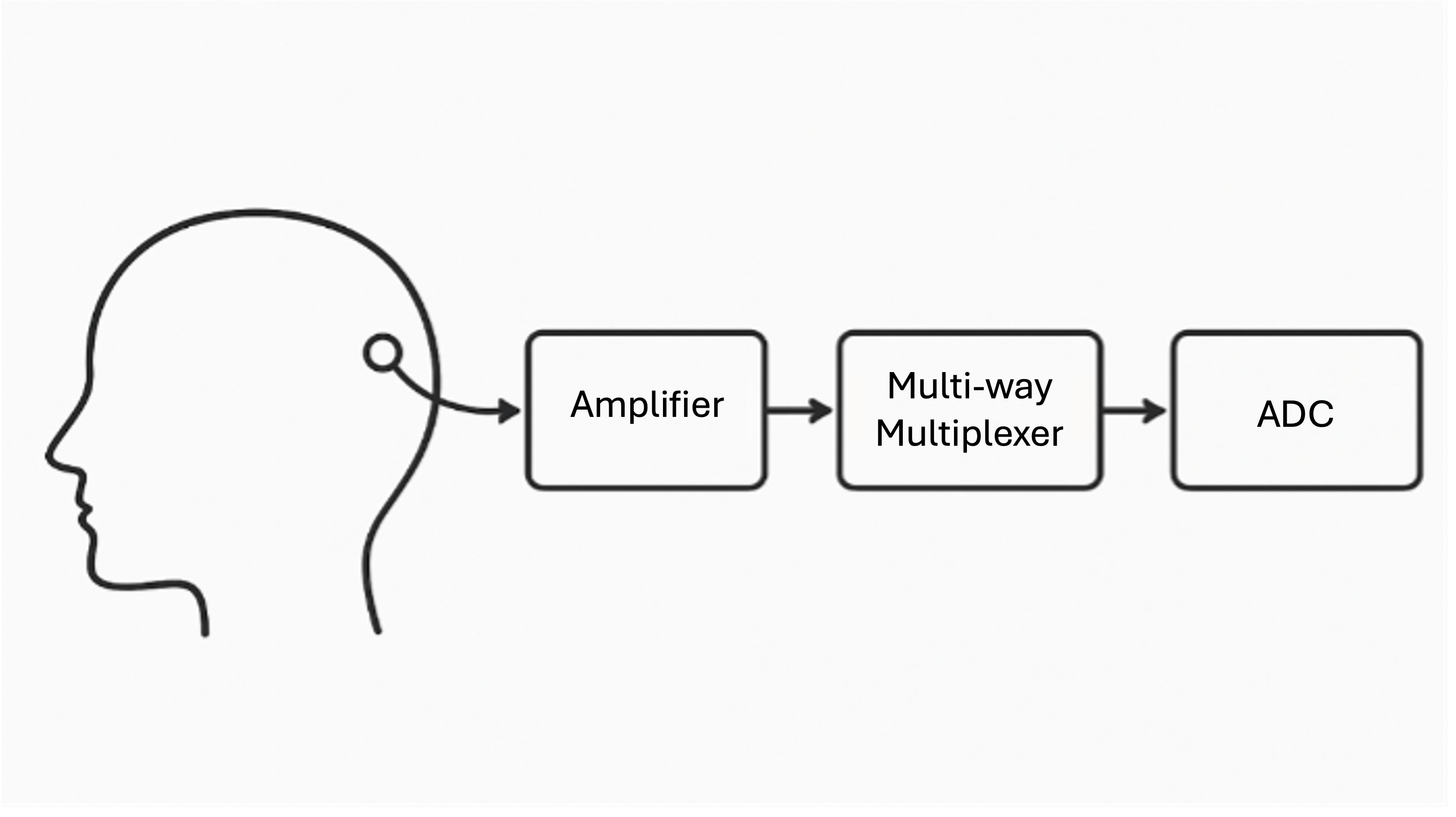}
    \caption{EEG signal acquisition process}
\end{figure}

Traditional EEG signal acquisition architectures mostly use multi-channel designs to achieve synchronous monitoring of different brain areas. Among them, time-division multiplexing technology is a common method that reduces the occupation of hardware resources by sampling multi-channel signals on the time axis and sharing a single analog-to-digital converter. For example, the classic time-division multiplexing architecture includes multiple low-noise amplifiers and an analog multiplexer. After each channel signal is amplified by the low-noise amplifier, it is transmitted to the analog-to-digital converter in sequence. However, when the number of channels increases significantly (such as more than dozens), the time-division multiplexing architecture faces certain technical bottlenecks. In order to avoid signal loss, the sampling rate of the analog-to-digital converter needs to be increased, which will lead to a significant increase in power consumption; at the same time, the crosstalk noise in the analog multiplexer may reduce the signal quality.

In order to solve these problems, frequency division multiplexing technology was introduced, which modulates the signals of each channel to different frequencies and then merges them into a single signal line, and demodulates them through a single analog-to-digital converter. This method not only improves the dynamic range of the system, but also reduces power consumption and redundant hardware requirements, and improves the overall efficiency of the acquisition system. In addition, the fast multiplexing technology proposed in recent years further optimizes the signal acquisition process. This technology significantly reduces the number of amplifiers by directly multiplexing signals at the electrode end instead of multiplexing after amplification, making ultra-high density recording possible, and providing a new technical path for future high-channel EEG acquisition systems.

According to the way of handling DC offset, the analog front end (AFE) of EEG acquisition is mainly divided into two categories: AC coupling and DC coupling, each with its own advantages and limitations. The AC-coupled analog front end uses a capacitor at the input of the amplifier to block the DC offset voltage generated by the electrode-tissue interface, which usually comes from the contact between the electrode material and the skin. The setting of the low cutoff frequency depends on the capacitor and pseudo-resistance in the feedback loop, which is usually less than 1 Hz to ensure that the low-frequency components of the EEG signal are retained. However, since the EEG signal is extremely weak and easily interfered by 1/f noise, chopping technology is widely used to reduce noise. For example, a typical AC-coupled chopper instrumentation amplifier can effectively reduce 1/f noise and amplifier offset through signal modulation at the input stage and secondary chopping at the output stage. The measured common-mode rejection ratio can exceed 120 dB, and the input-referred noise density can be as low as 57 nV/rtHz. Although this type of design can improve signal quality, the input capacitor and feedback capacitor occupy a large chip area, which limits its application in multi-channel or implantable systems. In addition, the existence of chopping spike noise requires additional filtering circuits to eliminate it, which increases the complexity of the design.

In contrast, DC-coupled analog front ends handle DC offsets by removing input capacitors and integrating low-pass filters or servo loops in the feedback loop. This design approach enables a more compact chip layout and is particularly suitable for invasive EEG recordings such as cortical electroencephalography. Some high-performance bioamplifiers use active integrators to achieve high-pass filtering, combined with small Miller capacitors and equivalent resistors of specific structures, to significantly reduce chip area while maintaining high input impedance. In addition, advanced dual mixed-signal servo loop architectures combine digital low-pass filters with digital-to-analog converters to dynamically cancel offsets, reducing chip area to 0.013 mm² while supporting synchronous acquisition of local field potentials and spike signals. However, DC-coupled schemes are sensitive to the nonlinear characteristics of the circuit, especially the nonlinear effects of pseudo-resistances, which may affect system stability and require additional calibration mechanisms in high-offset environments.

In recent years, in order to further improve the efficiency of EEG acquisition systems, module reuse and sparsity utilization technologies have become research hotspots. The module reuse method based on successive approximation registers integrates data conversion and signal processing functions (such as finite impulse response filtering and feature extraction) in the same module, thereby reducing hardware redundancy. For example, this method can simultaneously realize dot product calculation and signal digitization, which is particularly suitable for neural vector analysis systems. In addition, low-power analog-to-time conversion technology can directly convert electrode signals into time-encoded pulses, eliminating the need for a preamplifier and further reducing system power consumption. These technologies are particularly important in high-channel density scenarios, which help improve the integration and energy efficiency of the acquisition system.

At present, EEG acquisition systems are widely used in non-invasive recording (such as scalp EEG) and invasive recording (such as cortical electroencephalogram), showing great potential in the fields of brain-computer interface, epilepsy monitoring and sleep research. However, in practical applications, key issues such as low power consumption (less than 100 microwatts per channel), small area (less than 1 square millimeter per channel), high input impedance and low noise still need to be solved. For example, some advanced EEG acquisition systems have achieved parallel recording of more than 10,000 channels, greatly improving the temporal and spatial resolution, but long-term stability is still limited by electrode drift and brain movement interference. In response to the needs of multimodal data acquisition, reconfigurable analog front-end design has become a future trend, and the system can adapt to different types of signals by dynamically adjusting gain and bandwidth.

The development of EEG acquisition technology mainly presents the following trends: First, the realization of higher channel density. Whole-brain neural recording needs to support tens of thousands of channels, relying on miniaturized electrodes and intelligent algorithms (such as motion correction) to improve signal stability. Secondly, multimodal integration will become the mainstream. Reconfigurable analog front ends will support the acquisition of multiple signals such as voltage and current, and balance power consumption and chip area through hardware optimization. Finally, the analog-to-information conversion mode will be widely used, taking advantage of the sparsity of the signal, directly outputting decision information through event-driven sampling and feature extraction, thereby reducing the energy consumption of the analog-to-digital converter.

\subsection{EEG signal transmission}

EEG signal transmission technology aims to capture and transmit the brain's electrical activity wirelessly to support medical diagnosis, neuroscience research, brain-computer interface development, and other innovative applications. Traditional wired systems limit the patient's mobility due to cables, affecting the naturalness of data collection and the comfort of long-term use. The emergence of wireless transmission technology has significantly improved this problem, allowing patients to move freely in their daily lives while allowing longer monitoring. The core of this technology is to transmit EEG signals from the acquisition device to the receiving device, such as a smartphone, computer, or medical system. Its development is closely related to the progress of biomedical engineering, communication technology, and material science. According to the use method and deployment location of the device, EEG signal transmission technology is mainly divided into two categories: wearable devices and implantable devices. The two have their own characteristics in communication methods, application scenarios, and technical implementation. In addition, the exploration of emerging technologies such as visible light communication and ultrasonic communication provides new directions for future technological breakthroughs.

Wearable EEG devices are non-invasive and usually come in the form of headbands, helmets, headphones or patches. They collect EEG signals through electrodes placed on the scalp. Such devices have a wide range of applications in the medical and consumer fields, such as sleep monitoring, brain health assessment, attention training, meditation assistance, and the development of brain-computer interfaces. Sleep monitoring can help diagnose sleep disorders such as insomnia or sleep apnea; brain-computer interfaces control external devices by analyzing EEG signals, such as helping paralyzed patients operate wheelchairs or computers. Studies have shown that wearable devices mainly use radio frequency technology for wireless data transmission, commonly using industrial, scientific and medical frequency bands, such as 2.4 GHz Bluetooth or Wi-Fi protocols, which are widely adopted because they do not require a license and have a mature hardware ecosystem.

Some typical wearable EEG devices on the market include EMOTIV Insight, Muse, OpenBCI and NeuroSky MindWave. EMOTIV Insight is a lightweight five-channel device designed for real-time EEG monitoring, which transmits data to smartphones or computers via Bluetooth. It is widely used in field research, emotion analysis and brain-computer interface development. Muse is a device focused on meditation and sleep monitoring. It is equipped with seven sensors and connects to mobile applications via Bluetooth to provide real-time feedback to help users relax or improve sleep quality. OpenBCI is an open source platform that supports up to 16 channels of EEG signal acquisition and can transmit data via Bluetooth or Wi-Fi. It is favored by researchers and developers for its flexibility and customizability. NeuroSky MindWave is a single-channel device with a simple design, mainly used for educational and entertainment purposes, such as controlling simple games through EEG signals.

\begin{figure}
    \centering
    \includegraphics[width=0.3\textwidth]{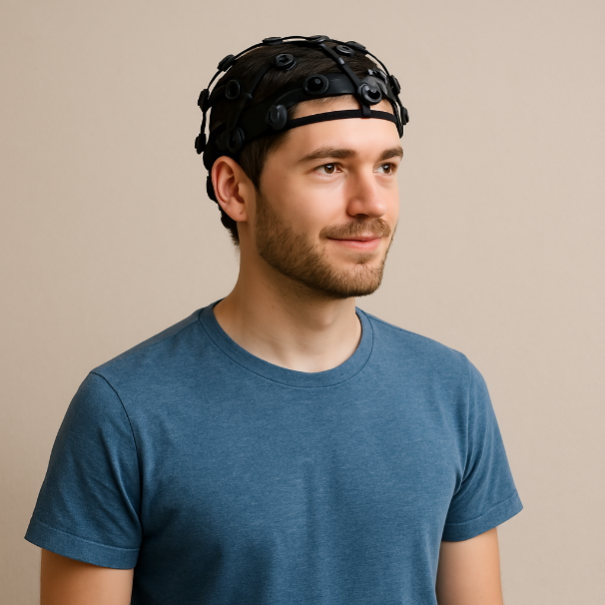}
    \caption{Schematic diagram of wearable EEG device}
\end{figure}

These wearable devices achieve real-time data transmission through wireless communication. Patients do not need to be fixed to bulky equipment and can be monitored for a long time at home, in the office or even in outdoor environments, which significantly improves the convenience and comfort of use. The advantages of industrial, scientific and medical frequency bands are their global versatility and low-cost hardware support, but there are also potential interference issues, especially in environments with dense wireless signals such as hospitals or cities. In addition, the battery life and miniaturization of the device are key challenges. In order to ensure wearing comfort, the device needs to be light and low power, so efficient protocols such as low-power Bluetooth are usually adopted. For example, this protocol minimizes power consumption while maintaining a stable connection, allowing the device to continue to operate for hours or even days. In addition, the contact quality of the scalp electrodes and signal noise suppression are also the focus of technical attention. Modern devices often use dry electrodes or active amplification technology to improve signal quality.

Unlike wearable devices, which are non-invasive, implantable devices need to be implanted in the body through surgery and are usually placed in the brain or skull to more accurately monitor and stimulate specific brain areas. Such devices play an irreplaceable role in the treatment of complex neurological diseases, such as Parkinson's disease, epilepsy, dystonia, chronic pain, and severe depression. They can directly record electrical activity in deep brain areas, capture signals that scalp devices cannot detect, or regulate abnormal neural activity through electrical stimulation, thereby alleviating symptoms and even improving the patient's quality of life. Implantable devices usually require wireless communication to interact with external devices to transmit monitoring data or receive control signals, such as adjusting stimulation parameters or uploading EEG recordings.

\begin{figure}
    \centering
    \includegraphics[width=0.4\textwidth]{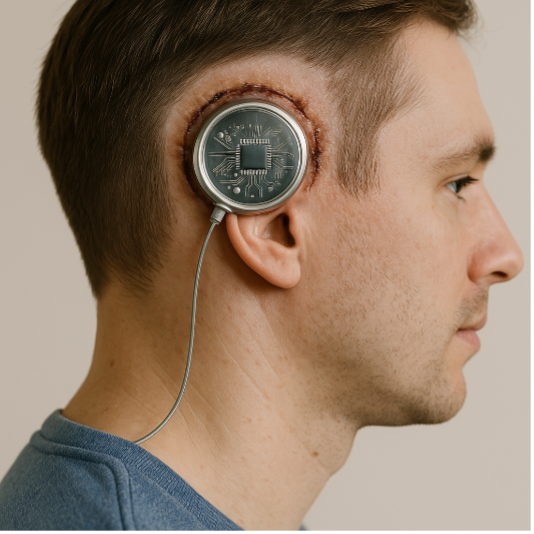}
    \caption{Schematic diagram of implanted device}
\end{figure}

Studies have shown that implantable devices mainly use radio frequency technology to communicate in the medical implant communication service band (402-405 MHz). This frequency band has good tissue penetration, low power requirements and safety, and is considered ideal for in-vivo communication. For example, Medtronic's Percept PC and Percept RC neurostimulators are representative devices for deep brain stimulation, used to treat Parkinson's disease and epilepsy. These devices not only provide electrical stimulation, but also record EEG signals, and communicate with external controllers through this frequency band, so that doctors can remotely adjust treatment plans. NeuroPace RNS System is a responsive neurostimulation system designed for patients with intractable epilepsy. It can monitor EEG signals in real time and provide electrical stimulation intervention immediately when abnormal activity is detected. It also relies on radio frequency communication technology.

In addition to RF technology, some implantable devices also use inductive coupling technology to transmit energy and data through electromagnetic induction between external coils and internal coils. This method performs well in short-distance communications and is often used in pacemakers or muscle stimulators. In EEG signal transmission, inductive coupling is often used in low data rate scenarios to help reduce the size and power consumption of implanted devices. However, its transmission distance is usually limited to within a few centimeters, and the data rate is low, which makes it difficult to meet applications that require high bandwidth, such as real-time transmission of high-resolution EEG data. Therefore, inductive coupling is often used in combination with RF technology to achieve more comprehensive functions.

The future development of EEG signal transmission technology will focus on improving performance, reducing costs, and expanding application scenarios to meet the diverse needs of the medical, scientific research, and consumer markets. Potential directions include multi-mode communication integration, such as combining radio frequency, ultrasound, and optical communications to improve flexibility; intelligent signal processing, such as using machine learning to optimize signal compression and denoising; and telemedicine applications, such as combining EEG devices with the Internet of Things to achieve remote diagnosis and real-time intervention. As technology continues to evolve, EEG signal transmission will bring more possibilities to healthcare, neuroscience research, and human-computer interaction.

\subsection{Classification and application of brain-computer interfaces}

Brain-computer interfaces can be classified according to different criteria, including classification based on the degree of invasiveness, signal type, and user interaction mode. These classifications can help researchers and engineers more comprehensively understand and optimize brain-computer interface systems so that they can play the greatest role in different application scenarios.

According to the degree of invasiveness, brain-computer interfaces can be divided into two types: invasive and non-invasive. Invasive brain-computer interfaces refer to systems that require surgical implantation of electrodes to directly record brain neural activity. This method usually implants electrodes on the surface of the cerebral cortex, or deep into the brain tissue to obtain high-resolution neural signals. Because the electrodes are in direct contact with the brain tissue, the signal quality of invasive brain-computer interfaces is high, the spatial resolution can reach the millimeter level, the frequency bandwidth is usually more than 200 Hz, and the signal amplitude is usually around 100 microvolts. This means that it can provide more accurate and stable signals than non-invasive brain-computer interfaces, and more precise control of external devices. In addition, the signals of invasive brain-computer interfaces are less affected by external noise and can support low-latency real-time control applications such as prosthetic control and neural restoration therapy.

The main applications of invasive brain-computer interfaces include: helping quadriplegic patients control mechanical prostheses through brain signals, enabling them to recover some autonomous activities; realizing neural prostheses, such as controlling robotic arms to grab objects through brain signals; and being used for brain-to-brain communication experiments, so that the neural activities of two people are directly interconnected, thus achieving non-verbal communication. These applications demonstrate the great potential of invasive brain-computer interfaces, especially in the fields of rehabilitation medicine and human-computer fusion. However, invasive brain-computer interfaces also face many challenges and limitations, such as high risks of surgery, including infection, tissue damage, and biocompatibility issues caused by long-term implantation. In addition, the high cost of manufacturing, implantation, and maintenance of the equipment limits its popularity among ordinary users. Despite this, research in recent years is exploring more minimally invasive electrode implantation methods, such as implanting microelectrodes through blood vessels to reduce the risks of invasive surgery.

In contrast, non-invasive brain-computer interfaces rely on external electrodes to measure brain activity, the most common method being electroencephalography. Electroencephalography records the brain's electrical activity by placing electrodes on the surface of the scalp to extract useful neural signals. Since this method does not require surgical implantation, it is low-cost, easy to use, and safer, so it is widely used in neuroscience research, brain-computer interaction systems, and commercial applications. For example, consumer-grade EEG devices on the market are relatively affordable and can meet different needs from academic research to personal health management.

The main technical features of non-invasive brain-computer interfaces include: high portability, some EEG devices have been designed as wearable headbands or wireless headphones, suitable for daily use; compared with invasive brain-computer interfaces, they are less expensive and suitable for large-scale applications. However, the quality of EEG signals is limited and is easily interfered by myoelectric noise (such as electrooculography, facial muscle movements), poor electrode contact, and environmental noise. Due to the obstruction of the scalp and skull, the spatial resolution of the EEG is low, usually only reaching the centimeter level, so its ability to decode fine brain activity is not as good as that of invasive brain-computer interfaces.

The main applications of non-invasive brain-computer interfaces include: cognitive status monitoring, such as detecting the user's attention level, fatigue status, and mood swings; brain-controlled games and entertainment devices, where users can control virtual characters or adjust the game environment through their thoughts; and smart home control, such as switching lights, TVs, or smart speakers through brain waves. Due to its non-invasiveness and operability, non-invasive brain-computer interfaces are gradually entering the consumer market and are being used in many fields such as education, medical rehabilitation, and human-computer interaction.

In addition to being classified by the degree of invasiveness, brain-computer interfaces can also be classified by the type of neural signals utilized. Among them, brain-computer interfaces based on event-related desynchronization and event-related synchronization are mainly used for motor imagery tasks. For example, when a user imagines the movement of his or her left or right hand in his or her mind, the system can parse the movement intention by identifying neural signals in specific brain areas, thereby achieving cursor control or mechanical prosthetic operation. Brain-computer interfaces based on event-related potentials utilize the brain's specific response to external stimuli (such as sound and flash), of which the P300 wave is the most commonly used signal. For example, the P300 speller allows users to spell words by looking at different letters, which is suitable for patients with ALS who cannot use keyboard input.

Brain-computer interfaces based on steady-state visual evoked potentials rely on the brain's response to light sources flickering at different frequencies and can be used for high-frequency interactive applications, such as quickly switching menus and controlling drones. The advantages of this method are strong signals, easy recognition, and low training costs, but long-term use may lead to visual fatigue. In addition, brain-computer interfaces based on slow cortical potentials are mainly used in clinical rehabilitation, such as helping locked-in syndrome patients communicate. Brain-computer interfaces based on sensorimotor rhythms are often used for tasks such as cursor control and wheelchair navigation, and have high control accuracy.

In recent years, hybrid brain-computer interface technology has gradually developed, combining multiple signal sources (such as EEG and near-infrared spectral imaging, EEG and magnetic resonance imaging) to improve the robustness and adaptability of the system. For example, the combination of EEG and near-infrared spectral imaging can simultaneously obtain high temporal resolution and high spatial resolution data, which is widely used in neurorehabilitation and cognitive assessment.

Brain-computer interfaces can also be classified according to the user's interaction method, mainly into active brain-computer interfaces, reactive brain-computer interfaces, and passive brain-computer interfaces. Active brain-computer interfaces require users to actively control brain activity, such as imagining limb movements to operate external devices. Reactive brain-computer interfaces are based on external stimuli, and the user's brain responds specifically to visual or auditory cues, such as steady-state visual evoked potentials. Passive brain-computer interfaces do not require active user operation, but the system autonomously monitors brain activity, such as for emotion detection and cognitive load assessment, and are widely used in scenarios such as driving safety monitoring and brain-computer interaction games..

\begin{figure}
    \centering
    \includegraphics[width=0.8\textwidth]{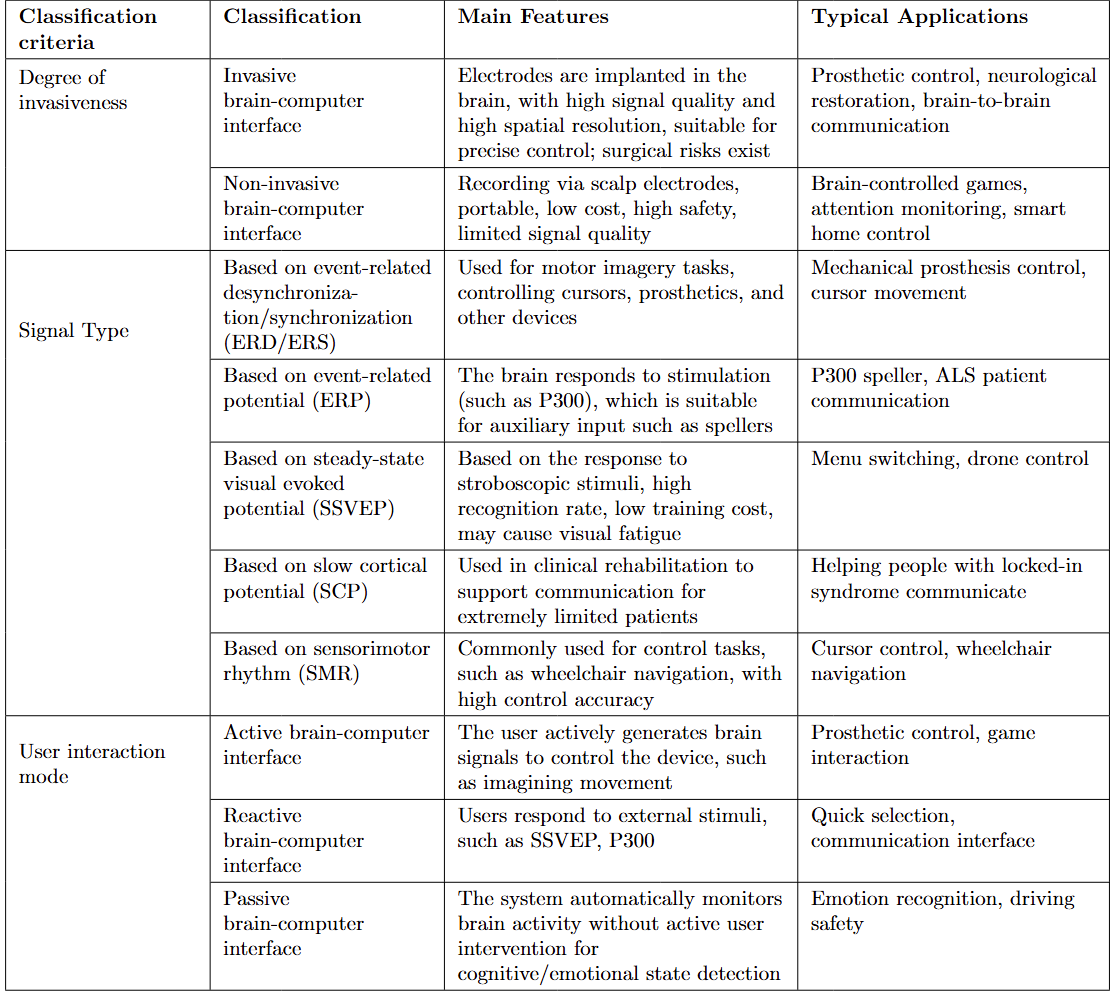}
    \caption{Classification of brain-computer interfaces}
\end{figure}

\section{Application of brain-computer interface in affective computing}

In recent years, brain-computer interface technology has gradually occupied an important position in emotional computing due to its unique capabilities. Brain-computer interface is a technology that directly extracts information from the brain's electrical activity signals and converts them into commands without relying on peripheral nerves or muscles. This feature makes brain-computer interface not only used for traditional neural control, but also as an effective tool for emotional computing, thereby achieving accurate perception of the user's emotional state.

The combination of affective computing and brain-computer interfaces has given rise to affective brain-computer interface systems, which can not only detect and analyze emotional state signals, but also use these signals to optimize the way people interact with computers. Therefore, the role of brain-computer interfaces is not limited to control functions, but also extends to the perception and feedback of user emotions. Especially in intelligent human-computer interaction systems, the enhancement of emotion recognition capabilities can significantly improve the naturalness of interaction, allowing machines to adapt to user needs more intelligently.

In brain-computer interface systems, EEG signals have become one of the most widely used biological signals due to their non-invasiveness and high sensitivity to emotional states. EEG signals are not only used in the medical field to diagnose neurological diseases such as epilepsy, brain tumors, and sleep disorders, but can also be used to assess an individual's mental and emotional state. Therefore, in scenarios where accurate assessment of individual emotions is required, brain-computer interface technology based on EEG signals has become a core tool for identifying emotional states. The advantage of this technology is that it can provide more intuitive and accurate results than traditional behavioral analysis methods, thereby improving the reliability of emotional computing.

The EEG-based emotion recognition system mainly includes a signal stimulation module and a signal processing module. The signal stimulation module involves the equipment used, the type of stimulation, and the data processing mode, while the signal processing module covers signal acquisition, preprocessing, feature extraction, feature selection, classification algorithm, and final performance evaluation. Many studies have been devoted to optimizing these two modules to improve the accuracy and stability of emotion recognition. By systematically analyzing and integrating existing research results, the main trends in the field can be summarized and information gaps that have not been fully explored can be revealed. This not only helps to optimize the existing system, but also provides a clear direction for future research.

\begin{figure}
    \centering
    \includegraphics[width=0.5\textwidth]{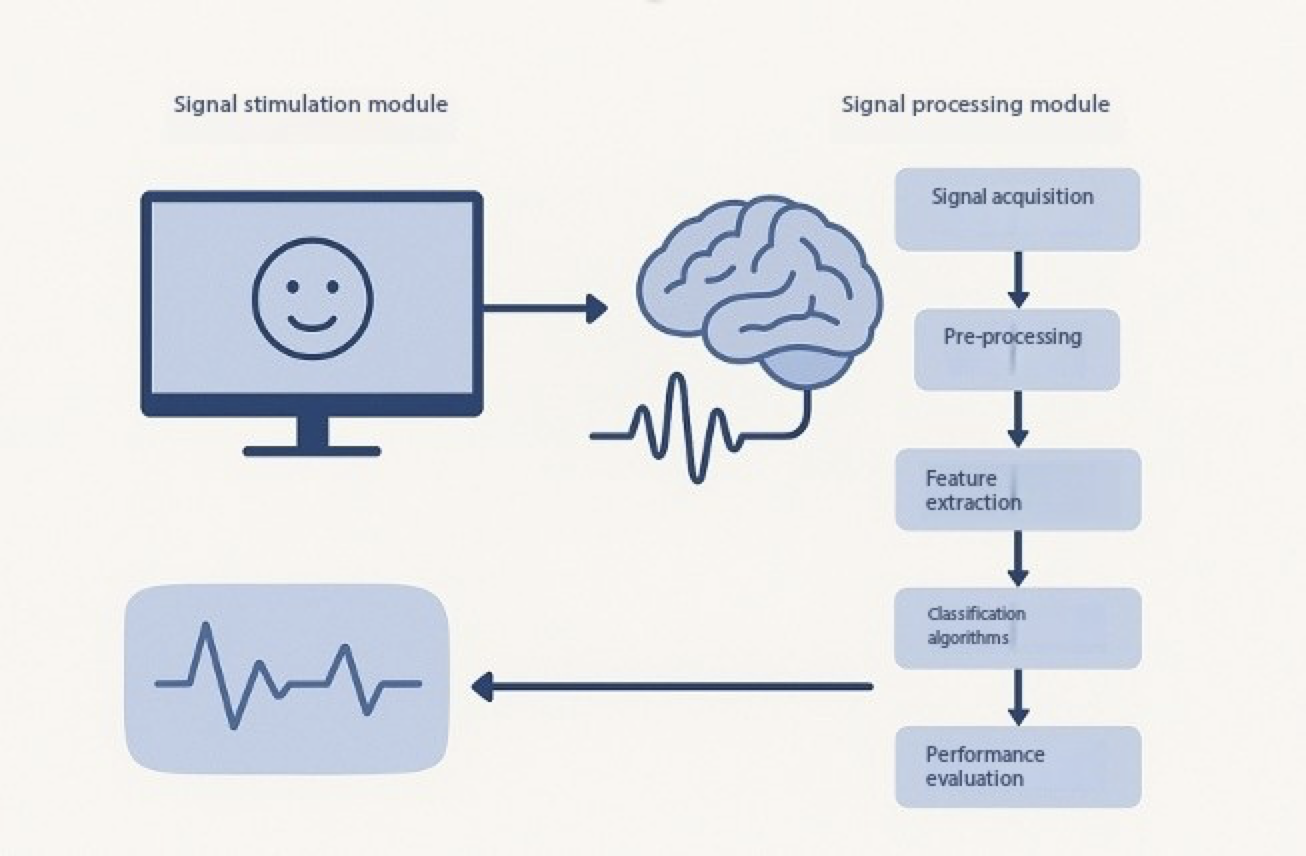}
    \caption{Emotion recognition system}
\end{figure}

In the development of affective computing, the recognition of EEG emotional states has become a core research direction. Emotion recognition based on EEG signals relies on signal changes in specific frequency bands, such as changes in activity in the frontal lobe and limbic system-related areas, to analyze the emotional state of an individual. Studies have shown that different emotional states will show significant differences in specific EEG bands, which makes EEG-based emotion classification possible. At the same time, with the advancement of machine learning and deep learning technologies, the accuracy of emotion recognition has also been significantly improved.

In addition, brain-computer interface technology also shows great potential in the field of intelligent robots. By applying brain-computer interface to robot interaction systems, robots can more accurately perceive the user's emotional state and make corresponding feedback. This combination can not only be used for medical rehabilitation and mental health intervention, but also in education, social companionship and other fields. For example, in assisting the emotional training of autistic children, brain-computer interface technology can help robots better understand children's emotional needs and provide a more humane interactive experience.

With the continuous advancement of affective computing technology, the role of brain-computer interfaces in emotion recognition and intelligent interaction will become increasingly important. The following chapters will further explore the specific applications of brain-computer interfaces in EEG emotional state recognition and robot interaction. First, we will focus on the emotional state recognition method based on EEG signals, analyze the EEG characteristics under different emotional states and related signal processing technologies. Subsequently, we will explore the practical application of brain-computer interfaces in robot interaction and show how they can enhance the robot's perception and response capabilities to human emotions. Through these in-depth analyses and examples, we will fully reveal the broad prospects of brain-computer interfaces in the field of affective computing.

\subsection{Identification of EEG emotional states}

As an important research direction in the field of affective computing, EEG emotion state recognition has its theoretical foundation based on the systematic cognition of the nature of emotion. Emotion, as a complex manifestation of human consciousness, is not only a direct reflection of the inner psychological state, but also an adaptive response mechanism of individuals to external environmental stimuli. This dynamic psychological and physiological phenomenon not only affects human decision-making processes and social interactions, but also forms a two-way regulatory relationship with physical and mental health status - positive emotions can effectively improve the quality of life and physiological functions, while long-term negative emotional states may lead to cognitive dysfunction and even induce organic diseases.

In the psychological theory system, emotion representation models are mainly divided into two theoretical schools: discrete emotion model and dimensional emotion model. The former regards emotion as a collection of several basic units, while the latter constructs a spatial mapping of emotional states through a multidimensional coordinate system. The theoretical origin of the discrete model can be traced back to Darwin's evolutionary view. On this basis, Tomkins proposed a classification system containing nine basic emotions, emphasizing the basic role of core emotions such as interest-excitement, surprise-fright in maintaining mental health. As a typical representative of this school, the Ekman model established six basic emotions with universal expression characteristics through cross-cultural research, including sadness, surprise, happiness, etc. These emotion units have core characteristics such as consistency of physiological expression and convergence of situational responses, which provides an important theoretical reference for subsequent emotion recognition research based on physiological signals.

However, with the development of cognitive neuroscience, researchers have found that discrete models have significant limitations in describing complex emotional states. Emotional experiences in real life often present multi-dimensional dynamic characteristics, which are difficult to fully represent with limited discrete categories. This theoretical breakthrough has given rise to the rise of dimensional models, which transform emotional states into continuous quantized spatial vectors by establishing a multi-dimensional emotional coordinate system. Russell's two-dimensional ring model is a typical representative. It uses two orthogonal dimensions, valence (pleasure level) and arousal (activation intensity), to construct an emotional space and divides the emotional state into four quadrants: high arousal positive valence areas correspond to positive activation states such as excitement and satisfaction; high arousal negative valence areas cover negative stress reactions such as tension and anxiety; low arousal negative valence areas represent inhibitory states such as depression and fatigue; low arousal positive valence areas include low activation positive states such as calmness and relaxation. This quantitative modeling method provides an operational mathematical framework for emotional feature extraction based on EEG signals.

As a key link in EEG emotion recognition research, the methodology of emotion induction experiment has evolved from subjective recall to standardized stimulation. Early studies mostly used subjective recall method to induce the target state by guiding the subjects to reproduce specific emotional memories, but this method has inherent defects such as poor state stability and difficulty in time synchronization. Modern experimental paradigms generally use international standardized stimulation systems, such as the International Affective Picture System (IAPS), which contains 1,200 visual stimulation materials labeled with valence-arousal, and the International Affective Digital Sound System (IADS), which provides 167 auditory stimulations calibrated in three dimensions. These standardized tools not only ensure the validity and reliability of experimental stimulation, but also significantly improve the comparability of data across studies. It is worth noting that current research is expanding from passive reception of stimulation to interactive situations, and constructing emotion induction models that are closer to real scenes through immersive environments such as virtual reality and somatosensory games. This paradigm innovation is of great value in improving the generalization ability of EEG emotion recognition models.

At the neural mechanism level, emotional states are closely related to brain electrophysiological activities. The asymmetric activation pattern of the prefrontal cortex is significantly correlated with the dimension of emotional valence. The enhanced activity of the left prefrontal cortex is usually accompanied by positive emotional experiences, while the dominant activation of the right side is associated with negative emotional states. Structures such as the amygdala and hippocampus of the limbic system play a core regulatory role in the dimension of emotional arousal, and their neural oscillation characteristics can be effectively captured through time-frequency analysis of EEG signals. These neurophysiological findings provide a theoretical basis for constructing an emotion recognition model based on multi-lead EEG signals. By integrating time domain, frequency domain and nonlinear features, researchers have developed a variety of effective feature extraction and pattern classification algorithms.

\begin{figure}
    \centering
    \includegraphics[width=0.4\textwidth]{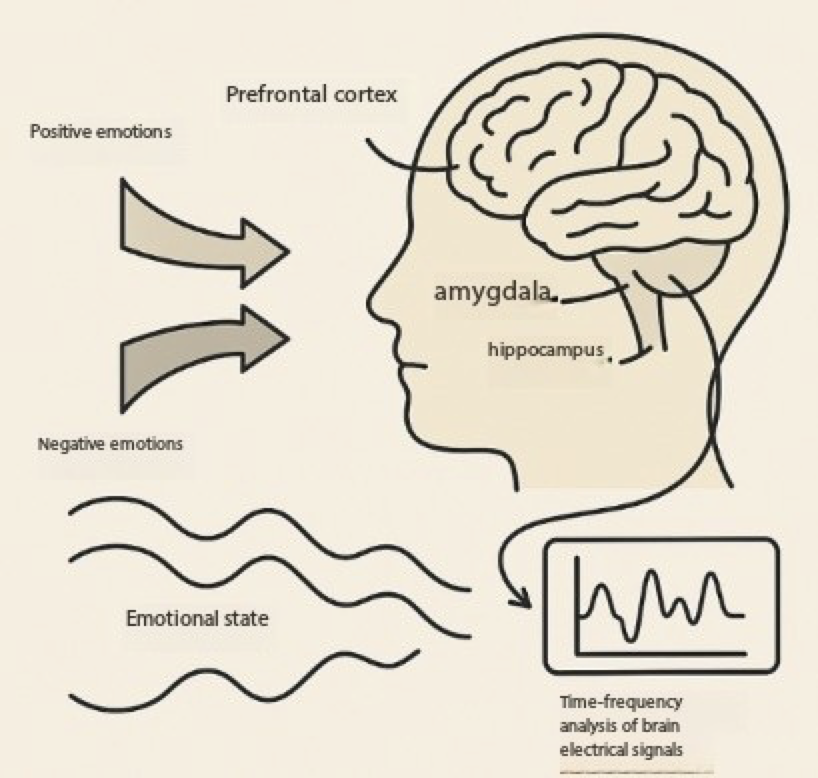}
    \caption{Schematic diagram of neural mechanism}
\end{figure}

The main challenges facing this field at present are the individual differences and environmental dependence of emotional states. There may be significant variations in the EEG response patterns of different subjects under the same stimulus, and there are also systematic differences in the expression of emotions in laboratory environments and real scenes. Solving these problems requires the development of more adaptive personalized modeling methods, and the establishment of a fusion analysis framework that includes multimodal data (such as EEG, physiological parameters, and behavioral performance). Future research trends will focus on dynamic emotional state tracking, cross-situational model transfer, and autonomous mining of emotional features based on deep learning. These breakthroughs will promote the practical application of brain-computer interface technology in the fields of mental health monitoring and intelligent interactive systems.

\subsection{Brain-computer interface and robot interaction}

An often underestimated but crucial component of brain-computer interface driving devices is how to convert the output of the neural signal decoder into a control signal suitable for driving the robotic device. The core of this conversion process lies in the design of the control strategy, which not only determines the efficiency of human-computer interaction, but also directly affects the practicality and user experience of the entire system. The current mainstream control strategies can be divided into two major paradigms: discrete and continuous. Each strategy has its unique advantages and applicable scenarios, and also faces different technical challenges.

The typical feature of discrete control strategy is that users can only send discrete instructions to external devices at intervals of a few seconds (usually 3-4 seconds). This control mode is naturally adaptable to brain-computer interfaces based on exogenous stimuli. For example, a system built with visual evoked potentials or steady-state visual evoked potentials essentially generates instructions by detecting the brain's response to specific discrete stimuli (such as flashing lights). Under this mechanism, the system structure determines that it can only generate intermittent control signals. In actual application scenarios, such discrete commands are often used to trigger robots to perform preset semi-autonomous actions, such as turning a wheelchair or a robotic arm to perform a grasping action. Although the information transmission rate of this mode is low (an average of 0.3 instructions per second), its advantage is that the instruction recognition has high certainty and stability.

In contrast, brain-computer interface systems based on endogenous neural activity (such as autonomous regulation of sensorimotor rhythms) theoretically have the potential to achieve continuous control. Such systems do not rely on external stimuli to trigger and can continuously decode the user's spontaneous neural activities such as motor imagery. However, in practical applications, most studies still tend to adopt a discretization processing strategy. This choice is mainly due to considerations of signal stability: the continuous probability output of the decoder is accumulated through a time integration algorithm (such as a moving average filter or an exponential smoothing filter), and valid instructions are triggered only when the confidence reaches a preset threshold. Although this approach sacrifices real-time performance, it significantly reduces the risk of false triggering caused by the non-stationarity of EEG signals. This compromise reflects the limitations of the current technical system in terms of signal decoding reliability.

Breaking through the shackles of discrete control modes and achieving true continuous interaction is an important direction for improving the performance of brain-controlled devices. Existing explorations are mainly carried out along two technical paths: one is to establish a direct mapping relationship between neural activity characteristics and robot motion parameters, and the other is to develop a new signal processing framework to improve the stability of decoding output. Typical cases of the former include directly mapping the neural oscillation characteristics corresponding to specific motion imagination patterns into multi-degree-of-freedom motion control signals for drones or robotic arms through linear or nonlinear transformation functions. This type of method has achieved continuous control at the six-degree-of-freedom level in a laboratory environment, but its strict requirements on signal decoding accuracy limit the scalability of actual application scenarios.

In terms of signal processing optimization, researchers have proposed a variety of innovative solutions. A team has developed a new control framework based on dynamic system theory, which significantly improved the stability of continuous control signals by modeling the temporal evolution characteristics of the decoder output. Another study used a composite filtering strategy, combined with the Savitzki-Golay filter and reverse bias correction technology, to effectively suppress noise interference and abnormal fluctuations in the classifier output. These technological breakthroughs enable healthy subjects to continuously control mobile robots to complete complex navigation tasks, marking the beginning of the continuous control strategy from theoretical verification to practical application.

The core challenge facing the current technology system is the limited set of available instructions. Even the most advanced non-invasive brain-computer interface system can hardly break through the upper limit of eight types of instructions that can be reliably identified. This has posed a significant constraint on the control of complex equipment that requires delicate operations (such as multifunctional exoskeletons or multi-joint robotic arms). To address this bottleneck, researchers have proposed an intelligent assistance enhancement strategy: on the one hand, through a preset behavior template library, discrete instructions are converted into a complex action sequence that can be autonomously executed by the robot; on the other hand, a semi-autonomous control framework for human-machine collaboration is constructed, enabling the robot to intelligently complete user instructions based on environmental perception information.

In terms of specific implementation, systems based on exogenous stimuli often adopt goal-oriented autonomous planning strategies. For example, after selecting the target position through the visual interface, the robot autonomously completes path planning and obstacle avoidance navigation. This "target selection-autonomous execution" mode has achieved good application results in intelligent wheelchair control. Systems based on endogenous signals tend to adopt hierarchical interaction design, expanding limited instructions into a multi-dimensional control space through a hierarchical menu system. Typical applications include controlling the robotic arm to approach the target object first, and then triggering a sequential operation process of grasping the action. It is worth noting that the latest research has begun to try to organically integrate the two control strategies, and simultaneously obtain discrete instructions and continuous adjustment signals through a hybrid brain-computer interface architecture, which provides new possibilities for breaking through the limitations of existing technologies. .

Although the current mainstream synchronous sequential control mode has achieved basic functions to a certain extent, its mechanized interaction process seriously restricts the naturalness and flexibility of human-machine collaboration. This rigid structure of "command sending-action execution" not only causes operation delays, but also makes it difficult for users to make dynamic adjustments in complex scenarios. To break through this limitation, researchers began to explore a new interaction paradigm that deeply integrates user intentions with machine intelligence, which marks an important leap from simple command transmission to intelligent collaborative decision-making in brain-controlled robot systems.

Among various brain-controlled robotic devices, the differences in application scenarios have a profound impact on the choice of control strategies. As the most common research object (accounting for 38.4\%), the control mode of telepresence mobile robots often prioritizes real-time feedback performance, so more than half of the systems adopt an interactive method based on exogenous stimuli such as steady-state visual evoked potentials. Such devices are usually equipped with environmental perception systems, and users set the macroscopic movement direction through discrete instructions, while micro-decisions such as obstacle avoidance are made autonomously by the machine. In contrast, electric wheelchairs (accounting for 22.1\%), as auxiliary equipment directly related to personal safety, emphasize stability in their control strategies. 57.9\% of the studies use exogenous paradigms to ensure the certainty of command recognition. It is worth noting that the robot arm (19.8\%) exhibits unique control characteristics in grasping operations: 52.9\% of the systems select the target object through visual evoked potentials, and nearly half of the solutions use motor imagery to achieve continuous trajectory control of the end effector. This mixed strategy reflects the high requirements of fine operations on the control dimension.

Wearable exoskeleton devices show obvious differentiation in control modes. 87.5\% of lower limb exoskeleton systems use self-paced psychological task control, which is in line with the physiological rhythm characteristics of walking initiation - users trigger the gait cycle through motor imagery, and the details of joint movement are automatically generated by the preset biomechanical model. Upper limb exoskeletons show the opposite trend, with 66.7\% of studies relying on exogenous paradigms. This is because their functions focus more on specific action triggers (such as grasping and releasing), which require clear discrete command confirmation. As an emerging control object, the three-dimensional spatial motion characteristics of quadcopters have prompted researchers to try hybrid control modes, using steady-state visual evoked potentials for flight mode switching, and mapping motor imagery to continuous adjustment of pitch and roll.

In-depth analysis of the distribution of control strategies shows that discrete control modes are absolutely dominant (93.0\%), which reflects the bottleneck of the current technical system in terms of signal decoding reliability. Even self-paced systems that theoretically support continuous control generally use a confidence threshold mechanism to discretize continuous outputs. Although this compromise improves operational safety, it sacrifices the fluency of human-computer interaction. A few breakthrough studies have made progress in the field of mobile robot navigation: by constructing a dynamic mapping model of motion imagery neural features and velocity vectors, combined with Kalman filtering for noise reduction, continuous trajectory tracking with centimeter-level accuracy is achieved. An innovative team has also developed an adaptive extraction algorithm for neural oscillation features that can analyze the user's intention to adjust the six-degree-of-freedom motion of the drone in real time, achieving a breakthrough of 5 continuous command updates per minute in a laboratory environment.

The introduction of a shared control framework provides a new approach to breaking through the limitations of discrete control. This framework positions the robot as an intelligent decision-making subject rather than a passive execution terminal. Typical applications include brain-controlled wheelchair systems: users trigger steering commands by imagining left and right movements, while the wheelchair autonomously calculates the steering radius and avoids obstacles. This hierarchical control architecture reduces the command transmission rate by 40\% while increasing task completion efficiency by 2.3 times. The advanced solution builds a dynamic weight allocation mechanism that automatically adjusts the human-machine control weight according to the complexity of the environment - giving users more freedom in open areas and enhancing the machine's autonomous obstacle avoidance function when entering narrow passages. Clinical trials have shown that this adaptive system reduces the operational error rate of patients with spinal cord injury by 68\%.

The integration of the error potential feedback mechanism has created a new dimension of two-way interaction between humans and machines. When the robot's behavior deviates from the user's expectations, the system corrects the motion trajectory in real time by detecting the error-related potentials. Breakthrough research has combined this neural feedback with reinforcement learning: when the robotic arm completes the grasping task, it continuously receives the user's subconscious error evaluation signals, and after 300-500 iterations, it can autonomously optimize the motion trajectory that meets individual preferences. Recent progress has shown that by decoding the user's subjective satisfaction evaluation of the robot's movements (such as fluency scores), the system can establish a personalized control model to achieve accurate reproduction of natural motion patterns in the field of prosthetic control.

The evolution of intelligent fusion strategies has driven brain-control systems toward cognitive collaboration. The latest experimental platform has been able to achieve multimodal intent analysis: in addition to traditional motor imagery signals, it simultaneously integrates neural features such as attention level and task urgency, allowing the robot to predict user needs. For example, when a neural oscillation pattern related to user anxiety is detected, the nursing robot will automatically reduce its movement speed and increase the frequency of environmental scanning. This forward-looking interaction mode upgrades the brain-computer interface from a simple control channel to a two-way cognitive interaction interface, laying the foundation for building a truly intelligent brain-controlled robot system.

% ------------------------------------------

\chapter{Emotional Computing of Physiological Electrical Signals}

\section{Types and characteristics of physiological electrical signals}

As an important research direction in the field of artificial intelligence and human-computer interaction, affective computing is committed to identifying, analyzing and simulating human emotional states through technical means. Among the many emotion recognition methods, analysis based on physiological electrical signals has gradually become the core technical path for emotion state detection due to its strong objectivity, high real-time performance and not easily affected by subjective disguise. The human body maintains life functions through complex bioelectric activities. These electrical signals not only carry physiological state information, but also have a deep correlation mechanism with emotional fluctuations. This chapter will systematically explain the main types of physiological electrical signals commonly used in affective computing and their characteristics, focusing on analyzing the generation mechanism of electrocardiogram signals and electromyogram signals, signal characteristics and their unique value in emotion recognition.

The generation of physiological electrical signals originates from the electrophysiological activities of human tissue cells. When the emotional state changes, the autonomic nervous system and the endocrine system will work together on various organ tissues to induce characteristic changes in electrical signals. This electro-chemical physiological response has clear timing characteristics and a quantifiable parameter system, providing a reliable biological information carrier for emotional computing. Compared with external behavioral signals such as facial expressions and voice intonation, physiological electrical signals have higher anti-interference and individual consistency, and can break through the limitations of cultural differences and subjective disguises to capture more essential emotional physiological reactions.

\begin{figure}
    \centering
    \includegraphics[width=0.5\textwidth]{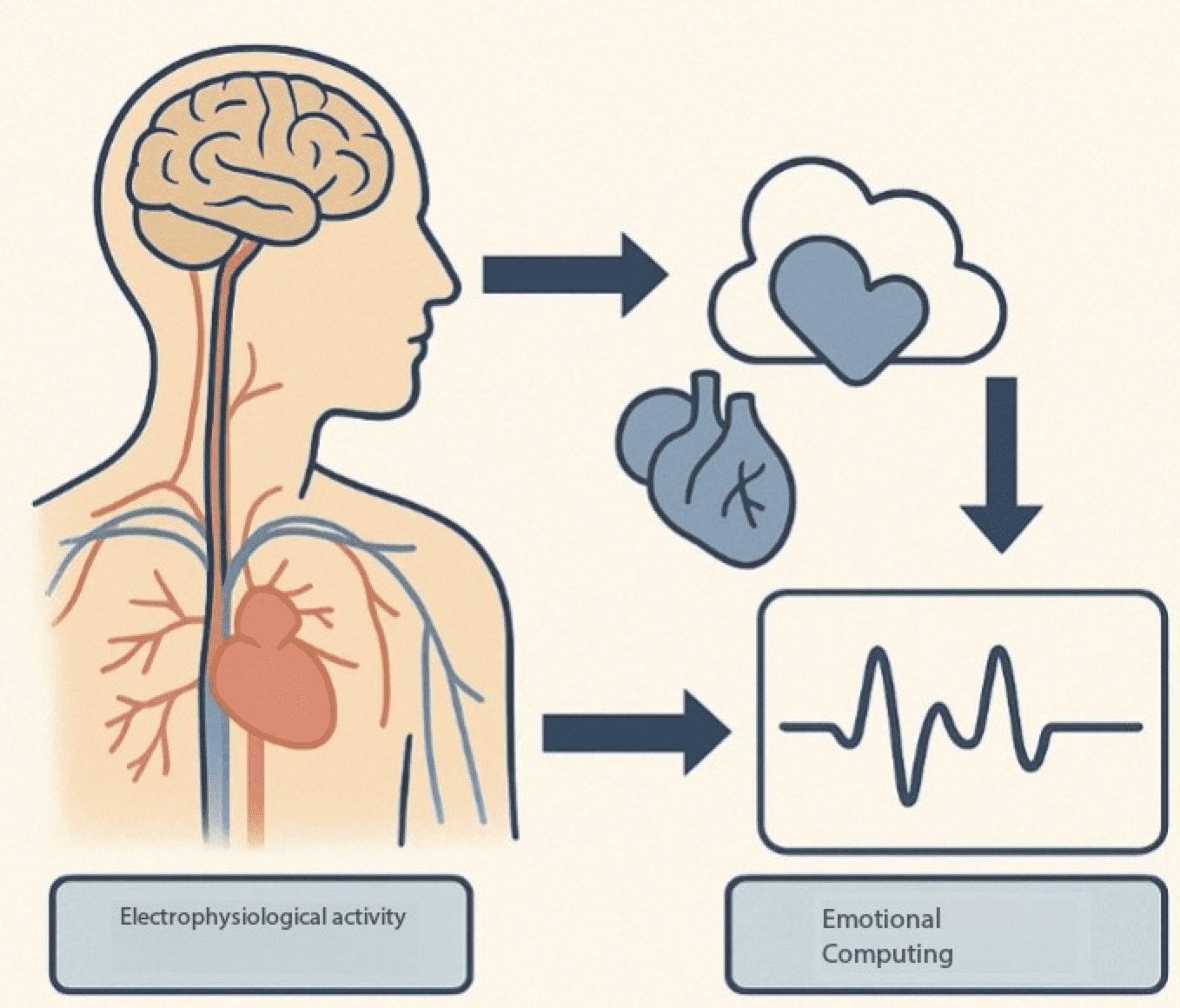}
    \caption{Schematic diagram of the generation of physiological electrical signals and emotion calculation}
\end{figure}

Among many physiological electrical signals, electrocardiogram signals have become the basic data type for emotional computing research because of their easy collection and rich physiological information content. As an organ directly regulated by the autonomic nervous system, the electrical activity of the heart is particularly sensitive to emotional changes. When an individual experiences emotional fluctuations, the balance between the sympathetic and parasympathetic nerves is broken, which directly changes the depolarization and repolarization process of myocardial cells, manifested as regular changes in the morphology, rhythm, and spectral characteristics of the ECG waveform. By analyzing the characteristic parameters such as RR interval variation, T wave amplitude, and ST segment displacement in the ECG signal, different emotional states such as anxiety, pleasure, and anger can be effectively identified. It is worth noting that the response of the ECG signal to emotional stimuli has a significant time delay characteristic, which is closely related to its conduction mechanism through neuro-humoral regulation. This physiological characteristic provides important time dimension information for building a dynamic emotion recognition model.

Unlike the electrocardiogram (ECG) that reflects the state of the autonomic nervous system, the electromyography (EMG) signal directly records the electrical activity characteristics of skeletal muscle fibers, providing a unique motor physiological dimension for affective computing. Emotional experience can trigger unconscious contraction reactions in motor systems such as facial expression muscles and limb muscles. Although such tiny electromyographic changes are difficult to detect with the naked eye, they can be accurately captured by highly sensitive surface electrodes. For example, when disgust is generated, the electromyographic activity of the corrugator muscles is significantly enhanced; while in a state of pleasure, the discharge frequency of the zygomatic major muscle will show a characteristic increase. The time-frequency domain feature analysis of the electromyographic signal (such as root mean square value, median frequency, and wavelet energy coefficient) can effectively distinguish active movement from emotion-induced muscle activity. This ability to distinguish plays a key role in avoiding motion artifact interference and improving the accuracy of emotion recognition. Especially in dynamic interactive scenarios such as virtual reality, the multi-channel synchronous monitoring technology of electromyographic signals can achieve a refined analysis of complex emotions. This chapter will be divided into two sections to explore the core characteristics of electrocardiogram and electromyographic signals. Through comparative studies of these two types of typical physiological electrical signals, readers will be able to establish a complete knowledge system of physiological computing and lay a theoretical foundation for discussing multimodal emotion computing models in subsequent chapters.

\subsection{Electrocardiogram (ECG) signal}

Electrocardiogram is a common method for recording the electrical activity of the heart. It measures the potential difference on the surface of the skin to obtain the depolarization and repolarization process of the heart in each cardiac cycle. The electrical activity of the heart is generated by the excitation of myocardial cells, forming unique electrical signals, which are captured by electronic recording equipment and presented as electrocardiograms. This technology has extremely high application value in clinical practice and is an important tool for the diagnosis of the cardiovascular system.

The ECG reflects the repetitive pattern of electrical depolarization and repolarization of the heart muscle with each beat. A typical ECG consists of five major waveforms, labeled P, Q, R, S, and T waves. In some cases, a U wave may also be present. The P wave represents depolarization of the atria and is the first visible waveform of cardiac activity. This is followed by the QRS complex, which primarily represents the depolarization of the ventricles. The T wave corresponds to the repolarization phase of the ventricles. The U wave is rare and is usually associated with late repolarization of the ventricles.

\begin{figure}
    \centering
    \includegraphics[width=0.5\textwidth]{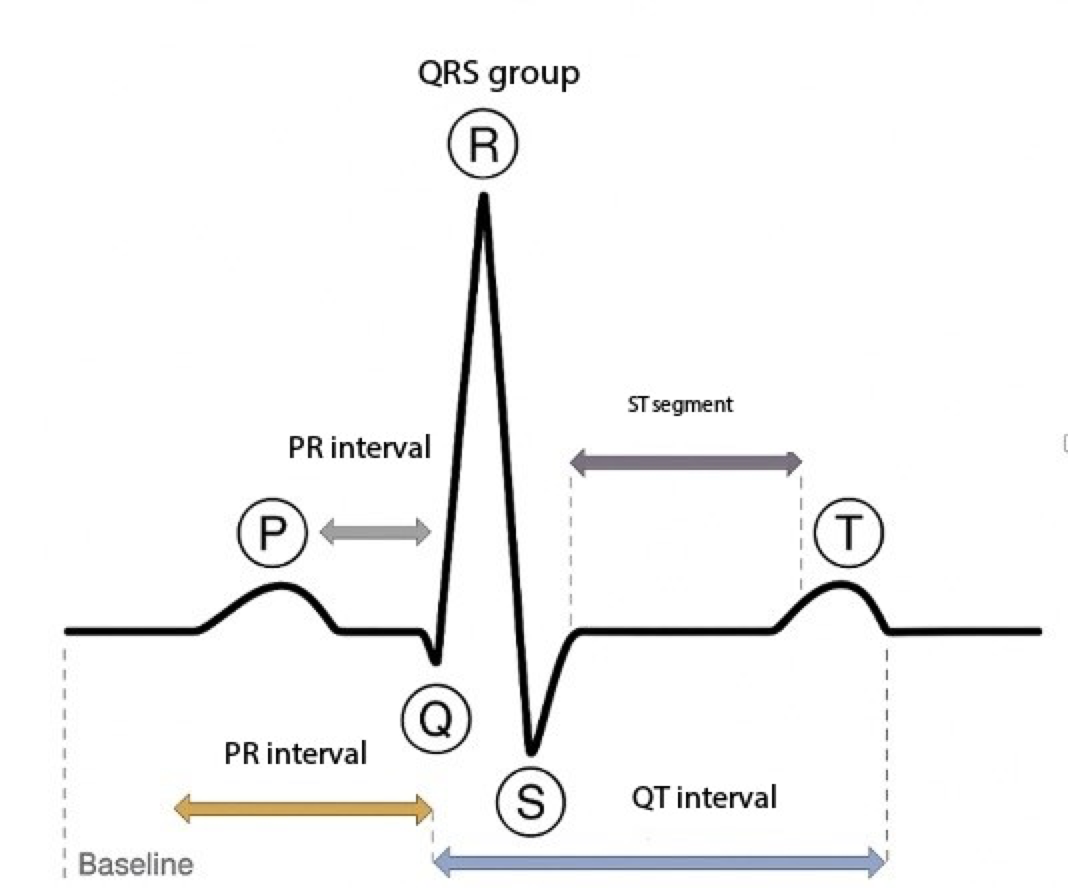}
    \caption{Electrocardiogram}
\end{figure}

The history of electrocardiogram can be traced back to 1903, when Dutch scientist Willem Einthoven designed a method to record cardiac action potentials using a galvanometer. He introduced P, Q, R, S, and T marks in the standard electrocardiogram and recorded the signal on paper with an ink pen. This primitive recording method had important scientific value at the time. Although modern technology has widely adopted digital equipment, traditional paper electrocardiogram recording is still retained in some clinical environments.

In a complete cardiac cycle, the \textbf{P wave} occurs due to atrium depolarization, which causes the atria to contract and pump blood into the ventricles. The P wave usually lasts about 90 milliseconds and has an amplitude of no more than $2.5 \times 10^{-4}$ volts. The following \textbf{QRS complex} reflects the depolarization of the ventricles and lasts about 80 milliseconds with an amplitude of up to 1 millivolt. The formation of this complex involves multiple steps, including depolarization of the septum and electrical signal conduction in the Purkinje fibers. The \textbf{T wave} that follows the QRS complex represents the repolarization of the ventricles and is usually smaller in amplitude than the QRS complex. The electrical signal of repolarization is conducted from the epicardium to the endocardium, forming the T wave.

Atrial repolarization is not usually seen on an ECG because it is smaller in amplitude and is masked by the QRS complex produced by ventricular depolarization. Although the atrial repolarization signal is not visible on a normal ECG, it is still an important part of the electrophysiological activity of the heart.

In actual measurement, ECG signals are collected through electrodes placed on the body surface. Different electrode positions provide different perspectives to fully evaluate the electrical activity of the heart. The earliest standard measurement method was to record through three-point limb electrodes, namely the left arm, right arm and left leg. The bipolar limb leads established by Einthoven include lead I (potential difference between left arm and right arm), lead II (potential difference between left leg and right arm) and lead III (potential difference between left leg and left arm).

As technology develops, another important measurement method has been introduced, namely the precordial leads. By placing six electrodes at specific locations on the chest, leads V1 to V6 are formed. The chest leads can provide detailed information about the anterior, lateral, and posterior walls of the ventricles, which helps diagnose heart diseases such as myocardial ischemia and myocardial infarction. This method is called Wilson precordial leads. The electrodes are arranged along both sides of the sternum and the fifth and sixth ribs on the left to capture the electrical activity of different areas of the heart.

In order to further improve the sensitivity of the measurement, Goldberg proposed the decoupling lead technology. By removing the common signal components from the limb leads and performing differential measurements, enhanced leads are formed. Combining limb leads, chest leads and enhanced leads, the standard 12-lead ECG recording has become the most commonly used ECG detection method in clinical practice. In addition, in order to reduce noise interference and enhance signal quality, modern ECG equipment generally uses differential amplifiers. By measuring the potential difference between the two electrodes and amplifying the signal changes, the differential amplifier can effectively suppress common-mode noise. This technology is particularly effective in eliminating environmental noise and equipment noise, making ECG recordings clearer and more reliable.

The ECG signal is an important tool for recording and analyzing the electrical activity of the heart. The standard twelve-lead ECG is the most common form of measurement, but in some cases, other minimally invasive electrode placement techniques are used. These techniques involve the use of a spring-loaded catheter with multiple electrode bundles that are placed into the left ventricle. Each bundle typically contains 16 electrodes, and up to 16 bundles can be deployed simultaneously. These electrodes fit snugly against the endocardial wall and directly record electrical activity close to depolarized cells. These types of measurements are often used for research purposes or to diagnose cases of complex cardiac pathology.

Electrode placement in the coronary arteries and ventricles is accomplished by inserting catheters equipped with electrodes. These catheters are usually equipped with two or three electrodes, and inserting multiple catheters simultaneously allows for more precise measurements. This approach provides greater spatial resolution, which aids clinicians in the detection and localization of cardiac abnormalities. The concept that the electrical activity of the heart can be described by a single current dipole was first proposed by Albert Einthoven. He proposed that the three-dimensional rotating field vector generated by the heart, the cardiac vector, could be used to represent the electrical activity of the heart. The cardiac vector reflects not only the direction of the current dipole, but also the magnitude of the dipole moment, which is the combined result of ion flow during polarization and repolarization in cardiac cells.

When building a mathematical model of the heart vector, it is usually assumed that the human body is a sphere and the origin of the heart vector is located at the center of the sphere. Although the human arm and left leg extend from the sphere, the model assumes that they are in the same plane and are equidistant from the center of the sphere. This simplification helps analyze the electrical activity of the heart. In addition, it is assumed that the electrical conduction in the sphere is uniform and isotropic to ensure the accuracy of the calculation.

In the traditional Einthoven lead system, the electrical activity of the heart is recorded through leads on the left hand, right hand, and left leg. Leads I, II, and III form a two-dimensional plane with an angle of 120°. Although these leads provide part of the information of the heart's vector, there is a direct mathematical dependence between them due to Kirchhoff's laws. This dependence limits the measurement results within the two-dimensional plane and cannot fully reveal the depolarization process of the heart.

To address this limitation, additional electrode placement methods can be used, such as Winston or Goldman leads. These methods provide more dimensional information by placing electrodes at different locations, thereby forming a complete three-dimensional cardiac vector. The formation of a three-dimensional vector allows the spatial characteristics of the cardiac electrical activity to be more accurately characterized. One of the main features of the electrocardiogram is the interpretation of hemodynamic phenomena, especially the systolic and diastolic cycles of the heart, that is, the heart rate. A complete cardiac cycle begins with atrial contraction, which corresponds to the P wave on the electrocardiogram. Subsequently, the ventricular contraction produces the QRS complex wave, which is the hallmark feature. The cardiac cycle ends with diastole, when both the atria and ventricles are in a relaxed state. After this, a new atrial contraction starts the cycle again. Due to this cyclical feature, the electrocardiogram plays an important role in assessing cardiac function and rhythm.

Generally speaking, the normal heart rate of an adult is about 75 beats per minute, which corresponds to a 0.8-second cycle on the electrocardiogram. In this cycle, atrial contraction usually lasts about 0.1 seconds, ventricular contraction lasts 0.3 seconds, and the remaining 0.4 seconds is the heart's diastole. The atrial diastole is 0.7 seconds, and the ventricular diastole is 0.5 seconds. This time distribution ensures the heart's effective pumping function.

The regulation of heart rate is influenced by many factors, including intrinsic and extrinsic mechanisms. The intrinsic mechanisms are mainly related to the spontaneous activity of the sinoatrial node. The excitability of the sinoatrial node can change due to factors such as stretching and temperature changes, which directly affects the heart rate. Increased temperature usually increases the heart rate, while decreased temperature slows it down.

The extrinsic mechanism mainly involves the regulation of the autonomic nervous system. The parasympathetic and sympathetic nervous systems regulate heart rate by releasing neurotransmitters such as acetylcholine or norepinephrine. The parasympathetic nerves slow down the heart rate through the vagus nerve, while the sympathetic nerves speed up the heart rate by releasing norepinephrine and adrenaline. In response to emergency situations, such as the "fight or flight" response, the secretion of adrenaline will significantly increase the heart rate to ensure that the body has an adequate blood supply.

\subsection{Electromyography (EMG) signal}

The recording of muscle electrical activity is an important biomedical measurement technology, which is widely used in medical diagnosis, exercise physiology, human-computer interaction and other fields. By recording muscle electrical activity, it is possible to evaluate muscle function, diagnose neuromuscular diseases, and assist in motor function rehabilitation training. In recent years, with the advancement of signal processing and machine learning technology, the application of muscle electrical signals in emotion recognition has gradually become a research hotspot.

There are two main methods for recording muscle electrical activity: surface electrodes and needle electrodes. Surface electrodes are a non-invasive method that detects electrical signals generated by muscle groups during contraction by attaching electrodes to the surface of the skin. These signals are actually the superposition of the electrical activities of multiple motor units. A motor unit is a functional unit consisting of a motor neuron and all the muscle fibers it controls. Since surface electrodes record the spatially weighted sum of the electrical activities of motor units, their signal characteristics are affected by multiple factors such as skin thickness, subcutaneous fat layer, and electrode position. Although the accuracy of surface electrodes is slightly lower than that of needle electrodes, they are particularly suitable for continuous monitoring and large-scale data acquisition due to their simple operation, high comfort, and non-invasive nature.

\begin{figure}
    \centering
    \includegraphics[width=0.5\textwidth]{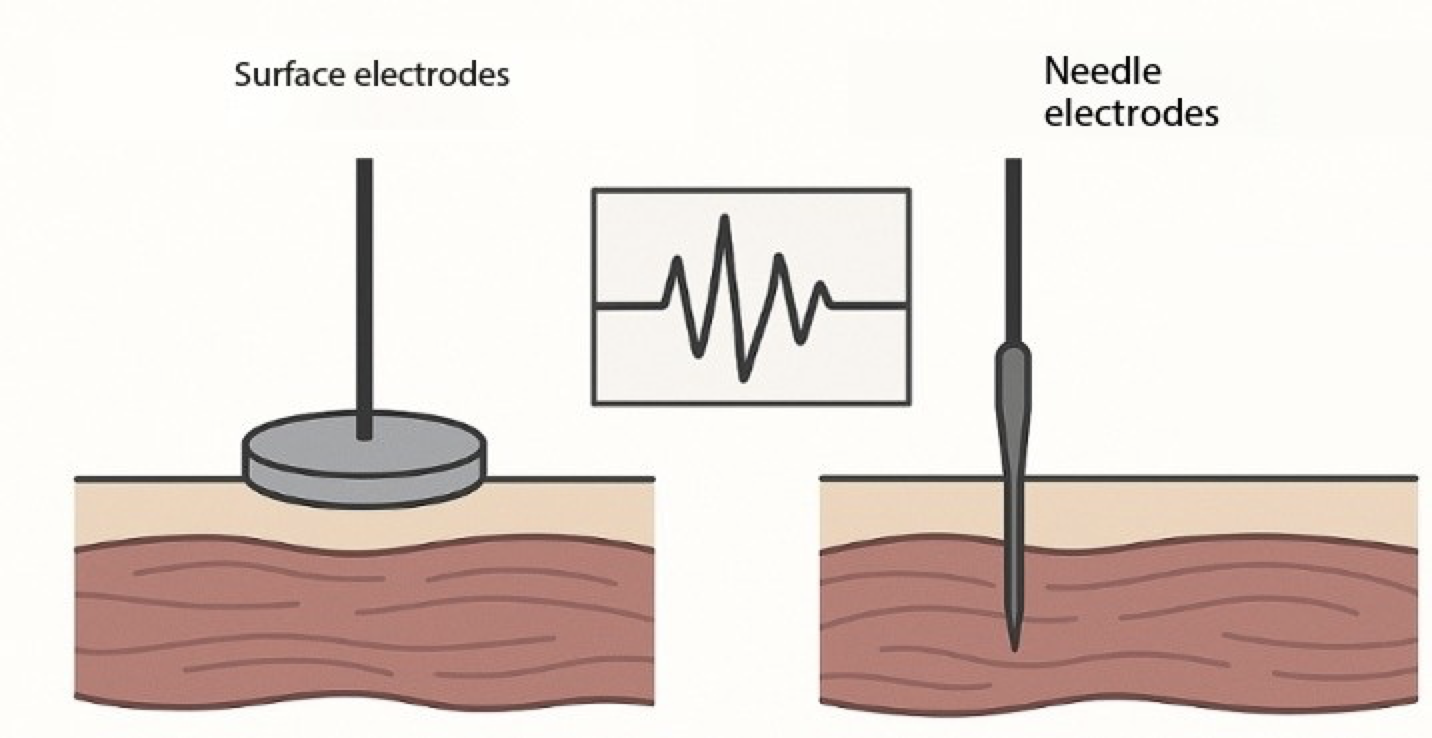}
    \caption{Schematic diagram of surface electrode and needle electrode}
\end{figure}

Needle electrodes are a common choice when more precise information about the electrical activity of muscle fibers is needed. This method records local electrical signals from specific muscle fibers by inserting a needle with an electrode directly into the muscle tissue. Needle electrodes include concentric electrodes and monopolar electrodes. Concentric electrodes consist of a central electrode and a ring-shaped outer electrode that can capture precise electrical signals from adjacent muscle fibers. This method is particularly useful for diagnosing neuromuscular diseases such as amyotrophic lateral sclerosis and myasthenia gravis.

The amplitude and duration of muscle electrical signals are usually within a specific range. The amplitude of electrical signals in healthy muscles is usually between 2 and 6 millivolts, and the duration is 5 to 8 milliseconds. Their typical waveform shows a three-phase potential, which is the electrophysiological characteristic of muscle fiber depolarization and repolarization. In the resting state, normal muscles do not produce significant electrical signals, but when contracted or externally stimulated, the amplitude and frequency of electrical signals will increase significantly. By analyzing these characteristics, the conductivity and excitability of muscles can be evaluated, thereby detecting the functional state of the neuromuscular system.

EMG signals are widely used in medicine. For example, in muscle strength assessment, the amplitude of the EMG signal is proportional to the force generated by muscle contraction. By measuring EMG signals, the strength of muscle contraction can be inferred, which can guide rehabilitation training and sports performance assessment. In addition, in the diagnosis of neurological diseases, the pattern of abnormal muscle fiber activity can be identified by analyzing the frequency and time domain characteristics of EMG signals.

It is worth noting that the application of electromyographic signals in emotion recognition has also attracted much attention. Changes in human emotions are often accompanied by subtle muscle movements, and these micro-movements are mainly manifested as changes in facial expression muscles. By recording the electrical signals of facial muscles through surface electrodes, researchers are able to extract feature information for emotion recognition and psychological state analysis. Emotion recognition systems based on electromyographic signals usually include steps such as signal acquisition, feature extraction, and classification. In the feature extraction stage, researchers usually use time domain, frequency domain, and time-frequency domain features, such as root mean square value, frequency mean, and wavelet transform coefficients. Subsequently, the extracted features are classified through machine learning models such as support vector machines, random forests, or deep neural networks to identify the emotional state of individuals.

In practical applications, EMG emotion recognition technology has been widely used in scenarios such as human-computer interaction and mental health monitoring. For example, intelligent interactive systems can use EMG signals to identify the user's emotional state and then make corresponding feedback, such as adjusting the voice assistant's tone or providing personalized content recommendations. In the field of mental health, EMG signals can be used to detect the physiological manifestations of psychological states such as anxiety and depression, and assist psychologists in early screening and diagnosis.

In addition, emotion recognition combined with multimodal data is also a current research trend. The combination of electromyographic signals with physiological signals such as facial expressions, voice, and heart rate can significantly improve the accuracy and robustness of emotion recognition. Multimodal emotion recognition systems can comprehensively analyze physiological and behavioral data of different dimensions to form more comprehensive emotion assessment results.

\section{Physiological signal processing methods}

In the field of affective computing, physiological signals, as an objective, continuous and difficult-to-forge information carrier, have received widespread attention in recent years. The key to emotion recognition is how to extract effective features that reflect the individual's emotional state from massive and complex physiological data, and then achieve efficient and accurate emotion classification. Based on this background, this chapter will systematically explain the physiological signal processing methods for emotion recognition, and build a complete emotion recognition system by introducing key steps such as data preprocessing, feature extraction, and classification algorithms. Emotion is an important and complex part of human cognitive activities. Its physiological basis is manifested in a variety of signals such as heart rate, skin electricity, and brain waves. These signals will change significantly under different emotional states. Therefore, how to fully mine the emotional information in the signal under the influence of noise interference and individual differences has become a core issue in emotion recognition research.

At the beginning of this chapter, we first need to preprocess the collected physiological signals reasonably. The collection process of physiological signals is often subject to various interferences such as environmental noise, instrument errors, and motion artifacts. These interferences will not only reduce the signal quality, but may also seriously affect the subsequent feature extraction and emotion classification work. By performing a series of preprocessing operations such as filtering, denoising, and normalization on the signal, the signal-to-noise ratio of the signal can be effectively improved, making the subsequent processing stage more targeted and effective. The preprocessing process is not limited to traditional filtering methods, but can also combine a variety of technical means such as time domain, frequency domain, and time-frequency domain to perform multi-scale and multi-angle analysis of complex signals, so as to better capture the weak changes and potential patterns in the signal. This part will lay a solid foundation for subsequent content, and also provide theoretical support and practical guidance for understanding the dynamic characteristics of signals.

After preprocessing, feature extraction of physiological signals becomes an indispensable step in emotion recognition. Each emotional state corresponds to specific physiological changes. How to extract effective features from the preprocessed signal is the key to achieving high-precision emotion classification. Feature extraction methods can rely on traditional statistical features, time domain and frequency domain features, or use advanced signal analysis methods such as wavelet transform and empirical mode decomposition to mine local features and instantaneous changes in signals. In addition, with the continuous development of deep learning technology, automatic feature learning methods based on convolutional neural networks and recurrent neural networks are gradually applied to the processing of physiological signals. These methods can automatically capture the hidden high-order nonlinear relationships and complex spatiotemporal dynamic features in the signal, thereby greatly improving the robustness and accuracy of the emotion recognition system. The work of feature extraction is not just a simple conversion of data, but a deep internal logic analysis of physiological signals, providing rich and accurate information support for subsequent emotion classification.

Physiological signal classification is the key step to classify the data after feature extraction according to different emotional states. The selection and design of the classifier directly affects the overall performance of the emotion recognition system. Traditional classification methods such as support vector machines, decision trees, and random forests have achieved good results to a certain extent due to their high generalization ability and adaptability to small sample data. However, with the continuous increase in data volume and the complexity of emotional states, the application of deep learning models, especially long short-term memory networks (LSTM) and convolutional neural networks (CNN) in physiological signal classification has received more and more attention. These models can not only make full use of large-scale data for end-to-end learning, but also maintain high accuracy and robustness when facing high-dimensional and nonlinear data. In the classification stage, how to deal with the fusion of multimodal information, how to achieve a unified description of emotional states between individuals, and how to effectively train and optimize the model have become hot topics and difficulties in current research. By combining traditional methods and deep learning methods, this chapter strives to build an emotion recognition framework with excellent comprehensive performance and strong adaptability, and provide a practical technical solution for practical applications.

\begin{figure}
    \centering
    \includegraphics[width=0.5\textwidth]{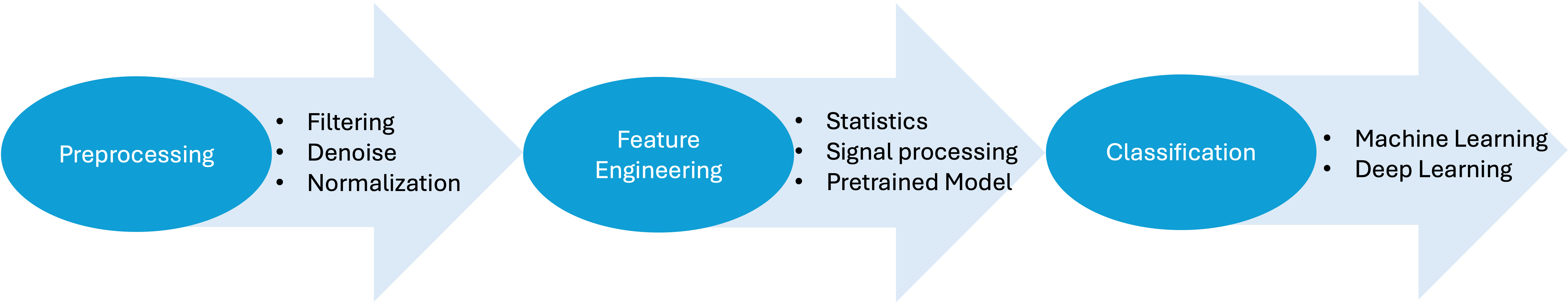}
    \caption{Physiological signal classification flow chart}
\end{figure}

The content of this chapter aims to provide readers with a systematic and complete set of physiological signal processing methods for emotion recognition. From the basic steps of signal preprocessing to the key technologies of feature extraction, and then to the model selection and optimization in the classification stage, each part is closely connected and together constitutes the core module of the emotion recognition system. Through a detailed discussion of each link of physiological signal processing, it can not only help readers deeply understand the basic principles of emotion recognition, but also provide a comprehensive reference for subsequent application practice and theoretical research. In the future, with the continuous advancement of data acquisition technology, computing power and algorithm theory, the application prospects of physiological signals in emotion recognition will be broader. The mutual cooperation and synergy between various sub-modules will further promote the development of the field of emotional computing and promote the in-depth application of emotional intelligence in multiple fields such as medical care, mental health, and intelligent interaction.

In the following chapters, we will discuss in detail the three steps of preprocessing, feature extraction, and classification. First, the physiological signal preprocessing section will introduce data acquisition and noise suppression technology, and explore how to improve data quality through signal filtering, artifact removal, and normalization. Then, in the physiological signal feature extraction section, we will focus on how to extract effective features that can reflect emotional states from preprocessed signals, and compare and analyze traditional methods with emerging technologies. Finally, the physiological signal classification section will introduce in detail the application of various classification algorithms in emotion recognition, analyze the advantages and disadvantages of different models and applicable scenarios, and explore the innovative application of deep learning technology in this field. Through the detailed elaboration of these three sections, we hope to build a research framework for emotion recognition that combines theory and practice for readers, and provide strong theoretical support and technical references for in-depth exploration of related fields in the future.

\subsection{Physiological signal preprocessing}

The problems caused by artifacts in biomedical signals are wide-ranging and varied, and their potential degradation effect on the performance of state-of-the-art signal processing algorithms is high. The magnitude of the noise removal problem and its importance are fully reflected in the scope of this chapter. By analyzing the various types of artifacts that contaminate biomedical signals and exploring filtering techniques to remove them without degrading the quality of the signal of interest, we can gain a deeper understanding of the challenges and solutions in this area. If a subject coughs or twists during ECG acquisition, the electromyogram associated with such activities will constitute interference or artifacts. In adult patients, this physiological interference can be minimized through strict guidance and self-control, however, this solution may not be applicable to infants and children. An interesting example of physiological interference is when the mother's ECG appears together with the fetus's ECG, the latter of which is the real signal of interest. In this case, external control is neither feasible nor ideal, and researchers are forced to develop innovative solutions to extract the desired signal.

Since most biomedical signals are weak at their source, high amplification factors of several hundred to several thousand times may be required. Electronic noise in the instrument amplifier is also amplified along with the desired signal. Although the thermal component of the noise can be reduced by cooling the equipment to very low temperatures, this step may not be practical or cost-prohibitive in most applications. Low-noise power supplies and modern electronic amplifiers with high input impedance, high common-mode rejection ratio, and high power supply rejection ratio are necessary when acquiring biomedical signals. Our environment is filled with natural and man-made electromagnetic waves. Electromagnetic waves broadcast by radio and television stations, as well as radiated by fluorescent lighting equipment, computer monitors, and other systems used in the laboratory or work environment, are picked up by cables, equipment, and connectors. 50 Hz or 60 Hz power waveforms are notorious for the many ways they can mix with and contaminate the signal of interest. This interference can be considered interference caused by the experimental environment. In most cases, simple electromagnetic shielding of cables and grounding of the equipment chassis can reduce electromagnetic and power interference. For experiments dealing with extremely weak signals, it may be necessary to use a metal mesh shielding cage to contain the subject and the instrument.

The ECG is a relatively strong signal with an easily identifiable waveform, and most types of interference affecting its signal can be removed by bandpass filtering. However, other signals with less identifiable waveforms and wider bandwidths may not be suitable for simple filtering procedures. For signals such as event-related potentials or somatosensory evoked potentials, the noise level may be much higher than the signal level, making the latter difficult to discern in a single recording. Before attempting to filter or preprocess the signal, it is important to have a good understanding of the noise processes involved. In chest lead ECG signals, low-frequency artifacts and baseline wandering can be caused by coughing or breathing with large chest movements, or by movement of the arms or legs when acquiring limb lead ECGs. Electrogastrograms are a common source of artifacts in chest lead ECGs. Low-frequency artifacts can also be caused by poor electrode contact or electrode polarization. Baseline wandering can also sometimes be caused by temperature changes and offsets in the instrument and amplifier. ECG signals with low-frequency artifacts can make isoelectric analysis of the ST segment difficult, and large baseline wandering can even cause positive or negative peaks in the ECG to be truncated by the amplifier or analog-to-digital converter.

\begin{figure}
    \centering
    \includegraphics[width=0.5\textwidth]{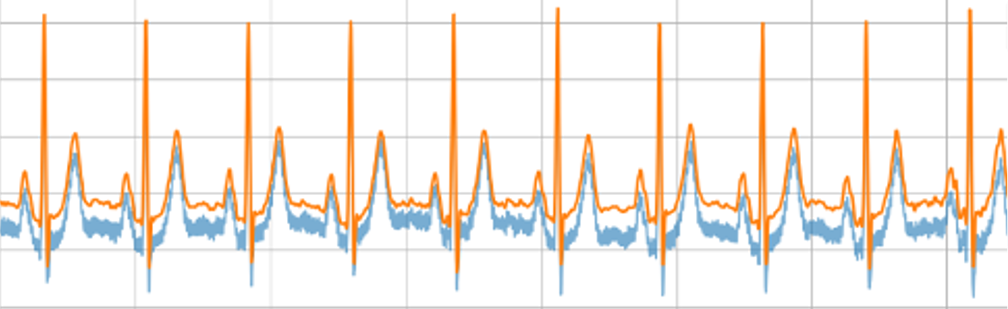}
    \caption{ECG with baseline drift before and after processing}
\end{figure}

The most common periodic artifacts in biomedical signals are 50 Hz or 60 Hz power supply interference. If the power supply waveform is not a pure sine wave due to distortion or truncation, harmonics of its fundamental frequency may also appear. If the interference is a periodic waveform but not a sine wave, such as a rectangular pulse, the harmonics will also be apparent. Power supply interference may be difficult to detect visually in signals with undefined waveforms such as phonocardiograms or electromyograms; however, if the interference occurs in signals with well-defined waveforms such as electrocardiograms or carotid pulse signals, it is easy to observe. In either case, the power spectrum of the signal should clearly indicate the presence of power supply interference, which appears as pulses or spikes at 50 Hz or 60 Hz; if harmonics are present, additional spikes will appear at integer multiples of the fundamental frequency. An ECG signal with 60 Hz interference will show the regular or periodic structure of the interference superimposed on the ECG waveform, and its power spectrum will clearly show a spike at 60 Hz, and additional spikes at 180 Hz and 300 Hz, representing the third and fifth harmonics, respectively. The recommended sampling rate for ECG signals is 500 Hz, but a higher sampling rate of 1000 Hz was used in this case because the ECG was recorded along with the phonocardiogram as a reference signal. A wider bandwidth also helps to better show artifacts and filtering effects. The bandwidth of interest for ECG signals is usually in the range of 0.05 - 100 Hz, including the 60 Hz component, so simple low-pass filtering is not suitable for removing power supply interference. Low-pass filtering the ECG below 60 Hz may smooth and blur the QRS complex, while affecting the PQ and ST segments. The ideal solution is to remove the 60 Hz component without sacrificing any other components.

In the study of emotion recognition, the preprocessing of physiological signals is a crucial step, because the original signals are often interfered by various noises, which may mask key emotional features. In order to extract meaningful patterns from these signals, we need to adopt appropriate signal processing techniques, among which time domain filters have attracted much attention due to their unique advantages. A notable feature of time domain filtering is that it can process the signal directly without in-depth analysis of the frequency characteristics of the signal and noise, at least in most cases without the need for spectral characterization in a direct way. In addition, compared with frequency domain filtering, time domain processing usually has a faster computational speed, which is particularly important in real-time emotion recognition applications.

Among the many time-domain techniques, synchronous averaging is a very effective method, especially for situations where the signal or event of interest can be acquired multiple times. The core of this method is to solve a common problem: when the frequency range of the signal and the noise overlap, traditional linear filters often have difficulty in effectively separating them. Synchronous averaging addresses this challenge by taking advantage of the repetitiveness of the signal. For example, when dealing with event-related potentials or somatosensory evoked potentials, we can record the corresponding signal segments multiple times by repeatedly applying the stimulus. Then, using the time when the stimulus occurred as a reference point, these segments are aligned and averaged. This method can also be applied to ECG signals by detecting the position of the QRS complex in the heart cycle, aligning the waveforms, and averaging to filter out noise. As long as the noise is random and has a zero average value and is not correlated with the signal itself, the averaging process can significantly improve the signal-to-noise ratio. In other words, as the number of averages increases, the impact of noise gradually decreases, while the clarity of the signal continues to increase. This improvement is not linear, but rather increases proportionally with the square root of the number of acquisitions, so the more signal segments acquired, the better the final signal-to-noise ratio is.

In addition to ECG and event-related potentials, synchronous averaging can also be extended to the processing of other physiological signals, such as electromyographic signals. Electromyographic signals record the potential changes during muscle activity and are often used to study facial expressions or body movements in emotional expression. Since electromyographic signals are easily affected by environmental noise, power supply interference, and other physiological activities, direct analysis is often difficult. Through synchronous averaging, we can use the repeatability of specific muscle movements to enhance signal quality. For example, in the experiment, the subjects are asked to repeat a certain expression movement (such as smiling or frowning) many times, and the corresponding electromyographic signals are recorded each time. Then, these signal fragments are aligned with the moment when the movement starts as the reference point and the average is taken. In this way, random noise is effectively weakened, while the muscle activity patterns related to expression are retained and highlighted. This method is particularly suitable for studying subtle emotional clues such as micro-expressions because it can extract reliable signal features in a noisy background.

In practice, the process of synchronous averaging requires several key steps. First, we need to obtain multiple repeated signal records. These signals may come from reactions triggered by external stimuli or naturally repeated events in quasi-periodic signals. Taking electromyographic signals as an example, the reference point for alignment can be determined by detecting the starting point of muscle contraction. Secondly, the signal part corresponding to the event is extracted and accumulated in a buffer. It should be noted that the duration of different signal segments may not be exactly the same, so the reference point should be used as the benchmark when aligning, and there is no need to force the uniformity of the end of the segment. Finally, the accumulated result is divided by the number of signal segments to obtain the averaged signal. This method relies on the precise alignment of the repeated parts of the signal. If the alignment is deviated, the signal features may become blurred on the time axis, thus affecting the accuracy of the analysis.

Taking visual evoked potential as an example, a single acquired signal often contains a lot of noise, making it difficult to directly distinguish key features. However, when we average the responses to multiple flash stimuli, the main peaks and valleys in the signal gradually emerge. For example, in normal visual evoked potentials in adults, the first positive peak usually appears about 120 milliseconds after the stimulus, often called P120, while the negative valleys before and after it are marked N80 and N145, respectively. By averaging multiple recordings, the latency of these features can be accurately measured, and the results usually fall within the normal range. This improvement in clarity is crucial for emotion recognition, because emotional states may be reflected by small changes in these latencies.

In the processing of ECG signals, synchronous averaging also demonstrates its powerful capabilities. Suppose we are faced with a noisy signal containing multiple heartbeats. We can first extract a typical QRS complex from the signal as a template, and then find the position of each heartbeat through template matching technology. This matching process is achieved by calculating the similarity between the template and the signal. Positions with high similarity usually correspond to the appearance points of the QRS complex. After determining these positions, we align and average the corresponding signal fragments, thereby significantly suppressing the impact of noise. The final average waveform not only removes random interference, but also retains the typical characteristics of the heartbeat, providing a reliable basis for subsequent sentiment analysis.

For the application of EMG signals, the potential of synchronous averaging is also worth exploring. For example, when studying emotions such as anger or happiness, subtle contraction patterns of facial muscles may be reflected through EMG signals. By recording the EMG response of the same expression multiple times and averaging it based on the starting point of muscle activity, we can filter out irrelevant noise and highlight the muscle activity characteristics related to emotion. This method is not only applicable to facial EMG signals, but can also be extended to the analysis of body movements, such as EMG changes in actions such as clenching a fist or waving a hand, providing more possibilities for multimodal emotion recognition.

Another major advantage of synchronous averaging is that it does not rely on frequency domain filtering, so the frequency content of the signal is not lost. Unlike traditional low-pass filters or moving average filters, this method preserves the integrity of the signal and avoids the distortion of characteristics caused by filtering. In addition, synchronous averaging can effectively suppress some structured noise, such as power supply interference. As long as the phase of the interference is different in each signal recording, the averaging process will weaken it. To fully utilize this property, the repetition rate of the stimulus should be as far as possible from the power supply frequency. For example, when collecting visual evoked potentials, the frequency of the flash stimulus can be set to 2.1 times per second instead of a value directly related to the power supply frequency of 50 Hz or 60 Hz.

In the preprocessing of physiological signals for emotion recognition, the diversity of time domain techniques provides us with a wealth of tools to deal with noise problems in different scenarios. When we cannot apply synchronous averaging techniques by acquiring signals multiple times, how to effectively remove random noise in a single signal record becomes a key issue that needs to be solved. At this time, the moving average filter, as a classic time domain method, can fill this gap well. Its core idea is to weaken the influence of noise through temporal smoothing. This method assumes that the characteristics of the signal have a certain stability in the time dimension, so the average value of the signal over a short period of time can be used to approximate its true form. Unlike synchronous averaging, which requires multiple copies of the signal, the moving average filter relies only on a single signal, and generates a smoothed output by sliding a window along the time axis and performing a weighted average of the samples in the window. Because this window is constantly moving, it can provide a new estimate at each time point of the signal, hence the name "moving average".

The implementation of a moving average filter is very intuitive. It calculates the output by taking a weighted combination of the current value of the signal and several previous values. This weighting process is determined by a set of filter coefficients, which not only reflect the importance of each sample, but also implicitly normalize the number of samples. For example, a simple and commonly used form is the Hanning filter, which smoothes by taking a specific weighted average of the current sample, the previous sample, and the previous two samples. The design of this filter is inspired by the early practice of signal processing and can effectively reduce high-frequency noise without over-complication. Imagine that when we are faced with a piece of electromyographic signal mixed with sharp fluctuations, the Hanning filter is like a gentle iron, ironing out those abrupt noise peaks while trying to preserve the core features related to muscle activity in the signal. This smoothing effect is particularly important in emotion recognition because it helps us more clearly observe the physiological changes related to emotional expression, such as the subtle rise and fall of facial muscles when smiling or frowning.

However, the moving average filter is not a panacea. Its smoothing characteristics can remove noise but may also blur the rapidly changing details in the signal. For example, in an electrocardiogram signal, the steep peak of the QRS complex may become less obvious due to over-smoothing, thus affecting subsequent sentiment analysis. Therefore, in practical applications, we need to reasonably select the filter window length and weighting method based on the characteristics of the signal and the needs of emotion recognition. For electromyographic signals, since their frequency range is usually wide and emotion-related features may be hidden in rapid muscle contractions, shorter windows may be more suitable to avoid losing key information.

\begin{figure}
    \centering
    \includegraphics[width=0.5\textwidth]{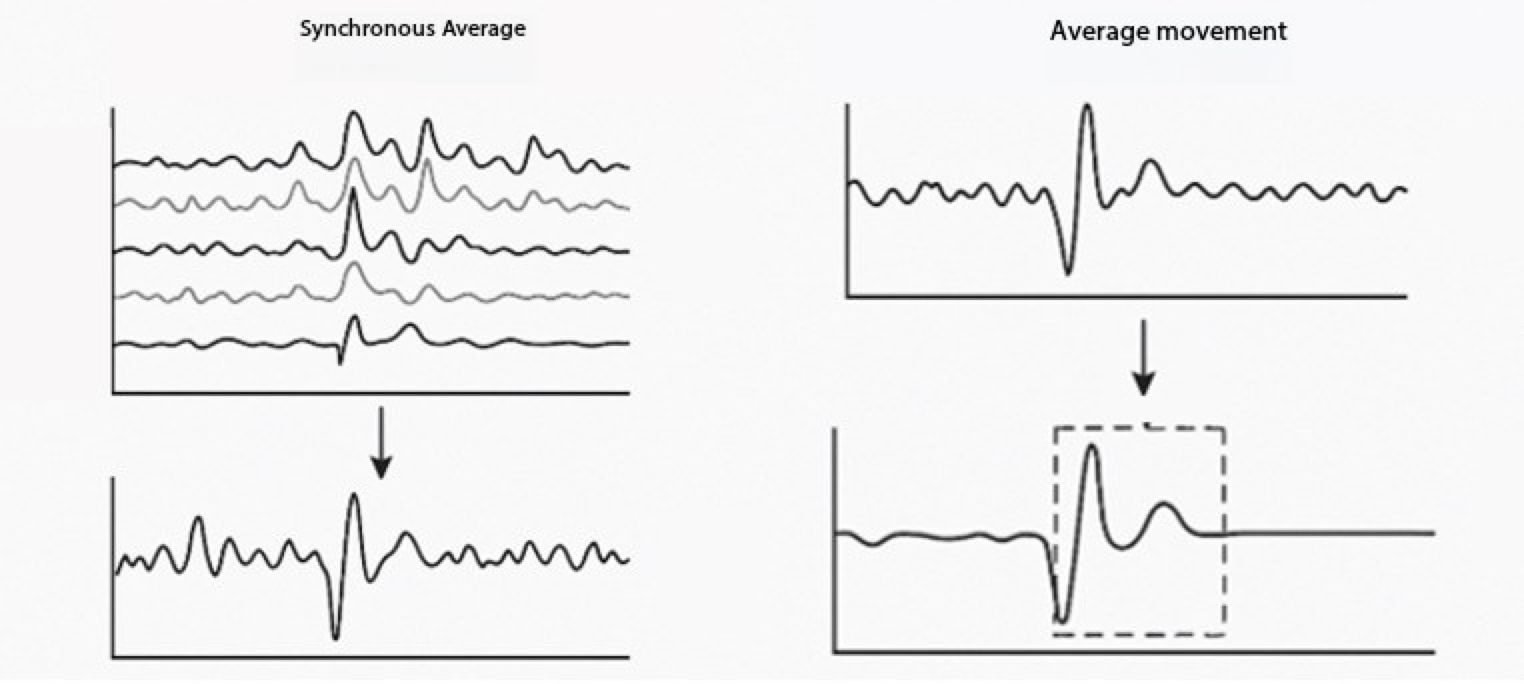}
    \caption{Schematic diagram of synchronous average and moving average filtering}
\end{figure}

In addition to smoothing high-frequency noise, physiological signal preprocessing also needs to deal with the interference of low-frequency artifacts, such as baseline drift, which is common in ECG signals. This drift is usually caused by breathing, body movement, or poor electrode contact, which will introduce slow fluctuations in the signal and directly affect the extraction of emotional features. At this time, derivative-based operators become a powerful time domain tool. The essence of the derivative operation is to capture the rate of change of the signal. It has no response to the constant components in the signal, but is extremely sensitive to the rapidly changing parts. Imagine that when we apply a derivative operation to an ECG signal with baseline drift, the slowly fluctuating low-frequency components will be greatly weakened, while high-frequency features such as the QRS complex will be highlighted. This feature makes the derivative operation similar to a natural high-pass filter, which can effectively remove the DC component and suppress low-frequency interference.

In practice, derivatives can be implemented by differential methods in digital signal processing. For example, the simplest form is to calculate the difference between the current sample and the previous sample. This first-order differential operation can quickly reflect the local change trend of the signal. If the high-frequency component needs to be further enhanced, we can apply the differential twice in succession to obtain the second-order derivative, which has the effect of measuring the acceleration of the signal change. This method is also promising in the processing of electromyographic signals. For example, when analyzing angry expressions, facial muscles may show rapid and violent contractions. The second-order derivative can help us capture the dynamic characteristics of these moments, thereby revealing hidden emotional clues. However, the derivative operation also has its limitations. Due to its amplification of high-frequency components, the noise in the signal may be enhanced together, especially in the first-order difference, where this effect is particularly obvious. To alleviate this problem, we can adopt an improved method, namely the three-point central difference. This method can suppress low-frequency artifacts and control the amplification of high-frequency noise to a certain extent by combining the information of the current sample with the two previous and next samples.

Taking the electrocardiogram signal as an example, when we are faced with a waveform contaminated by low-frequency baseline drift, both the first-order difference and the three-point center difference can significantly remove the drift effect. However, the results show that while both methods highlight the QRS complex, they also inevitably weaken the morphology of the P wave and the T wave, and even make the entire waveform look unlike a typical electrocardiogram signal. This reminds us to be cautious when using derivative operations, especially in emotion recognition. If the goal is to preserve the complete morphology of the signal to analyze features such as heart rate variability, the derivative operation may not be the best choice. However, it still has its unique value in certain specific tasks, such as detecting the exact location of the QRS complex, which is crucial for subsequent synchronous averaging or other analysis steps.

In contrast, frequency domain filters offer a completely different approach. If time domain filters directly trim and polish the "time story" of the signal, frequency domain filters decompose the signal into different frequency chapters and then edit these chapters in a targeted manner. Although moving average filters and derivative operations in the time domain are simple and efficient, their frequency responses are often indirectly generated byproducts rather than the focus of the initial design. For example, when the moving average filter suppresses high-frequency noise, its effect on the frequency axis is not very sharp, and it usually only shows significant attenuation near a specific frequency. Although the high-pass characteristics of the derivative operation are clear, the law of its gain changing with frequency is also relatively rough and lacks precise control.

The advantage of frequency domain filters is that they allow us to tailor the frequency characteristics of the signal according to the specific needs of emotion recognition. Whether it is low-pass filtering to retain the slow changes of the signal, high-pass filtering to highlight fast dynamics, or even band-pass or band-stop filtering to focus on a specific frequency band, we can achieve it through frequency domain design. For example, in the processing of electromyographic signals, if we know that the muscle activity related to emotion is mainly concentrated in a certain frequency range, we can design a bandpass filter to accurately extract this part of information while removing irrelevant low-frequency drift and high-frequency noise. This flexibility is particularly important in multimodal emotion recognition, because different physiological signals (such as electrocardiogram, electromyography, and EEG) may require completely different frequency processing strategies.

In practice, frequency domain filters can be implemented in a variety of ways. We can first perform a Fourier transform on the signal, adjust its frequency components in the frequency domain, and then return to the time domain through an inverse transform; or we can directly design an equivalent time domain filter and apply it to the signal samples. Classic design methods include Butterworth filters, Chebyshev filters, elliptic filters, and Bessel filters, each of which focuses on smoothness, transition steepness, and phase characteristics. For example, the Butterworth filter is known for its smooth passband response, which is very suitable for sentiment analysis that needs to preserve the natural form of the signal; while the elliptic filter is known for its steep transition band, which can quickly separate the target frequency band from the noise in the electromyographic signal.

\subsection{Physiological signal feature extraction}

One of the most important time-domain features of an ECG signal is the duration of the cardiac cycle. Typically, the length of the cardiac cycle is derived by measuring the time span from one R wave to the next. Other relevant features include the duration of individual waves, such as the duration of the QRS complex, and the time interval between waves, such as the time interval between the T wave and the P wave. The measurement of this interval is particularly important because it reflects the relationship between the cardiac repolarization phenomenon and the sinus node pulse rate.

The duration of the QRS complex is an important time domain feature. It is usually identified by its unique shape and relatively stable time constant. On the other hand, the time interval between the T wave and the P wave also has significant clinical significance, which reveals the separation of two important events in the electrical activity of the heart. In the electrocardiogram, the QRS complex is mainly manifested as a high-frequency area, while the P wave and T wave are mainly composed of low-frequency components. The ST segment is limited in time and mainly contains low-frequency content.

There are significant differences in the frequency content of normal and abnormal ECGs. A normal heart rate is usually between 60 and 100 beats per minute, while a patient with atrial fibrillation may have a heart rate of more than 200 beats per minute. In addition to changes in frequency, the slopes of the heart's depolarization and repolarization processes also change in pathological conditions. These changes require a wider frequency bandwidth to fully describe the different phenomena. In general, a standard ECG can be basically described by the first eight harmonics of the Fourier transform. Fourier analysis provides a basic representation of the frequency domain characteristics of the ECG. For seemingly normal ECGs, frequency analysis usually covers the range of 0 to 100 Hz. When detecting arrhythmias, the frequency analysis may need to be extended to 200 Hz. It is worth noting that as the frequency increases, noise gradually dominates the spectrum, reducing the effectiveness of the additional information.

\begin{figure}
    \centering
    \includegraphics[width=0.4\textwidth]{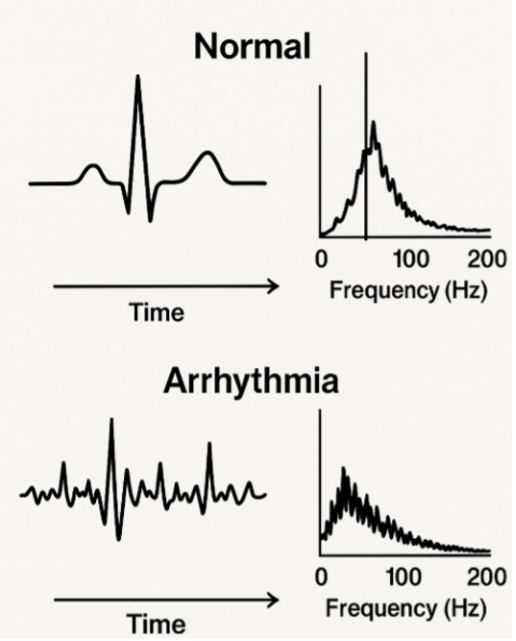}
    \caption{Schematic diagram of normal and abnormal ECG frequency comparison}
\end{figure}

In addition, the frequency effects measured by electrodes do not always reflect the true spectrum of cellular activity. Different types of noise, including respiratory signals and electromyographic signals of skeletal muscles, often interfere with the measurement results. For this reason, filters are usually used to remove these noises, such as applying notch filters to remove device and capacitor noise. Respiration signals are one of the common artifacts. Since breathing involves the movement of the chest and diaphragm, the amplitude of the electromyographic signals generated is large. Although the respiratory frequency is much lower than the heart rate, the displacement of the electrodes will cause modulation effects in the spectral content of the signal. Monitoring the breathing process through a flow probe and subtracting the relevant signal from the electrocardiographic signal is an effective noise reduction method.

Wavelet analysis is also widely used in ECG signal processing. Unlike traditional Fourier transform, wavelet transform can provide localized information in the time-frequency domain, which helps to capture short and complex events in ECG signals. It reveals the characteristics of the signal at different time scales by convolving the signal with the mother wavelet. Wavelet analysis is particularly suitable for detecting the pattern and positioning of QRS complex waves, and can also quantify the amplitude and scaling of these waves. Daubechies and Coiflet wavelets are commonly used tools in ECG analysis.

In the analysis of physiological signals, EMG signals are also an important research object. Since EMG signals reflect the electrical activity of muscle fibers, they are often used to evaluate muscle function, diagnose neuromuscular diseases, and study muscle fatigue. In order to extract effective features from EMG signals, time domain, frequency domain, and time-frequency domain analysis methods are usually used.

A commonly used time domain feature is the root mean square value, which is used to represent the power level of the signal. The root mean square value provides a quantification of the strength of muscle contraction, which in turn reflects the ability of the muscle to generate force. In addition, the average rectified value is also often used in EMG analysis. It describes the average value of the absolute changes in the signal and is used to evaluate the smoothness or non-smoothness of the signal. In the frequency domain, EMG signals are usually concentrated between 50 and 500 Hz, and their main energy distribution is between 70 and 300 Hz. By analyzing the spectral characteristics of the EMG, the changes in muscles under contraction and fatigue states can be effectively distinguished. During muscle fatigue, the power spectrum usually shifts from high frequency to low frequency, and this feature provides a strong basis for evaluating muscle function.

The application of wavelet analysis in EMG signals also shows significant advantages. Through wavelet decomposition, the characteristic changes of EMG signals at different time scales can be detected. For example, wavelet analysis can identify different types of contraction patterns during isometric contractions and detect the difference between fast contractions and slow contractions. In addition, wavelet analysis can also be used to evaluate the recruitment status of motor units. Detecting the response delay of motor units through short-time Fourier transform or wavelet transform can help identify abnormalities in neuromuscular function.

\subsection{Physiological signal classification}

The classification of ECG signals is mainly used to identify abnormal states of the heart, and its methods can be divided into traditional methods and machine learning-based methods. Traditional methods usually rely on the extraction of time and frequency domain features of the signal, such as the morphology of the QRS complex, the RR interval, and heart rate variability. Through these features, researchers can use threshold judgment or rule systems for classification. A classic example is the Pan-Tompkins algorithm, which extracts QRS complex features through steps such as bandpass filtering, differentiation, square and moving window integration, and then realizes the classification of heartbeats. This method is simple and intuitive, with low computational complexity, and is suitable for some basic classification tasks. However, with the increase in data complexity and the increase in classification requirements, the limitations of traditional methods have gradually emerged, especially when dealing with diverse and noisy ECG signals.

In recent years, the rapid development of machine learning technology has brought new possibilities for ECG signal classification. Support vector machine is a common machine learning algorithm that distinguishes normal heartbeats from abnormal heartbeats by finding the optimal hyperplane. Studies have shown that the classification accuracy of support vector machine can be significantly improved when time domain and frequency domain features are used in combination. Random forest, as an integrated learning method, can effectively process high-dimensional feature data by constructing multiple decision trees and integrating their results for classification, and has good robustness and generalization ability. In addition, with the rise of deep learning technology, the application of neural networks in ECG signal classification is increasing. Convolutional neural networks can automatically extract the spatial features of signals, while recurrent neural networks are good at processing time series data. Researchers have found that combining these two networks into a hybrid model can further improve the classification performance. For example, a study proposed a hybrid model based on convolutional neural networks and recurrent neural networks for automatic detection of arrhythmias. The model can directly extract features from the original signal and classify it, achieving a high accuracy rate.

In actual research, the methods of ECG signal classification are constantly innovating. For example, some scholars use transfer learning technology to migrate the model pre-trained on a large-scale dataset to a small sample dataset, thereby improving classification performance. This method is particularly suitable for scenarios with limited data resources. In addition, for applications with high real-time requirements, such as wearable ECG monitoring devices, researchers are also exploring lightweight models to reduce computational complexity while ensuring accuracy.

The classification of electromyographic signals mainly serves motion recognition and muscle function assessment, and its methods can also be divided into traditional methods and machine learning-based methods. Traditional methods usually rely on the extraction of time domain features, such as root mean square, zero crossing rate, and wavelength, and judge the activity state of the muscle through threshold judgment or pattern matching. For example, the threshold method based on root mean square features can quickly detect whether the muscle is in an active state. The advantage of this method is that it is simple to implement and has a small amount of calculation, but it often performs poorly in complex motion recognition or multi-channel signal processing.

Similar to ECG signals, machine learning techniques have also been widely used in EMG signal classification. Support vector machines perform well in multi-class gesture recognition, especially when combined with time domain and frequency domain features, they can effectively distinguish different gesture patterns. The K nearest neighbor algorithm is an instance-based learning method that classifies by calculating the distance between the sample to be classified and the training sample. It is particularly suitable for classification tasks with small sample data sets. In recent years, deep learning models have gradually dominated the classification of EMG signals. Convolutional neural networks can automatically extract the spatial features of signals, while long short-term memory networks are good at processing time series data. Studies have shown that combining these two networks into a hybrid model can significantly improve the accuracy and real-time performance of EMG signal classification. For example, a study proposed a gesture recognition method based on multi-channel EMG signals, which uses convolutional neural networks to simultaneously process signal data from multiple channels to achieve high-precision gesture classification.

In practical applications, research on EMG signal classification is also making continuous breakthroughs. For example, some scholars have designed a real-time EMG signal classification system based on embedded systems, which uses a lightweight machine learning model to achieve low-power operation and high real-time gesture recognition. This system is particularly suitable for scenarios that require instant feedback, such as prosthetic control. In addition, there are studies that further improve the recognition accuracy of complex movements through the analysis of multi-channel EMG signals combined with deep learning models, providing new technical support for the field of human-computer interaction.

When selecting a classification algorithm for ECG and EMG signals, multiple factors need to be considered, including accuracy, real-time performance, computational complexity, and the characteristics of the data set. Deep learning models usually have higher accuracy, but require a large amount of data for training and have high computing resource requirements, so they are more suitable for scenarios with sufficient data and good hardware conditions. Traditional methods and shallow machine learning models, such as support vector machines and random forests, often perform better on small sample data sets and have lower computational complexity, making them suitable for applications with high real-time requirements, such as wearable devices or embedded systems. In addition, characteristics of the data set, such as noise level and class imbalance, will also affect the choice of algorithm. For data with high noise, robust algorithms (such as random forests) may have more advantages; for data with class imbalance, data enhancement or transfer learning techniques may be needed to improve performance.

In recent years, researchers have proposed many innovative classification methods, which have promoted the development of ECG and EMG signal analysis. In the field of ECG signals, some studies have achieved automatic arrhythmia detection through deep learning, and used a hybrid model of convolutional neural networks and recurrent neural networks to directly extract features from the original signals and classify them, significantly improving the diagnostic efficiency. In addition, the application of transfer learning also provides new ideas for the classification of small sample data sets. By using pre-trained models, researchers can obtain better classification results when data is limited. In the field of electromyographic signals, multi-channel signal processing and real-time classification are research hotspots. For example, some studies have achieved high-precision recognition of complex gestures by processing multi-channel electromyographic signals through convolutional neural networks; and lightweight classification models based on embedded systems provide practical solutions for prosthetic control and rehabilitation training.

Physiological signal classification is a challenging and promising research field. The classification methods of ECG and EMG signals have evolved from traditional feature extraction to automatic feature learning based on machine learning, which has greatly improved the classification performance. Traditional methods still have a place for their simplicity and efficiency, while machine learning methods, especially deep learning models, have become mainstream with their powerful feature extraction capabilities and high accuracy. In practical applications, researchers continue to promote technological progress through innovative algorithms and system designs, such as using transfer learning to solve small sample problems or developing lightweight models to meet real-time requirements. In the future, with the advancement of data acquisition technology and the improvement of computing power, the research on physiological signal classification can further explore cutting-edge technologies such as small sample learning to adapt to small sample scenarios, while paying attention to the interpretability and real-time performance of the algorithm, bringing greater breakthroughs to fields such as medical diagnosis and human-computer interaction.

\section{Sentiment Analysis of Physiological Electrical Signals}

As the direct output of the human autonomic nervous system, physiological electrical signals can reflect the dynamic changes of emotions with millisecond accuracy, providing an objective biological basis for the quantification and identification of emotional states. This chapter will systematically explain the application framework of physiological electrical signals in emotion analysis, starting from the correlation mechanism between physiological signals and emotional states, explore the key technical paths of emotional computing, and lay a theoretical foundation for building an accurate and explainable emotion recognition model.

The human body's emotional experience is essentially an integrated response of the nervous system to external stimuli, a process achieved through the synergistic action of the autonomic nervous system and the central nervous system. When an individual experiences emotions such as pleasure, fear, and sadness, the balance between the sympathetic and parasympathetic nerves changes, which in turn triggers synchronous fluctuations in physiological indicators such as heart rate, skin conductance, and electromyographic activity. For example, anxiety is often accompanied by an acceleration of heart rate and an increase in skin conductance, while a state of relaxation is associated with a slowing of breathing rhythm and an increase in the energy of EEG alpha waves. These physiological electrical signals can not only distinguish the valence (positive/negative) and arousal (high/low) of emotions, but also reveal the temporal evolution characteristics of emotions. Taking the electrocardiogram signal as an example, the high-frequency component of its RR interval variability is closely related to parasympathetic nerve activity and can be used as a biomarker of emotion regulation ability; while the phase response of skin electrical activity is highly synchronized with emotional arousal events and is widely used to detect instantaneous emotional fluctuations.

It is worth noting that the multimodal characteristics of physiological electrical signals provide complementary information for emotion analysis. Electroencephalography (EEG) reflects the emotional load in cognitive processing through cortical neural oscillations, electromyography (EMG) captures muscle motor unit potentials related to facial micro-expressions, and electrooculography (EOG) associates attention allocation with emotional salience assessment. This coupled analysis of multi-channel signals can break through the information limitations of a single modality and build a more robust emotion representation system. However, the relationship between physiological signals and emotional states is not a simple linear correspondence. Individual differences, environmental interference, and the subjectivity of emotional experience all introduce complexity. For example, the same fear stimulus may cause significant differences in the degree of sympathetic nerve activation in different subjects, which is closely related to the individual's emotion regulation strategy, genetic traits, and even cultural background. Therefore, the establishment of a universal emotion-physiology mapping model requires taking into account both group rules and personalized adaptation.

\begin{figure}
    \centering
    \includegraphics[width=0.5\textwidth]{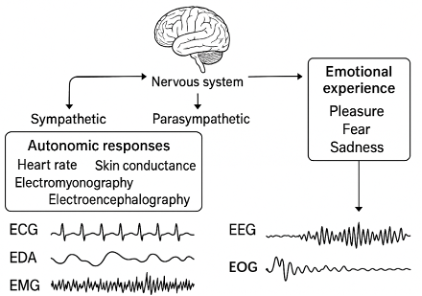}
    \caption{Multimodal physiological electrical signals}
\end{figure}

Emotional analysis of physiological electrical signals is essentially a cross-modal mapping problem of "signal → feature → emotion". Its technical implementation involves the deep integration of signal preprocessing, feature engineering and machine learning algorithms. In the signal preprocessing stage, it is necessary to design noise reduction schemes based on the physical characteristics of different physiological modalities: for example, wavelet transform is used to eliminate power frequency interference in ECG signals, and independent component analysis is used to separate eye movement artifacts in EEG. Feature extraction needs to take into account the time domain, frequency domain and nonlinear pattern, such as extracting the slope characteristics of tonic/phasic components from EDA signals, or quantifying the complexity of HRV signals through fuzzy entropy. These features constitute a multi-dimensional characterization of emotional states, but their high dimensionality and redundancy also require researchers to use algorithms such as principal component analysis (PCA) and maximum correlation minimum redundancy to optimize features.

In recent years, breakthroughs in deep learning technology have opened up new paths for end-to-end emotion computing. Convolutional neural networks can learn spatiotemporal sensitive feature representations directly from raw physiological signals, while long short-term memory networks are good at capturing emotion-related temporal dependencies. More cutting-edge cross-modal fusion architectures (such as Transformer) integrate multi-channel physiological signals through self-attention mechanisms, showing significant advantages in cross-subject emotion recognition tasks. However, the interpretability of the model remains a key bottleneck restricting clinical applications. To this end, researchers have begun to introduce physiological prior knowledge to constrain model structures, such as using heart rate variability spectral features as input gating signals for LSTM, or using graph neural networks to model the functional coupling relationship between ECG and EEG. This hybrid paradigm of "data-driven + knowledge-guided" is expected to enhance the traceability of its physiological significance while maintaining model performance.

This chapter will systematically discuss the core issues of emotion analysis of physiological electrical signals. First, in the section on the relationship between physiological signals and emotional states, the biological mechanisms of autonomic nervous activity and emotional response will be deeply analyzed, and the complementary representation capabilities of multimodal physiological signals will be explained in combination with classical emotion theory, and the impact of individual differences and situational factors on emotion-physiology associations will be discussed. Subsequently, the section on emotion computing methods for physiological electrical signals will systematically sort out the technical chain of emotion computing, compare the advantages and disadvantages of traditional machine learning and deep learning paradigms, and focus on discussing solutions to key issues such as signal noise reduction, feature optimization, and model generalization.

\subsection{Relationship between physiological signals and emotional states}

The relationship between physiological signals and emotional states has always been an important research topic in the fields of psychophysiology and affective computing. This relationship not only reveals the deep mechanism of human emotions, but also provides broad prospects for technological applications. Studies have found that physiological signals such as heart rate variability, electroencephalogram, skin conductivity, and breathing rate can reflect the emotional state of an individual to a certain extent. For example, when a person is in a highly excited mood, the heart rate may increase and the skin conductivity will also increase; and changes in specific frequency bands in the electroencephalogram, such as alpha wave activity in the frontal lobe area, may be related to the experience of positive emotions. However, this association is not simple and direct. It is deeply affected by individual differences, situational factors, and the complexity of the physiological signals themselves. Therefore, understanding and utilizing this relationship is full of potential, but also accompanied by many challenges.

From a historical perspective, the connection between emotions and physiological signals has long attracted the attention of scholars. In the late 19th century, William James proposed an influential view that emotions are essentially the perception of physiological changes in the body. This theory captured the public's imagination at the time and inspired widespread discussion about the nature of emotions. However, as research deepened, subsequent scholars found that physiological reactions do not always correspond one-to-one to specific emotions. For example, blocking visceral feedback does not completely eliminate emotional experience, and the intervention of cognitive factors can also significantly change an individual's emotional state. This finding shows that the generation and experience of emotions is a multidimensional process, and relying solely on physiological signals may not be enough to fully explain its complexity. Nevertheless, James's theory laid the foundation for later research, prompting scientists to begin exploring more specific physiological indicators, such as whether changes in skin conductivity can reveal hidden psychological activities. This exploration even gave birth to the prototype of lie detection technology. Although its scientific validity was questioned due to methodological issues in the mid-20th century, it undoubtedly promoted the further development of emotion research.

In terms of theoretical framework, modern emotion research mainly relies on two models to understand the relationship between physiological signals and emotional states. One is the discrete emotion model, which believes that humans have several basic emotions, such as happiness, sadness, anger, fear, surprise and disgust, and these emotions are universal and independent. The other is the dimensional model, which places emotions in a continuous coordinate system and uses the two dimensions of arousal and pleasure to describe the intensity and positive and negative nature of emotions. For example, high arousal and high pleasure may correspond to excitement or joy, while low arousal and low pleasure may point to fatigue or depression. These models provide researchers with a clear framework, making the measurement of physiological signals and the interpretation of emotional states more systematic. In addition, some scholars have proposed the concept of "core emotion", believing that the basis of emotion is a primitive positive and negative feeling or a contrast between vitality and laziness, and this basic emotion may be expressed through brain activity and physiological signals. This view further emphasizes that physiological signals may be better at capturing the overall characteristics of emotions rather than those more delicate or philosophical emotional experiences.

In actual measurements, physiological signals mainly come from the autonomic nervous system and the central nervous system. Heart rate variability reflects the changes in heartbeat intervals and is usually regulated by the sympathetic and parasympathetic nervous systems. Studies have found that when individuals are under stress or negative emotions, heart rate variability tends to decrease, while it increases in a relaxed or positive emotional state. Skin conductivity reflects the degree of emotional arousal by measuring sweat gland activity. For example, skin conductance increases significantly when one is afraid or excited. As a representative of the central nervous system, the electroencephalogram can capture the electrical activity of the cerebral cortex. Changes in different frequency bands are closely related to emotional states. For example, the enhancement of alpha waves in the frontal lobe may be associated with calm or positive emotions, while the activity of beta waves may indicate anxiety or tension. In addition, breathing rate also changes with emotional state. For example, breathing becomes rapid and shallow when anxious, and tends to be slow and deep when relaxed. The measurement of these signals has benefited from technological advances in recent years, especially the popularity of wearable devices. Devices such as smart bracelets and heart rate monitors enable researchers to collect data in real time in natural environments, greatly expanding the breadth and depth of research.

With the development of technology, the research on emotion recognition has ushered in new breakthroughs. In particular, the application of machine learning and deep learning technologies has significantly improved the accuracy of extracting emotional information from physiological signals. For example, by applying convolutional neural networks to the analysis of EEG data, researchers have achieved an emotion classification accuracy of more than 90\% on some data sets, which can effectively distinguish positive, neutral, and negative emotions. In addition, more and more studies have begun to adopt multimodal methods, combining multiple signals such as EEG, ECG, and skin conductivity to more comprehensively capture the characteristics of emotional states. The advantage of this method is that it not only relies on the representation of a single signal, but also constructs a more three-dimensional emotional portrait through the fusion of multi-dimensional data. With the continuous improvement of wearable devices, real-time emotion monitoring has gradually moved from the laboratory to daily life, bringing new possibilities to areas such as mental health monitoring and human-computer interaction.

Despite this, this field still faces many challenges. First, individual differences are an issue that is difficult to ignore. The same emotion, such as anger, may trigger completely different physiological reactions in different people. Some people may have a sharp increase in heart rate, while others may show significant changes in skin conductance. This difference makes it difficult for a universal emotion recognition model to be applicable to everyone. Secondly, data labeling is also a big problem. Emotional states are usually dynamic, especially in long-term monitoring. How to accurately label the emotional state at each moment is a huge burden for researchers. In addition, the gap between the experimental environment and real life also limits the promotion of research results. Although well-controlled stimulation conditions in the laboratory facilitate data collection, it is difficult to fully simulate the complex and changeable emotional experience in the natural environment.

Personalized models are considered to be an effective means of dealing with individual differences. Through transfer learning or adaptive algorithms, the system can be adjusted according to each person's physiological baseline, thereby improving the accuracy of recognition. At the same time, advances in unsupervised learning and semi-supervised learning techniques are also expected to reduce reliance on large amounts of labeled data, allowing models to extract meaningful patterns from unlabeled raw data. In addition, the further development of wearable devices will push emotion recognition into the "wild." These devices are not only lightweight and easy to use, but can also covertly record a variety of physiological signals, such as skin conductivity, body temperature, and heart rate, providing researchers with a more realistic and richer source of data. At the same time, the application of virtual reality technology also provides new possibilities for emotion induction. Through immersive scene design, researchers can more directly induce target emotions, thereby improving the quality of data.

The relationship between physiological signals and emotional states is a research field full of complexity but great potential. From historical theoretical exploration to modern technological breakthroughs, this topic has achieved remarkable results in the fields of psychophysiology and affective computing. Although problems such as individual differences, data annotation, and experimental paradigms still exist, with the continuous advancement of personalized models, wearable devices, and machine learning technology, future emotion recognition systems are expected to achieve higher accuracy and practicality. It is expected that in the next decade, this field will usher in greater development, not only to more accurately capture the dynamic changes of emotional states, but also to play an important role in multiple scenarios such as mental health monitoring, educational assistance, and human-computer interaction. By deeply understanding the relationship between physiological signals and emotions, we may be able to better interpret the inner world of human beings and open up new paths for the integration of technology and humanities.

\begin{figure}
    \centering
    \includegraphics[width=0.5\textwidth]{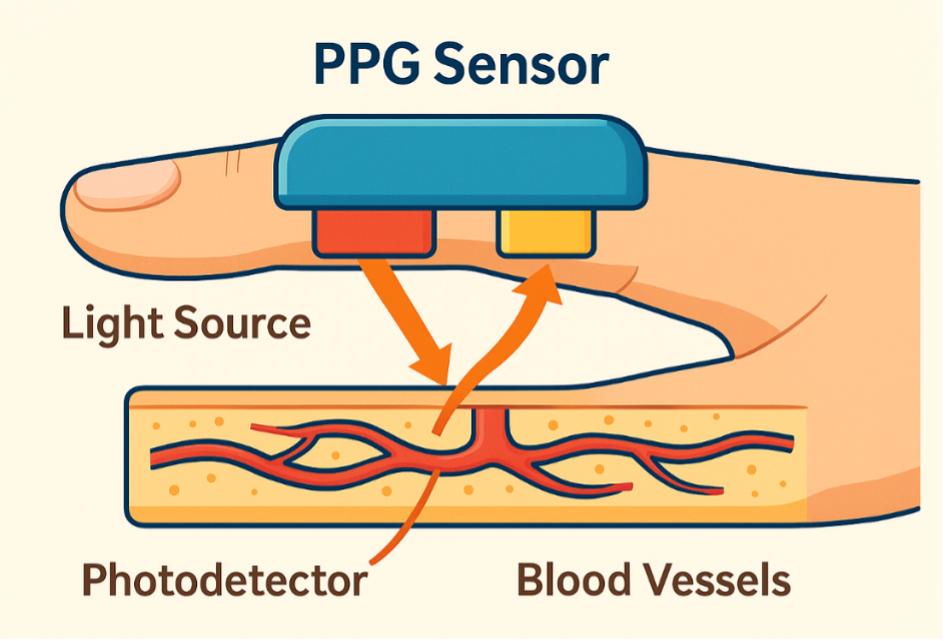}
    \caption{Photoplethysmography acquisition diagram}
\end{figure}

Reduced heart rate variability is associated with a variety of psychiatric disorders, such as depression, anxiety, and alcohol use disorder. Since heart rate can analyze the activity of sympathetic and parasympathetic nerves through time series, it is widely used to detect emotional states. However, heart rate is highly dependent on the physical state at the time of measurement, so further feature extraction is needed in the emotion classification task. In related studies, an experiment selected five key features to identify five emotions: sadness, anger, fear, happiness, and relaxation. The support vector machine classifier was used for emotion recognition, and the average accuracy finally achieved reached 56.9\%. Another study adopted a multimodal approach to combine multiple physiological signals such as blood volume pulse, electromyography, skin electrodermal activity, skin temperature, and respiration for emotion classification. The statistical classifier was trained by support vector machine and Fisher linear discriminant method, and features were extracted from multiple subjects and compared. The results showed that the average classification accuracy of Fisher linear discriminant method was 28.83\%, while the accuracy of support vector machine classifier reached 46.5\%.

Electrodermal activity is another physiological signal commonly used in affective computing. By placing two electrodes on the fingers to measure the conductivity of the skin, the emotional state of the individual can be obtained. Since the exocrine glands in the palm are highly sensitive to emotional changes, measuring the galvanic response of the skin by applying a small voltage can effectively capture emotional fluctuations. In a relaxed state, skin conductivity decreases, while it increases under stress or tension. Related research uses smart gloves to collect electrodermal activity data and induce emotions based on the International Affective Picture System Database. Statistical analysis methods are used to extract features and classify the data into different emotional arousal levels, and finally an average classification accuracy of 64.32\% is obtained.

Breathing pattern is also one of the important physiological features in affective computing. Breathing signals can be measured by resistive wire strain gauges, photoplethysmography, or foam-based pressure sensors, and can be integrated into smart textiles for convenient monitoring. Studies have found that a decrease in breathing rate usually indicates a state of relaxation, while deep and fast breathing may mean happiness or anger. In addition, irregular changes in breathing are often associated with negative emotions. For example, shallow and fast breathing may indicate that an individual is in a state of tension or fear, while shallow and slow breathing patterns may be associated with depression. A study used a deep learning algorithm to analyze breathing patterns and classified them based on an emotional database. The recognition accuracy of valence and arousal levels reached 73.06\% and 80.78\%, respectively.

Changes in skin temperature can also be used for emotion recognition. When measured at the fingertips, vasodilation causes the fingertips to warm up when relaxed, while vasoconstriction causes them to cool down when stressed or anxious. Similar to skin electrodermal activity detection, skin temperature sensors can also be integrated into wearable devices such as gloves to enable mobile monitoring. Studies have shown that measuring skin temperature with smart gloves and extracting features in combination with specific emotional stimuli (such as watching movies or pictures) can effectively classify different emotional states. Experimental results show that positive emotions are easier to identify than negative emotions, and the accuracy of emotion classification using skin temperature features can reach 89.29\%.

Electromyography, as another physiological signal, can provide rich information about muscle activity. Its measurement usually requires electrodes to be attached to the surface of the skin to record the electrical signals generated when the muscles contract. In emotional computing, electromyography is often used to detect specific emotion-related muscle activities. For example, the tension of the upper trapezius muscle in the neck can be used to assess mental stress, while facial muscle activity can be used to analyze changes in facial expressions. Studies have found that high-frequency signals generated when muscles contract are usually associated with an increase in emotional valence. A study combined galvanic skin response with electromyography signals to construct an emotion recognition system, in which galvanic skin response is related to the level of emotional arousal, while electromyography signals are used to assess emotional valence. Experiments have shown that the system can effectively identify the emotional state of an individual, providing strong support for multimodal emotional computing.

Facial recognition, speech recognition, and cardiac-related methods are able to measure emotions independently and identify them on two dimensions: valence and arousal. Among them, breathing signals play a special role in emotion recognition, and their recognition capabilities are usually limited to specific emotional states, such as panic, fear, concentration, or depression, and cannot cover a wider range of emotion categories. Skin electrodermal activity and skin temperature are mainly used to detect arousal levels, while electromyographic signals are mainly used to classify valence levels. Therefore, in order to recognize the full range of emotional states, it is recommended to combine multiple physiological signals for multimodal emotion calculation. The advantage of this method is that it can improve the accuracy of emotion recognition, although the setting of multimodal fusion is more complicated.

For clinical research, static measurements of EEG, EMG, ECG, or pulse volume changes may be more suitable because these methods can provide high accuracy and enable diverse emotion classification. For example, a research system achieved an accuracy of 89.73\% in emotional arousal recognition and 63.76\% in valence classification. When the classification bandwidth was increased, that is, the emotion search range was expanded, the arousal accuracy increased to 96.58\% and the valence accuracy increased to 89.93\%. However, due to the high sensitivity of EEG signals, the system is susceptible to motion artifacts. Another study optimized the emotion recognition system for racing drivers, using facial EMG, ECG, respiratory signals, and skin electrical activity to achieve an overall accuracy of 79.3\% for five emotional states (high stress, low stress, disappointment, excitement, and neutrality). This system is particularly suitable for scenarios where emotions need to be recognized in dynamic motion environments.

In situations where mobile emotion measurement is required, such as elderly care, ECG signals may be a better solution. Some smart wearable devices, such as the smart T-shirt developed by a company, can measure ECG signals and extract heart rate variability features. Through ECG signals and heart rate variability features, five emotional states can be distinguished with an accuracy of 56.9\%. This method can be widely used for emotion monitoring in daily life and outdoor activities. For home environments, medical practices or psychological assessments, remote and non-invasive emotion recognition methods are better choices. For example, you can interact with virtual images on smart devices or install cameras in the room for continuous emotion monitoring. Facial recognition technology has made significant progress in emotion recognition, and an algorithm can achieve an accuracy of up to 89\% for seven emotional states. Speech recognition can also be used for emotion classification, but its accuracy drops significantly when the number of emotion categories increases. For example, when only happy and sad emotions are recognized, the accuracy can reach 80.46\%, but when six emotions are recognized, the accuracy drops to 49\%.

In the calculation of emotions from physiological signals, the biggest challenge is that the measurement dimensions of different signals are limited. For example, electromyographic signals are mainly used for valence classification, while skin electrodermal activity and skin temperature are only used to measure the arousal level. These signals can be fused to achieve a more complete description of emotions. However, heart rate variability, respiratory signals, and pulse volume changes can measure both valence and arousal dimensions at the same time, so pulse volume changes may become an important parameter for future research. Studies have shown that the correlation between pulse rate variability measured using photoplethysmography and heart rate variability measured using electrocardiogram signals exceeds 95\%. If the reliability of wrist pulse volume change measurement can be further verified, it is expected that emotion recognition can be performed solely on smart watches in the future.

In terms of experimental settings, different studies used different methods. For example, one study compared two experimental situations: one was to let participants enter an active situation by playing video games, and the other was to enter a passive situation by watching visual stimuli. Both experiments were conducted in a laboratory environment. Another study used a mobile wearable device to record ECG signals. During the experiment, participants watched a video and remained seated. The use of this mobile wearable device may be one of the reasons for the moderate accuracy. In addition, a study used a portable EEG device to measure EEG signals in a laboratory environment, but because EEG sensors are extremely sensitive to vibration and movement, it is not advisable to use them in a dynamic environment.

Research on facial recognition often focuses on the effects of lighting conditions. As long as there is sufficient lighting, it may not make much difference whether the measurement is in a real environment or a laboratory environment. Another study on ECG signals and facial physiological signals was conducted in a quiet room, with subjects sitting in a comfortable chair and playing music with headphones. The study showed that the accuracy of emotion measurement based solely on ECG signals is relatively low. In addition, a study on speech recognition used professional actors to directly record speech data. The results showed that the accuracy of the actors in conveying emotions may have a greater impact on the final recognition results.

Research on multimodal emotion recognition shows that the fusion of different signals can improve the overall accuracy. For example, some studies have combined facial recognition, speech recognition, and gesture analysis in emotion computing, and the experimental data were recorded simultaneously on multiple participants. Studies have shown that multimodal data fusion can significantly improve the accuracy of emotion prediction. Another study used a sensor kit to record multimodal physiological signals in a laboratory environment. However, the intuitiveness and naturalness of this method still need to be optimized, so its applicability in mobile environments may be low. In addition, a study on racing drivers used a multi-sensor wearable device, which included a head-mounted facial electromyography sensor, a chest electrocardiogram and breathing sensor, a glove-embedded skin electrodermal activity sensor, and a data acquisition and communication module. The system is designed to assess the basic emotional state of racing drivers in a high-intensity environment.

Regarding the mobility of emotion recognition, with the development of smart wearable devices, such as smart watches or smart fabrics, non-invasive emotion recognition has become more widely used in daily life. For example, a smart wristband has developed a method to measure photoplethysmography signals and skin electrodermal activity from the wrist. These wearable devices can provide a more convenient way to monitor emotions, which is more practical than traditional fixed measurement methods. With the further development of technology, future emotion recognition systems may be more intelligent and applicable to a variety of daily situations.

% ---------------------------------

\part{Social Robots and Emotional Interaction}

\chapter{Design Principles for Social Robots}

\section{Interaction Model of Social Robots}

The interaction framework of the social robot adopts a three-layer architecture of "perception-decision-feedback", which can effectively simulate the information processing mechanism of humans in the social process, thereby achieving a more humanized interaction experience.

\subsection{Perception layer: integrated multimodal sensors}

The perception layer is the first gateway for social robots to contact the outside world. It is responsible for collecting various information from the environment and users. By integrating multiple sensors, the robot can obtain rich and comprehensive perception data, providing a solid foundation for subsequent decision-making and feedback.

Visual sensors are an important part of the perception layer. Take the Intel RealSense D455 multimodal kit as an example. It has multiple functions such as depth perception, RGB camera and infrared sensor. Among them, the depth perception accuracy is 4 meters.ZAxis error is less than2\%This high-precisionDepth perception enables robots to accurately perceive the three-dimensional structure of the surrounding environment, realize object recognition, scene understanding, and capture the user's body movements. For example, the robot can judge the distance between the user and itself through depth information, so as to adjust its behavior and tone of voice and maintain an appropriate social distance. RGB cameras can capture color images for face recognition, expression recognition, and lip reading recognition. Through face recognition technology, robots can remember the appearance characteristics of different users and realize personalized greetings and communication; expression recognition allows robots to perceive the user's emotional state, such as happiness, sadness, anger, etc., so as to make corresponding emotional responses in conversations; lip reading recognition is of great significance in noisy environments, and can assist voice recognition to improve the accuracy of information acquisition. Infrared sensors play an advantage in low-light conditions, and can help robots perceive the user's movements and presence even in dark environments, ensuring that robots can work normally under various lighting conditions.

\begin{figure}
    \centering
    \includegraphics[width=0.5\textwidth]{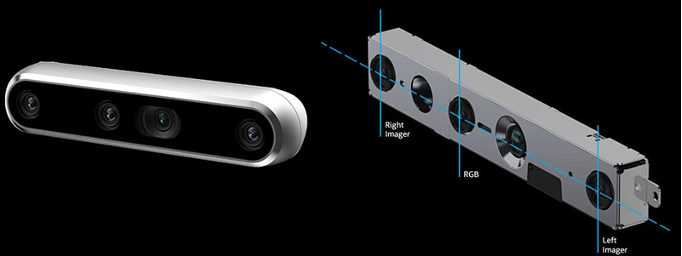}
    \caption{Appearance of Intel RealSense D455 depth camera}
\end{figure}

The voice sensor is also a key part of the perception layer. It can capture the user's voice signal and convert it into text information through voice recognition technology. In the process of voice recognition, many factors need to be considered, such as the speed, tone, accent and background noise of the voice. In order to improve the accuracy of voice recognition, social robots usually use advanced voice processing algorithms and deep learning models to train and optimize a large amount of voice data. For example, some social robots can recognize multiple languages and dialects to meet the needs of different regions and user groups. At the same time, the voice sensor also has a voice wake-up function.(e.g. "Hey Siri"), users can activate the robot through specific wake-up words and communicate with it conveniently and quickly.

Tactile sensors add tactile perception capabilities to social robots. By installing tactile sensors on the robot's shell, hands or other contact parts, the robot can sense the user's touch strength, position and pattern. For example, when a user gently touches the robot's head, the robot can interpret it as a friendly and intimate behavior and give corresponding positive feedback, such as making a pleasant sound or a happy expression. Tactile sensors can also be used to detect the robot's own state, such as whether it has been collided or squeezed, so as to take corresponding protective measures to avoid damage. At present, some robot platforms with flexible electronic skin (E-skin), such as HuggieBot 3.0, can achieve multi-point tactile distributed perception, and even identify the user's touch behavior intention through a "tactile map".

\begin{figure}
    \centering
    \includegraphics[width=0.5\textwidth]{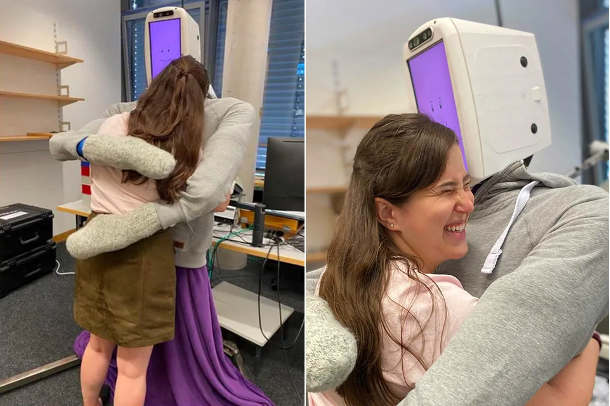}
    \caption{HuggieBot 3.0 with flexible electronic skin}
\end{figure}

In addition, the perception layer can also integrate other types of sensors, such as environmental sensors (temperature, humidity, light, etc.) and olfactory sensors, to obtain more comprehensive environmental information and further enrich the robot's perception capabilities. For example, environmental sensors can help the robot adjust its operating state according to the ambient temperature and humidity, such as automatically turning on the heat dissipation function in a high temperature and high humidity environment; olfactory sensors can detect odors in the environment to determine whether there are abnormal situations, such as fire or harmful gas leakage.In medical care and home service robots, the integrated environmental perception system not only improves the interactive intelligence, but also enhances the safety protection function.

\subsection{Decision-making layer: interaction strategy optimization}

The decision layer plays a core role in the interaction framework of social robots. Its main task is to conduct in-depth analysis and processing based on the rich information collected by the perception layer, and generate appropriate dialogue strategies and behavioral decisions. This process combines a variety of cutting-edge technologies to ensure that the robot can make natural, intelligent and human-desirable responses.

Deep learning technology plays an important role in the formulation of dialogue strategies. By building complex neural network models, robots can extract features and recognize patterns from large amounts of data transmitted from the perception layer. For example, deep neural networks can analyze user voice and text input to identify semantic information, emotional tendencies, and intentions. These models can capture subtle patterns and relationships in conversations by learning from massive amounts of data, thereby providing robots with more accurate decision-making basis. For example, intelligent customer service robots can track user goals in multiple consecutive rounds of conversations and maintain continuity at the semantic level to avoid the phenomenon of "answering questions that are not asked". In the optimization of dialogue strategies, deep learning models can not only understand the user's current input, but also combine the conversation history to predict the user's possible subsequent reactions. This ability enables robots to generate more coherent and forward-looking dialogue strategies. For example, when a user asks about the weather, the robot can not only provide current weather information, but also proactively provide relevant suggestions based on the conversation history and the user's interest preferences, such as whether to bring rain gear or whether it is suitable for outdoor activities.

In recent years, the rapid development of big model technology has brought new possibilities to the decision-making layer of social robots. Models represented by DeepSeek big model and BERT semantic encoding have powerful language understanding and generation capabilities through pre-training on massive text data. These models can handle complex language tasks such as question and answer, text generation and semantic understanding, providing social robots with more intelligent conversation capabilities. The application of big model technology enables social robots to better understand users' intentions and emotions. For example, the AI virtual human emotional companionship system "Virtual Partner" has achieved super-human emotional companionship with its solid self-developed big model capabilities. This technology enables robots to generate empathetic responses based on the user's emotional state and enhance the user's emotional experience.

In addition, large model technology can also help robots maintain consistency in multiple rounds of conversations. By understanding the context of the conversation, the robot can generate more coherent and natural responses. For example, Google DeepMind's Talker-Reasoner architecture divides the conversation into two parts: the fast Talker and the thoughtful Reasoner.Inspired the design strategy of "layered response" in human-computer interaction,This architecture enables the robot to perform complex reasoning and planning while responding quickly to users.

\begin{figure}
    \centering
    \includegraphics[width=0.5\textwidth]{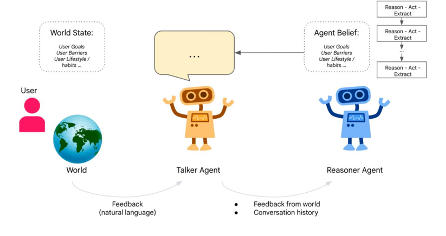}
    \caption{Diagram of the Talker-Reasoner method}
\end{figure}

In multimodal interaction, the decision layer needs to process information from different modalities, such as vision, speech, and touch. The application of large model technology in this field has also made significant progress. For example, multimodal large models can process visual and language information at the same time, allowing robots to combine users' expressions and body movements in conversations to generate more natural and emotional responses. This multimodal interaction capability not only improves the robot's conversation quality, but also enhances its adaptability in complex environments. In social scenarios, robots can judge users' emotional states by observing their expressions and movements, and adjust their conversation strategies accordingly. This capability enables robots to better integrate into human social environments and provide more personalized services.For example, after integrating image-text alignment models such as CLIP, the robot can recognize the content of picture books and combine it with voice explanations in child-accompanying scenarios, and even recognize the child’s smile or sleepy expression, and adjust the rhythm of interaction in a timely manner.

Affective computing technology also plays an important role in the decision-making layer. By analyzing the emotional information in the user's voice tone, expression and text, the robot can generate emotional responses and enhance the user's emotional experience. For example, when the user shows sadness, the robot can provide emotional support through comforting words and warm tone. The application of model technology in affective computing has also made significant progress. For example, some emotion recognition models can accurately identify the user's emotional state by analyzing the user's voice and text, and generate corresponding emotional responses. This technology enables robots to better understand the user's emotional needs and provide more considerate services.For example, the Kismet robot developed by the MIT Media Lab can generate corresponding facial expressions and voice responses by analyzing human facial expressions and voice features, thereby achieving natural interaction with humans.

\begin{figure}
    \centering
    \includegraphics[width=0.5\textwidth]{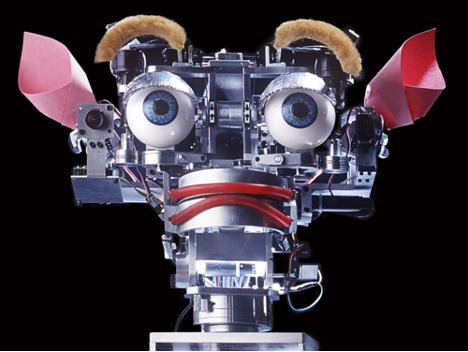}
    \caption{Robot Kismet}
\end{figure}

\subsection{Feedback layer: multi-channel emotional expression coordination mechanism}

The feedback layer is an important part of the emotional communication and interaction between social robots and users. It is responsible for conveying the robot's decisions and intentions to users through multiple channels, including expressions, voice, body movements, etc. In order to make the robot's feedback more natural, coordinated and emotional, it is necessary to establish a multi-channel emotional expression coordination mechanism.

In terms of facial expression feedback, social robots can display various expressions, such as smiles, frowns, and blinks, through screens, LED lights, or other display devices. These expressions should match the robot's emotional state and the content of the conversation to enhance the effect of emotional expression. For example, when the robot expresses happiness, it can show a big smile accompanied by bright eye flashes; when the robot expresses confusion, it can frown and reveal confusion in its eyes. At the same time, the change of expression should have a certain fluency and naturalness, avoiding abrupt switching, so that users can feel that the robot's emotions are real and coherent. For example, the Emo robot, which is equipped with 26 precision actuators, can predict and imitate human facial expressions, including smiles, within 840 milliseconds. It also has a high-resolution camera in its pupils to track the eyes of the conversation partner, and is equipped with an artificial intelligence model that can predict and respond to human expressions and learn complex emotions frame by frame from sample videos.

\begin{figure}
    \centering
    \includegraphics[width=0.5\textwidth]{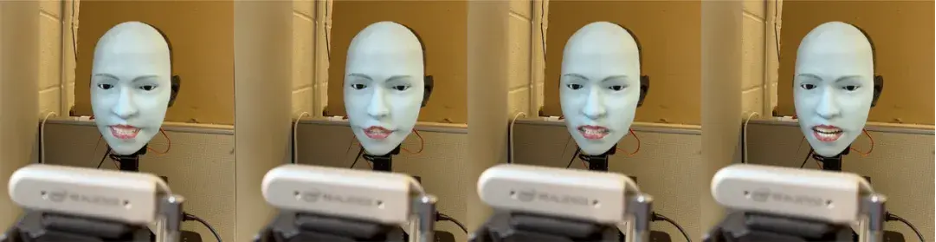}
    \caption{Emo robot learning facial expressions}
\end{figure}

Voice feedback is one of the main channels for social robots to communicate with users. In voice feedback, emotional expression is mainly reflected in intonation, speaking speed, and volume. For example, when the robot expresses excitement, it can increase its intonation and speaking speed, and its voice will be louder; when the robot expresses comfort, it can lower its intonation and speaking speed, and its voice will be softer. In addition, the timbre and pronunciation of voice can also be used to express emotions, such as using warm and friendly timbre to express friendliness, and using clear and accurate pronunciation to express professionalism. In order to achieve emotional voice feedback, social robots usually use speech synthesis technology, and combine it with emotional voice databases for training and optimization to make the generated voice more emotional.The EveR series of robots developed by the Korea Institute of Science and Technology (KIST) have the ability to express emotions through voice feedback. When EveR-2 Muse sang the Korean lyrical song "I Will Close My Eyes For You", it was able to convey sadness by adjusting the tone, speed and volume of its voice, allowing the audience to feel the emotional connotation of the song. This ability to convey emotions through voice characteristics makes social robots more vivid and expressive when interacting with humans.

\begin{figure}
    \centering
    \includegraphics[width=0.5\textwidth]{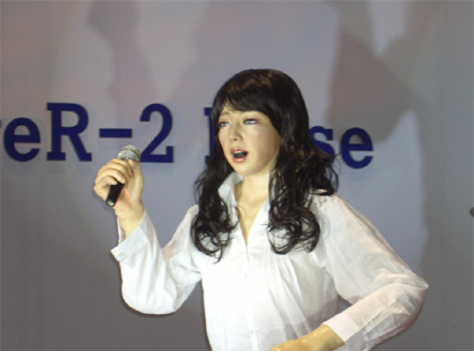}
    \caption{EveR-2 robot has the ability to lip sync}
\end{figure}

Body movement feedback adds more dimensions to the emotional expression of social robots. By controlling the movements of the robot's arms, head, body and other parts, various emotions and intentions can be conveyed. For example, when the robot expresses welcome, it can extend its arms to make a hugging gesture; when the robot expresses rejection, it can shake its head and retract its arms. The design of body movements should be consistent with human social habits and cultural norms, and avoid using movements that may cause misunderstanding or discomfort. At the same time, body movements should be coordinated with facial expressions and voice feedback to form an overall emotional expression system. For example, when the robot expresses apology, it can lower its head, speak in a low voice and bow slightly. These movements, facial expressions and voices work together to convey a sincere apology.The Pepper robot developed by SoftBank Robotics can express a variety of emotional states by coordinating its arm movements, head movements and voice intonation. In its interactions with autistic patients, Pepper can show emotional expressions such as happy dancing or angry fist pumping through dance, gestures and facial expressions, helping patients understand the connection between emotions and body language and learn how to express their emotions with body language.

\begin{figure}
    \centering
    \includegraphics[width=0.5\textwidth]{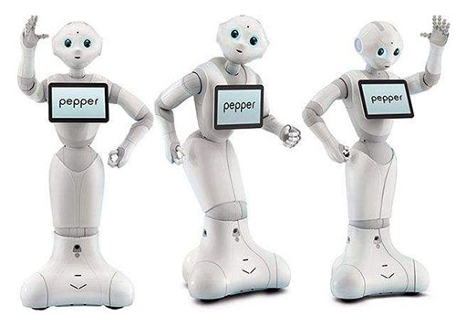}
    \caption{Pepper robot developed by SoftBank Robotics}
\end{figure}

In order to achieve the coordination of multi-channel emotional expression, social robots need to establish a unified emotional model and expression rules. Based on psychological theory, the emotional model can divide emotions into basic emotions (such as happiness, sadness, anger, fear, etc.) and compound emotions (such as surprise, anxiety, disgust, etc.), and define the expression form of each emotion on different channels. For example, based on the PAD emotional model (Pleasure-Arousal-Dominance), emotions with high pleasure are usually expressed as positive emotions, such as happiness; emotions with high arousal are expressed as strong emotions, such as anger or excitement; and emotions with high dominance are expressed as confidence and control, such as confidence or relaxation. The expression rules specify how to coordinate the feedback of each channel in different dialogue scenarios and emotional states to achieve the best emotional expression effect. For example, when the robot is having a pleasant chat with the user, it can use smiling expressions, brisk voices, and relaxed body movements to convey happy emotions; when the robot provides advice to the user, it can use focused expressions, steady voices, and slightly forward body movements to express seriousness and concern.

\begin{figure}
    \centering
    \includegraphics[width=0.5\textwidth]{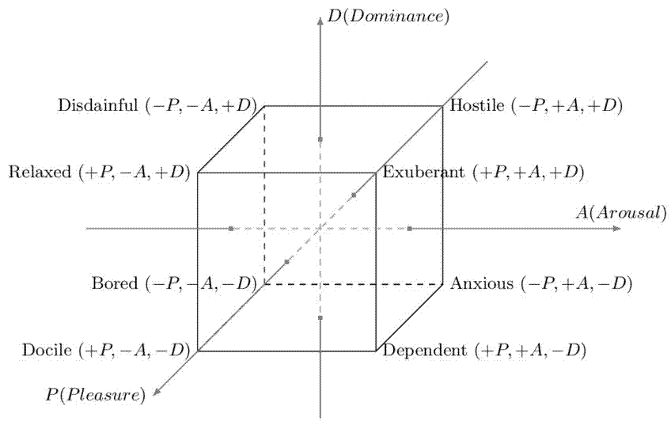}
    \caption{The PAD emotional model}
\end{figure}

In addition, the feedback layer can also enhance emotional expression through environmental interaction. For example, the robot can control the surrounding lighting, music and other environmental factors to create an atmosphere that matches the emotion. When the robot expresses romance, it can dim the lights and play soft music; when the robot expresses vitality, it can turn on bright lights and play cheerful music. Through interaction with the environment, the robot's feedback is richer and more vivid, which can better attract the user's attention and trigger emotional resonance. At the same time,Some service robots already have the ability to link with smart home systems and support dynamic adjustment of environmental feedback to adapt to different intentions.

The three-layer interaction framework of "perception-decision-feedback" provides a comprehensive and systematic solution for the design of social robots. Through the perception layer integrating multiple sensors, social robots can obtain rich and accurate environmental and user information; the decision layer based on reinforcement learning enables the robot to intelligently process information and generate appropriate dialogue strategies; and the feedback layer of the multi-channel emotional expression coordination mechanism makes the robot's feedback more natural, coordinated and emotional. The continuous improvement and development of this interaction framework will promote the widespread application of social robots in various fields and bring more convenience and fun to human social life.

\subsection{Design principles of social robots}

In the design of social robots, the design of interactive behaviors is crucial, as it directly affects the communication effect and user experience between users and robots. Among them, the Proxemics principle and the Turn-taking mechanism are two key aspects, which regulate the interactive behaviors of robots from the spatial and temporal dimensions, making them more natural, comfortable and efficient.

\subsubsection{Proxemics principle}

\textbf{Definition and Importance of Proxemics}
The Proxemics principle was proposed by American anthropologist Edward T. Hall in the 1950s. It studies the use and significance of spatial distance between people in the process of communication. In the design of social robots, following the Proxemics principle can help robots better understand human social habits, adjust their own movement trajectory and position, and thus create a communication atmosphere that makes users feel comfortable and natural. If robots can properly apply the Proxemics principle, it can not only improve user acceptance and satisfaction, but also enhance the adaptability and practicality of robots in social situations.

\textbf{Specific Application of Proxemics Principle}

\begin{enumerate}
    \item \textbf{Social distance:} 
    According to the Proxemics principle, the distance between people can be divided into different types, mainly including intimate distance, personal distance, social distance and public distance. In social situations, the distance between the robot and the user is usually kept between 0.45 and 1.2 meters, which is considered to be the social distance. Within this distance range, users usually feel more comfortable and natural, and will not feel that the robot is too close and intrusive, nor will the communication effect and intimacy be affected because of the distance.
    
    Intimate distance refers to a very close distance, usually within 0.45 meters. This distance is mainly used between close relationships, such as family members and lovers. In the interaction between social robots and users, unless it is a specific emotional companionship scenario, intimate distance is generally not used to avoid making users feel uncomfortable. Personal distance is between intimate distance and social distance, about 0.45-1.2 meters. This distance is suitable for friends, colleagues and other familiar people. After establishing a certain degree of trust and familiarity with the user, the social robot can appropriately adjust to the personal distance to enhance the affinity and intimacy of communication.
    
    Public distance refers to a longer distance, usually more than 3.6 meters, which is mainly used in public speeches, performances, etc. In large-scale events or speech scenes where social robots participate, the robots need to reasonably adjust their position and volume according to the size of the venue and the distribution of the audience to ensure that the information can be clearly conveyed to all audiences.
    \begin{figure}
        \centering
        \includegraphics[width=0.5\textwidth]{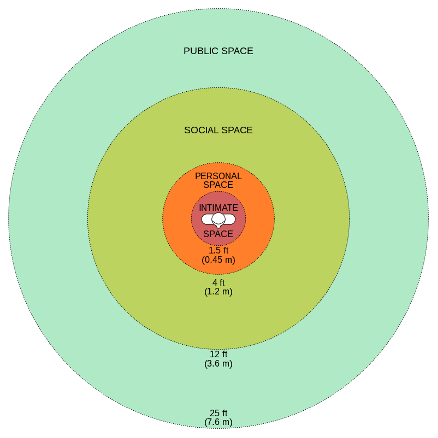}
        \caption{Proxemics principle interpersonal distance chart}
    \end{figure}
    
    \item \textbf{Adjustment of robot's movement trajectory:} 
    In order to better follow the Proxemics principle, social robots need to have flexible mobility and be able to adjust their movement trajectory according to different social scenarios and user needs. For example, in a small meeting, the robot can move to different social distances in a timely manner according to the location of the participants and the topic of discussion. When discussing more formal and serious topics, the robot can maintain a social distance to show professionalism and respect; when discussing more relaxed and casual topics, the robot can move closer and enter a personal distance to enhance the affinity and interactivity of the communication.
    
    In addition, robots also need to consider the user's personal space and privacy needs. During the movement process, avoid sudden and large movements to avoid scaring the user or invading the user's personal space. At the same time, the robot can use some non-verbal signals, such as nodding and smiling, to indicate its intention to move in advance, so that the user has enough time and psychological preparation.
    
\end{enumerate}

\subsubsection{Case Analysis of Proxemics Principle}

\begin{enumerate}
    \item \textbf{Application of service robots in hotels:} 
    In hotel scenarios, social robots often need to communicate and interact with guests at close range. For example, when checking in at the front desk, the robot can maintain a social distance to demonstrate professionalism and courtesy. When guests need help carrying luggage or being guided to the room, the robot can adjust to the personal distance in a timely manner according to the guest's movements and expressions to provide more intimate and thoughtful service. In this process, the robot not only improves the efficiency and quality of service by flexibly applying the Proxemics principle, but also enhances guest satisfaction and loyalty.
    \begin{figure}
        \centering
        \includegraphics[width=0.5\textwidth]{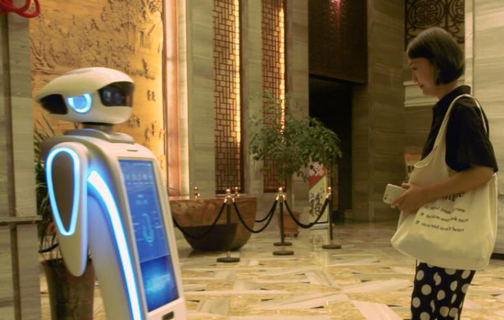}
        \caption{Hotel service robot}
    \end{figure}
    \item \textbf{Application of educational robots in classrooms:} 
    In a classroom environment, educational robots need to interact with students effectively. When explaining knowledge points, the robot can maintain a social distance to ensure that all students can clearly hear and see the teaching content. When conducting one-on-one tutoring or answering questions with students, the robot can get closer and enter a personal distance to enhance the affinity and interactivity of communication. By properly applying the Proxemics principle, educational robots can better stimulate students' interest and enthusiasm in learning and improve teaching effectiveness.
    \begin{figure}
        \centering
        \includegraphics[width=0.5\textwidth]{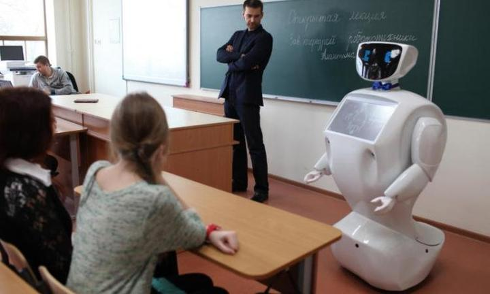}
        \caption{Educational robot in classroom}
    \end{figure}

\end{enumerate}

\subsubsection{Turn-taking mechanism}

\begin{enumerate}
    \item \textbf{Definition and Importance of Turn-taking Mechanism:}
    Turn-taking mechanism refers to the rules and strategies for how participants take turns speaking during a conversation. In daily human communication, turn-taking is a natural and tacit mechanism that ensures the fluency and effectiveness of the conversation. For social robots, mastering and applying the turn-taking mechanism is also crucial. Through reasonable turn-taking, the robot can better understand the user's intentions and conversation rhythm, avoid interrupting the user or being silent for a long time, thereby improving the naturalness and affinity of the conversation.
    \item \textbf{Implementation of Turn-taking Mechanism:}
    Voice Activity Detection (VAD) is a commonly used turn-taking implementation technology that is used to detect pauses and silences in voice signals. In the design of social robots, through VAD technology, the robot can monitor the user's voice signal in real time, determine whether the user has finished speaking, and then decide when to respond. For example, when the user is speaking, the robot can detect the pause in the voice through VAD technology. If the pause time exceeds a certain threshold, the robot can assume that the user has finished speaking and then respond.
    \begin{figure}
        \centering
        \includegraphics[width=0.5\textwidth]{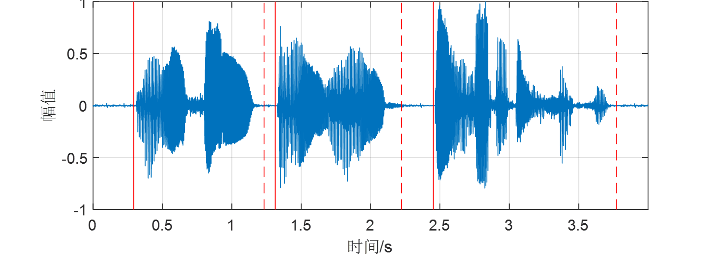}
        \caption{Voice signal endpoint detection (VAD)}
    \end{figure}
    The accuracy and reliability of VAD technology are crucial to the implementation of the Turn-taking mechanism. In order to improve the performance of VAD, researchers usually use a variety of signal processing and machine learning algorithms, such as energy threshold-based methods, probability model-based methods, etc. These methods can effectively identify pauses and silence segments in speech signals and maintain a high detection accuracy even in noisy environments.
    Also, eye contact is also a way to achieve turn-taking. As an important non-verbal signal in human communication, it can convey rich information and emotions. In the design of social robots, through eye contact technology, robots can maintain appropriate eye contact with users and enhance the naturalness and affinity of the conversation.For example, the Domo robot developed by the Massachusetts Institute of Technology (MIT) is equipped with human-like eyes and built-in cameras to perceive its surroundings. These eyes are designed so that humans can easily understand the robot's gaze direction, thereby enhancing the naturalness and effectiveness of human-robot interaction.at the same time,When the user is speaking, the robot can capture the user’s eyes through the camera and determine whether the user’s attention is focused on itself, so as to better understand the user’s intentions and the rhythm of the conversation.
    \begin{figure}
        \centering
        \includegraphics[width=0.5\textwidth]{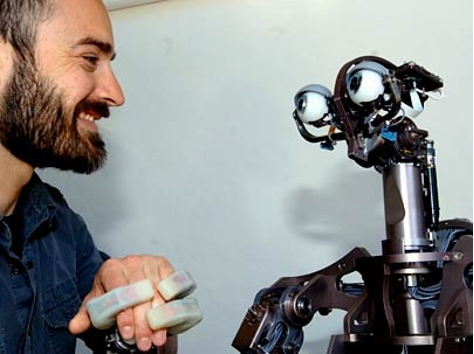}
        \caption{Domo robot with human-like eyes}
    \end{figure}

\end{enumerate}

\subsubsection{Case Analysis of Turn-taking Mechanism}
\begin{enumerate}
    \item \textbf{Application of intelligent customer service robots:} 
    In intelligent customer service scenarios, social robots need to have efficient conversations with users and answer their questions and needs. By using the Turn-taking mechanism, the robot can better understand the user's intentions and avoid interrupting the user or being silent for a long time. For example, when the user is describing a problem, the robot can detect the pause in the voice through VAD technology to determine whether the user has finished speaking, and then decide when to respond. At the same time, the robot can also maintain appropriate eye contact with the user through eye contact technology to enhance the naturalness and affinity of the conversation. This efficient Turn-taking mechanism not only improves the service quality of the customer service robot, but also improves user satisfaction and experience.

    \item \textbf{Application of virtual assistants in home scenarios: }
    In family scenarios, virtual assistants need to communicate and interact with family members on a daily basis. By using the Turn-taking mechanism, virtual assistants can better integrate into family life and provide more intimate and thoughtful services. For example, when family members are discussing family affairs, virtual assistants can participate in the conversation in a timely manner and provide relevant suggestions and information through VAD technology and eye contact technology. This natural Turn-taking mechanism not only enhances the practicality of virtual assistants, but also improves the interactive experience of family members.
\end{enumerate}

\subsubsection{The synergy between the Proxemics principle and the Turn-taking mechanism}

\begin{enumerate}
    \item \textbf{Coordination between space and time:}
    The Proxemics principle and the Turn-taking mechanism regulate the interactive behavior of social robots from the spatial and temporal dimensions respectively. In practical applications, the two need to cooperate with each other to create a natural and comfortable communication atmosphere. For example, in a social gathering, the robot can adjust the distance with the user through the Proxemics principle to keep it within the social distance range; at the same time, through the Turn-taking mechanism, the robot can have a natural conversation with the user, speak and respond in a timely manner. This synergy of space and time not only improves the robot's interactive effect, but also enhances the user's experience and satisfaction.

    \item \textbf{Synergy between emotion and cognition:}
    The Proxemics principle and the Turn-taking mechanism also involve the synergy of emotion and cognition. By properly applying the Proxemics principle, the robot can create a space environment that makes users feel comfortable and safe, thereby alleviating users' tension and anxiety. At the same time, through an effective Turn-taking mechanism, the robot can better understand the user's intentions and emotions and provide more intimate and thoughtful services. This synergy of emotion and cognition not only improves the robot's interaction quality, but also enhances the user's trust and dependence on the robot.

    \item \textbf{Synergy between Culture and Individual Differences:}
    In different cultural contexts, the application of the Proxemics principle and the Turn-taking mechanism will also be different. For example, in some cultures, people prefer to maintain a closer social distance, while in other cultures, people prefer to maintain a longer social distance. Social robots need to flexibly adjust their interactive behaviors according to different cultural backgrounds to adapt to the needs and habits of different users. At the same time, for individual differences, such as the user's personality, age, gender, etc., the robot also needs to be personalized to provide more intimate and thoughtful services.
    
    In summary, the Proxemics principle and the Turn-taking mechanism are two important aspects in the design of social robot interaction behavior. By reasonably applying the Proxemics principle, the robot can adjust its own movement trajectory and position to create a comfortable and natural communication atmosphere for users. Through an effective Turn-taking mechanism, the robot can better understand the user's intentions and conversation rhythm, and achieve natural and smooth conversations. The synergy of the two not only improves the robot's interaction effect, but also enhances the user's experience and satisfaction. In the future development, with the continuous advancement of technology and the continuous expansion of application scenarios, the Proxemics principle and the Turn-taking mechanism will play a more important role in the design of social robots, bringing more convenience and fun to human social life.
\end{enumerate}

\section{Personalization of Social Robots}

\subsection{The Importance of Personalized Design}
Personalized design is extremely important in improving the user acceptance and trust of social robots. By giving robots unique personality traits, the interactive experience and emotional connection between users and robots can be significantly enhanced.

Many studies and practical application cases have fully demonstrated the positive effects of personalized design. For example, the "extroverted" personality design of the Pepper robot has significantly increased its acceptance in the service industry by 37\%. This data shows that when robots show distinct personality traits, users are more willing to interact with them and are more inclined to trust and rely on these robots.The differences in preferences for personalized robots among different user groups are becoming increasingly apparent. For example, children prefer lively and humorous personalities, while the elderly prefer calm and caring personalities, which has been verified in medical care and education scenarios.From a psychological perspective, personalized design can enhance users' identification and emotional connection with robots. When robots show personality traits similar to humans, users are more likely to resonate and feel close to them, which promotes more positive interactions and improves the robot's use effect and user experience.

From a technical perspective, the implementation path of personalized design continues to expand. In 2018, Mila and Facebook jointly published a paper proposing a deep learning framework based on the persona-CHAT dataset (generated by artificially simulated multi-type personality dialogues), which verified that the injection of personality traits significantly improved the dialogue effect. In the same year, the neural model developed by the MIT research team used a multi-level attention mechanism to perform IDB encoding of movie character dialogues, achieving effective extraction of personality vectors. The model performed well in the task of dialogue role classification. In the direction of multimodal fusion, a 2024 review study pointed out that a multimodal system that integrates text, voice and visual information can enhance the expression of personality traits (see Figure 7-14 for a schematic diagram of a multimodal architecture), while emphasizing the use of explainable design to enhance users' understanding of robot behavior. This direction provides technical support for personalized services in scenarios such as medical care and education. In the field of cultural personality reconstruction, the historical figure simulation technology proposed in the 2024 study constructs a personality model by parsing specific character corpora (such as the poetry and literature of Li Bai and Su Dongpo), and its derived cultural guide robots have been successfully applied in museum scenarios. The ChatHaruhi project (Figure 7-14) conducted MBTI personality assessments on 32 fictional characters based on a large-scale pre-trained language model. The results showed that their personalities matched human perception by 82.8\%, confirming the effectiveness of this technology in restoring complex personalities. In terms of strengthening social attributes, a 2019 study confirmed that robots with social participation designs have significantly improved their acceptance in human-robot interaction (HRI), and that emotional processing mechanisms can enhance the effectiveness of persuasive strategies. This finding provides an important basis for the commercial application of service robots.

\begin{figure}
    \centering
    \includegraphics[width=0.5\textwidth]{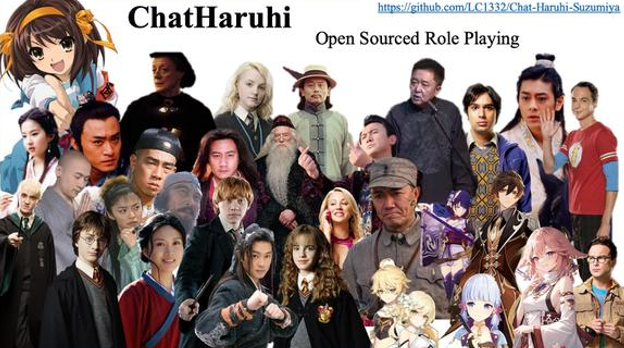}
    \caption{The ChatHaruhi library creates 32 chatbot roles}
\end{figure}

\subsection{Methods and Practices of Personalized Design}
Personalization design plays a vital role in the development of social robots. By giving robots unique personality traits, users' acceptance and trust can be significantly improved, thereby enhancing the interactive experience between users and robots. The following will explore the methods and practices of personality design, including five-factor model (Big Five) mapping, dynamic personality adjustment, personalized dialogue generation technology based on deep learning, neural models and multi-level attention mechanisms, personalized personality feature integration in multimodal dialogue systems, personality simulation of historical figures and story characters, application of large-scale pre-trained language models, social participation and emotional processing, as well as cognitive architecture and human-inspired design.

\subsubsection{Big Five Mapping}
The Five Factor Model is a commonly used personality classification model in psychology, which includes five dimensions: Openness, Conscientiousness, Extraversion, Agreeableness, and Neuroticism. By converting these dimensions into behavioral parameters, robots can show different personality traits. For example, a robot with high openness may show more creativity and curiosity, while a robot with high conscientiousness may be more reliable and organized.

In practical applications, the five-factor model can be used to design and evaluate the personality characteristics of social robots. For example, researchers can evaluate the personality characteristics of users through questionnaires or behavioral observations, and then set corresponding behavioral parameters for the robot based on these characteristics. This mapping method can not only help robots better adapt to the needs and preferences of users, but also improve users' identification and trust in robots. For example, in the field of elderly care, by evaluating the personality characteristics of the elderly, corresponding behavioral parameters can be set for nursing robots, so that they can better interact with the elderly and provide more considerate care services.

\begin{figure}
    \centering
    \includegraphics[width=0.5\textwidth]{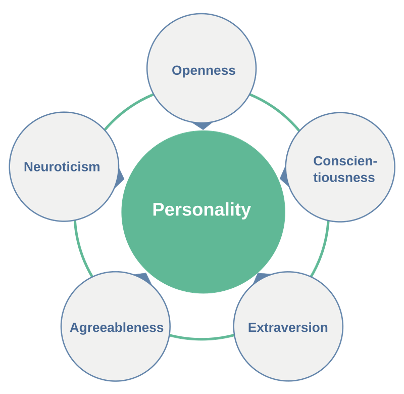}
    \caption{Big Five personality classification model}
\end{figure}

\subsubsection{Dynamic Personality Adjustment}
Dynamic personality adjustment is an important development direction of personalized design. By adjusting the robot's behavior and personality traits in real time, the robot can better adapt to the needs and preferences of different users. For example, the DynPer framework proposed by Zhejiang University can adjust the robot's behavior in real time according to the user's personality. This dynamic adjustment capability enables the robot to provide more personalized services, thereby improving user satisfaction and experience.

In practical applications, dynamic personality adjustment can be achieved through a variety of technologies. For example, through machine learning algorithms, robots can analyze the user's conversation content and behavior patterns, and thus adjust their own behavior and personality traits in real time. This dynamic adjustment capability can not only improve the robot's adaptability and flexibility, but also enhance the interactive experience between users and robots. For example, in the field of education, teaching robots can adjust teaching content and methods in real time according to students' learning progress and interests, providing more personalized teaching services.

\subsubsection{Personalized dialogue generation technology based on deep learning}
Personalized dialogue generation technology based on deep learning is one of the important means of personalized design. Through big data training, robots can show specific personality traits, thereby improving the quality of conversation and user experience. For example, in 2018, a paper jointly published by Mila and Facebook proposed a method to give chatbots personalized personality traits through deep learning models and big data. The method uses the persona-CHAT dataset, which is a dataset in which workers play different personalities and generate conversations. Studies have shown that when chatbots are given personality information, the conversation effect is significantly improved.

In practical applications, personalized dialogue generation technology based on deep learning can be implemented in a variety of ways. For example, by using pre-trained language models such as GPT-3 or BERT, robots can generate more natural and personalized dialogue content. This technology can not only improve the robot's conversational ability, but also enhance the emotional connection between users and robots. For example, in the field of customer service, chatbots can generate more intimate and personalized responses based on users' emotions and needs, thereby improving user satisfaction.

\subsubsection{Neural Model and Multi-level Attention Mechanism}
Neural models and multi-level attention mechanisms are another important means of personalized design. By learning the embedding of character images from conversations, robots can better capture and model personality traits, thereby improving the quality of conversations and user experience. For example, in 2018, a research team from the Massachusetts Institute of Technology (MIT) proposed a neural model that can learn the embedding of character images, that is, the embedding of personality, from conversations. The model converts the IDB encoded text of the conversation into a vector through a multi-level attention mechanism and a memory network to capture the aspects of different movie characters. In addition, the study also uses text descriptions as prior knowledge to improve model performance. Experimental results show that the model performs well in the task of dialogue role classification.

In practical applications, neural models and multi-level attention mechanisms can be implemented in a variety of ways. For example, by using deep neural networks, robots can analyze conversation content and extract feature vectors related to personality traits. This technology can not only improve the robot's conversation ability, but also enhance the emotional connection between users and robots. For example, in the entertainment field, chatbots can analyze the user's conversation content and generate replies that are more in line with the user's interests and preferences, thereby improving the user's entertainment experience.

\subsubsection{Personalized Personality Feature Integration in Multimodal Dialogue Systems}
In multimodal dialogue systems, the integration of personalized personality features is one of the important directions of personality design. By combining text, voice, and visual information, robots can show richer and more personalized personality features, thereby improving user satisfaction and experience. For example, a review article in 2024 discussed how to integrate personalized personality features in multimodal dialogue systems. The article also emphasized the importance of improving the transparency and explainability of the personality features of the dialogue agent so that users can understand the robot's behavior.

In practical applications, the integration of personalized personality features in multimodal dialogue systems can be achieved through a variety of methods. For example, by using multimodal data fusion technology, the robot can analyze the user's voice, expression, and body movements to better understand the user's emotions and needs. This technology can not only improve the robot's adaptability and flexibility, but also enhance the emotional connection between the user and the robot. For example, in the medical field, nursing robots can provide more intimate and personalized nursing services by analyzing the patient's voice, expression, and body movements.

\subsubsection{Personality Simulation of Historical Figures and Story Characters}
Personality simulation of historical figures and story characters is another important direction of personalized design. By simulating historical figures or story characters, robots can provide a richer and more personalized experience, thereby improving user satisfaction and experience. For example, research in 2024 proposed methods for developing chatbots that can simulate historical figures or story characters. These robots are designed to provide educational, entertainment, or cultural experiences, further promoting the development of personalized dialogue strategies.

In practical applications, personality simulation of historical figures and story characters can be achieved through a variety of methods. For example, by using natural language processing technology, robots can generate dialogue content that matches historical figures or story characters. This technology can not only improve the robot's conversation ability, but also enhance the emotional connection between users and robots. For example, in the field of education, teaching robots can provide more vivid and interesting teaching content and improve students' learning interest by simulating historical figures or story characters.

\subsubsection{Application of Large-scale Pre-trained Language Models}
The application of large-scale pre-trained language models is one of the important means of personalized design. By using large-scale pre-trained language models, robots can generate more natural and personalized conversation content, thereby improving user satisfaction and experience. For example, a 2023 study conducted personality assessments on 32 character chatbots created by the ChatHaruhi library based on the Big Five personality and MBTI dimensions. The study found that modern character chatbots based on large-scale pre-trained language models can effectively portray the personality traits of the corresponding characters and have a high degree of consistency with human-perceived personality (82.8\%). In addition, the study also proposed potential strategies for shaping the personality of chatbots.

In practical applications, the application of large-scale pre-trained language models can be achieved in a variety of ways. For example, by using pre-trained language models such as GPT-3 or BERT, robots can generate more natural and personalized conversation content. This technology can not only improve the robot's conversation ability, but also enhance the emotional connection between users and robots. For example, in the entertainment field, chatbots can generate more vivid and interesting conversation content by using large-scale pre-trained language models, thereby improving users' entertainment experience.

\subsubsection{Social Engagement and Emotional Processing}
Social engagement and emotional processing are important aspects of personification design. By designing robots with socially engaging personalities, acceptance in human-robot interaction (HRI) scenarios can be improved, thereby enhancing the emotional connection between users and robots. For example, a 2019 study explored how designing robots with socially engaging personalities can improve acceptance in human-robot interaction (HRI) scenarios. The study found that emotional processing plays an important role in persuasion when social robots use persuasive strategies.

In practical applications, social engagement and emotional processing can be achieved in a variety of ways. For example, by using emotion recognition technology, robots can analyze the user's emotional state and generate corresponding emotional responses. This technology can not only improve the adaptability and flexibility of robots, but also enhance the emotional connection between users and robots. For example, in the field of customer service, chatbots can analyze the user's emotional state and generate more intimate and personalized responses to improve user satisfaction.

\subsubsection{Cognitive Architecture and Human-inspired Design}

Cognitive architecture and human-inspired design are important theoretical and methodological supports for personalized design. By designing and implementing human-inspired cognitive architecture, robots can exhibit more natural and humanistic behaviors, thereby improving user satisfaction and experience. For example, a 2023 paper discussed in detail the design, implementation, and testing of a human-inspired cognitive architecture for social robots. The architecture aims to give social humanoid robots the ability to perform human social behaviors and virtual robot emotions. Tests have shown that this approach performs well in successful application cases.

In practical applications, cognitive architecture and human-inspired design can be implemented in a variety of ways. For example, by using deep learning and reinforcement learning techniques, robots can learn and imitate human cognitive and behavioral patterns, thereby showing more natural and humane behaviors. This technology can not only improve the adaptability and flexibility of robots, but also enhance the emotional connection between users and robots. For example, in the field of education, teaching robots can provide more vivid and interesting teaching content and improve students' interest in learning by learning and imitating human cognitive and behavioral patterns.

% --------------------------

\chapter{Emotional Interaction Technology of Social Robots}
With the continuous development of social robot technology, emotional interaction technology has become an important part of improving the human-computer interaction experience. Emotional interaction not only enables robots to understand the user's emotional state, but also respond accordingly to these emotional states, thereby enhancing user satisfaction and trust. This section will explore the implementation mechanism of emotional interaction, including emotion recognition technology and emotion response technology.

\section{Implementation Mechanism of Emotional Interaction  }

\subsection{Emotion Recognition Technology}

Emotion recognition technology is the basis of emotional interaction. It identifies the user's emotional state by analyzing multiple information such as the user's speech, expression, voice and physiological signals. In recent years, multimodal fusion and context perception have become important development directions of emotion recognition technology.

Multimodal fusion refers to the comprehensive analysis of information from different modalities to improve the accuracy of emotion recognition. The EmoFusion system developed by Tsinghua University is a typical example. The system achieves efficient recognition of the user's emotional state by weighting multimodal features through the attention mechanism. Studies have shown that the accuracy of the EmoFusion system in emotion recognition tasks has reached 92.1\%. In the EmoFusion system, information from multiple modalities such as speech, facial expressions, and physiological signals are input into the model at the same time. Through the attention mechanism, the system can automatically select the most useful features for emotion recognition, thereby improving the accuracy of recognition. For example, when a user is speaking, the system not only analyzes the tone and speed of his voice, but also combines the user's facial expressions and body language to comprehensively judge his emotional state. This cross-modal fusion method makes emotion recognition more comprehensive and accurate.

Context-aware technology corrects emotional judgments by combining conversation history and scene information. Amazon's Alexa is a successful example of context-aware application. Alexa's Contextual ASR (contextual automatic speech recognition) system can dynamically adjust its understanding of the user's voice based on the user's conversation history and current scene information, thereby more accurately identifying the user's emotional state. For example, when a user is talking to Alexa, if the user has previously mentioned an unpleasant event, Alexa can adjust its understanding of subsequent voices based on this context information and identify the user's emotional changes. This context-aware capability enables Alexa to be more intelligent and humane in emotional interactions.

\begin{figure}
    \centering
    \includegraphics[width=0.5\textwidth]{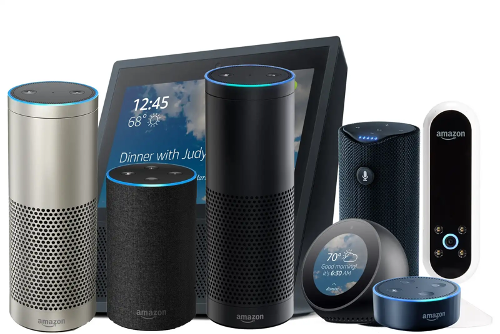}
    \caption{Alexa, Amazon’s intelligent voice assistant}
\end{figure}

\subsection{Emotion Response Technology}

Emotional response technology is an important part of the field of affective computing. Its core lies in expressing emotions through verbal and nonverbal behaviors and providing feedback during the interaction process, thereby improving emotional understanding and empathy. With the rapid development of artificial intelligence and interactive technology, emotional response technology has been widely used in many fields, including human-computer interaction, intelligent customer service, virtual assistants, and mental health support.

The theoretical basis of emotional response technology mainly comes from the intersection of psychology, cognitive science and artificial intelligence. Psychologists provide theoretical support for emotional response technology by studying the mechanisms of human emotional expression and understanding. For example, Paul Ekman's emotion theory points out that humans have six basic emotions (happiness, sadness, anger, surprise, fear and disgust), which can be expressed through non-verbal behaviors such as facial expressions, voice intonation and body language. Emotional response technology can recognize and respond to human emotions by simulating these emotional expression mechanisms.

\begin{figure}
    \centering
    \includegraphics[width=0.5\textwidth]{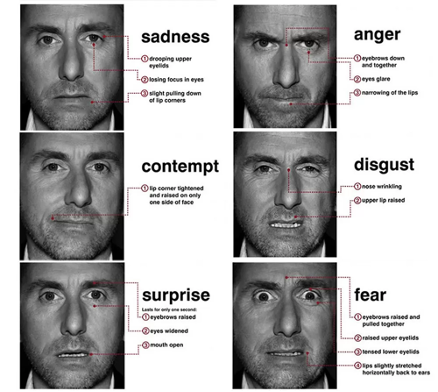}
    \caption{Paul Ekman's Emotion Theory}
\end{figure}

Emotional response technology mainly achieves emotional expression and feedback through verbal and non-verbal behaviors. Verbal behaviors include voice intonation, language content, and dialogue structure, while non-verbal behaviors include facial expressions, body language, and spatial distance. In practical applications, emotional response technology can be implemented in many ways, such as voice generation, visual feedback, or action response.

Emotional response technology has broad application prospects in the field of human-computer interaction. For example, in intelligent customer service, emotional response technology can understand the user's emotional state and provide corresponding emotional support through voice recognition and natural language processing technology. In the field of virtual assistants, emotional response technology can enhance the user's interactive experience through the simulation of facial expressions and body language. In addition, emotional response technology can also be applied to mental health support to help users relieve stress and anxiety.

Evaluating the effect of emotional response technology is the key to ensuring its practical application value. Evaluation methods usually include indicators such as user satisfaction surveys, emotion recognition accuracy, and interaction fluency. Through these evaluation methods, researchers can understand the performance of emotional response technology in different scenarios and further optimize its performance.

\section{Application Case Analysis of Emotional Interaction}
Emotional interaction technology has shown great potential in many fields. The following will list its typical application cases from different fields to deeply analyze the actual value and effectiveness of emotional interaction.

\subsection{Education}
New York public elementary schools have introduced Moxie robots to assist in teaching, focusing on educational support for children with autism. With its emotional interaction function, the robot provides emotional companionship and guidance for children with autism. Through interactive communication and emotional feedback with children, it effectively improves the social skills of children with autism by up to 41\%. This case shows that emotionally interactive robots can provide strong support for the education of special children, helping them to better develop their social skills and integrate into social life.

\begin{figure}
    \centering
    \includegraphics[width=0.5\textwidth]{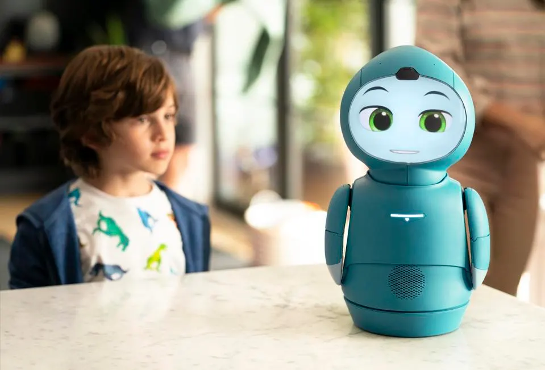}
    \caption{Moxie robot}
\end{figure}

\subsection{Medical field}
Paro seal robots have been used in medical scenarios to care for patients with Alzheimer's disease, and have achieved remarkable results. It interacts with patients through tactile comfort, which can effectively reduce the anxiety level of patients with Alzheimer's disease, and the clinically verified effectiveness is as high as 68\%. Paro robots interact with patients with their cute appearance and docile behavior, giving them emotional comfort, reducing their psychological burden, and improving their quality of life. At the same time, they also provide auxiliary support for the nursing work of medical staff.

\begin{figure}
    \centering
    \includegraphics[width=0.5\textwidth]{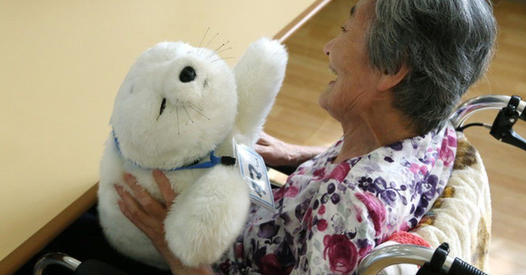}
    \caption{Paro seal robot}
\end{figure}

\subsection{Service Industry}
Retail industry: Affective computing technology is applied to retail service robots, enabling them to keenly perceive changes in customers' emotions. When customers express dissatisfaction, robots can adjust their service methods in a timely manner, such as changing the tone of communication, providing more considerate solutions, etc., thereby effectively improving customer satisfaction and winning better reputation and more customer loyalty for retailers.

Family services: In a family environment, affective computing technology helps robots interact better with the elderly. Robots can identify the emotional state of the elderly and take appropriate measures when they are depressed, such as playing music they like, telling interesting stories, etc., to relieve their unpleasant emotions, provide emotional companionship for them, alleviate their loneliness, and improve their happiness in their later years.

Nursing homes: Affective computing technology plays a positive role in nursing homes, helping robots reduce cognitive impairment and loneliness in Alzheimer's patients. Robots can stimulate brain activity and slow cognitive decline by interacting with patients emotionally, such as reminiscing about the past and playing simple games. At the same time, they can provide emotional companionship and care to patients, improving their quality of life.

\subsection{Entertainment and social fields}
Affective computing technology is applied to social robots, enabling them to recognize and respond to human emotions. The robot can generate corresponding responses based on the user's emotional state. For example, when the user is in a happy state, the robot can share interesting topics or jokes to further enhance the user's pleasure; when the user is depressed, the robot can give comfort and encouragement. In addition, social robots can also create personalized articles and poems to meet users' needs for emotional expression and cultural entertainment, adding new colors to people's social and entertainment lives.

In the field of social media, emotional computing technology enables robots to imitate human behavior and become important participants in the production of information content, the dissemination of opinions, and the guidance of public opinion. Robots can generate content that meets user needs based on users' emotional tendencies and interests, guide the direction of topic discussions, influence the direction of public opinion, and inject new vitality into the development of social media.

\begin{figure}
    \centering
    \includegraphics[width=0.5\textwidth]{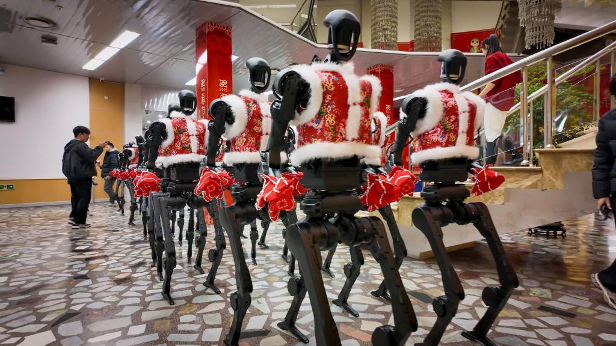}
    \caption{Unitree H1, a humanoid robot from Yushu Technology, made its debut on the Spring Festival Gala}
\end{figure}

\subsection{Military and Security Field}
In military scenarios, affective computing technology plays an important role. On the one hand, it can be used for morale assessment and psychological diagnosis. By analyzing the emotional state of soldiers, psychological problems of soldiers can be discovered in a timely manner, providing scientific basis for military commanders to take corresponding measures, enhance the scientificity and accuracy of combat command, and improve the combat effectiveness of the army.

In terms of border inspection and security monitoring, affective computing technology helps robots identify potential security threats. Robots can analyze people's emotions, behaviors and other characteristics to determine whether they have security risks, thereby improving security capabilities and contributing to maintaining national security and social stability.

\begin{figure}
    \centering
    \includegraphics[width=0.5\textwidth]{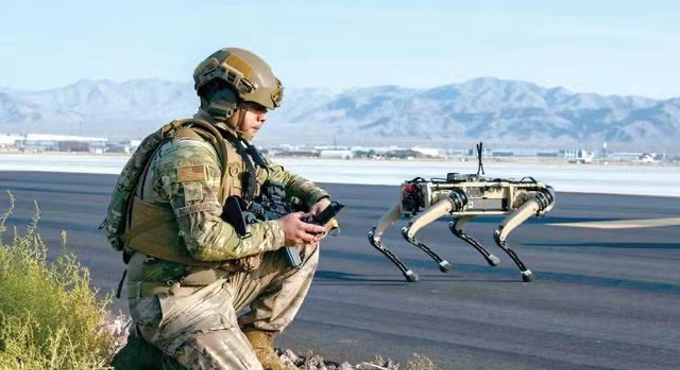}
    \caption{Quadruped bionic robots enter the PLA equipment sequence}
\end{figure}

\subsection{Other fields}
Recruitment field: Affective computing technology can be applied to the interview stage of the recruitment process. Robots can assist recruiters in understanding job seekers more comprehensively by analyzing the emotional state and language expression of job seekers, improve the accuracy and efficiency of recruitment, and provide technical support for enterprises to select suitable talents.

Driving assistance: In driving scenarios, affective computing technology enables robots to monitor the driver's emotional state. When the driver is tired or distracted, the robot can issue prompt reminders, such as sound alarms, adjusting the in-car environment, etc., to ensure driving safety and reduce the occurrence of traffic accidents.

% ------------------------------

\chapter{Current Challenges}

\section{Technical Challenges}
As an emerging technology field, affective computing aims to capture subtle changes in human emotions through technical means and convert them into computable signals, thereby providing support for multiple fields such as medical care, education, and entertainment. However, achieving this goal is far from easy, because the expression of human emotions has highly complex multimodal characteristics. Emotions are not presented in a single way, but rather integrate facial expressions, voice intonation, body posture, text content, and even physiological signals such as heart rate or skin galvanic response. For example, when a person is angry, he may frown, speak louder, and have a faster heartbeat, but these manifestations vary greatly depending on cultural background or individual differences. This multimodal characteristic requires the system to not only process a single data source, but also integrate multiple information, and this process itself is accompanied by many technical obstacles. At the same time, the accuracy and reliability of the affective computing system directly determine its practical value in real-world scenarios, especially in high-risk fields such as medical diagnosis or educational counseling, where any wrong judgment may lead to serious consequences. However, the current system still seems to be unable to cope with problems such as insufficient generalization ability, data bias, and model transparency. These challenges not only limit the development of technology, but also lay the foundation for the discussion in this chapter. This chapter will explore the difficulties of multimodal emotion recognition and the core issues of accuracy and reliability of emotion computing to reveal the current state of technology and future improvement directions.

\begin{figure}
    \centering
    \includegraphics[width=0.5\textwidth]{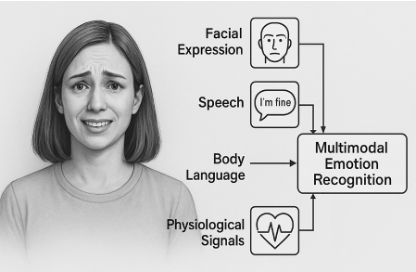}
    \caption{Schematic diagram of the concept of affective computing}
\end{figure}

As one of the most challenging links in affective computing, multimodal emotion recognition aims to improve the accuracy and stability of emotion recognition by fusing multiple input signals. Facial expressions are generally considered to be the most intuitive manifestation of emotions. For example, a smile is often associated with happiness, while a frown may indicate anger or confusion. However, it is far from enough to judge emotions based solely on facial expressions, because human emotional expressions are often multi-layered. For example, when a person says "I'm fine", a calm tone may mask inner anxiety, while a trembling voice or clenched fists may reveal true feelings. In addition, changes in voice pitch, speed of speech, and subtle body movements all play an important role in emotional expression. Furthermore, physiological signals such as heart rate or galvanic skin response can provide deeper emotional insights, but the collection of these signals requires professional equipment and is easily affected by the external environment, such as interference during exercise or poor contact of the equipment. How to effectively integrate these signals from different sources has become a core problem facing multimodal emotion recognition.

The complexity of integrating multimodal data is reflected in all aspects of technical implementation. A common fusion strategy is to merge all raw data before classification. This method can capture the correlation between different modes, such as the coordinated changes of facial expressions and voice intonation. However, this method has extremely high requirements for data quality and time synchronization. Imagine that facial expressions may switch quickly in a few milliseconds, and the ups and downs of voice intonation may last for several seconds. How to accurately align these signals in the time dimension is a headache. Another method is to extract features and make preliminary judgments in each mode separately before aggregating the results. This method has low computational cost and is suitable for scenarios with limited resources, but it often ignores the deep interactive information between modes. Studies have shown that both methods have their own advantages and disadvantages, but in dynamic and noisy real-world environments, the system often finds it difficult to maintain stable performance. For example, in a cafe full of background noise, the microphone may not be able to capture speech clearly, and the dim light may make it difficult for the camera to recognize facial details. The noise and data loss in these real-world scenarios further amplify the technical difficulty.

\begin{figure}
    \centering
    \includegraphics[width=0.5\textwidth]{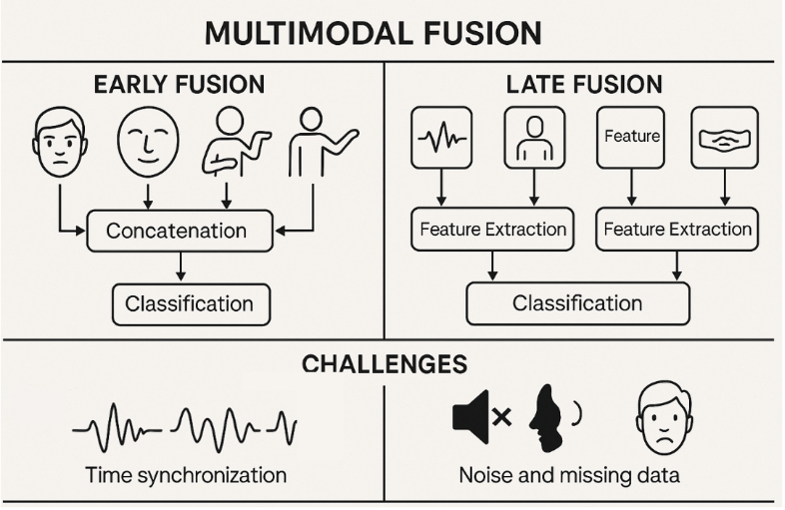}
    \caption{Multimodal data integration methods and challenges}
\end{figure}

The complexity of real-world applications is also reflected in the diversity and limitations of the data itself. Existing emotion recognition systems may perform well in the laboratory, but often fail in real-world scenarios. For example, a system that works well in a quiet conference room may fail in a crowded public place due to environmental interference. In addition, the current datasets used to train models often have significant defects. Many datasets are mainly based on recordings of Western people, with limited sample sizes and usually collected under controlled conditions, which makes it difficult to reflect the diversity of emotional expressions in the real world. For example, emotional expressions in Asian culture may be more restrained, while African culture may tend to be more extroverted. This cross-cultural difference causes the model to perform poorly in global applications. Even more problematic is that emotion annotation itself is highly subjective. The same segment of speech with a rapid tone may be labeled as "angry" by one annotator, while another may think it is "excited". This inconsistency introduces noise to the training data and directly affects the performance of the model. In recent years, researchers have tried to alleviate this problem by using unlabeled data through semi-supervised learning or unsupervised learning, but the actual effects of these methods still need to be further verified in larger-scale experiments.

\begin{figure}
    \centering
    \includegraphics[width=0.5\textwidth]{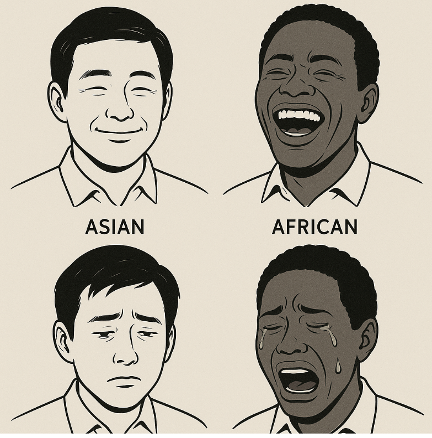}
    \caption{Differences in emotional expression in different cultures}
\end{figure}

Technological advances have brought new hope for multimodal emotion recognition. Deep learning techniques, such as transformer-based models, are able to process time series data and capture cross-modal dependencies, while graph neural networks excel at modeling complex interactions between modalities. These tools have great potential in theory, but face resource bottlenecks in practice. They require a lot of computing power and high-quality annotated data, and the current situation of data scarcity and high computational costs limits their widespread application. In addition, performance differences in cross-cultural scenarios further highlight the current dilemma. A model trained on North American users' smile data may not accurately recognize the more implicit expressions of happiness from East Asian users. This cultural bias not only affects the fairness of the system, but also limits its potential for promotion in the global market. To solve this problem, it is necessary to build more inclusive datasets, but this involves high costs, strict privacy regulations, and the complexity of cultural differences, which is difficult to achieve in the short term.

The accuracy and reliability of affective computing systems are key to their implementation. Unlike tasks such as object recognition or speech transcription, emotion recognition lacks clear objective criteria. An object can be clearly classified as a "cat" or a "dog," but emotional states such as "happy" or "sad" often vary from person to person and are deeply affected by context. For example, a person may smile when receiving a gift, but may be disappointed rather than happy. This subjectivity and ambiguity make it difficult for systems to achieve high accuracy, especially in scenarios that require real-time responses, such as virtual assistants or customer service robots. Insufficient generalization ability is the primary issue affecting accuracy. Many models perform well on benchmark datasets, but are vulnerable in dynamic environments. For example, a speech emotion recognition system trained in a quiet environment may not be able to adapt to the effects of user fatigue, stress, or environmental noise. This performance degradation stems from the difference in distribution between training data and actual scenarios. Although domain adaptation techniques can partially alleviate this problem, it still requires additional labeled data and computing resources.

Data bias is also an important factor affecting reliability. If the training data mainly comes from a specific population, such as young white men, then the system's emotional judgments on other groups, such as the elderly or non-Western users, may be significantly inaccurate. In medical health monitoring, this bias may lead to misdiagnosis, such as misjudging the anxiety of elderly patients as calmness, thereby delaying treatment. In addition, the current mainstream deep learning models are often regarded as "black boxes" and their prediction processes lack transparency. For example, a system may judge that a user is "angry" based on voice tone and facial expressions, but cannot explain why "frustration" or "anxiety" is excluded. In high-risk areas such as medical decision-making or psychological counseling, this opacity may reduce the trust of professionals because they need clear evidence to support system output. Explainable artificial intelligence technology has received attention in recent years, such as highlighting key features through attention mechanisms, or reconstructing decision-making processes through generative models, but how to effectively apply these methods to multimodal and time-dependent emotional data remains an unsolved mystery.

The lack of uniformity in evaluation standards further exacerbates the reliability problem. Different studies may use different performance indicators, such as accuracy, F1 score, or mean square error, making it difficult to compare the performance of models in different scenarios. For example, speech emotion recognition may focus more on classification accuracy, while physiological signal analysis may focus more on the prediction consistency of time series. This difference not only makes it difficult for researchers to judge which method is better, but also hinders the establishment of technical standards. To solve this problem, the industry and academia need to work together to develop a unified emotion evaluation framework, but this involves the coordination of interests of multiple parties and is difficult to achieve in the short term. The circular dilemma of data scarcity and bias is the core obstacle to accuracy and reliability. Improving model performance requires more diverse data, but data collection is limited by privacy regulations and ethical constraints. For example, collecting users' facial videos or physiological signals requires explicit consent, and many users are reluctant to participate due to privacy concerns. Even if data is collected, how to ensure its representativeness and fairness is also a problem. Researchers try to make up for the shortcomings through synthetic data or data augmentation technology, but synthetic data often cannot fully reflect the complexity of real emotions, and its effect still needs to be improved.

The technical challenges currently faced by affective computing include both data-level limitations and algorithm design difficulties. The complexity of multimodal emotion recognition and the lack of accuracy and reliability of affective computing together constitute the bottleneck of technological development. Through in-depth analysis of these issues, this chapter not only reveals the gap between theory and practice of affective computing, but also provides a direction for thinking about future technological breakthroughs. Only by making comprehensive progress in data quality, algorithm innovation, and ethical norms can affective computing truly realize its potential in real-world scenarios.

\subsection{Challenges of multimodal emotion recognition}

As an important research direction in the field of affective computing in recent years, multimodal emotion recognition aims to integrate information from different physiological modalities, such as electroencephalogram, electrocardiogram, galvanic skin response, facial expressions, and eye movement data, to obtain complementary features, thereby significantly improving the performance of emotion recognition. Compared with the limitations of a single modality, multimodal fusion can effectively alleviate the weaknesses caused by single signal noise, missing or insufficient expression ability. By organically combining the characteristics of different modalities, researchers hope to build a more robust and accurate emotion recognition system to adapt to the complex and changeable human emotional expression.

In the study of multimodal emotion recognition, the way to fuse different modal information has become one of the key issues. For example, some studies have constructed a new emotion discrimination space by analyzing physiological signals. Specifically, the researchers used EEG signals as auxiliary information, combined with discriminative canonical correlation analysis techniques, extracted features from a variety of physiological data, and finally trained the emotion recognizer through machine learning methods. This method makes full use of the unique advantages of EEG signals in capturing brain activity and makes up for the shortcomings of other modalities in deep emotional expression. Similar ideas have also been adopted by other researchers. They trained feedforward neural networks by fusing multiple signals and found that the fused model performed significantly better than the results of relying on a single signal when recognizing various emotions. The reason for this performance improvement is that different modal signals are complementary in emotional expression. For example, facial expressions may more intuitively reflect emotional categories, while physiological signals can more accurately capture the intensity or arousal level of emotions.

In the practice of feature fusion, researchers have also explored a variety of specific implementation methods. For example, some studies have spliced facial expression images with corresponding EEG signal features to generate a higher-dimensional feature map, and trained emotion recognition models on this basis. Experimental results show that this combination of multimodal features significantly outperforms single modality features in performance. In addition, some studies have attempted to hybridly fuse EEG, galvanic skin response, and facial data, first estimating the arousal level by integrating EEG and galvanic skin response features, and then incorporating facial data into the final fusion process. This gradual fusion method makes full use of the advantages of each modality in different emotional dimensions, such as the sensitivity of galvanic skin response to states of tension or excitement, and the ability of facial expressions to represent specific emotional categories.

In addition to feature-level fusion, decision-level fusion has also become an important strategy in multimodal emotion recognition. The researchers proposed an emotion recognition framework based on decision fusion, training independent classifiers at the electrocardiogram, electromyogram, and skin conductance levels, and then integrating the outputs of these classifiers by majority voting to determine the final emotion category. The advantage of this method is that it allows each modality to function independently, while reducing the impact of single-modality misjudgment through collective decision-making. Similar ideas have also been applied to the fusion of multi-domain features, such as integrating features in the time domain, frequency domain, and wavelet domain to identify combinations that can stably characterize the underlying signal characteristics, thereby improving the classification effect. This multi-domain fusion generates the final feature vector through splicing, providing the model with more comprehensive emotion information.

In practical applications, researchers have also tried to process physiological signals by weighted average fusion to classify valence and arousal levels. Experimental results show that this method performs slightly better than valence in the classification of arousal levels, which may be related to the fact that physiological signals are more sensitive to intensity changes than to the representation of positive and negative emotions. In addition, there are studies that combine EEG data with the audiovisual features of videos and optimize the feature space through dimensionality reduction technology to improve the computational efficiency and recognition accuracy of the model. These methods show that multimodal fusion not only needs to pay attention to the way signals are integrated, but also needs to consider how to retain key information in high-dimensional data.

\begin{figure}
    \centering
    \includegraphics[width=0.5\textwidth]{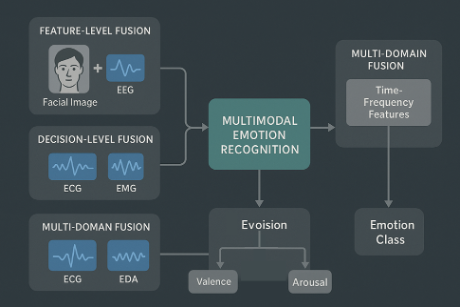}
    \caption{Schematic diagram of feature fusion method}
\end{figure}

The introduction of deep learning technology has further promoted the development of multimodal emotion recognition. For example, a study developed a late fusion method based on deep convolutional neural networks to estimate positive and negative emotion scores. The researchers designed two fusion strategies: one is to train a neural network model for each modality and then integrate their outputs; the other is to calculate the average probability of the pre-trained model on different emotion categories and select the category with the highest probability as the final result. This late fusion method makes full use of the powerful ability of deep learning in feature extraction, while adapting to the needs of different signal combinations through flexible fusion strategies. Similarly, there are studies that convert electrocardiogram signals into image form, generate frequency domain and time-frequency domain features through Fourier transform and wavelet transform, and then extract multi-domain features by neural network, and finally improve classification accuracy through decision fusion. This combination of multimodality and multi-domain provides a new idea for emotion recognition.

Although multimodal fusion has shown significant advantages in theory and experiments, it still faces many challenges in its implementation. First, how to effectively integrate information from different modalities is a complex issue. Feature-level fusion concatenates the features of each modality into a single vector. Although it is simple and intuitive, it may lead to information redundancy or conflict when faced with large heterogeneity between modalities. Although decision-level fusion can preserve the independence of each modality, its performance is highly dependent on the quality of each classifier and the design of the fusion rules. Studies have shown that there is no consensus on the pros and cons of feature-level fusion and decision-level fusion. For example, in some experiments, feature-level fusion has a slightly higher accuracy in emotion classification than decision-level fusion, while in other scenarios, decision-level fusion performs better due to its adaptability to differences between modalities.

In addition, in real-world applications, not all modal signals used in the training phase may be available in the testing phase. For example, both EEG and eye movement data may be used during training, while only eye movement data is available during testing. This modality missing problem poses a challenge to the robustness of multimodal emotion recognition. Researchers have tried to solve this problem through methods such as transfer learning, proving that even if the test data only contains some modalities, the model can still use the multimodal information in the training phase to achieve good performance. This approach not only improves the practicality of the system, but also expands its application potential in real-world scenarios, such as emotion-based brain-computer interface technology.

The difficulty of multimodal emotion recognition lies not only in how to design an efficient fusion strategy, but also in how to deal with practical challenges such as data heterogeneity, modality loss, and computational complexity. By combining the diversity of physiological signals with the powerful capabilities of deep learning technology, future research is expected to further break through these bottlenecks and lay the foundation for building a more intelligent and humane emotion recognition system.

\subsection{Accuracy and Reliability of Affective Computing}
Although sentiment computing has made significant progress, its accuracy and reliability still face many challenges, especially the generalization ability of models in practical applications. This chapter will discuss the accuracy and reliability of sentiment recognition, focusing on its core challenge - designing models that can generalize well on unseen data or new data sets, and explore the main obstacles such as limited data set samples, non-standard data segmentation techniques, and data variability between individuals, while proposing possible solutions.

The core goal of emotion recognition is to accurately judge the emotional state of an individual by analyzing multimodal data such as facial expressions, voice intonation, and physiological signals. However, the excellent performance of the model in the laboratory environment is often difficult to directly migrate to the real scene. One of the roots of this problem lies in the limitations of training data. The currently available emotion datasets are usually small in scale, and the number of samples is insufficient to cover a wide range of population characteristics and emotional expressions. For example, some datasets may only contain subjects from a specific cultural background or age group, which makes the model prone to performance degradation when facing unseen subjects with different backgrounds. In addition, emotional expression is highly individual, and even the same emotion may be expressed in completely different ways by different people. This inter-individual data variability further exacerbates the difficulty of model generalization, making the accuracy and reliability of emotion recognition systems in cross-individual applications questionable.

% \begin{table}[htbp]
%     \footnotesize
    % \centering
%     \caption{Some Imbalanced Sentiment Datasets}
%     \label{tab:imbalanced_datasets}
%     \renewcommand{\arraystretch}{0.8}
%     \setlength{\tabcolsep}{4pt}
%     \begin{tabular}{clllp{2.5cm}}
%         \toprule
%         \textbf{No.} & \textbf{Dataset Name} & \textbf{Sample Size} & \textbf{Class Distribution} & \textbf{Imbalance Characteristics} \\
%         \midrule
%         1 & IMDb sentiment classification & 50,000 & Positive: 25,000 / Negative: 25,000 / Neutral: None & Balanced binary classification, lacks neutral sentiment common in real-world scenarios \\
%         2 & SST-2 (Stanford Sentiment Treebank) & 67,349 & Positive / Negative (biased toward positive) & No neutral category; some samples tend to be overly positive \\
%         3 & GoEmotions (Google) & 58,000+ & 27 sentiment categories (incl. neutral, highly imbalanced) & Few samples in rare emotion categories; highly long-tailed distribution \\
%         4 & SemEval-2017 Task 4 (Twitter Sentiment) & 48,000 & Positive, Neutral, Negative (more neutral) & High neutral ratio; short text increases emotional ambiguity \\
%         5 & MELD (Multi-modal EmotionLines Dataset) & 13,000 & Joy, Sadness, Anger, Fear, Disgust, Neutral, etc. & Very few samples in low-frequency emotions such as fear and disgust \\
%         \bottomrule
%     \end{tabular}
% \end{table}

\begin{figure}
    \centering
    \includegraphics[width=0.8\textwidth]{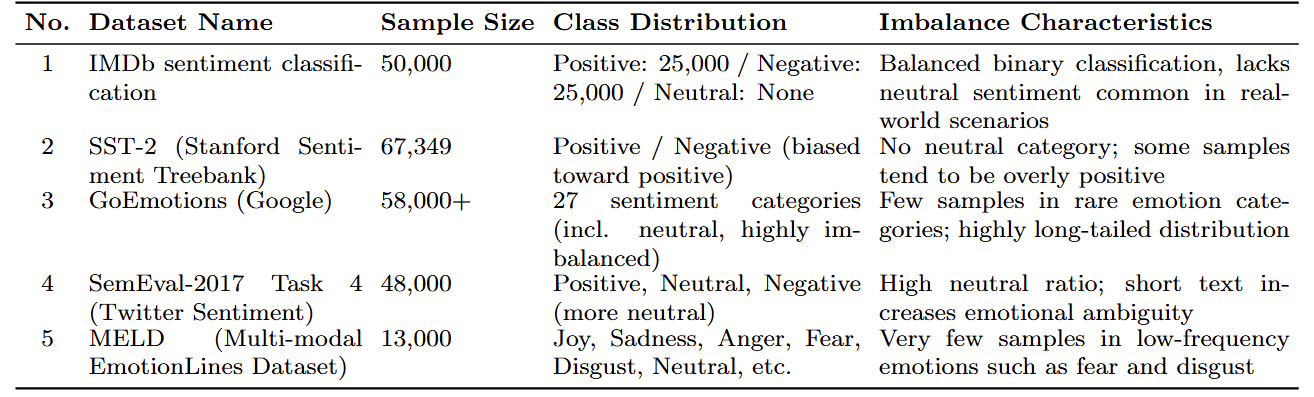}
    \caption{Some Imbalanced Sentiment Datasets}
\end{figure}

In order to meet these challenges, researchers have conducted extensive explorations in data segmentation techniques. Common data segmentation methods are mainly divided into two types: subject-dependent and subject-independent. The former uses the data of the same batch of subjects in the training and testing stages. Although it can achieve high accuracy in a specific group, its generalization ability is weak because the model is too dependent on the characteristics of the subjects seen during training. In contrast, the subject-independent segmentation technique completely separates the subjects in the training set and the test set, thereby simulating the situation in which the model faces completely new subjects in real scenarios. The research in Section 7 shows that models trained under the subject-independent setting are more likely to achieve good generalization effects. This is because this method forces the model to learn more general emotional features rather than the expression habits of specific individuals. However, although the subject-independent method is theoretically closer to the actual application needs, the performance of existing recognition models in this setting is still not ideal, especially in terms of real-time and intelligence. For example, when the model needs to quickly process unseen emotional data in a dynamic environment, its prediction results often have large deviations and even fail to meet the minimum requirements of practical applications.

The reasons for this phenomenon can be attributed to many aspects. First, the limited sample size of the data set is a long-standing problem in the study of affective computing. Since the collection of emotional data requires the voluntary participation of the subjects and involves privacy protection and ethical issues, the construction of large-scale high-quality data sets faces many difficulties. The insufficient sample size directly limits the model's ability to learn diverse features, especially in scenarios independent of the subjects, the model may not be able to fully capture the universal laws of emotional expression. Secondly, non-standard data segmentation techniques further exacerbate the complexity of the problem. In the field of affective computing, different research teams often adopt different segmentation strategies, such as segmentation by time series, segmentation by emotional category, or random segmentation. This inconsistency makes the evaluation of model performance lack a unified standard and makes it difficult to make fair comparisons between different studies. In addition, inter-individual variability is not only reflected in the style of emotional expression, but may also be affected by the external environment (such as light, noise) and internal state (such as fatigue, health status), all of which may become noise that interferes with the accuracy of the model.

% \begin{table}[htbp]
%     \footnotesize
%     \centering
%     \caption{Segmentation Strategy Examples and Descriptions}
%     \label{tab:segmentation_strategies}
%     \renewcommand{\arraystretch}{0.8}
%     \setlength{\tabcolsep}{4pt}
%     \begin{tabular}{cllp{2.5cm}}
%         \toprule
%         \textbf{No.} & \textbf{Strategy} & \textbf{Paper Title} & \textbf{Description} \\
%         \midrule
%         1 & Time-based Split & \textit{EEG-based Emotional Recognition using CNN} & Data is split chronologically; models simulate prediction of future data. Large performance fluctuations observed (50\%–90\%). \\
%         2 & Emotion Category Split & \textit{Emotion Detection with Multi-class Multi-label CNN (Medium)} & Multi-label classification by emotion categories to address imbalance. Commonly used in datasets like GoEmotions. \\
%         3 & Random Split & \textit{Facial Emotion Recognition using CNN (ScholarWorks SJSU)} & Random splitting (e.g., 70:30) may cause data leakage, leading to overestimated performance. \\
%         4 & Subject/Speaker-Independent Split & \textit{SER with Speaker-Independent Evaluation} & Ensures that speakers in training and testing sets do not overlap, matching real-world use cases but often reduces performance. \\
%         5 & Predefined Split & \textit{Personalized Emotion Recognition using Past-Present-Future Splits (ACL Anthology)} & Manually defined splits simulate realistic usage, such as using a time sliding window across past, present, and future data. \\
%         \bottomrule
%     \end{tabular}
% \end{table}

\begin{figure}
    \centering
    \includegraphics[width=0.8\textwidth]{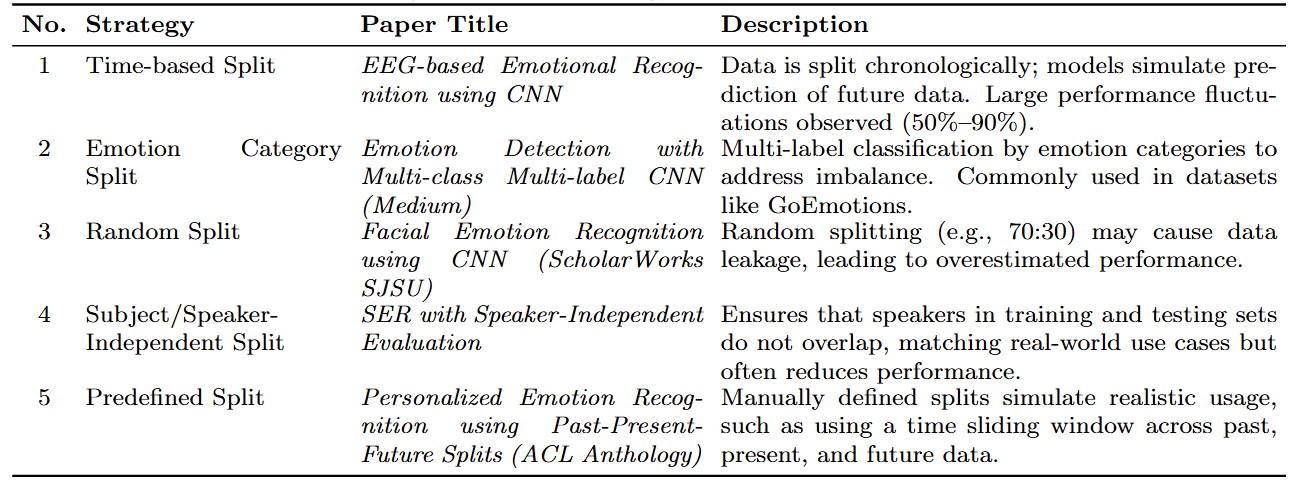}
    \caption{Segmentation Strategy Examples and Descriptions}
\end{figure}

In order to improve the generalization ability of emotion recognition models, an intuitive solution is to train and validate them on a large number of subjects. By increasing the number and diversity of subjects, the model can be exposed to a wider range of emotional expression patterns, so that it can better adapt to unseen data when testing. For example, if an emotion recognition system is trained on a dataset containing hundreds of subjects of different ages, genders, and cultural backgrounds, its prediction accuracy when facing new users will be significantly better than that of a model trained on data from only dozens of people. However, the implementation of this solution is limited by the availability of datasets. In practice, researchers often need to rely on existing public datasets, and the scale and diversity of these datasets usually cannot fully meet the needs. To make up for this deficiency, data augmentation technology has become a widely tried means. Data augmentation transforms the original data (such as rotation, scaling, adding noise, etc.) to artificially generate more training samples, thereby alleviating the problem of insufficient sample size to a certain extent. However, data augmentation is not a panacea. If the enhancement process is not careful enough, it may lead to an imbalance in data distribution. For example, some emotion categories are over-enhanced while samples of other categories are still scarce, which will eventually cause the model to overfit during training. That is, the model is too sensitive to the specific patterns of the enhanced data and cannot be generalized to real scenarios.

In addition to the challenges at the data level, the intelligence level of the model design itself also has a profound impact on the accuracy and reliability of emotional computing. Existing subject-independent recognition models often lack sufficient adaptability when facing complex real-world scenarios. For example, in real-time applications, emotion recognition systems need to process multimodal inputs (such as facial expressions and voice) simultaneously and give accurate judgments in a short time. However, current models usually perform well on a single modality, and are prone to information conflicts or improper weight distribution when multimodal fusion occurs. In addition, the dynamic nature and contextual dependence of emotions themselves also increase the difficulty of recognition. Human emotions are not static, but evolve over time and environment. For example, a person's anger after being criticized may quickly turn into frustration, and this rapidly changing emotional conversion places higher demands on the model's real-time tracking capabilities. Most existing models are trained based on static data and lack the ability to model the dynamic evolution of emotions, which greatly reduces their reliability in continuous tasks.

In order to solve the above problems, future research can start from multiple directions. First, in terms of data collection, larger and more diverse emotion datasets can be constructed through cross-institutional cooperation or crowdsourcing. For example, combining mobile devices and wearable technologies to collect users' emotional data in natural environments can not only increase the sample size, but also improve the authenticity of the data. Secondly, in model design, more advanced deep learning techniques such as transfer learning and meta-learning can be introduced. Transfer learning can effectively make up for the shortcomings of small sample datasets by utilizing models pre-trained on large-scale general datasets and then fine-tuning them for specific emotional tasks. Meta-learning, on the other hand, enables the model to quickly adapt to new tasks and new subjects through the method of "learning how to learn", thereby improving generalization ability. In addition, the improvement of multimodal fusion technology is also a key direction. By designing smarter fusion strategies (such as attention mechanisms), the model can better integrate information from different modalities and avoid the interference of single modal noise.

In terms of data enhancement, future work needs to pay more attention to the rationality of enhancement strategies. For example, enhancement methods can be designed based on the semantic characteristics of emotional expression rather than simply applying general transformations. For example, for facial expression data, new expression samples can be generated by simulating muscle movements rather than relying solely on geometric transformations. This semantically driven enhancement method can generate data that is closer to the true distribution, thereby reducing the risk of overfitting. In addition, in order to cope with inter-individual variability, researchers can also explore hybrid methods that combine personalized modeling with general modeling. For example, a general model is used for rough emotion recognition in the initial stage, and then the model parameters are gradually adjusted according to the user's specific data to form a personalized recognition system. This method can not only ensure the basic performance of the model on unseen data, but also improve its adaptability to specific users.

The accuracy and reliability of emotion computing are the core issues in current research, and the key lies in improving the generalization ability of the model on unseen data. Limited data set samples, non-standard data segmentation techniques, and inter-individual data variability are the main obstacles to achieving this goal. These problems can be alleviated to a certain extent by training models on a large number of subjects, carefully applying data augmentation techniques, and improving model design. However, the complexity of emotion recognition determines that this field still requires long-term exploration and innovation. Future development requires not only breakthroughs in technology, but also coordinated progress in data collection, ethical norms, and application scenario design. Only in this way can emotion computing truly move towards the goal of high accuracy and high reliability and bring more profound value to human society.

\section{Ethical and social challenges}

With the rapid development and widespread penetration of technology, the ethical and social challenges caused by affective computing have gradually surfaced and become an issue that cannot be ignored. These challenges not only stem from the limitations of the technology itself, but also touch upon the core value system of human society at a deeper level, including the definition of privacy rights, the reshaping of trust relationships, and the profound transformation of the interaction model between technology and humans. As a technology that aims to perceive, understand and respond to human emotions, affective computing has exciting application prospects, but at the same time, it also brings unprecedented ethical dilemmas and social reflections. This chapter will deeply analyze the complexity of affective computing at the ethical and social levels, focusing on two core issues: privacy and data security issues, and ethical considerations of human-computer relationships. Through a systematic discussion of these issues, we hope to outline a more responsible and sustainable framework for the future development of affective computing to ensure that technological progress not only serves efficiency and convenience, but also protects human dignity and well-being.

The core of affective computing is to capture and interpret human emotional states through technical means. This process relies on the real-time collection and analysis of user physiological signals (such as heart rate, galvanic skin response), facial expressions, voice intonation and even text content. This capability provides a broad space for personalized services, psychological support and optimization of human-computer interaction. However, this "emotional insight" is also accompanied by profound ethical concerns. When a device can detect the fluctuations in our hearts earlier than we can, and even "read" our unspoken emotions, can our inner world still remain independent and private? When these emotional data are stored, transmitted or even used for commercial purposes, who will ensure their security and legality? Furthermore, when machines evolve from simple tools to beings that can express "empathy" or "care", how should the boundaries of the relationship between humans and machines be defined? These issues not only concern the pros and cons of technical design, but also directly challenge the ethical bottom line of human society, prompting us to rethink the tension between technological progress and social values. This chapter will start with the issues of privacy and data security, exploring the high sensitivity of emotional data and its potential risks; then turn to the ethical considerations of the human-computer relationship, analyzing how technology reshapes human emotional dependence and social connections. Through the interweaving analysis of these two major topics, we will reveal how emotional computing, while promoting technological innovation, profoundly affects our lifestyles and moral concepts.

\begin{figure}
    \centering
    \includegraphics[width=0.5\textwidth]{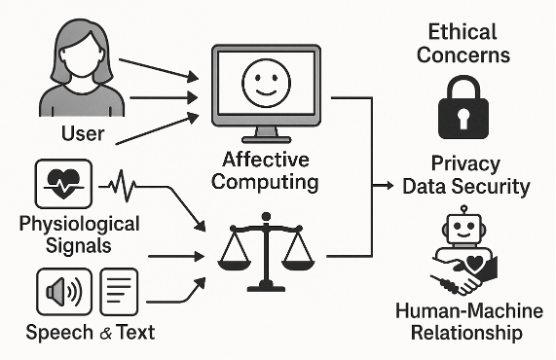}
    \caption{Diagram of potential risks in emotional computing}
\end{figure}

In the practical application of affective computing, privacy and data security undoubtedly constitute one of the most urgent and complex ethical challenges. The operation of affective computing systems depends on the deep mining of users' emotional states, which requires them to collect and process a large amount of highly personalized data. For example, an application for mental health monitoring may need to record the user's heart rate changes, emotional ups and downs in voice, and subtle differences in facial expressions in real time. These data not only reflect the user's current emotional state, but may also reveal deeper information, such as mental health status, living habits, and even details of social relationships. Compared with traditional personal information (such as name or address), emotional data is more sensitive because it directly points to the individual's inner world. Once this data is abused or leaked, the consequences may far exceed the scope of conventional privacy violations. For example, insurance companies may use emotional data to assess the psychological stability of policyholders, thereby adjusting premium standards or directly refusing insurance; employers may judge their work performance or loyalty by analyzing employees' emotional fluctuations. This potential risk of surveillance and discrimination makes privacy protection an unavoidable issue in the development of affective computing, and the commercial use of data further amplifies this challenge.

The vulnerability of data security has added new complexity to the privacy issue. Emotional computing systems often run on the cloud or connected devices, which means that users' emotional data needs to be transmitted between multiple nodes. However, current network security technology is far from impeccable, and hacker attacks and data leaks occur frequently. According to a global study in 2023, more than 60\% of connected devices have encountered security threats, and emotional computing devices are often the primary target of attackers due to the unique value and high sensitivity of their data. What is more worrying is that many developers of emotional computing applications did not fully consider the necessity of data protection at the beginning of the design. For example, some smart speakers continue to record and upload voice data containing emotional clues without the user's explicit consent. This "secret collection" behavior not only infringes on the user's right to know and autonomy, but also lays hidden dangers for data abuse. In addition, the commercialization trend of emotional data has further blurred the boundaries of privacy. In the digital economy, data is hailed as the "new oil", and emotional data is favored because it can accurately portray user needs. Advertisers may use this data to push more targeted advertisements, and social platforms may optimize content recommendation algorithms by analyzing user emotions. However, such commercialization often comes at the expense of user privacy. Users may not be aware of the purpose for which their emotional data is used, let alone having substantial control over it.

Although the EU's General Data Protection Regulation (GDPR) provides a legal framework for data privacy, its scope of application and enforcement vary significantly around the world, especially in the emerging field of affective computing, where the lag of relevant regulations is particularly obvious. How to find a balance between technological innovation and privacy protection has become an ethical problem that needs to be solved urgently. To meet this challenge, researchers and developers need to incorporate the concept of "privacy first" into technology design. For example, edge computing technology can be used to limit the processing of emotional data to the user's device and avoid uploading it to the cloud; or differential privacy technology can be used to add noise to data analysis to protect individual identities. In addition, transparency and user authorization are also crucial. The affective computing system should inform users of the purpose of data collection and scope of use in a clear and easy-to-understand manner, and provide a simple exit mechanism. However, the improvement of technical means is only part of the solution. The comprehensive protection of privacy and data security still depends on the improvement of laws, the establishment of social norms, and the improvement of public awareness. Only when users can truly understand and control their own emotional data can affective computing move forward steadily on the ethical track and not become a tool for privacy erosion.

\begin{table}[htbp]
    \centering
    \caption{Some Laws Related to Data Privacy Protection}
    \label{tab:data_privacy_laws}
    \scriptsize
    \renewcommand{\arraystretch}{1.2}
    \setlength{\tabcolsep}{2pt}
    \begin{tabular}{@{}p{1cm}p{2.2cm}p{1.2cm}p{2.8cm}p{2.8cm}p{3.5cm}@{}}
    \toprule
    \textbf{No.} & \textbf{Title of Act} & \textbf{Country / Region} & \textbf{Main Content} & \textbf{User Rights} & \textbf{Penalties} \\
    \midrule
    1 & The United States Privacy Rights Act (APRA) & USA (proposed), 2024 & Unify privacy standards nationwide, restrict data transfer, and emphasize data minimization & Access, correction, deletion, export, opt-out & Enforcement by FTC and state attorneys general; users can file lawsuits \\
    2 & California Consumer Privacy Act (CCPA) & USA (California), 2020 & Protect California residents’ data privacy and limit data sales & Know, access, delete, and refuse to sell data & Maximum compensation of \$750 per breach \\
    3 & Personal Information Protection Law (PIPL) & China, 2021.11.1 & Standardize personal data processing and cross-border flow management & Access, correction, deletion, portability, withdrawal of consent & Max fine: 50 million yuan or 5\% of prior year's turnover \\
    4 & Data Security Law & China, 2021.9.1 & Enforce data classification, grading protection, and security review & — & High administrative penalties \\
    5 & General Data Protection Regulation (GDPR) & EU, 2018.5.25 & Transparent and strict standards for personal data protection with informed consent & Access, rectification, erasure, data portability, objection to processing & Max fine: €20 million or 4\% of annual turnover \\
    \bottomrule
    \end{tabular}
\end{table}

If privacy and data security are the external challenges facing affective computing, then the ethical considerations of the human-machine relationship constitute its internal core dilemma. As affective computing technology gives machines the ability to recognize, simulate, and even respond to human emotions, the boundary between humans and machines is becoming increasingly blurred. This change has not only changed the way we interact with technology, but also has a profound impact on human emotional dependence, social interaction, and self-cognition. When a device can "sense" our sadness through voice tone or facial expressions, and comfort us with a warm tone, will we gradually regard it as a source of emotional support? Will the deepening of this relationship weaken the real connection between people? These issues are not only about the realization of technical functions, but also prompt us to re-examine the ethics of the human-machine relationship in the context of affective computing, as well as the potential impact of technology on the structure of human society.

First, emotional computing may lead to excessive human dependence on machines. Studies have shown that when machines show human-like empathy, users tend to be more likely to trust them and even develop emotional attachment to them. For example, a 2024 experiment found that 40\% of people with depression who used emotional computing assistants said they would rather talk to their assistants than communicate with family or friends. This phenomenon not only highlights the potential of emotional computing in the field of psychological support, but also raises deep ethical concerns: If people rely more and more on machines to meet their emotional needs, will they reduce interactions with real humans, thereby exacerbating social isolation? Furthermore, when the emotional response of the machine is merely an algorithm-based simulation rather than real understanding and resonance, will this "false empathy" mislead users and make them unknowingly fall into an unreal emotional relationship? The deepening of this dependence may relieve the emotional pressure of individuals in the short term, but in the long run, it may weaken the most precious emotional connections in human society.

Secondly, the deepening of the human-machine relationship also brings about the ethical dilemma of responsibility attribution. In the application of affective computing, the decision-making of machines is often based on complex algorithms and data models, but these models are not perfect. For example, an educational robot may determine that a student has lost interest in a course by analyzing his or her facial expressions and voice, and adjust the teaching content. However, if the algorithm misjudges the student's emotions, resulting in poor teaching results or even psychological stress, who should be held responsible for this? Is it the developer, the user, or the machine itself? The traditional responsibility allocation framework is stretched when facing affective computing, because the machine is neither a completely autonomous subject nor a simple tool. This ambiguity not only challenges the existing legal system, but also requires us to redefine the power and responsibility relationship between people and technology at the ethical level. In addition, when affective computing is used in more complex scenarios, such as medical diagnosis or legal consultation, the issue of responsibility attribution will become more difficult, further highlighting the urgency of ethical considerations.

In addition, affective computing may also have a profound impact on human self-cognition. When machines can identify our emotions more accurately than we can ourselves, will we gradually lose control of our own emotions? For example, a smart bracelet may prompt the user that he is "angry" through heart rate and skin electrical response before the user is aware of the anger. Although this externalized emotional interpretation is helpful for self-management, it may also make users dependent on the judgment of the machine, thereby weakening the ability of self-reflection. In more extreme cases, when affective computing is used to manipulate user emotions, such as stimulating the desire to buy through specific music or interface design, its ethical boundaries will become more blurred. Does this manipulation infringe on the user's autonomy? How can we find a balance between technological empowerment and ethical constraints? These issues are not only about the mental health of individuals, but also about the potential threat of technology to human free will.

To address the ethical challenges of human-machine relationships, researchers have proposed a "human-centered" design concept, emphasizing that technology should always serve human needs rather than take over the role of the guest. For example, affective computing systems can set clear boundaries of interaction to avoid over-simulating human emotions, thereby reminding users that they are tools rather than partners. At the same time, it is also crucial to educate the public to understand the limitations of affective computing. Only when users realize that the machine's emotional response is algorithm-driven rather than real emotional resonance can they treat this relationship more rationally. In addition, interdisciplinary collaboration—including the participation of ethicists, psychologists, and sociologists—will help develop more comprehensive ethical guidelines to ensure that affective computing does not undermine human autonomy and social connections while enhancing human-machine interaction. Through these efforts, we can establish a healthy balance between technology and humanity and prevent affective computing from becoming a double-edged sword.

The ethical and social challenges of affective computing are a multi-dimensional, cross-domain issue, covering privacy and data security, ethical considerations of human-machine relationships, and many other aspects. These challenges are not stumbling blocks to technological development, but opportunities to promote its maturity. By incorporating privacy protection principles into the design, strengthening user control, and clarifying the ethical boundaries of human-machine relationships, we can find a harmonious coexistence between technological innovation and social responsibility. The ultimate goal of affective computing should not only be a technological breakthrough, but also to improve the quality and meaning of life by understanding and responding to human emotions. However, the realization of this goal is inseparable from the joint efforts of technology developers, policymakers, and the public. Only when ethical and social challenges are properly addressed can affective computing truly become an emotional bridge connecting humans and technology, rather than a controversial ethical minefield. In the future, as technology evolves further, the answers to these questions may become clearer, but before that, we must welcome this change with an open mind and a cautious attitude to ensure that affective computing gives machines "emotional intelligence" while not losing the warmth and bottom line of human society.

\subsection{Privacy and data security issues}
With the in-depth development of affective computing technology, privacy and data security issues have gradually surfaced and become a challenge that needs to be solved urgently. Especially in the field of highly sensitive data such as brain-computer interfaces, the risk of privacy leakage not only threatens the personal information security of users, but may also cause broader social and ethical issues. This article will conduct an in-depth discussion on the privacy and data security issues in affective computing, analyze its potential risks, technical response strategies and future development directions.

The core of affective computing is to extract emotional information from a variety of physiological and behavioral signals. Taking brain-computer interface as an example, many studies use EEG data to capture the user's emotional state. These data are usually processed by methods such as transfer learning to improve the adaptability and accuracy of the model. However, EEG signals not only contain emotional information, but also carry a lot of other private information, such as the user's identity characteristics, health status, and psychological state. The richness of this information makes EEG data uniquely valuable in emotion recognition, but also brings significant privacy risks. Studies have shown that users can be identified with extremely high accuracy through only a few seconds of EEG data. This high-precision user identification capability means that once the data is improperly used or leaked, users may face the risk of identity exposure, abuse of health information, and even psychological manipulation. Similar situations are not limited to the field of brain-computer interface, but are also prevalent in other branches of affective computing. For example, emotion analysis based on facial expressions may leak the user's identity or emotional preferences, while voice emotion recognition may expose the user's psychological state or personal habits.

As privacy issues become increasingly prominent, governments and institutions around the world have begun to strengthen the protection of personal data through legislation. Europe's General Data Protection Regulation, which came into effect on May 25, 2018, clearly stipulates the strict standards that companies must follow when collecting, processing and storing user data, giving users greater control over their data. Similarly, China's Personal Information Protection Law, which was implemented on January 11, 2021, also puts forward clear requirements for the collection and use of personal information. The introduction of these laws reflects the global emphasis on data privacy protection, and also sets a higher compliance threshold for the development and application of affective computing technology. Whether it is a brain-computer interface or other affective computing systems, developers must fully consider these legal requirements in the technical design to ensure the legality and security of user data.

In order to meet the privacy challenge, researchers have proposed a variety of technical strategies to build privacy-preserving affective computing systems. Among them, cryptographic methods are an important solution. Cryptography ensures that data is not directly accessed during transmission and processing through technologies such as homomorphic encryption, secure multi-party computing, and secure processors. For example, in the study of driver drowsiness detection based on EEG, researchers used secure multi-party computing technology to achieve the same analysis results as unencrypted data while protecting user privacy. This method provides a feasible path for privacy protection under the premise of controllable computational costs. However, the application of cryptography is not without limitations, and its computational complexity may limit its use in large-scale real-time affective computing tasks.

Another common privacy protection strategy is data perturbation. The perturbation method reduces the extractability of sensitive information in the data while maintaining the utility of emotion recognition by adding noise to the original data or performing data transformation. Differential privacy technology is a typical representative of this strategy. It uses mathematical means to ensure that even if an attacker obtains the processed data, the original information cannot be accurately inferred. In addition, data reconstruction technology can also transform the original signal into a form that is difficult to reverse parse, thereby further enhancing privacy protection. This method is widely used in brain-computer interfaces and speech emotion recognition. For example, in speech emotion analysis, researchers can mask the user's identity information by adding background noise or changing the pitch characteristics while retaining the core features of emotional expression. The advantage of the perturbation method lies in its flexibility and low computational overhead, but its challenge lies in how to find a balance between privacy protection and data utility.

\begin{figure}
    \centering
    \includegraphics[width=0.5\textwidth]{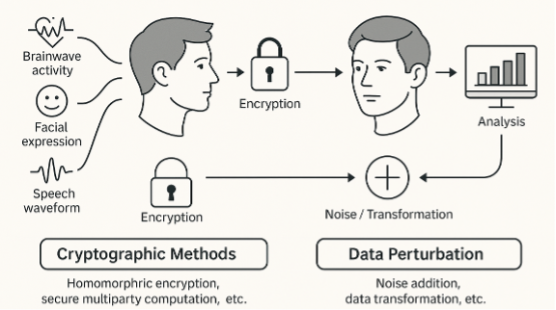}
    \caption{Privacy protection concept diagram}
\end{figure}

Machine learning-assisted systems provide another approach to privacy protection. Such systems can not only be used for emotion recognition itself, but can also help users better understand privacy policies and provide risk warnings in data usage decisions. For example, by designing intelligent privacy management tools, users can clearly understand potential privacy risks before authorizing data use, so that they can make more informed choices. Although this method does not directly act on the data itself, it adds a layer of security to the emotional computing system by enhancing users' privacy awareness and control capabilities.

Although the above strategies have alleviated the privacy issue to a certain extent, the field of affective computing still faces many unresolved challenges. Taking brain-computer interface as an example, recent studies have proposed an unsupervised multi-source decentralized migration method to protect the privacy of EEG data in offline scenarios. This method achieves high classification accuracy and avoids direct sharing of sensitive data by using the parameters or prediction results of the source model instead of the original data for transfer learning. Experiments on specific datasets have verified the effectiveness of this method. However, compared with the accuracy of emotion recognition, the issue of privacy protection has not received enough attention in research and application. Many developers prefer to pursue the improvement of technical performance and ignore the importance of data security. This unbalanced attention may cause the system to face greater risks in actual deployment.

In addition to privacy leaks, affective computing systems also face security threats from adversarial attacks. Adversarial attacks trick machine learning models into outputting incorrect results by introducing tiny perturbations into the input data. These perturbations are often so subtle that they are difficult for humans to detect, but they are enough to cause serious interference to the system. In the field of brain-computer interfaces, adversarial attacks are particularly prominent. For example, researchers designed a specific attack method for an EEG-based regression model, which successfully changed the estimated results of the driver's drowsiness level by generating tiny perturbations, with a success rate of nearly 100\%. Similar problems also exist in other affective computing scenarios. For example, facial expression recognition models may misjudge the user's emotional state due to adversarial samples, and voice emotion analysis systems may draw incorrect conclusions due to noise interference.

The potential consequences of adversarial attacks should not be underestimated. In extreme scenarios of brain-computer interfaces, hackers may directly affect the user's emotions and cognitive functions by invading the system, or even manipulate the motor cortex to induce unexpected behavior. For example, in military applications, opponents may use adversarial attacks to send false instructions to operators, causing misoperation or emotional confusion. Although this possibility is still theoretical speculation, it has attracted great attention from security experts. A report pointed out that the abuse of affective computing technology may provide malicious actors with a direct way into the human brain, thereby creating chaos or causing more serious consequences.

Despite the growing threat of adversarial attacks, research on defenses against affective computing systems has lagged behind. At present, researchers have proposed some potential defense strategies to improve the adversarial robustness of the system. Data modification is a common method that adjusts the training data or test data to enhance the model's ability to resist interference. For example, adversarial training introduces adversarial samples during the training process so that the model learns to recognize and resist similar attacks. In addition, data compression and randomization techniques can also weaken the impact of perturbations during the testing phase. Another strategy is model modification, which directly improves the robustness of the model through techniques such as regularization or defense distillation. Regularization reduces the sensitivity of the model to small perturbations by limiting its complexity, while defense distillation generates a more stable model structure through multi-stage training. In addition, the application of auxiliary tools also provides new ideas for defense. For example, using additional machine learning modules to detect adversarial samples can issue warnings or block abnormal inputs in time when an attack occurs.

However, the application of these defense strategies in the field of affective computing is still in the exploratory stage. Taking brain-computer interfaces as an example, there is currently no research on adversarial defenses specifically targeting emotion recognition systems. This is in stark contrast to the rapid pace of technological development in this field. In the future, as affective computing becomes more popular in the fields of healthcare, education, and entertainment, the importance of privacy and security issues will become more prominent. Developers need to find a balance between technological innovation and risk control to ensure that the system can not only accurately understand users' emotions, but also effectively protect their privacy and security.

The issue of privacy and data security in affective computing is a multi-dimensional and multi-level challenge. From brain-computer interfaces to facial expression analysis to speech emotion recognition, each technology is accompanied by unique privacy risks and security threats. Cryptography, data perturbation, and machine learning-assisted systems provide technical support for privacy protection, while data modification, model modification, and auxiliary tools open up paths for defense against adversarial attacks. However, the implementation of these solutions still needs to overcome many difficulties, including computing costs, utility loss, and legal compliance. In the future, the field of affective computing needs to work together in technology research and development, policy formulation, and user education to build an ecosystem that is both smart and secure. Only in this way can this technology benefit mankind while minimizing potential risks.

\subsection{Ethical considerations of human-computer relationships}
Emotion recognition systems rely on a large amount of personal data, including facial images, voice samples, electroencephalogram (EEG) data, and other biological signals. The collection and storage of this data raises serious privacy issues. If a user's emotional data is collected, stored, or shared without permission, it may lead to personal privacy leaks or even be used to manipulate user behavior. For example, companies can use sentiment analysis data to optimize advertising and influence consumer decisions in a more targeted manner, while government agencies may use emotion recognition technology for social monitoring.

In order to protect user privacy, strict data collection and storage specifications must be established. First, the collection of emotional data should be based on explicit user consent (opt-in) rather than default collection. Second, data anonymization and de-identification can reduce the risk of user identity leakage. Finally, the use of encryption technology and secure storage solutions such as blockchain can help improve the security of emotional data and reduce the possibility of malicious attacks.

With the development of emotion recognition systems, machines can not only understand human emotions, but also adjust their own behavior based on the analysis results to make them more "human". For example, virtual assistants can provide comfort when users express sadness, and social robots can adjust their interaction methods based on human emotional feedback. This highly intelligent emotional interaction enhances the intimacy of the human-machine relationship, but it also raises some ethical issues.

\begin{figure}
    \centering
    \includegraphics[width=0.5\textwidth]{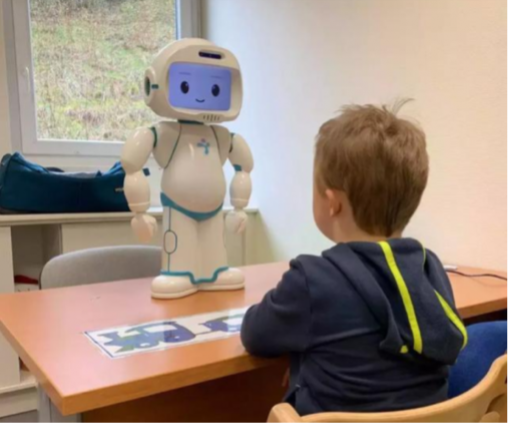}
    \caption{Qtrobot, developed by LuxAI, is a social robot designed for children with autism spectrum disorder (ASD).}
\end{figure}

First, users may become overly dependent on emotion recognition systems, viewing them as true "friends" or "psychological supporters." This dependence may weaken emotional communication between people and even affect mental health. Second, the machine's emotion recognition ability is essentially based on data pattern analysis, rather than truly understanding or experiencing emotions. In other words, the machine is "simulating" human emotions, but it does not have true empathy. If users have a wrong perception of the machine's "emotional response," it may affect their emotional regulation ability and social behavior.

To address these issues, developers need to be transparent in the design of emotion recognition systems to avoid misleading users into thinking that machines have real emotional awareness. In addition, users should be encouraged to use emotion recognition technology as an auxiliary tool rather than a substitute for human relationships.

Emotion recognition systems are usually based on machine learning models, and the source of training data for these models determines their applicability to different populations. However, due to data imbalance, emotion recognition systems may show bias. For example, some studies have found that existing emotion recognition algorithms may have different accuracy rates when identifying emotions of different races, which may lead to discriminatory treatment of certain groups.

This bias not only affects individual user experience, but may also lead to social injustice in certain application scenarios. For example, in the recruitment process, if a company uses emotion recognition technology to assess the emotional stability of applicants, and the system has biased emotion recognition for certain groups, it may lead to employment discrimination. For another example, in the judicial system, if sentiment analysis is used to assess the credibility of suspects, and the technology performs inconsistently across different races and cultural backgrounds, it may affect judicial justice.

In order to reduce the bias of emotion recognition systems, researchers should ensure the diversity of training data and develop fairness optimization algorithms. In addition, in practical applications, when using emotion recognition systems for decision-making, other human supervision mechanisms should be combined to reduce the negative impact of algorithmic bias.

The transparency and explainability of AI system decisions have always been important topics in the ethical discussion of artificial intelligence. In the field of emotion recognition, users usually cannot know how the system makes a certain emotional judgment. For example, a virtual assistant may judge that the user is angry based on the tone of the user's voice, but the user does not know the specific basis of the analysis. This "black box" problem reduces the user's trust in the system and may lead to misjudgment.

To improve transparency, emotion recognition systems should provide clear explanations of their decisions. For example, the system can show users the main basis for its emotion analysis, such as facial expression changes, voice characteristics, or heart rate fluctuations. In addition, the use of explainable AI (XAI) methods can help users understand the system's analysis logic, thereby improving trust and controllability.

The use of emotion recognition technology may have a profound impact on the user's mental health. First, emotion tracking may cause users to pay too much attention to their own emotions, thereby exacerbating anxiety. For example, if a person finds that their emotion analysis data shows "persistent stress" or "chronic anxiety", it may strengthen their perception of negative emotions and increase their psychological burden.

Secondly, emotion recognition technology may be used for emotional manipulation. Advertisers can use emotion data to precisely place advertisements to maximize the user's emotion-driven consumption behavior; social media platforms can optimize information flow through emotion analysis, making it easier for users to immerse themselves in it and even form emotional dependence.

Therefore, when designing an emotion recognition system, ethical boundaries should be set to avoid using technology for emotional manipulation. For example, it should avoid collecting and analyzing emotional data without informing the user, and ensure that the user can choose whether to use the relevant function. In addition, psychology experts can be introduced to participate in system design to evaluate the potential impact of technology on the user's mental health.

% ====================================ch10

\chapter{Future Development Trends}

\section{Technology Development Trends}
The rapid development of emotion recognition technology is first of all due to the deep integration of deep learning and emotion recognition. As one of the core technologies in the field of artificial intelligence, deep learning has proven its powerful capabilities in many fields such as image recognition, speech processing and natural language understanding. In the field of emotion recognition, the application of deep learning has also brought revolutionary changes. By building a complex neural network model, the system can automatically extract emotional features from massive data without relying on traditional manual feature engineering. For example, convolutional neural networks (CNNs) perform particularly well in facial expression recognition. They can capture subtle changes in human facial muscles and even distinguish micro-expressions such as "slightly raised corners of the mouth" or "slight wrinkles between eyebrows". With the continuous expansion of the data set size and the continuous optimization of the model structure, future CNN models will be able to more accurately identify emotional states in complex scenarios, such as maintaining high accuracy in low light or expression occlusion. At the same time, recurrent neural networks (RNNs) and their variants, such as long short-term memory networks (LSTMs) and gated recurrent units (GRUs), have shown unique advantages in processing speech emotions and text emotions. They are good at analyzing time series data and can capture the ups and downs in speech intonation or the emotional ups and downs in text sentences. Imagine that the smart assistant of the future can not only understand what you say, but also judge whether you are happy or frustrated by the tone of your voice. This ability will undoubtedly greatly improve the naturalness of the interaction. What's more exciting is that the emergence of the Transformer model has brought new possibilities for emotion recognition. Transformer initially shined in the field of natural language processing, but its powerful attention mechanism also makes it promising in multimodal emotion recognition. The Transformer model of the future will be able to seamlessly integrate multiple information such as text, voice and images. For example, by analyzing the conversation content, the speaker's tone and facial expressions in a video, the speaker's true emotions can be comprehensively judged. This multimodal fusion capability will bring a qualitative leap in the accuracy and application scenarios of emotion recognition.

However, the expression of human emotions is never single-dimensional, and judging emotions solely by facial expressions or voice is often not comprehensive enough. This leads to another important trend - the rise of multimodal emotion recognition. Emotions are complex, and they may be expressed through facial expressions, voice intonation, body movements, and even physiological signals (such as heart rate or skin galvanic response) at the same time. Single-modal recognition systems are easily affected by noise or lack of context, while multimodal emotion recognition significantly improves the robustness and accuracy of the system by integrating multiple sources of information. In the future, multimodal emotion recognition systems will be more mature and able to intelligently fuse data from different modalities. For example, an intelligent customer service system may analyze the user's voice, facial expressions, and input text at the same time to determine whether the user is satisfied with the service and adjust the response strategy accordingly. The key to achieving this goal lies in the advancement of data fusion technology. In the future, fusion algorithms will no longer simply splice data from each modality together, but will automatically explore the correlation and complementarity between different modalities through deep learning. For example, tremors in voice may echo tension in facial expressions. In addition, cross-modal learning will also become a highlight. This approach uses the knowledge of one modality to enhance the recognition capabilities of another modality. For example, by training the model with voice data, it can better understand the emotional clues in facial expressions. It is worth mentioning that one challenge facing multimodal emotion recognition is the high cost of data annotation. To solve this problem, unsupervised learning and semi-supervised learning will play an important role in the future. These methods can make full use of unlabeled natural data, such as videos and texts on social media, so that the system can still learn emotional patterns in the absence of manual annotation. It can be foreseen that as multimodal emotion recognition technology matures, we will see more intelligent systems that can "read minds". They can not only perceive our explicit emotions, but also capture the subtle emotions hidden under multiple expressions.

\begin{figure}
    \centering
    \includegraphics[width=0.2\textwidth]{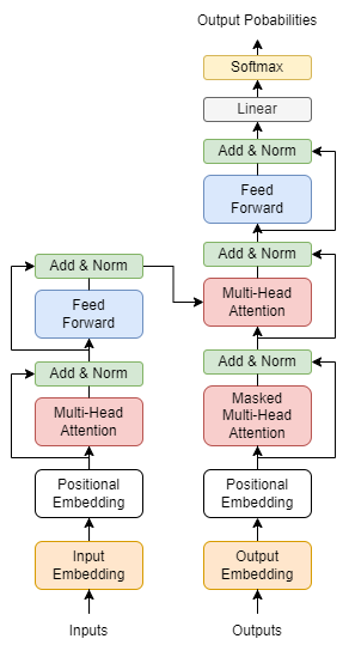}
    \caption{Multimodal emotion recognition system concept diagram}
\end{figure}

At the same time, the real-time and imperceptible nature of emotion recognition is also a top priority for future development. With the rapid development of the Internet of Things and edge computing technologies, emotion recognition systems will gradually get rid of their dependence on cloud computing and instead realize real-time processing on local devices. This shift will greatly improve the system's response speed while reducing the latency of data transmission. For example, future smartphones may have built-in emotion recognition modules that analyze users' voices and facial expressions in real time and actively suggest taking a break when users feel tired. This imperceptible emotion recognition means that the system can complete emotion analysis without the user noticing, thereby providing a more natural service. The support of edge computing technology is the basis for achieving this goal. By deploying lightweight emotion recognition models on edge devices, the system can complete processing without uploading sensitive data to the cloud, which not only improves efficiency but also enhances privacy protection. In addition, the development of embedded systems will enable emotion recognition technology to be integrated into more everyday devices, such as mirrors in smart homes that can adjust the lighting atmosphere by observing your expression, or cars that analyze the driver's emotional state to remind them to pay attention to safety. However, the privacy issues brought about by imperceptibility cannot be ignored. Future emotion recognition systems need to incorporate privacy protection mechanisms in their technical design, such as using federated learning or differential privacy technologies to ensure that users' emotional data will not be abused. It is conceivable that in the near future, our devices will be like a caring friend, silently understanding our emotions without disturbing us and providing just the right amount of help.

Personalization and adaptability of recognition is another trend worth paying attention to. Everyone has a unique style of expressing emotions. For example, some people raise their voices when they are angry, while others may become silent. General emotion recognition models often have difficulty adapting to such individual differences, and future systems will be able to automatically adjust recognition strategies based on the characteristics of users. Transfer learning will play a key role in this process, allowing the model to transfer knowledge learned from one user to another, so as to quickly adapt to the new user's emotional expression habits. For example, an emotion recognition system may be pre-trained on a large dataset, and then fine-tune the model through brief interactions with a user to make it more consistent with the user's emotional pattern. Online learning further enhances the system's adaptability. This method allows the system to continue learning during operation, such as by observing the user's reactions in different situations, and gradually optimizing the judgment of the user's emotions. Furthermore, the introduction of user profiling technology will enable the emotion recognition system to combine the user's background information (such as age, gender, cultural habits) and behavior patterns to build a more comprehensive emotion understanding framework. Imagine that in the future, intelligent assistants can not only recognize your emotions, but also provide personalized responses based on your personality and preferences - when you are depressed, it may play your favorite music instead of the stereotyped suggestion that you "relax." This personalized emotion recognition will greatly enhance the user experience and make technology truly integrated into our lives.

The development of technology is inseparable from the deep integration with other branches of artificial intelligence. As an important part of artificial intelligence, emotion recognition will co-evolve with technologies such as natural language processing and computer vision to jointly promote the intelligence of human-computer interaction. For example, the emergence of emotional agents will enable machines to not only perceive and understand emotions, but also express emotions in an appropriate way. A virtual assistant with emotional intelligence may comfort you with a warm tone when you are lost, or show the same enthusiasm when you are excited. This "two-way emotional interaction" will make the human-computer relationship closer to communication between people. In addition, the combination of emotion recognition and natural language processing will enable machines to generate emotional language, such as adjusting the tone and wording of conversations according to the user's emotions. And the integration with computer vision allows the system to perceive emotions more accurately through visual clues, such as by analyzing the posture and expression of the characters in the video to judge their emotional state. It can be foreseen that the deep integration of emotion recognition and artificial intelligence will give birth to a series of more intelligent and more "human" interactive systems.

After discussing the technical development trends of emotion recognition, it is not difficult to find that behind all this, there are two key driving forces: the progress of artificial intelligence and the innovation of interactive technology. In the following part of this chapter, we will further explore the "progress of artificial intelligence" and "innovation of interactive technology", analyze how they inject new vitality into emotion recognition, and look forward to their coordinated development in the future.

\subsection{Progress of Artificial Intelligence}
As the core driving force in the field of emotion recognition, the technological progress of artificial intelligence (AI) not only provides strong support for emotional computing, but also points out the direction for future development trends. With the continuous evolution of technology, emotion recognition is gradually moving from a simple classification task of a single modality to a complex system of multimodality, real-time, imperceptible, personalized and emotional intelligence. The progress of artificial intelligence is reflected in the innovation of deep learning models, the deepening of multimodal fusion, the application of edge computing, the optimization of personalized strategies, and the rise of emotional intelligence. This section will explore these emerging technologies in depth, analyze their specific implementation methods, and look forward to how they will shape the future of emotion recognition. At the same time, we will also pay attention to privacy protection and ethical considerations to ensure that technological development is coordinated with social needs.

The breakthrough of artificial intelligence in the field of emotion recognition first stems from the continuous evolution of deep learning technology. In the past decade, convolutional neural networks (CNNs) have dominated facial expression recognition. Through multi-layer convolution and pooling operations, they can automatically extract emotional features from images, such as a raised corner of the mouth indicating pleasure and a frown indicating anger. However, with the increase in task complexity, such as the need to recognize micro-expressions or mixed emotions, traditional CNNs have gradually revealed their limitations in processing high-dimensional data and capturing subtle changes. To this end, researchers have begun to explore more expressive model architectures, among which the rise of the Transformer model is particularly noteworthy. Transformer initially achieved success in the field of natural language processing. Its core lies in the self-attention mechanism, which can capture long-range dependencies in data. In emotion recognition, this feature makes it particularly suitable for analyzing dynamic changes in emotional expressions. For example, in speech emotion recognition, Transformer can extract subtle ups and downs in tone from a speech signal, distinguish whether the speaker is calm or excited, and even perceive hidden anxiety without changing the speaking speed. In the future, researchers plan to combine Transformer with CNN to build a multimodal emotion recognition system. For example, a system may receive a user's facial video, voice signal, and text input at the same time, and integrate this information through the Transformer's attention mechanism to determine whether the user is using a sarcastic tone to cover up his true emotions. This multimodal fusion not only improves recognition accuracy, but also makes it possible to understand complex emotions, such as identifying patients' contradictory psychology in psychological counseling scenarios.

At the same time, graph neural networks (GNNs), as another emerging deep learning model, also show unique advantages in emotion recognition. GNNs are good at processing data with non-Euclidean structures, such as the spatial relationship between facial feature points or the propagation of emotions in social interactions. In facial expression recognition, researchers can capture dynamic changes by constructing a relationship graph between feature points. For example, the combination of drooping eyes and drooping mouth corners may indicate sadness, and this dynamic relationship is difficult for traditional CNNs to model directly. In the future, the application of GNNs will be further expanded, such as combining physiological signals (such as heart rate or skin electrical response) to construct multidimensional emotion maps, so as to more comprehensively understand the user's emotional state. In order to improve the efficiency of GNNs, researchers are developing lightweight architectures that can run on resource-constrained devices, which lays the foundation for real-time emotion recognition.

\begin{figure}
    \centering
    \includegraphics[width=0.8\textwidth]{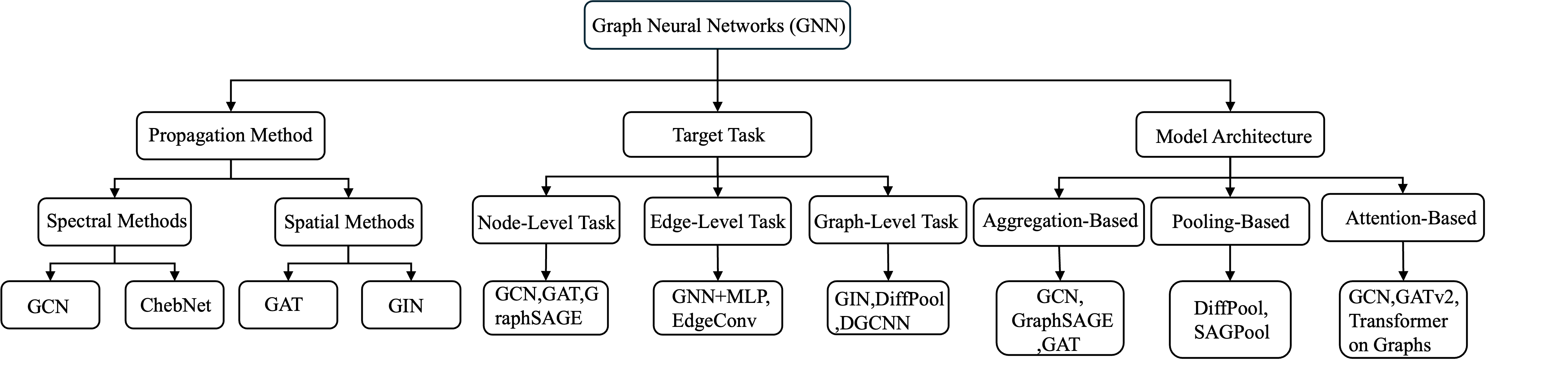}
    \caption{Graph neural network for emotion recognition}
\end{figure}

The future of emotion recognition depends not only on the innovation of model architecture, but also on the deep integration of multimodal information. Human emotions are multidimensional, and a single modality such as facial expression or voice often cannot fully reflect the inner state. For example, a person may smile but have a low voice, and this contradiction requires a combination of multiple information sources to accurately interpret. Multimodal emotion recognition significantly improves the robustness and accuracy of the system by integrating facial expressions, voice, text, and physiological signals. Advances in artificial intelligence are pushing this field to new heights, among which cross-modal learning is a key direction. Cross-modal learning uses the knowledge of one modality to enhance the recognition ability of another modality, such as assisting the judgment of facial expressions through changes in voice intonation. This method is particularly suitable for scenarios where data is scarce. For example, in some cultures, emotional expression is more restrained, and when single-modal data is not enough to train a reliable model, cross-modal learning can make up for the deficiency through rich data from other modalities. In the future, researchers will develop more efficient cross-modal algorithms, such as using generative adversarial networks (GANs) to generate data of missing modalities, thereby achieving seamless integration between different modalities. For example, a system might generate virtual facial expressions from a user’s speech for use in training models or enhancing real-time interactions.

Attention mechanisms also play a crucial role in multimodal emotion recognition. Traditional fusion methods usually treat all modalities equally, but in fact the importance of each modality may be different in different situations. For example, in a noisy environment, facial expressions may be more reliable than voice, while text information may become the main clue when a video call is interrupted. The attention mechanism can dynamically weight the importance of different modalities according to the context, so as to capture emotional clues more accurately. The future development trend is to develop smarter attention networks, such as dynamic attention mechanisms, which can adjust weights in real time according to the user's context (such as fatigue or concentration). For example, an intelligent education system may pay more attention to the student's eyes and gestures when he or she appears confused, rather than the voice disturbed by background noise. This refined fusion will make the emotion recognition system closer to human intuition.

\begin{figure}
    \centering
    \includegraphics[width=0.3\textwidth]{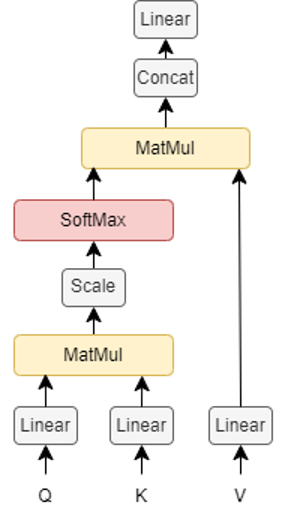}
    \caption{Transformer model multi-head attention mechanism}
\end{figure}

With the rise of the Internet of Things and edge computing technologies, real-time and imperceptible emotion recognition has become possible. The traditional cloud computing model requires data to be transmitted to the server for processing, which not only increases latency but may also raise privacy concerns. Edge computing delegates data processing capabilities to local devices, such as smartphones or smart cameras, so that emotion recognition can be completed in milliseconds. For example, a smart watch with a built-in emotion recognition module can monitor the user's heart rate and voice in real time, judge their stress level, and remind the user to take a deep breath through vibration when anxiety is detected. In the future, researchers will focus on developing lightweight models, such as compressing complex deep learning models to a size suitable for edge devices through model pruning or knowledge distillation technology. At the same time, imperceptible emotion recognition will further enhance the user experience. This technology allows the system to complete emotion analysis without the user's awareness, such as smart homes monitoring the emotions of family members through cameras and microphones, and automatically adjusting lights and music to relieve tension. To achieve this goal, researchers are developing more advanced sensors, such as high-resolution thermal imagers, to capture subtle emotional signals caused by changes in facial blood flow, as well as smarter algorithms to ensure that the analysis process is completely transparent to users.

\begin{figure}
    \centering
    \includegraphics[width=0.3\textwidth]{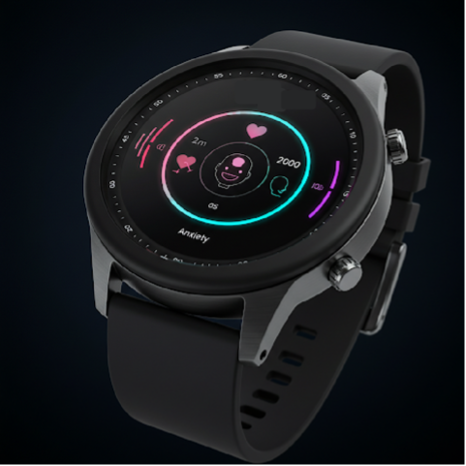}
    \caption{Smartwatch with built-in emotion recognition module}
\end{figure}

The personalization and adaptability of emotion recognition are also important directions for future development. Everyone has a unique style of expressing emotions. For example, some people are used to expressing joy with exaggerated expressions, while others tend to be more restrained. General emotion recognition models often have difficulty adapting to such individual differences, resulting in inaccurate recognition results. Transfer learning provides a solution that allows the model to be pre-trained on a large dataset and then fine-tuned through brief interactions with specific users. For example, a virtual assistant may quickly learn the user's emotional patterns through a few minutes of conversation when it is first used, such as distinguishing whether the user's "tears in laughter" is happy or sad. In the future, transfer learning will be further optimized to reduce the amount of data required for personalization, such as through the few-shot learning technology, user-specific models can be generated with only a few interactions. In addition, online learning will enable the system to continuously optimize during operation. For example, a companion robot can gradually improve its understanding of the user's emotions by observing the user's reactions in different situations over a long period of time, such as learning to distinguish whether the user's silence when tired requires rest or simply does not want to talk. This adaptive ability will make the emotion recognition system more flexible and adapt to the dynamic changes in the user's emotional expression.

The rise of emotional agents marks another leap forward in emotion recognition technology. Emotional agents can not only perceive and understand emotions, but also express them in an appropriate way, thus achieving "two-way emotional interaction". For example, a virtual assistant can comfort a user with a warm tone when the user is upset, or show the same enthusiasm when the user is excited. This ability relies on the development of emotion generation technology, such as generating emotional responses through natural language processing, or generating facial expressions that match emotions through computer vision. In the future, emotional agents will play an important role in virtual reality (VR) and the metaverse. In virtual environments, emotional agents can monitor users' immersion and emotional reactions in real time and dynamically adjust content. For example, in a virtual meeting, when the system detects that the user is tired, the emotional agent may suggest a pause or adjust the rhythm, or even show a caring expression through the virtual avatar. This real and vivid interaction will make the virtual experience more humane.

However, the widespread use of emotion recognition technology also brings privacy and ethical challenges. Users' emotional data is highly sensitive and once leaked, it may be used for improper purposes, such as manipulating emotions or commercial marketing. Federated learning and differential privacy technologies offer hope for solving this problem. Federated learning allows models to be trained without sharing original data. For example, multiple devices can process user data locally and only upload model updates to the server to protect privacy. Differential privacy ensures that even if the data is intercepted, it cannot be traced back to an individual by adding noise to the data. In the future, these technologies will be widely integrated into emotion recognition systems. For example, a smart speaker may complete all emotion analysis locally and only use anonymized statistical data for model optimization. At the same time, the establishment of an ethical framework is crucial. Researchers need to clarify the boundaries of technology use, such as prohibiting the use of emotion recognition for surveillance without consent, and develop a responsibility attribution mechanism to ensure that technology development complies with social norms.

Advances in artificial intelligence are injecting new vitality into emotion recognition technology. From the innovation of deep learning models such as Transformer and GNN, to the deepening of multimodal fusion and cross-modal learning, to the realization of edge computing, imperceptibility, personalization and emotional intelligence, these technologies will jointly promote emotion recognition to a higher level. At the same time, the integration of privacy protection and ethical considerations will ensure that the technology is on the right track. It can be foreseen that the emotion recognition system of the future will be more intelligent, accurate and natural, which will not only change the way we interact with machines, but also have a profound impact in education, medical care, entertainment and other fields. Machines will no longer be cold tools, but partners that can understand our emotions and accompany us in our growth. The continuous breakthroughs in artificial intelligence will gradually make this vision a reality.

\subsection{Innovation of interactive technology}

In the field of affective computing, innovations in interactive technologies are driving the development of emotion recognition at an unprecedented pace. With the rise of emerging technologies such as virtual reality (VR), augmented reality (AR), brain-computer interface (BCI), tactile feedback technology, and the metaverse, emotion recognition is shifting from traditional screen- and voice-based interaction to a more immersive, intuitive, and multi-dimensional experience. The advancement of these technologies not only provides new application scenarios for emotion recognition, but also greatly enriches the possibilities of emotional expression and perception. This section will explore in depth how these innovative technologies are combined with emotion recognition technology, analyze their future development trends, and look forward to their far-reaching impact on the field of human-computer interaction.

Virtual reality technology brings users into a completely virtual environment through head-mounted displays and controllers, opening up new frontiers for the research and application of emotion recognition. The immersive nature of VR allows users to move freely in the virtual world, while emotion recognition technology can capture users' emotional responses in real time through a variety of sensors. For example, in a virtual social scene, a user may have a conversation with a virtual character. The system can analyze the user's facial expressions through a camera, detect voice intonation through a microphone, and even capture gestures through a handle to determine the user's emotional state - whether it is excitement, tension or boredom. This real-time emotional feedback provides a basis for optimizing the interactive experience: virtual characters can adjust their behavior according to the user's emotions, such as giving a comforting response when the user shows anxiety, or speeding up the conversation when the user is excited. This technology has shown potential in the field of psychotherapy, where patients can face fear or trauma in a safe virtual environment, while the system monitors their mental state through emotion recognition to help therapists adjust intervention strategies.

In the future, the combination of VR and emotion recognition will be more in-depth. With the advancement of sensor technology, the system will be able to monitor more physiological signals, such as heart rate, galvanic skin response, and eye tracking, to more accurately assess the user's immersion and emotional involvement. For example, in a virtual adventure game, when the system detects that the user's heart rate is accelerated and the pupil is dilated, it can be inferred that the user is in a state of tension or excitement, and dynamically adjust the difficulty or atmosphere of the scene-such as increasing the tension of the background music or introducing new challenges. In addition, emotion recognition in VR can also be deeply integrated with brain-computer interface technology. By analyzing the user's brain wave signal (EEG), the system can perceive his emotional changes in real time, such as from calm to frustration, and adjust the virtual content accordingly. For example, in educational VR applications, if students show fatigue during the learning process, the system can automatically switch to more relaxing content or introduce interactive games to re-stimulate their interest. It can be foreseen that with the portability of VR hardware and the improvement of computing power, emotion recognition will be seamlessly embedded in the virtual experience, and it may even be possible to directly manipulate the emotional expression of the virtual environment through thoughts, such as users can make the virtual avatar smile just by "wanting to smile". This highly personalized and intuitive interaction will greatly enhance the immersion and practicality of VR, and promote its widespread application in entertainment, education, medical and other fields.

\begin{figure}
    \centering
    \includegraphics[width=0.5\textwidth]{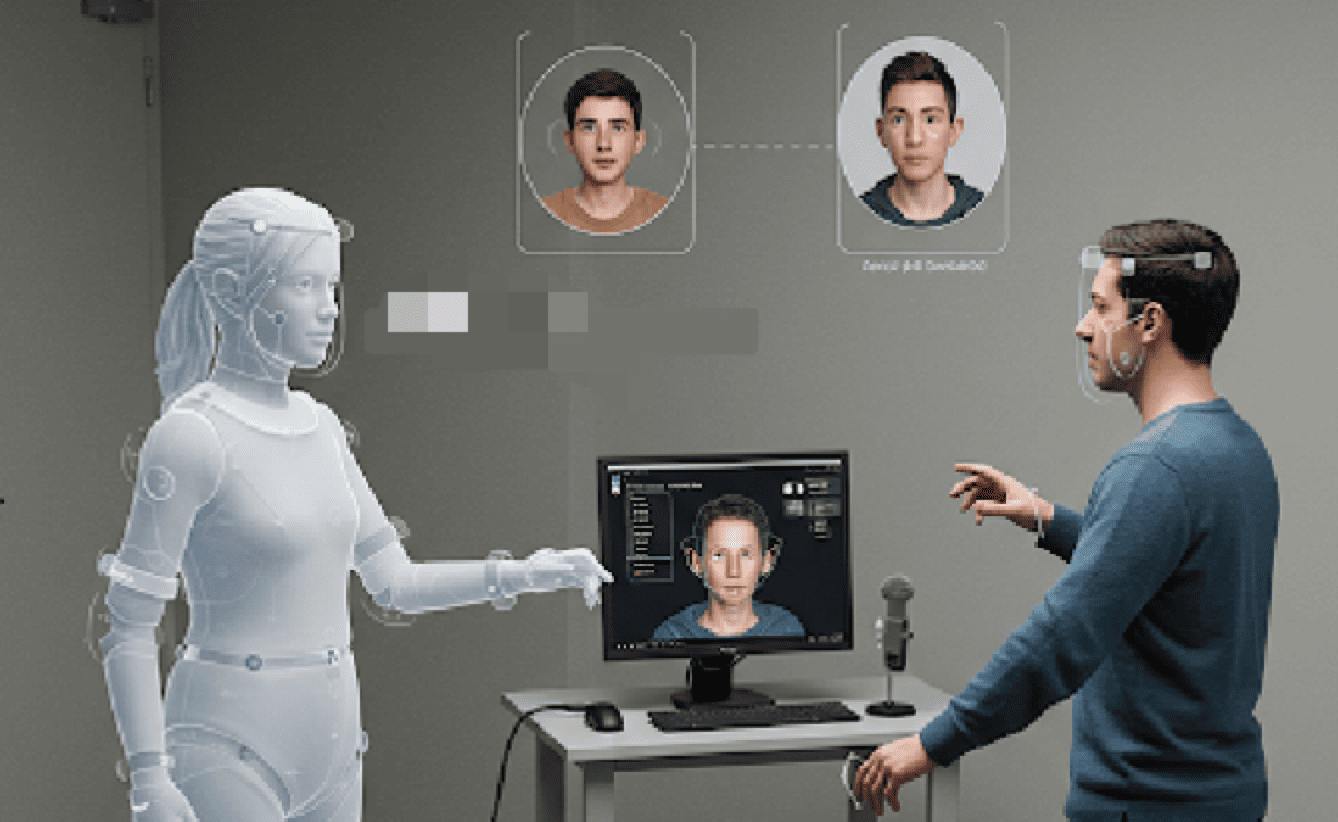}
    \caption{Virtual interaction}
\end{figure}

Unlike VR, augmented reality technology overlays virtual information onto the real world, providing users with a hybrid interactive experience. In the field of emotion recognition, the application potential of AR technology lies in enhancing the experience of interpersonal communication and environmental perception. Imagine an AR pair of glasses with a built-in camera and microphone that can analyze the facial expressions and voice intonation of both parties in the conversation in real time and display emotional cues in the user's field of view. For example, in business negotiations, AR glasses can detect subtle changes in the other party's tone or hesitation in the eyes, and remind the user in the form of text or icons: "The other party may feel uneasy" or "The other party is interested in your proposal." This technology can not only improve communication efficiency, but also help users better understand the emotions of others in scenarios such as education and training, cross-cultural communication, etc.

In the future, AR technology will further integrate with emotion recognition, showing a trend of being more intelligent and personalized. With the maturity of eye tracking and head posture recognition technology, AR systems will be able to analyze users' attention and emotional response to specific objects. For example, in a museum visit scene, when a user stares at an exhibit for a long time and smiles, the system can judge that he is interested in it and automatically push relevant background stories or interactive content. In addition, AR can also be combined with tactile feedback technology to provide users with multi-sensory experiences related to emotions through smart gloves or tactile vests. In a virtual shopping scene, when a user touches a virtual product, the system can adjust the intensity and mode of tactile feedback according to his emotional response-if the user shows excitement, the gloves may simulate stronger texture feedback; if the user hesitates, a softer touch is provided to encourage decision-making. This multimodal interaction will make the AR experience more real and vivid. Looking to the future, with the development of 5G networks and edge computing, AR devices will achieve lower latency and higher computing power, and the real-time and accuracy of emotion recognition will be greatly improved, which may give rise to new application scenarios. For example, in telemedicine, doctors use AR glasses to perceive patients' emotions in real time and adjust treatment plans.

Brain-computer interface technology provides a revolutionary method for emotion recognition by directly connecting the human brain with external devices. Traditional BCI systems mainly rely on electroencephalogram (EEG) signals to monitor the user's brain activity, while in the field of emotion recognition, this technology can decode the user's emotional state such as concentration, relaxation, excitement or fatigue in real time. For example, in an intelligent education system, BCI can monitor the student's attention level in class. When it detects that the student is distracted or tired, the system can automatically adjust the difficulty of the teaching content or introduce interactive links to re-attract their attention. The high real-time and accuracy of this technology gives it a unique advantage in personalized interaction.

\begin{figure}
    \centering
    \includegraphics[width=0.5\textwidth]{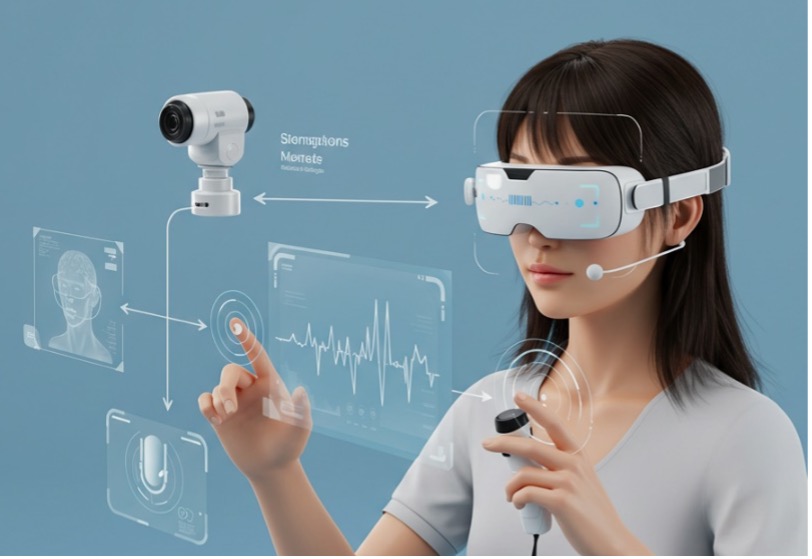}
    \caption{Brain-computer interface device interaction}
\end{figure}

In the future, the combination of BCI and emotion recognition will move towards a deeper level of intelligence. Through the analysis of brain wave signals by deep learning algorithms, the system can not only recognize basic emotions, but also capture more complex emotional changes, such as the transition from expectation to disappointment, and even predict the user's emotional trends. For example, in a virtual reality game, when BCI detects that the user is tired of repetitive tasks, the system can automatically introduce new challenges or rewards to maintain their interest. In addition, BCI technology can also work with VR and AR technologies to create a true "mind interaction". In a virtual environment, users may be able to control the character's emotional expression or scene changes with just their thoughts, such as "thinking about calmness" can make the virtual world soft and peaceful. This technology may also extend to daily life. For example, in a smart home scenario, after the BCI system senses the user's low mood, it automatically adjusts the light color or plays soothing music to improve his mood. It can be foreseen that with the popularization of non-invasive BCI devices and the advancement of signal processing technology, emotion recognition will shift from external behavior analysis to direct decoding of internal psychological states, bringing unprecedented depth and breadth to human-computer interaction.

As an emerging means of interaction, tactile feedback technology provides users with a multi-sensory experience beyond vision and hearing through vibration, pressure or temperature changes. In the field of emotion recognition, this technology can significantly enhance the effect of emotional expression and perception. For example, a smart watch can convey the user's emotional state through different vibration patterns - fast and strong vibrations indicate excitement, and slow and soft vibrations indicate calmness. This technology is particularly useful in remote communication, such as a long-distance couple "transmitting heartbeats" through tactile devices, allowing the other party to feel each other's emotional fluctuations.

In the future, tactile feedback technology will be more closely integrated with emotion recognition, showing a highly personalized and dynamic trend. The system can adjust the mode and intensity of tactile feedback in real time by analyzing the user's emotional state. For example, in a virtual reality training scenario, when the user shows tension when completing a task, the tactile vest can simulate a gentle massage to relieve stress; when the user succeeds, it provides a strong pat on the back to enhance the sense of achievement. In addition, tactile feedback can also be deeply integrated with VR and AR technologies to bring users a more realistic sensory experience in a virtual environment. For example, in a virtual tourism scenario, when a user touches the sand on a virtual beach with a tactile glove, the system can adjust the fineness and temperature of the touch according to their emotional response - if the user feels happy, the warmth is enhanced; if the user is bored, the touch of waves is introduced to stimulate interest. With the development of material science and micro-sensor technology, the portability and accuracy of tactile devices will be greatly improved. In the future, tactile feedback may realize the customized design of "emotional touch", such as tailoring unique touch patterns for different users' emotional preferences, so as to play a greater role in emotional companionship, education and entertainment.

As a virtual world concept that integrates virtual reality, augmented reality and artificial intelligence, the Metaverse provides a broad stage for emotion recognition technology. In the Metaverse, users interact with others through virtual avatars, and emotion recognition technology can monitor the user's emotional state in real time through multimodal data - facial expressions, voice intonation, gestures and even physiological signals - and dynamically reflect it on the avatar. For example, when a user feels happy in the Metaverse, his avatar will naturally smile and speed up the pace of movement; when the user feels frustrated, the avatar may bow his head or speak in a low tone. This kind of emotional mapping not only enhances the user's sense of immersion and substitution, but also makes virtual social interaction more real and vivid.

In the future, emotion recognition in the metaverse will develop in a more intelligent and ecological direction. With the advancement of artificial intelligence technology, the system will be able to infer the deep emotional state and personality traits of users by analyzing their behavioral trajectories and interaction patterns in the metaverse. For example, a user who is often active in virtual parties may be identified as an extrovert, and the system will recommend more social activities for him; while a user who prefers to explore alone may receive personalized task push. In addition, emotion recognition in the metaverse can also be combined with the AI-driven behavior of virtual characters to create more autonomous and emotionally intelligent NPCs (non-player characters). These virtual characters can adjust their reactions in real time according to the user's emotional state, such as trying to appease the user when he is angry and celebrating together when the user is excited. It is conceivable that with the development of blockchain and distributed computing technology, the metaverse will form a decentralized emotional interaction ecology, and the user's emotional data may be stored and autonomously controlled in encrypted form to customize personalized virtual experiences. In the future, emotion recognition technology may become one of the core pillars of the metaverse, promoting the deep integration of the virtual world and the real world at the emotional level.

Innovations in interactive technologies such as virtual reality, augmented reality, brain-computer interface, tactile feedback, and the metaverse have brought unprecedented opportunities for emotion recognition. These technologies not only expand the application scenarios of emotion recognition, but also greatly enrich the ways of emotional expression and perception through multimodal data fusion and real-time feedback. In the future, with the continuous advancement of sensor technology, artificial intelligence, and computing power, emotion recognition will be more accurate, real-time, and personalized, and it may be possible to achieve seamless decoding from external behavior to internal psychological state, and even predict users' emotional trends and actively optimize the interactive experience. For example, a system that integrates VR, BCI, and tactile feedback may automatically generate a virtual forest scene when the user feels stressed, and help the user regain calm through mind-controlled relaxing music and warm tactile feedback.

\section{Social and cultural influences}
In the future development trend of affective computing, emotion recognition technology is changing the way we interact with the world at an astonishing rate. From the daily use of smart devices to complex social systems, this technology not only improves the ability of machines to understand human emotions, but also has a profound impact on our social structure, cultural customs and ethical concepts. As emotion recognition technology is increasingly integrated into daily life, its impact on the social and cultural level has become an issue that cannot be ignored. This section will discuss the enhancement of social interaction, the adaptability of cultural customs, and the challenges of ethical privacy, and provide an introduction for the subsequent discussion of the social adaptability of human-machine integration and the role of robots in cross-cultural communication.

The core of emotion recognition technology is to give machines the ability to perceive and respond to human emotions, which is first reflected in the optimization of social interaction. In traditional face-to-face communication, people rely on language, facial expressions and body movements to convey emotions, and the emergence of emotion recognition technology has injected new possibilities into this process. For example, in the field of education, teachers can monitor students' learning status in real time through systems equipped with emotion recognition functions. Studies have shown that students' emotions such as interest, confusion or anxiety directly affect their learning effects. When the system detects that students are frustrated by the difficulty of the course, teachers can adjust the teaching rhythm or provide additional support in time, thereby creating a more inclusive and efficient learning environment. This technology not only improves the quality of education, but also opens up new paths for personalized teaching. In the medical field, emotion recognition technology also shows great potential. Doctors can use this technology to monitor patients' emotional fluctuations, especially in the field of mental health. For example, the emotional changes of patients with depression are often subtle and difficult to detect, and the emotion recognition system can assist medical staff to detect problems earlier and provide intervention by analyzing the patient's voice tone or facial expressions. In addition, for special groups such as children with autism, emotion recognition technology can also help them learn emotional expression and understanding through simulated interaction and feedback, thereby improving social skills. These applications show that emotion recognition technology is reshaping the way people communicate with each other, making social interactions more sensitive and humane.

\begin{figure}
    \centering
    \includegraphics[width=0.5\textwidth]{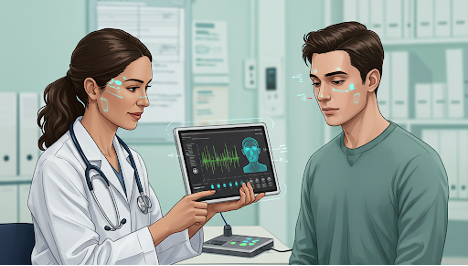}
    \caption{Emotion recognition technology in education and medical care}
\end{figure}

However, the impact of this technology on social interaction is not entirely positive. As emotion recognition devices become more popular, people may gradually rely on technology to interpret the emotions of others, while neglecting their natural observation and empathy abilities. Especially among the younger generation, long-term use of emotion recognition tools may weaken their experience of interacting with real humans and lead to the degradation of social skills. In addition, the accuracy of the technology has also become a major concern. If the emotion recognition system misinterprets the emotional state, such as misjudging a neutral expression as anger, it may cause misunderstandings or even conflicts, especially in high-risk scenarios such as court trials or public security monitoring. Therefore, when promoting the integration of emotion recognition technology into social interactions, its reliability must be ensured and the final judgment of humans must be retained to avoid excessive intervention of technology in interpersonal relationships.

The change in social interaction caused by emotion recognition technology is only one aspect of its impact. Its impact on cultural customs is also worthy of attention. The expression and understanding of emotions vary significantly in different cultures, which brings complexity to the design and application of technology. For example, in Eastern culture, people tend to express emotions implicitly and facial expressions may be more restrained, while in Western culture, direct emotional expression is more common. An emotion recognition system trained on data from a single culture may not accurately interpret emotional signals from other cultures. For example, in Japanese culture, a smile may represent politeness rather than happiness, while in a Western context, the same expression is often interpreted as a positive emotion. If the technology fails to recognize these nuances, it can lead to misunderstandings and even conflicts in cross-cultural interactions. To meet this challenge, researchers are developing cross-cultural emotion recognition models to improve the universality of the system by building a dataset containing multicultural emotional expressions.

In addition, some studies have proposed using transfer learning technology to allow the system to quickly adapt to new cultural environments with fewer samples. For example, the system can automatically adjust the algorithm based on the user's language or geographic location to more accurately identify emotional patterns in local culture.

Cultural diversity not only requires technology to be adaptable, but also brings ethical considerations in design. In some cultures, emotions are considered private, and unauthorized emotional monitoring may be considered offensive. For example, in European countries that emphasize personal privacy, the collection of emotional data is strictly regulated by the General Data Protection Regulation (GDPR).

A system that analyzes the user's emotional state without their consent may be resisted for violating laws and cultural norms. Therefore, developers of emotion recognition technology must have a deep understanding of the customs and values of the target culture to ensure that the system operates under the premise of respecting cultural differences. For example, when designing smart assistants for multicultural markets, user customization options can be added to allow users to choose whether to enable emotion recognition functions and the scope of data use. This cultural sensitivity not only helps promote technology, but also avoids social disputes caused by cultural misunderstandings.

When discussing the impact of emotion recognition technology on society and culture, ethical and privacy issues are undoubtedly the focus that cannot be avoided. As a highly sensitive personal information, the collection and use of emotion data has aroused widespread concern. First of all, emotion recognition systems usually need to obtain multimodal data of users, such as facial images, voice and text. The leakage of this data may lead to serious privacy crises. For example, users' facial expression data may be used for unauthorized identity identification or abused by advertisers to push customized emotional marketing content.

Imagine a scenario where a user cries in front of a smart TV, and the system records this emotional state and sells it to a third party, which then pushes ads for antidepressants. This kind of emotional exploitation without consent not only violates privacy, but can also cause psychological burden on users.

Secondly, the potential bias of emotion recognition technology may also exacerbate social injustice. Due to the limitations of training data, the system may not accurately recognize the emotional expressions of certain groups. For example, studies have found that some commercial emotion recognition models have biased judgments on the facial expressions of ethnic minorities, which may lead to unfair treatment in scenarios such as recruitment or security checks.

To address this problem, researchers are exploring fairness-enhancing algorithms to reduce bias through data enhancement and model correction. However, the fairness of technology not only relies on algorithm improvement, but also requires a diverse development team and extensive social participation to ensure that the system reflects the needs of different cultures and populations.

What is even more worrying is that emotion recognition technology may be used for emotional manipulation. In the business and political fields, criminals may use technology to analyze users' emotional weaknesses and push precise advertisements or guide public opinion. For example, social media platforms may use emotion recognition technology to identify users' anxiety and then push inflammatory content to manipulate their behavior.

This practice not only threatens individual autonomy, but may also pose a risk to social stability. Therefore, society urgently needs to establish a legal and ethical framework to regulate the collection and use of emotional data. For example, clear policies can be formulated to require companies to obtain informed consent from users when using emotion recognition technology and to undergo regular independent audits to ensure that the technology is not abused.

When exploring the social and cultural impact of emotion recognition technology, we naturally touch upon its role in human-machine integration. With the rapid development of robotics and artificial intelligence technology, human-machine interaction is becoming more frequent and in-depth, and emotion recognition technology, as the key to human-machine interaction, can significantly enhance the emotional intelligence of machines, making them better adapted to the needs of human society. In home scenarios, service robots can perceive users' emotions through emotion recognition and provide personalized companionship. For example, when a user feels tired due to work pressure, the robot may play soothing music or take the initiative to initiate a relaxing conversation; when a user appears lonely, the robot can accompany him in entertainment activities.

This emotional sensitivity not only improves the user experience, but also alleviates social problems to a certain extent, such as the loneliness caused by aging. However, human-machine integration is not without challenges. Humans need to adapt to coexisting with emotionally intelligent machines and learn to maintain the authenticity of emotional communication with the assistance of technology. At the same time, robots must also follow social norms to avoid causing discomfort due to excessive intervention. For example, an overly proactive emotional robot may be seen as an invasion of privacy, so its design should focus on respecting emotional boundaries.

These issues will be further elaborated in the subsequent section “Social Adaptability of Human-Machine Integration” to explore how humans and machines can achieve harmonious coexistence in society.

The social and cultural impact of emotion recognition technology also extends to the field of cross-cultural communication. In today's globalized world, interactions between different cultures are becoming increasingly frequent, and emotion recognition technology makes it possible for robots to play a bridging role in this scenario. For example, in international business negotiations, emotion recognition systems can analyze the emotional state of participants in real time, helping translators to convey emotional colors more accurately and avoid misunderstandings caused by cultural differences.

In the field of education, language learning robots can adjust teaching strategies through emotion recognition. For example, when students feel frustrated by language barriers, the robot can slow down the speech or provide encouragement, thereby improving learning outcomes.

In tourism scenarios, tour guide robots can adjust their interpretation style according to tourists' emotional reactions and provide more considerate services. However, to realize these applications, emotion recognition technology must have a high degree of cultural adaptability. Researchers are developing algorithms that can recognize multicultural emotional expressions and ensure the robustness of the system in multilingual environments.

In addition, robots need to demonstrate cultural humility in cross-cultural communication, that is, to acknowledge the limitations of their own knowledge and seek human assistance when necessary. These issues will be explored in depth in the section "Robots' Role in Cross-Cultural Communication" to reveal how emotion recognition technology can help promote cultural integration in globalization.

The social and cultural impact of emotion recognition technology is multidimensional and far-reaching. It brings new opportunities to human society by optimizing social interactions, adapting to cultural differences, and responding to ethical challenges, but it also raises new questions. Future development needs to find a balance between technological innovation and social responsibility to ensure that emotion recognition technology does not undermine individual rights and social harmony while improving the quality of life. Subsequent sections of this chapter will further focus on the social adaptability of human-machine integration and the role of robots in cross-cultural communication, in order to comprehensively analyze the potential and challenges of emotion recognition technology in the future society. Through these discussions, we hope to present readers with a clear and profound picture of how this technology will shape the future of mankind.

\subsection{Social adaptability of human-machine integration}
Since its advent in the late 1990s, emotion recognition technology has experienced a leap from basic theoretical research to widespread application. With the advancement of artificial intelligence and machine learning technology, this technology has gradually moved from the laboratory to the market and penetrated into all aspects of human life. Initially, emotion recognition was limited to simple emotion classification, such as distinguishing between joy, anger, sadness and happiness through facial expressions; but now it has developed into a complex system that can comprehensively analyze multimodal data such as voice intonation and physiological signals. The evolution of this technology not only reflects the improvement of computing power, but also reflects the expectation of humans to give "humanized" characteristics to machine intelligence.

In the future society, human-machine integration will become a new normal. The interaction between humans and machines will no longer be limited to the cold functional task execution, but will evolve into an emotional and dynamic coexistence relationship. Imagine that when you return home tired, an intelligent assistant can not only turn on the lights for you, but also sense your low mood, actively play soothing music and say "You have worked hard today, do you need to talk?" The realization of this scenario is the embodiment of emotion recognition technology as a bridge connecting human emotions and machine intelligence. However, the social adaptability of technology is not achieved overnight. It needs to be constantly adjusted in practical applications and coordinated with human culture, psychology and social norms.

\begin{figure}
    \centering
    \includegraphics[width=0.5\textwidth]{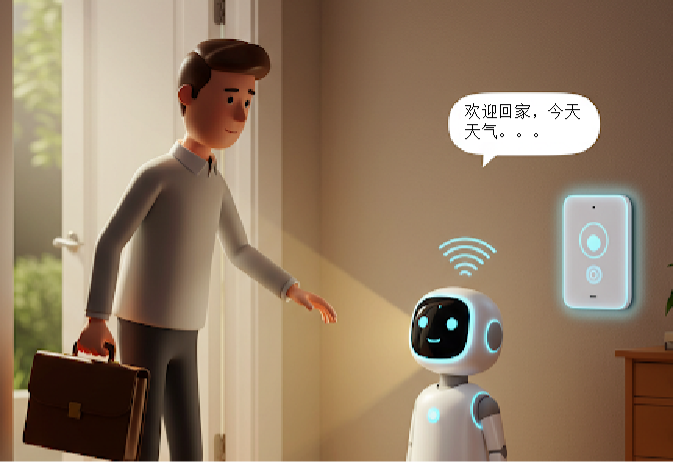}
    \caption{Human-machine integration}
\end{figure}

In the home environment, emotionally intelligent robots are gradually becoming an important tool for companionship and support. Take the Pepper robot developed by SoftBank Corporation of Japan as an example. This robot is equipped with an advanced emotion recognition system that can capture the user's facial expressions through a camera and analyze the voice tone through a microphone to judge the user's emotional state. When Pepper detects that the user is depressed, it may take the initiative to initiate a conversation, such as "You look a little unhappy, is there anything I can help you with?" At the same time, it can also play music or tell jokes according to the user's preferences to relieve emotions. This personalized emotional interaction is particularly important for elderly people living alone. Studies have shown that emotionally intelligent robots can significantly improve mental health in elderly care. For example, a systematic review by Pu et al. (2019) found that elderly people living alone who interacted with emotional robots had about 30\% fewer symptoms of loneliness and depression, showing the potential of technology in improving the quality of life. In addition, emotion recognition technology can also be combined with smart home systems to form a more comprehensive emotional support network. For example, when the system detects that a family member is in a state of anxiety for a long time, it can automatically adjust the color temperature of the light, reduce indoor noise, and even remind other family members to pay attention to this situation. This seamless emotional care not only strengthens the connection between family members, but also makes machines an indispensable part of family life.

In the field of education, emotion recognition technology provides new tools for personalized teaching. The Tega robot developed by the Massachusetts Institute of Technology is a typical example. This robot monitors students' learning status in real time by analyzing their facial expressions, voice intonation and body movements. For example, when Tega finds that students are confused or frustrated because the math problems are too difficult, it will pause the teaching content and provide encouraging words, such as "It's okay, let's try it together!" or adjust the teaching strategy to break down the problem steps to reduce the difficulty. Gordon et al. (2015) showed that students who used Tega to assist in teaching had an increase of about 20\% in participation and performance in second language learning, thanks to the robot's accurate response to students' emotions. Emotion recognition technology can also play a role in large-scale online education. For example, by analyzing students' camera images and voice feedback, the system can identify which students are bored or confused about the course content, thereby providing real-time suggestions to teachers and optimizing classroom interactions. This technology not only improves teaching effectiveness, but also makes students feel cared for and understood, thereby enhancing their motivation to learn.

In the medical field, emotion recognition technology provides a new way for the treatment and rehabilitation of special populations. Taking the social training of autistic children as an example, the Milo robot developed by the University of Cambridge in the UK can simulate human emotional expressions and help children learn to recognize and express emotions through interactive games. Milo's facial screen can display different expressions, such as smiling or frowning, and guide children to imitate or respond through voice prompts. For example, when Milo says "I'm happy, can you give me a smiley face?", the child needs to try to imitate the smile, and Milo will give real-time feedback based on the child's expression. The experiment of Scassellati et al. (2012) showed that after 12 weeks of Milo training, autistic children's scores in social skills tests increased by about 25\%, and their emotional understanding ability also improved significantly. In addition, emotion recognition technology also shows potential in the field of mental health. For example, some smart devices can detect depression or anxiety symptoms by analyzing the user's voice patterns and physiological signals (such as heart rate), and promptly remind the user to seek professional help. This technology not only provides patients with the opportunity for early intervention, but also reduces the burden on the medical system.

The appearance design of emotionally intelligent machines directly affects the user's first impression and emotional response. According to the "uncanny valley" theory (Mori, 1970), when the appearance of a robot is highly similar to that of a human but not completely identical, people tend to feel uneasy or even fearful. For example, although the Atlas robot of Boston Dynamics is very advanced in terms of functionality, its overly realistic humanoid appearance may make users feel uncomfortable. In contrast, the Pepper robot of Japan's SoftBank adopts a cartoon-like design. Its small body and rounded facial features make it more friendly, and users are more likely to regard it as a "friend" rather than a threat. Studies have shown that the acceptance of appearance design is closely related to the psychological comfort of users (Bartneck et al., 2009), which suggests that developers need to find a balance between similarity and acceptability.

Behavior patterns and the naturalness of emotional expression are also key factors affecting acceptance. If the robot's emotional expression is too mechanical or exaggerated, users may feel unreal or even disgusted. For example, if an emotional robot says "Don't be sad" in a stiff tone when the user is sad, it may be counterproductive. On the contrary, conveying emotions through subtle facial expressions and soft tones often wins the favor of users. Take the Pepper robot as an example. When it expresses concern, it tilts its head slightly and slows down its speech. This delicate behavioral design significantly enhances the user's sense of trust. Research shows that the naturalness of emotional expression is positively correlated with user acceptance (Breazeal, 2003), which means that future technology development needs to pay more attention to the delicacy of behavioral simulation.

\begin{figure}
    \centering
    \includegraphics[width=0.5\textwidth]{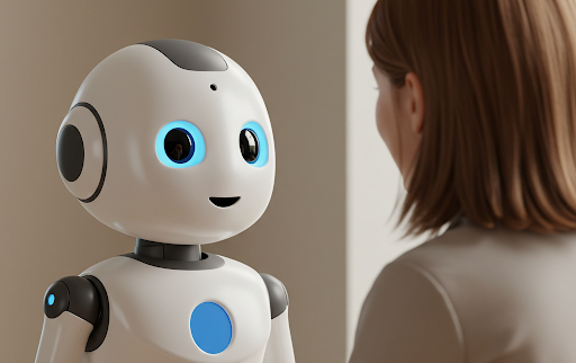}
    \caption{Robot emotional expression}
\end{figure}

The impact of cultural background on acceptance cannot be ignored. There are significant differences in preferences for emotional expression among different cultures. For example, in Eastern culture, people tend to be more accepting of robots that express emotions in a more subtle way. Japanese users may prefer a low-key and gentle interaction like Pepper, while in Western culture, direct and enthusiastic emotional expression may be more popular. American users may expect robots to show stronger personalities, such as humor or a straightforward conversational style. A cross-cultural study by Trovato et al. (2013) found that Egyptian users were less accepting of emotional robots than Japanese users, which was related to differences in trust in technological intervention in the culture. Therefore, the development of emotion recognition technology needs to incorporate cultural adaptability to ensure that robots can adjust their interaction methods according to the cultural characteristics of the target market. A specific example is the feedback from users on emotional robots in different cultural backgrounds. In Japan, the deployment of Pepper robots in homes and commercial places has been widely successful, with users considering them as "family members" as high as 60\%. In Europe, some users expressed discomfort with Pepper's "too friendly" behavior and believed that it was too intrusive. This difference highlights the importance of cultural adaptability. Cross-cultural research further shows that technology developers need to optimize robots’ emotional expression strategies through user research and localized design to improve their acceptance in the global market (Trovato et al., 2013).

There are multiple challenges to the adaptability of emotionally intelligent robots in social norms, one of which is the privacy issue caused by excessive intervention. For example, if a household robot proactively provides emotional support when the user does not explicitly request it, such as "You look tired today, do you want to take a break?" It may be regarded as an invasion of private space. To address this problem, designers have proposed the concept of "emotional humility", that is, robots should maintain appropriate restraint when interacting with humans and not try to dominate emotional exchanges (Calo, 2015). For example, when a robot detects that the user is emotionally abnormal, it can first ask "Do you need my help?" instead of directly intervening. This design not only respects the user's autonomy, but also meets the general expectation of society for privacy.

Emotion recognition technology relies on the collection of a large amount of personal data, including facial images, voice features, and physiological signals. The leakage of this data may lead to a serious crisis of trust. For example, if an emotion recognition system leaks a user's depression data to a third-party advertiser, the user may face harassment or even psychological harm from precision marketing. To address this challenge, researchers are developing a technical framework based on federated learning and differential privacy. Federated learning allows data to be processed on local devices without uploading to the cloud, while differential privacy protects the anonymity of individual data by adding noise (McMahan et al., 2017). These technologies can maximize user privacy while ensuring the accuracy of emotion recognition.

The application of emotion recognition technology may also cause ethical dilemmas, especially in medical scenarios. For example, if an emotional robot misjudges a patient's emotional state and recommends a treatment plan based on it, it may lead to deviations in medical decision-making. Fiske et al. (2019) pointed out that over-reliance on emotion recognition technology in psychotherapy may weaken the judgment of human doctors and even cause patients to over-trust technology. To avoid such problems, technology deployment must follow strict ethical standards, such as requiring all key decisions to be reviewed by human doctors and ensuring that technology is used only as an auxiliary tool rather than a dominant one. In addition, it is also crucial to establish a transparent user consent mechanism, and users should be clearly aware of how the technology works and its limitations.

On the technical level, the rise of multimodal emotion recognition technology will significantly improve the accuracy and robustness of the system. By integrating multiple information sources such as vision (facial expressions), hearing (voice intonation) and physiological signals (heart rate, skin galvanic response), the system can capture the emotional state of humans more comprehensively. For example, an intelligent assistant may judge that the user is in a state of tension by analyzing the user's frowning movements, low tone of voice and accelerated heart rate, and provide relaxation suggestions accordingly. Poria et al. (2017) pointed out that multimodal fusion can increase the accuracy of emotion recognition from 70\% of single modality to more than 85\%.

\begin{figure}
    \centering
    \includegraphics[width=0.5\textwidth]{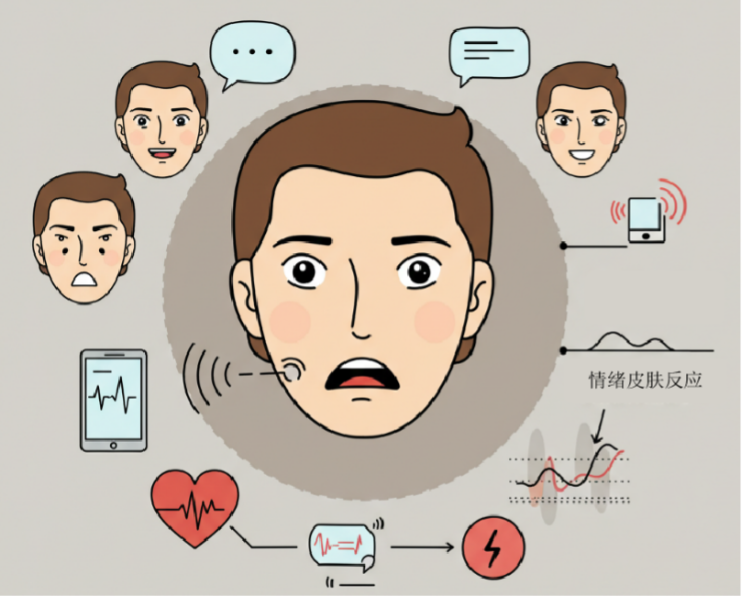}
    \caption{Multimodal emotion technology}
\end{figure}

In addition, personalization and situational adaptability will also become the focus of technological development. Through deep learning and reinforcement learning, robots can dynamically adjust emotional expression according to the user's personality traits, living habits and specific scenarios. For example, an extroverted user may like the robot's enthusiastic interaction, while an introverted user prefers low-key companionship. This personalized design will further enhance the user experience (Breazeal, 2003). At the same time, the application of edge computing and 5G technology will make the emotion recognition system more efficient, and data processing can be completed locally, thereby reducing latency and enhancing privacy protection.

At the social level, emotion recognition technology is expected to play a greater role in public services and business. In public services, emotionally intelligent robots can be used as psychological support tools. For example, in rescue work after natural disasters, robots can use emotion recognition technology to identify the psychological state of victims, provide targeted emotional comfort or guide them to accept professional help (Bethel \& Murphy, 2010). In the business field, emotion recognition technology has been applied to customer service. For example, Microsoft's AFFDEX system can optimize service processes by analyzing customers' facial expressions. When customer dissatisfaction is detected, the system automatically reminds customer service staff to adjust their tone or provide discounts (McDuff et al., 2016). These applications will further promote the popularization of technology in society.

The widespread application of technology requires a sound legal and ethical framework as a guarantee. At the policy level, the government can formulate regulations requiring companies to obtain explicit consent from users when deploying emotion recognition technology and to undergo regular third-party audits to prevent data abuse (Crawford \& Calo, 2016). At the same time, public education is crucial. Popularizing knowledge about affective computing through school courses or community activities can help people understand the functions and limitations of technology, thereby reducing misunderstandings or resistance to technology (Picard, 2000). For example, a simple popular science video can show the public how emotion recognition can improve life while reminding them of privacy risks.

The social adaptability of emotion recognition technology in human-machine integration is a multi-dimensional and complex topic. This article demonstrates the potential of technology by analyzing its applications in the fields of family, education, and medicine; reveals the key factors for the popularization of technology by exploring the impact of appearance design, behavioral patterns, and cultural background on human acceptance; proposes solutions for the coordination of technology and society by discussing privacy protection, social norms, and ethical dilemmas; and finally, describes a blueprint for future development by looking forward to technological progress and social applications.

Emotion recognition technology opens up new possibilities for human-machine interaction. From accompanying lonely elderly people to training autistic children to optimizing customer service, it is gradually changing the way we live. However, the social adaptability of technology is not without challenges. Privacy risks, ethical dilemmas and cultural differences may become obstacles to its popularization. Future development needs to seek a balance between technological innovation and social responsibility, and ensure that technology can improve the quality of life while maintaining individual rights and social harmony through interdisciplinary cooperation and continuous exploration.

\subsection{The role of robots in cross-cultural communication}

In the tide of globalization, cross-cultural communication has become an indispensable part of human society. With the rapid development of emotion recognition technology, robots are gradually transforming from simple tools to intelligent agents that can perceive, understand and respond to human emotions, especially in the complex scenario of cross-cultural communication, where their role becomes increasingly important. Robots can not only serve as tools for language translation, but also understand the emotional expressions of people in different cultural backgrounds through emotion recognition technology, thus playing a key role in cultural mediation, emotional bridges and personalized services. This section will explore the role of robots in cross-cultural communication in depth, analyzing their potential and limitations in cultural adaptability, technical challenges, ethical considerations and future development trends. By combining specific examples and research results, we will fully reveal how emotion recognition technology can help robots cope with the complex needs of cross-cultural communication in a globalized society, and look forward to its broad prospects for future development.

In today's world, cultural diversity brings rich layers to communication, but also increases the possibility of misunderstanding. The core of emotion recognition technology is to detect and interpret human emotional states through facial expressions, voice intonation, body movements, and even physiological signals such as heart rate or skin galvanic response. However, when this technology is applied to cross-cultural scenarios, its complexity increases significantly. For example, in Japanese culture, a smile may be just a polite social signal rather than a direct expression of happiness; while in Western culture, the same smile is often interpreted as a manifestation of positive emotions. This difference is not limited to facial expressions, but also reflected in the tone of language, body posture, and the subtleties of social norms. An emotion recognition system trained on data from a single culture, if it has not been cross-culturally adjusted, may make wrong judgments when facing users from different cultures, or even cause conflicts. Therefore, the first task of robots in cross-cultural communication is to understand and adapt to the diversity of these emotional expressions.

Imagine a real scenario: In an international business negotiation, a Chinese negotiator and an American negotiator are discussing the terms of cooperation. The Chinese negotiator is accustomed to expressing dissatisfaction in an indirect way, such as a slight frown or pause, while the American negotiator may be more inclined to speak bluntly. A robot equipped with emotion recognition can play the role of an emotional translator if it can monitor the emotional state of both parties in real time. It can analyze the micro-expressions and intonation of the Chinese negotiator to prompt the American negotiator of the possible dissatisfaction of the other party, and at the same time convey the intention of the American party to the Chinese negotiator by adjusting its own language and tone. This ability not only relies on technology, but also requires the robot to have deep cultural knowledge and situational awareness. Research shows that systems based on multimodal emotion recognition (combining facial, voice and text analysis) have significantly improved their accuracy in cross-cultural scenarios. For example, the multimodal emotion computing framework proposed by Poria et al. (2017) improves the robot's performance in identifying complex emotions by about 15\% by fusing visual and auditory signals. This lays a technical foundation for the practical application of robots in cross-cultural communication.

The application scenarios of robots in cross-cultural communication are far more than business negotiations. In the field of education, robots can be used as an auxiliary tool for cultural teaching to help students understand and experience the emotional expressions of different cultures. For example, a language learning robot can help students learn how to express emotions appropriately in a specific culture while mastering the language by simulating the enthusiastic tone of the French or the more restrained expression of the Germans. Researcher Baylor (2009) found that students who used virtual agents (such as robots) for emotional teaching scored 20\% higher in cross-cultural communication ability tests, indicating that robots can not only impart language knowledge, but also improve students' cultural adaptability through emotional simulation. In tourism scenarios, tour guide robots can dynamically adjust services based on tourists' emotional reactions. For example, when a group of tourists from the Middle East showed fatigue while visiting a museum, the robot could analyze their facial expressions and speech speed, slow down the pace of the explanation or suggest a break, thereby improving the tourists' experience. In the medical field, robots can provide emotional support to patients across language and cultural barriers. Research by Scassellati et al. (2012) showed that robots equipped with emotion recognition capabilities can play a role in cross-cultural therapy for children with autism by adjusting the interaction method to suit the needs of families from different cultural backgrounds.

However, the application of robots in cross-cultural communication is not without challenges. The primary technical difficulty lies in the demand for data. Emotion recognition models require a large amount of diverse data to cover the emotional expression patterns of different cultures, but collecting and labeling this data is a time-consuming and expensive task. For example, to train a model that can recognize implicit emotions in East Asian culture, thousands of hours of video and audio data may be required, and the acquisition of this data is often restricted by privacy and ethics. To address this problem, researchers are exploring unsupervised learning and transfer learning techniques. For example, Chen et al. (2017) proposed a deep learning method for cross-cultural emotion recognition, which enables the system to quickly adapt to new cultural environments by pre-training the model on Western cultural data and then fine-tuning it with a small amount of East Asian data. The accuracy of this method reached more than 85\% in the test, showing its potential in data-scarce scenarios. In addition, the introduction of meta-learning also provides robots with the ability to "learn how to learn", enabling them to quickly adjust their emotion recognition strategies through a small amount of interaction when they come into contact with new users.

\begin{figure}
    \centering
    \includegraphics[width=0.5\textwidth]{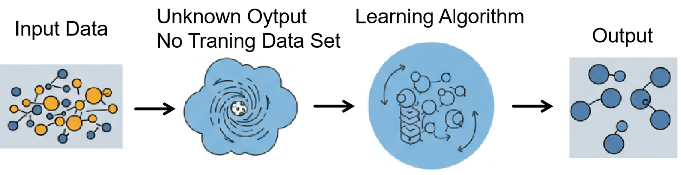}
    \caption{Unsupervised learning process}
\end{figure}

In addition to data issues, robots also need to have a high degree of contextual awareness in cross-cultural communication. The same emotion may have completely different meanings in different scenarios. For example, in a formal business meeting, a smile may indicate friendliness, while in a family gathering, it may be sarcastic. If robots cannot adjust their behavior according to the context, misunderstandings may occur. To address this problem, researchers are developing behavioral decision-making systems based on reinforcement learning. Such systems allow robots to continuously optimize their behavioral strategies through real-time interaction with users. For example, a service robot may find through trial and error that nodding and whispering responses are more popular when interacting with Japanese users, while more exaggerated gestures and enthusiastic tones are required when interacting with Italian users. This dynamic adjustment capability enables robots to maintain efficient communication in different cultural environments.

The integration of language and culture is another key challenge. Traditional translation systems tend to focus only on the literal meaning of the text, while ignoring the influence of emotions and cultural context. For example, "I'm fine" may be a polite perfunctory expression in English, but may be misunderstood as sincere satisfaction in some cultures. The combination of emotion recognition technology and natural language processing provides a solution to this problem. An emotional translation system can adjust the tone and wording of the translation by analyzing the speaker's intonation and expression. For example, when a Spanish user angrily says "¡Todo está bien!" (Everything is fine), the system can adjust the translation to "It's all fine, but I'm upset!" based on his emotional state, thereby more accurately conveying his intention. Research by Schuller and Batliner (2013) shows that this emotion-enhanced translation system increased user satisfaction by about 30\% in cross-cultural tests, showing its potential in improving communication fluency.

However, technological advances also bring ethical and privacy challenges. The collection and analysis of emotional data inevitably involves the privacy rights of users. In cross-cultural scenarios, the definition and protection standards of privacy vary from culture to culture. For example, in Europe, the General Data Protection Regulation (GDPR) has strict restrictions on the processing of personal data, while in other regions, relevant regulations may be more relaxed. A globally applicable robot system that fails to comply with local laws may lead to legal disputes or even user boycotts. In addition, the behavior of robots in cross-cultural communication may also cause ethical disputes. For example, if a robot fails to correctly identify cultural differences, it may inadvertently offend users. For example, in some Middle Eastern cultures, direct eye contact is seen as disrespectful, while in Western cultures it may be a symbol of respect. A cross-cultural study by Trovato et al. (2013) found that the acceptance of robot greeting behavior by Japanese and Egyptian users differed by as much as 40\%, which shows the importance of cultural sensitivity in robot design. To address these issues, developers need to introduce "cultural humility" in the design stage, that is, let the robot acknowledge the limitations of its own knowledge and seek human assistance when uncertain. For example, when encountering complex cultural situations, the robot can proactively ask the user: "Is my response appropriate?" This strategy can not only reduce misunderstandings, but also enhance users' trust in the robot.

\begin{figure}
    \centering
    \includegraphics[width=0.5\textwidth]{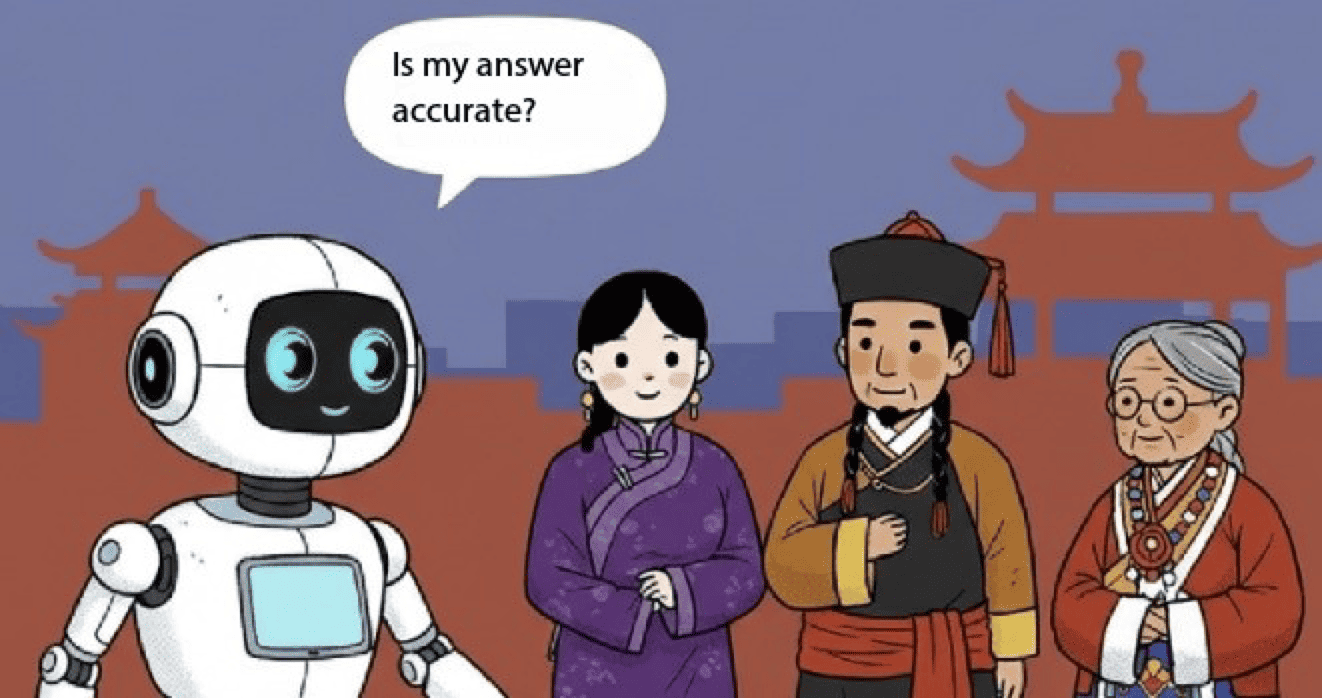}
    \caption{Robot “Cultural Humility”}
\end{figure}

Another ethical issue is the risk of cultural bias. If the emotion recognition model uses biased data during training, it may misjudge or discriminate against certain cultures. For example, Rhue (2018) found that some commercial emotion recognition systems have low accuracy in identifying the emotions of African-American users, which may be due to the lack of diversity in the training data. To address this problem, researchers are developing fairness-enhancing algorithms to reduce bias through data enhancement and model correction. For example, the method proposed by Buolamwini and Gebru (2018) improved the fairness of emotion recognition by about 10\% by balancing the data distribution of different populations. At the same time, diverse development teams and broad social participation are also considered important guarantees to ensure the fairness of technology. Future robot design requires joint efforts at the technical and social levels to avoid exacerbating cultural stereotypes.

Looking ahead, with the acceleration of globalization and the advancement of artificial intelligence technology, the role of robots in cross-cultural communication will usher in new development opportunities. First, emotion recognition technology will be further integrated with multimodal analysis technology to form a more comprehensive emotion understanding system. This system can comprehensively analyze the user's facial expressions, voice intonation, body movements and physiological signals, thereby achieving higher accuracy in complex scenarios. For example, a multimodal robot can determine whether the user is in a state of stress by detecting changes in the user's heart rate and fluctuations in tone, and choose the appropriate comfort method based on the cultural background. In Western culture, it may directly ask "Are you okay?", while in East Asian culture, it may express concern through indirect means (such as offering a cup of tea). This multimodal fusion technology has made breakthroughs in the laboratory. Poria et al. (2017) showed that its emotion recognition accuracy can reach more than 90\%, laying the foundation for future practical applications.

Secondly, the cultural adaptability of robots will be significantly enhanced. With the development of transfer learning and meta-learning techniques, robots will be able to quickly adapt to new cultural environments. For example, a service robot can learn the cultural preferences of users and adjust their behavior patterns through several interactions with them. Imagine a scenario: when a robot first enters an Indian home, it quickly learns to interact with users by nodding and responding softly by observing the user's body language and tone of voice. This adaptive ability will enable robots to provide personalized emotional support on a global scale. In addition, the establishment of a global emotional expression dataset will also promote this trend. Future research may create a database of hundreds of cultural emotional patterns for robots to learn and call, thereby achieving true cultural universality.

The application areas of robots in cross-cultural communication will also continue to expand. In international cooperation, robots can act as cultural mediators to help team members understand each other's cultural differences and promote collaborative efficiency. For example, in a team composed of Chinese, German and Brazilian members, robots can analyze the emotional expressions of each member to indicate potential misunderstandings and make communication suggestions. In transnational education, robots can act as virtual mentors to help students experience the emotional expressions of different cultures. For example, an American student learning Chinese can understand the emotional meaning of "face" in Chinese culture by interacting with robots. In the field of global health, robots can serve as telemedicine assistants to help doctors overcome language and cultural barriers and provide psychological support to patients. For example, during the COVID-19 pandemic, a robot equipped with emotion recognition can monitor patients' anxiety levels through video calls and provide personalized comfort based on their cultural background.

At the same time, the improvement of ethical and legal frameworks will provide guarantees for the widespread application of robots. In the future, the international community may formulate unified standards and guidelines to ensure that the design and use of robots in cross-cultural communication meet ethical requirements. For example, a global robot ethics convention may require all emotion recognition systems to disclose their data sources and algorithm logic to enhance transparency. At the same time, governments will also strengthen privacy protection legislation to ensure that user data is not abused. The public's awareness and acceptance of emotion recognition technology will also gradually increase. Picard (2000) pointed out that as people's understanding of emotional computing deepens, their concerns about privacy may gradually turn into expectations for the benefits of technology. This increase in social acceptance will lay the foundation for the popularization of robots around the world.

Driven by technological innovation, robots of the future may go beyond current imagination and become true "cultural empaths". They can not only recognize emotions, but also actively participate in cultural exchanges by learning and simulating behavioral patterns of different cultures. For example, a robot may imitate gestures or intonations in its culture when interacting with users, thereby narrowing the emotional distance. This ability will enable robots to play a more active role in a globalized society, transforming from passive tools to active cultural participants. In addition, with breakthroughs in quantum computing and neural network technology, robots' emotion recognition capabilities may reach a level close to that of humans. For example, an emotion recognition system based on quantum algorithms may analyze millions of data points in milliseconds to achieve real-time cross-cultural emotional understanding.

However, future development is also accompanied by new challenges. Rapid technological advances may exacerbate the digital divide, making it difficult for regions with limited resources to enjoy the convenience brought by robots. In addition, over-reliance on robots in cross-cultural communication may weaken human cultural adaptability and lead to the phenomenon of "cultural laziness". To avoid these problems, future robot design needs to find a balance between technological innovation and social impact. For example, robots can be designed as "assistants" rather than "substitutes", encouraging humans to actively participate in cross-cultural communication by providing advice rather than direct decision-making.

The role of robots in cross-cultural communication is an area full of opportunities and challenges. Emotion recognition technology gives them the ability to perceive and respond to human emotions, enabling them to play the role of cultural mediator and emotional bridge in a globalized society. However, the application of technology needs to fully consider factors such as cultural differences, technical limitations, ethical issues and privacy protection to ensure that it does not undermine individual rights and social harmony while promoting cross-cultural understanding. Future development will rely on the coordinated efforts of technological innovation, social adaptation and ethical governance. With the continuous expansion of multimodal technology, cultural adaptability and application fields, robots will provide more personalized and intelligent emotional support and services on a global scale. Ultimately, they may become not only a tool for cross-cultural communication, but also an important link to connect different cultures and enhance human understanding.

% --------------ref

\end{document}